\documentclass[aps,prd,nofootinbib,twocolumn,superscriptaddress,preprintnumbers,balancelastpage,longbibliography,nobibnotes,preprintnumbers]{revtex4-1}
\usepackage{graphicx,epsfig,psfrag,bm,amssymb,adjustbox}
\usepackage{dcolumn}
\usepackage{bm}
\usepackage[dvipsnames]{xcolor}
\usepackage{braket}
\usepackage{mathrsfs,amsfonts,hepunits,color}
\usepackage{hyperref}
\usepackage{comment}
\usepackage{physics}
\usepackage{subcaption}
\usepackage[labelfont=bf,
   justification=justified,
   format=plain]{caption} 

\usepackage{lineno}

\newcommand{\sqms}{\affiliation{Superconducting Quantum Materials and Systems Center (SQMS), Fermi National Accelerator Laboratory, Batavia, IL 60510, USA}}

\begin{document}

\title{{\small Report of the Snowmass 2021 Theory Frontier Topical Group on}
\\
Quantum Information Science for HEP}
\author{Simon~Catterall}
\thanks{Editor and topical group convener}
\email{smcatter@syr.edu}
\affiliation{Department of Physics, Syracuse University, Syracuse NY 13244, USA}

\author{Roni~Harnik}
\thanks{Editor and topical group convener}
\email{roni@fnal.gov}
\affiliation{Theory Division, Fermi National Accelerator Laboratory, Batavia, IL 60510, USA}
\sqms

\author{Veronika~E.~Hubeny}
\thanks{Editor and topical group convener}
\email{vhubeny@ucdavis.edu}
\affiliation{Center for Quantum Mathematics and Physics (QMAP)\\ 
Department of Physics \& Astronomy, University of California, Davis CA, USA}

\author{Christian~W.~Bauer}
\affiliation{Ernest Orlando Lawrence Berkeley National Laboratory, 
\& Berkeley Center for Theoretical Physics, \\ Department of Physics, University of California, Berkeley, CA 94720, USA}

\author{Asher~Berlin}
\affiliation{Theory Division, Fermi National Accelerator Laboratory, Batavia, IL 60510, USA}

\author{Zohreh~Davoudi}
\affiliation{Maryland Center for Fundamental Physics and Department of Physics, University of Maryland, College Park, MD 20742, USA\\
Institute for Robust Quantum Simulation, University of Maryland, College Park, Maryland 20742, USA}

\author{Thomas~Faulkner}
\affiliation{Department of Physics, University of Illinois Urbana-Champaign, Urbana, IL 61801, USA}

\author{Thomas~Hartman}
\affiliation{Department of Physics, Cornell University, Ithaca NY, USA}

\author{Matthew~Headrick}
\affiliation{Martin Fisher School of Physics, Brandeis University, Waltham MA, USA}

\author{Yonatan~F.~Kahn}
\affiliation{Department of Physics, University of Illinois Urbana-Champaign, Urbana, IL 61801, USA}
\affiliation{Illinois Center for Advanced Studies of the Universe, University of Illinois Urbana-Champaign, Urbana, IL 61801, USA}
\sqms

\author{Henry~Lamm}
\affiliation{Theory Division, Fermi National Accelerator Laboratory, Batavia, IL 60510, USA}

\author{Yannick~Meurice}
\affiliation{Department of Physics and Astronomy, University of Iowa, Iowa City, Iowa 52242, USA}

\author{Surjeet~Rajendran}
\affiliation{Department of Physics and Astronomy, The Johns Hopkins University, Baltimore, MD 21218, USA}

\author{Mukund~Rangamani}
\affiliation{Center for Quantum Mathematics and Physics (QMAP)\\ 
Department of Physics \& Astronomy, University of California, Davis CA, USA}

\author{Brian~Swingle}
\affiliation{Martin Fisher School of Physics, Brandeis University, Waltham MA, USA}

\date{\today}

\begin{abstract}
We summarise current and future applications of quantum information science to theoretical
high energy physics. Three main themes are identified and discussed; quantum simulation, quantum sensors and
formal aspects of the connection between quantum information and gravity. Within these themes, there are important research questions and opportunities to address them in the years and decades ahead.   Efforts in developing a diverse quantum workforce are also discussed. This work
summarises the subtopical area {\it Quantum Information for HEP TF10} which forms part of the
Theory Frontier report for the Snowmass 2021 planning process.
\begin{center}
---\\
\emph{Submitted to the  Proceedings of the US Community Study}\\ 
\emph{on the Future of Particle Physics (Snowmass 2021)}\\ 
\end{center}
\end{abstract}


\maketitle

\tableofcontents

\section{Executive Summary}
Ideas and methods of quantum information science have
found wide application to theoretical high energy
physics in recent years. This report divides those aspects into three main themes; quantum
simulation, quantum sensors and the application of ideas in quantum information to formal
aspects of quantum field theories (QFTs) and gravity.

Understanding the theoretical aspects of quantum information in quantum gravity and quantum field theories has catalyzed vibrant cross-disciplinary collaborations (e.g. between the lattice gauge theory and quantum gravity communities) and revealed new phenomena and organizational principles.
There are many types of avenues for exploration, ranging from 
the concrete (e.g.\ what happens dynamically in a particular system), 
to organizational (e.g.\ what attribute of a given system is relevant or useful in characterizing it),
to exploratory (e.g.\ how does these various attributes' behavior depend on the system),
to mathematical (e.g.\ how does one actually define a certain quantity or meaningfully generalize it from its original formulation).
Although the central focus is different in each of these avenues, they are interlinked, and progress in one informs the others.  
This has opened up a vast and wonderful new theoretical playground.

Complementary to these formal developments there have been extensive efforts to develop the
framework and algorithms needed to simulate QFTs on quantum computers.
The difficulties in simulating quantum theories classically motivate simulation on a quantum computer~\cite{Feynman:1981tf,Jordan:2017lea,pqa_loi}.  Quantum simulation
has the potential to explore non-perturbative physics in theories and regimes that have previously been
inaccessible to classical computing. Examples include probing the real time dynamics of hadronic collisions and associated non-perturbative phenomena such as fragmentation and hadronization in particle colliders, the determination of the neutron
star equation of state from finite density QCD, measuring the viscosity of the QCD quark-gluon plasma,  or simulating chiral fermions in beyond-the-Standard-Model (BSM)
theories.

Current efforts are centered about understanding the best 
truncation schemes for gauge theories, and the development
of efficient classical Hamiltonian simulation methods such as tensor networks in higher dimensions including fermions and their role in developing efficient quantum simulation algorithms. Theoretical work in HEP is needed to identify the simplest paths that
will bring us closer to simulating the QFTs of interest to particle physicists. These efforts
should, in addition, provide interesting benchmarks for the quantum computing community as devices are being developed. Indeed, HEP theorists have developed active collaborations with quantum hardware developers both in universities and private companies.

The third strand of this report focuses on the development of quantum sensors. Here there is an opportunity, even in the near term, to leverage the technology that is developed for QIS to directly search for new physics, including searching for new particles, dark matter, and gravitational waves. Theorists have been instrumental in making interesting connections between quantum sensing technologies and fundamental physics. 

In addition to research at the HEP-QIS interface, there is a need to develop and train a workforce in that US which is diverse and well-versed in the tools of quantum information science. To this end, HEP theorists are involved in workforce development efforts~\cite{Bauer:2022hpo} which we will comment upon.

This synergy between QIS and HEP benefits both fields. HEP theorists will be needed to build new bridges and establish existing connections between fundamental physics and quantum science. The research directions presented in this summary are relatively new, but are ambitious. New surprises are surely ahead.

\section{Quantum Simulation for High Energy Physics}
There are a wide variety of problems in high energy physics that cannot be addressed
using classical computation. These include real time scattering processes, finite density
strongly interacting matter and theories of BSM physics that incorporate
supersymmetry or chiral fermions. Monte Carlo simulation of the
path integral representing these quantum systems fails as a computational
method because of the infamous sign problem. Hamiltonian approaches using
classical computation suffer from the fact that the relevant Hilbert spaces grow
exponentially with the number of degrees of freedom which rapidly renders them
beyond the reach of classical computers.
In contrast, a quantum computer is capable, in principle, of representing dynamics
on
such spaces with a number of quantum bits that increases only linearly with the 
number of degrees of freedom. Furthermore quantum operations ideally retain quantum correlations and effectively perform massively parallel computations on the encoded wavefunction.

However the practical
application of quantum computers to such problems faces several immediate problems. First,
the quantum field theories (QFTs) that describe such systems possess
an infinite numbers of degrees of freedom. Thus
before one can contemplate simulating such a system with a quantum computer one 
must first truncate the theory. In classical Monte Carlo
simulation one introduces a lattice in spacetime as part of
this process. For quantum computation one also needs 
to effect a truncation 
in the field space. This latter truncation or digitization should preserve
as much symmetry as possible. This is particularly true in the case of gauge theories where violations of
gauge symmetry have the potential to ruin unitarity and consistency of the theory.
Furthermore, the truncated theory should retain a simple local encoding
in terms of the elementary quantum bits or qubits that
form the building blocks of the quantum computer.

Recent work has also been devoted to the problem of simulating scattering from first principles.
It is believed that all relevant scattering and hadronization
processes can be simulated on a quantum computer using resources
that scale only polynomially with the system size \cite{Jordan:2011ne}. Quantum
simulations potentially offer an ab initio method for 
understanding these non-perturbative phenomena that are
crucial for interpreting the results of collider experiments \cite{Bauer:2021gup}.

Parton distribution functions (PDFs) 
involve a matrix element of two quark fields separated by a light-like direction, which can not be calculated using traditional lattice field theory techniques directly due to a sign problem, although several indirect methods have been successfully developed in recent years~\cite{Constantinou:2022yye}.
Quantum computers give rise to the possibility to compute the matrix element relevant for the PDFs directly and from first principles. Several proposals have been put forward
in recent years to demonstrate how PDFs can be accessed on a quantum computer \cite{Lamm:2019uyc,echevarria2021quantum,li2021partonic,mueller2020deeply,qian2021solving}.
At this stage, it is not clear what the realistic computational resource
requirements are for computing PDFs and hadronic tensor to given accuracy, as the
complete algorithms, including that needed for preparation of hadronic states in QCD on a quantum
computer, are either non-existing or premature. PDFs can also be computed using
effective field theories describing the collinear and soft physics in jet-like collider events~\cite{Bauer:2021gup}.
Another example of important long-distance effects in collider events is collinear radiation,
which traditionally is described by parton-shower algorithms see e.g., Ref.~\cite{Bauer:2006mk}. 
The very nature
of a probabilistic Markov-Chain algorithm makes including quantum-interference effects challenging,
since collider events typically contain a very large number of final-state particles. This problem, therefore, is also a suitable candidate for quantum simulation. A
quantum algorithm has been developed in Ref.~\cite{Bauer:2019qxa} that reproduces the regular parton shower,
while by computing all possible amplitudes at the same time, it also includes quantum-interference
effects, see also Refs.~\cite{williams2021quantum,gustafson2022collider} for more progress.

There are two related approaches to quantum simulation that have been explored which can be loosely
termed analog and digital simulation. In both cases one maps the physical
system of interest to hardware whose quantum dynamics is capable of solving or simulating the 
original theory. In the case of analog simulation the hardware is tailored to the problem
at hand and requires substantial modification to solve a different problem. The digital
case more closely corresponds to a classical computer with elementary unitary quantum gates
acting on quantum states represented by qubits, or qudits (in cases where the hardware allows for more than a binary representation). It is possible to show, in analogy with classical
logic, that any quantum computation can be carried out with just a handful of one and two
qubit or qudit gates.

A quantum circuit then plays the role of the program
that describes the precise sequence of gates and the configuration of qubits that
are needed to simulate a given Hamiltonian. Like in classical computing, it should be possible
to change the circuit 
relatively easily in a high level programming language to solve new problems in a manner which
is agnostic about the underlying hardware.
There are many types of hardware platform that are currently being investigated for
quantum simulation including ultra cold
atoms in optical lattices~\cite{Zohar:2015hwa,Bermudez:2010da,Laflamme:2015wma,Meurice:2011aji,Bazavov:2015kka,Tagliacozzo:2012df}, trapped ions~\cite{Davoudi:2019bhy,Monroe:2019asq,Nguyen:2021hyk} including a proposal to construct such a trapped ion
machine using a storage ring \cite{Brooksetal}, superconducting circuits on a chip~\cite{Martinez:2016yna,Lamm:2019bik,Raychowdhury:2018osk,PhysRevD.99.094503}, superconducting bulk cavities~\cite{Alam:2022crs, Kurkcuoglu:2021dnw} and configurable Rydberg atom arrays \cite{keesling,51atom,qedryd,celi,surace2020lattice,Meurice:2021pvj}. Gate fidelities, coherence times, the number of available qubits, and the feasibility for qudit encodings, all vary significantly across
these difference platforms and at this time it is not clear which type of platform(s)
will ultimately prove most effective for the quantum simulation of high energy physics
theories. 

\begin{figure*}[htbp]
  \centering
  \includegraphics[scale=0.525]{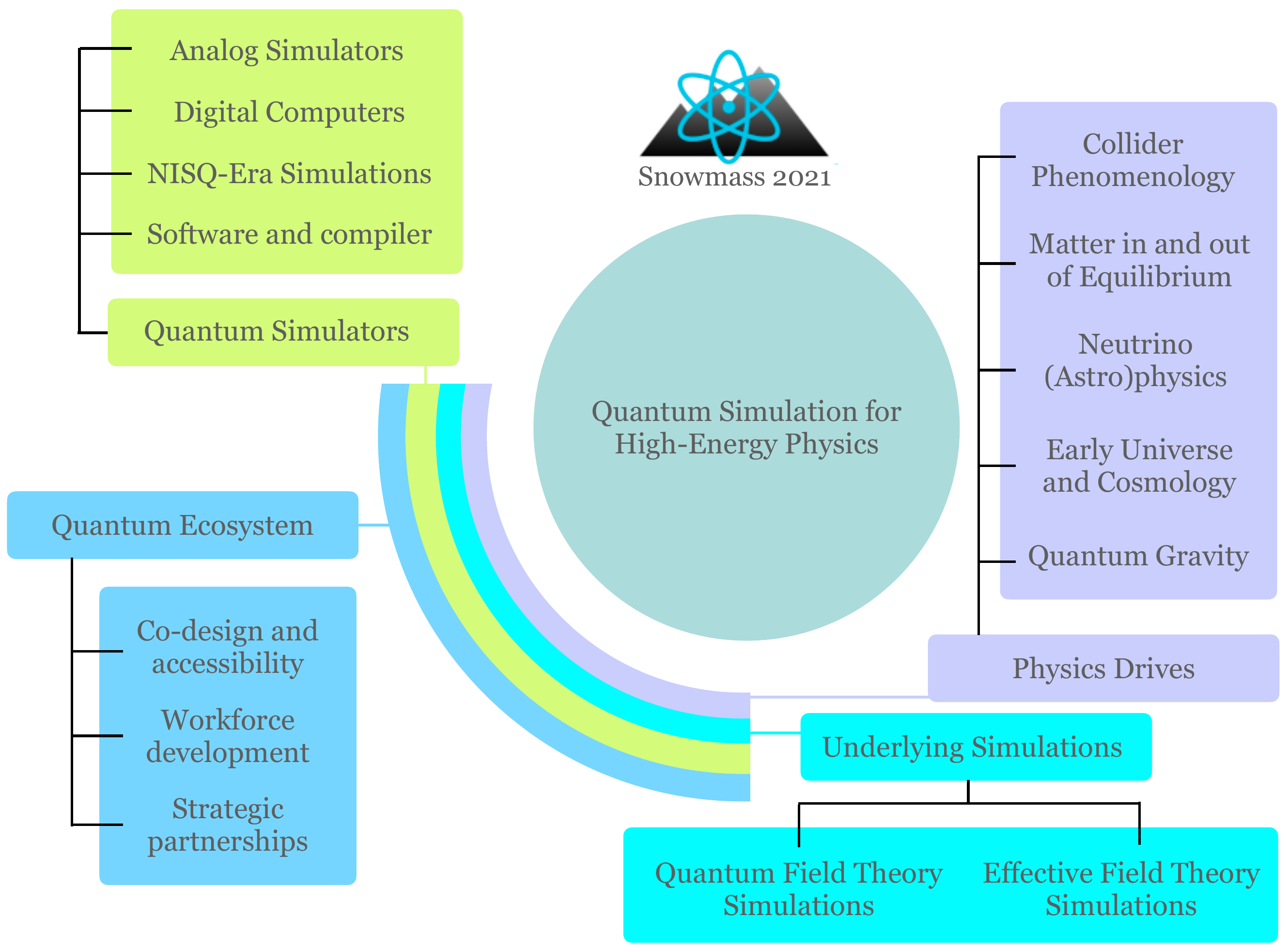}
  \caption{Physics drives, methodologies and strategies for quantum simulation
  in HEP, from~\cite{Bauer_Overview}.}
\label{chart}
\end{figure*}

Figure~\ref{chart} which is taken from a recent Snowmass white paper~\cite{Bauer_Overview}
gives an overview of the physics drives, methodologies and strategies that are being employed
in the application of quantum simulation to high energy physics.

A variety of approaches are currently being explored for Hamiltonian simulation such as the original
Kogut-Susskind formulation of Hamiltonian lattice gauge theory 
\cite{PhysRevD.11.395}, efforts to find optimal
bases for truncation such as those derived from tensor networks \cite{Liu:2013nsa,Meurice:2022xbk,Meurice:2020pxc}, pre-potential and loop-string-hadron formulations ~\cite{mathur2010n,Raychowdhury:2018osk}, quantum link models \cite{Wiese:2013uua,Wiese:2021djl},
and qubit regularizations \cite{Bhattacharya:2020gpm}.
A great deal of work is focused
on strategies to retain gauge invariance even in the presence of basis truncation,
correct for finite time step errors and reduce the effects of noise. A detailed review of these topics
can be found in a recent white paper~\cite{Bauer:2022hpo}.

Once one has fixed the truncation scheme it is straightforward to implement a discrete time evolution operator via Suzuki-Trotter decomposition which factorizes the unitary rotations associated with different non-commuting parts of the Hamiltonian. The optimal decomposition of terms to simulatable terms, the ordering of such terms, and the amount of errors one expects from digitization schemes constitute an area of research for both general Hamiltonians and the Hamiltonian of gauge theories of Standard Model, see e.g., Refs.~\cite{shaw2020quantum,ciavarella2021trailhead,paulson2021simulating,kan2021lattice,Nguyen:2021hyk}.
The basic elements of a digital circuit are single-qubit rotations (Pauli operators)
and CNOT gates acting on a pair of qubits which flip the target qubit when the control qubit is in the $\ket{1}$ state. 

The general aim of the field in recent years has been to explore
the different approaches to quantum simulation 
for a variety of simple QFTs with the goal 
of figuring out the best truncation schemes and evolution algorithms that
can ultimately be scaled to gauge theories including fermions
in (3+1) dimensions. It is
particularly important in this regard
to understand the limitations of what is possible on NISQ era machines
which possess limited numbers of qubits and no error correction.

Paralleling these developments
in quantum simulation of QFTs there is strong evidence (see section III) of a holographic connection
between strongly entangled quantum systems and gravity. This is visible in the original
AdS/CFT correspondence \cite{Mald}, the Ryu-Takayanagi proposal for entanglement entropy in holographic
systems \cite{ryu2006holographic} and applying ideas in quantum entanglement to 
black holes in AdS spacetime \cite{VanRaamsdonk:2010pw}. These topics and other
more recent work are summarized in a recent Snowmass white paper~\cite{Faulkner:2022mlp}. Efforts are also underway to
explore these ideas using classical and ultimately quantum
simulation of spin and gauge models on discrete tessellations
of hyperbolic space (Wick rotated AdS space) \cite{Asaduzzaman:2020hjl,Asaduzzaman:2021bcw}. 
Such models are also closely connected to MERA tensor network models that have
been proposed for building highly entangled ground states of many body
systems \cite{Evenbly} and play an important role
in certain proposals for quantum error correction \cite{Pastawski:2015qua}.

\subsection{Towards simulation of QCD and 
\\
other gauge theories}

Gauge theories are one of the pillars of the standard model. Strongly coupled gauge theories are a target for exploration with high performance computing, using
both classical methods - lattice QCD (LQCD) and, in the future, with quantum devices. However
to make use of future quantum computers
requires the development of new ways to represent QFTs that can be mapped to quantum
hardware and new algorithms to carry out simulations that go well beyond what has been done with classical computers.

A natural starting point for gauge theory is the Kogut-Susskind formulation of Hamiltonian lattice gauge theory  \cite{PhysRevD.11.395} (though there are alternatives as are described below). Even with a Hamiltonian one wishes to simulate, there are
interesting conceptual challenges need to be resolved to make such simulations feasible in the future, as well as technical problems at the interface of theory and design. HEP theorists are clearly critical to address the challenges in this new field. In some cases the decades of accumulated knowledge in LQCD can be leveraged, yet some problems
are inherently related to quantum simulation and require completely new ideas.

\paragraph*{\bf Discretization of space:} To turn the infinite-dimensional Hilbert space of quantum field theories into finite-dimensional system that quantum computers can simulate, one necessary step is to discretize continuous spatial dimensions into a finite-volume lattice~\cite{RevModPhys.55.775, PhysRevD.11.395, RevModPhys.51.659}, inevitably leading to discretization errors and finite-volume effects. Such discretization reduces spacetime symmetries and introduce new operators that modifies the nonperturbative renormalization. Investigations of approaching the infinite-volume and continuum limits of certain QFTs have recently emerged~\cite{PhysRevD.104.094519, PhysRevD.103.014506} and deserve more studies. 
Most gauge theory studies consider the Kogut-Susskind Hamiltonian~\cite{PhysRevD.11.395}, Hamiltonians with reduced discretization errors~\cite{Luo:1998dx,Carlsson:2001wp, Spitz:2018eps} will improve the convergence towards the infinite-volume and continuum theories, which allows quantum simulations at larger lattice spacings and smaller lattices for the same error. Recently, quantum simulations of these improved Hamiltonians has been initiated~\cite{Carena:2022kpg} where the corresponding time-evolution operators are derived together with the construction of the quantum circuits for a general gauge theory. 
Studies on the finite-volume technology in lattice calculations have been shown powerful to access physical observables \cite{Luscher:1986pf,Briceno:2017max, Davoudi:2018wgb, Bulava:2022ovd}, which will potentially enable the mapping between the finite and infinite-volume physics in quantum simulations of field theories.

\paragraph*{\bf Digitization of Fields:}
In addition to the use of
spatial lattices and discretization in time digital simulations require truncation of the local Hilbert space at each lattice point. Digitization represents the task of formulating, representing, and encoding truncations for QFTs suitable for digital quantum computers.  Some natural encodings exist for fermionic degrees of freedom~\cite{Jordan:1928wi,Bravyi2002FermionicQC,Chen:2018nog}. Further proposals discuss how to map lattice fermions (e.g. Wilson and staggered) onto these encodings~\cite{Muschik:2016tws} or use gauge symmetry to eliminate the fermions~\cite{Zohar:2018cwb,Zohar:2019ygc}. The relative merits of each approach are only beginning to be understood. The question of gauge boson digitization is murkier, with complicated tradeoffs~\cite{Hackett:2018cel,Alexandru:2019nsa,Singh:2019uwd,Singh:2019jog,Davoudi:2020yln,Bhattacharya:2020gpm,alexandru2022universality,Barata:2020jtq,Kreshchuk:2020kcz,Ji:2020kjk}. Digitizing may reduce symmetries -- either explicitly or through finite truncations~\cite{Zohar:2013zla}. Care must be taken as the regulated theory may not have the original theory as its continuum limit~\cite{Hasenfratz:2001iz,Caracciolo:2001jd,Hasenfratz:2000hd,PhysRevE.57.111,PhysRevE.94.022134,article}. A particularly illustrative example of the complications between truncations and renormalization can be found in Ref.~\cite{Wilson:1994fk}. Prominent proposals for digitization can be broadly classified~\cite{digi_loi} into: Casimir dynamics~\cite{Zohar:2012ay,Zohar:2012xf,Zohar:2013zla,Zohar:2014qma,Zohar:2015hwa,Zohar:2016iic,Klco:2019evd,Ciavarella:2021nmj} potentially with auxillary fields~\cite{Bender:2018rdp}, conformal truncation~\cite{Liu:2020eoa}, discrete groups (e.g. $D_4$ as a subgroup of SU(2) or $S_{1080}$ as a subgroup of SU(3))~\cite{Hackett:2018cel,Alexandru:2019nsa,Yamamoto:2020eqi,Ji:2020kjk,Haase:2020kaj}, dual variables~\cite{PhysRevD.99.114507,Bazavov:2015kka,Zhang:2018ufj,Unmuth-Yockey:2018ugm,Unmuth-Yockey:2018xak}, light-front kinematics~\cite{Kreshchuk:2020dla,Kreshchuk:2020aiq}, the pre-potential and loop-string-hadron formulation~\cite{mathur2010n,Raychowdhury:2018osk,Raychowdhury:2019iki,Davoudi:2020yln}, quantum link models~\cite{Wiese:2014rla,Luo:2019vmi,Brower:2020huh,Mathis:2020fuo}, and qubit regularization~\cite{Singh:2019jog,Singh:2019uwd,Buser:2020uzs}. In tensorial reformulations discussed in II.B, symmetry-preserving \cite{Meurice:2019ddf,Meurice:2020gcd} truncations are applied to character expansions and provide
controllable finite dimensional approximations. 

\paragraph*{\bf State preparation:}
Given a digitization, the next obstacle is initializing strongly-coupled quantum states in terms of fundamental fields. Much of the literature emphasized ground-state preparation~\cite{peruzzo2014variational,Kokail:2018eiw,Abrams:1998pd,nielsen2000quantum,PhysRevLett.117.010503,farhi2000quantum,Farhi472,Kaplan:2017ccd,Ciavarella:2021lel, Farrell:2022wyt, Atas:2021ext,Atas:2022dqm} but thermal and particle states have been investigated~\cite{2010PhRvL.105q0405B,Jordan:2011ne,Jordan:2011ci,Garcia-Alvarez:2014uda,Jordan:2014tma,Moosavian:2017tkv,Lamm:2018siq,Gustafson:2019mpk,Gustafson:2019vsd,Harmalkar:2020mpd,Gustafson:2020yfe,Jordan:2017lea,Klco:2019xro,PhysRevLett.108.080402,brandao2019finite,Clemente:2020lpr,motta2020determining,deJong:2021wsd,Gustafson:2021imb,Davoudi:2022uzo}. For methods which construct states using regulated theories, careful study of the renormalization needed to properly match onto the physical limit is required~\cite{Maiani:1990ca,Bruno:2020kyl}.

\paragraph*{\bf Time propagation:}
%
Propagating for a time $t$ requires the application of
a unitary operator $\mathcal U(t)=e^{-iHt}$,
which generically cannot be implemented exactly on a quantum computer and needs to be approximated.
A common method is trotterization, whereby $\mathcal{U}(t) \approx (\prod_i e^{iH_i t/N})^N \equiv (e^{-iH't/N})^N$ where $H = \sum_i H_i$ is a sum of $k$-local terms $H_i$, and $H'$ is an approximation of $H$ defined via this relation, which is generally hard to deduce.
The renormalization of lattice field theory in  Minkoswki spacetime due to trotterization is to introduce a temporal lattice spacing and new operators depending upon it \cite{PhysRevD.104.094519}. For some $H$, this allows for efficient simulations~\cite{Jordan:2011ne,Jordan:2011ci,Jordan:2017lea,Garcia-Alvarez:2014uda,Jordan:2014tma,Moosavian:2017tkv,Bender:2018rdp,haah2018quantum,Du:2020glq,PhysRevX.11.011020} and the errors from using $H'$ can be reduced via scale settings \cite{PhysRevD.104.094519}. Another way to reduce trotterization error might be using Hamiltonians with reduced lattice artifacts~\cite{Luo:1998dx,Carlsson:2001wp, Bulava:2022ovd, Carena:2022kpg}.
Other approximations of $\mathcal U(t)$ exist: QDRIFT~\cite{PhysRevLett.123.070503}, variational approaches~\cite{cirstoiu2020variational,gibbs2021longtime,yao2020adaptive}, Taylor series~\cite{PhysRevLett.114.090502}, use of nonlinear effects to increase the Trotter step \cite{ymbook,Gustafson:2019vsd} and qubitization~\cite{Low2019hamiltonian}. Initial resource comparisons have been performed for various gauge theories, see Refs.~\cite{shaw2020quantum,ciavarella2021trailhead,paulson2021simulating,kan2021lattice,Nguyen:2021hyk}.

\paragraph*{\bf Observables:}
It is straightforward to evaluate the expectation values of instantaneous Hermitian operators. Observables that depend on time-separated operators (such as parton distribution functions~\cite{Lamm:2019uyc,Kreshchuk:2020dla,Echevarria:2020wct}, particle decays~\cite{Ciavarella:2020vqm}, and viscosity~\cite{Cohen:2021imf}) are more challenging. Naively, the first measurement collapses the state, preventing further evolution.  Ways to overcome this have been proposed, including ancillary probe-and-control qubits~\cite{PhysRevLett.113.020505,Ortiz:2000gc,Lamm:2019bik,Lamm:2019uyc,Gustafson:2020yfe} and phase estimation~\cite{Abrams:1998pd,Roggero:2018hrn}. For time-separated matrix elements, it is still not known how to perform
nonperturbative renormalization like RI/SMOM~\cite{Martinelli:1994ty,Aoki:2007xm,Sturm:2009kb} on quantum computers.

\paragraph*{\bf Gauge invariance and errors:}
Noisy quantum devices can also be viewed as introducing new operators. The best-studied examples of this are related to gauge-violating operators~\cite{Stannigel:2013zka,Stryker:2018efp,Halimeh:2019svu,Lamm:2020jwv,Tran:2020azk,Halimeh:2020ecg,Halimeh:2020kyu,Halimeh:2020djb,Halimeh:2020xfd,VanDamme:2020rur,Kasper:2020owz,Halimeh:2021vzf}.  Which operators are introduced and which symmetries are broken are both hardware and digitization dependent. In tensorial reformulations discussed in II.B, the full integration over the gauge fields
provides manifestly gauge-invariant discrete reformulations.

\begin{center}
    ---
\end{center}

Analog quantum simulations provide another potentially promising path to quantum simulation of QFTs. Such simulators may naturally exhibit fermionic and bosonic degrees of freedom, or provide tunable interactions between larger local Hilbert. Nonetheless, one needs to engineer the interactions of these degrees of freedom to represent the dynamics of the QFT Hamiltonians of interest, which is generally a challenging task in current platforms. For example, simulating the dynamics of both Abelian and non-Abelian gauge theories in 2+1 and higher dimensions have proven hard~\cite{zohar2012simulating,zohar2011confinement,zohar2013simulating,tagliacozzo2013optical,ott2021scalable,gonzalez2022hardware} given either higher-body interactions or non-local interactions depending on the representation of the Hamiltonian. Some progress has been reported in recent years, but a first implementation of complete building blocks of a lattice gauge theory with high fidelity remains an important goal of the program in the coming decade. Rigorous error-bound analysis, error corrections, and verifiability will need to be developed for analog simulations as well. 

The role of hardware implementation and benchmarks in guiding the course of developments in quantum simulation cannot be overstated. Many recent experiments and implementations of a variety of QFT problems on a range of quantum platforms in both analog and digital modes (see e.g., Refs.~\cite{martinez2016real,Klco:2018kyo,Nguyen:2021hyk,kokail2019self,lu2019simulations,alam2021quantum,Klco:2019evd,Atas:2021ext,ciavarella2021trailhead,Ciavarella:2021lel,xu20213+,mildenberger2022probing,schweizer2019floquet,gorg2019realization,mil2019realizing,yang2020observation,Zhou:2021kdl} have generated a platform for communications and collaborations with experts in quantum hardware technology. It has also generated proposals and experiments dedicated to developing simulators suitable for QFTs, see e.g., Refs.~\cite{Hauke:2013jga,Davoudi:2019bhy,surace2020lattice,Macridin:2018gdw,davoudi2021towards,casanova2011quantum,Zohar:2012xf,Banerjee:2012xg,dasgupta2022cold,zohar2013quantum}. This is a critical path as otherwise theory and algorithmic developments will be disconnected from the reality of hardware. To achieve meaningful progress, a series of models from low-dimensional theories and simpler gauge groups need to be identified and progressively made more complex to follow, or ideally, guide, hardware developments. Finally, hybrid classical-quantum approaches to quantum simulation should be taken advantage of in both near and far terms~\cite{bender2018digital,davoudi2021towards,gonzalez2022hardware}. See Ref.~\cite{Bauer:2022hpo} for more references and context.

LQCD began around the time the computing became available to the scientific community and developed to its current mature state alongside the increase in computing power. These decades of experience can be leveraged as quantum computers develop. The challenges of simulating gauge theories, with an eye towards QCD, will be a rich area of research in the decades ahead.

\subsection{Tensor networks}

Tensor network methods are playing an 
increasingly important role in several branches of physics and in quantum information science.
In the context of HEP, tensor networks have appeared as ways to reformulate lattice gauge theory models to obtain a fully discrete formulation that is suitable 
for quantum computation and coarse graining methods. Tensor networks 
also provide tools to understand entanglement in conformal field theories and their connection to gravity \cite{Meurice:2022xbk}.

In the context of lattice gauge theory, tensors can be seen as the 
translationally invariant, local {\it building blocks} of exact 
discretizations of the path integral. 
They encode both the local and global symmetries of the original model. It is easy to design approximations (truncations) that preserve these symmetries and to design simplified 
models that should have the same correct universal continuum limit as the 
original model. Developing these building blocks and optimizing the approximations for NISQ machines and classical computers are important tasks for the near-term future. 
Tensor networks can also be used to perform 
classical simulations of quantum circuits.
This is useful for developing and testing quantum computing algorithms and quantum computational advantage.

Tensor networks have their origins in condensed matter physics where they provide a suitable
truncated basis for constructing 
the low lying energy eigenstates of strongly interacting many body systems. In particular matrix
product states and the DMRG algorithm provide very accurate ground state wavefunctions for one dimensional
systems \cite{PhysRevLett.69.2863,Banuls:2019bmf,Cirac:2020obd}. The annual lattice conferences have helped foster interactions among the communities involved
 \cite{mcbpos2013,ympos2013,kvapos2014,yspos2014,mcbpos2014,juypos2014,ympos2014} and
 the number of contributions has grown steadily with the years \cite{mcbpos2018}.
Current applications to particle physics include
the Schwinger and Thirring models and gauge theories in $(2+1)$ dimensions - see
\cite{Banuls:2019rao} and references therein. The basic
idea is to express the ground state wavefunction as the trace of a product of local
tensors which are functions of a set of parameters. Varying these parameters
allows for a good estimation of the ground state.

\begin{figure}[htbp]
  \centering
  \includegraphics[width=\hsize]{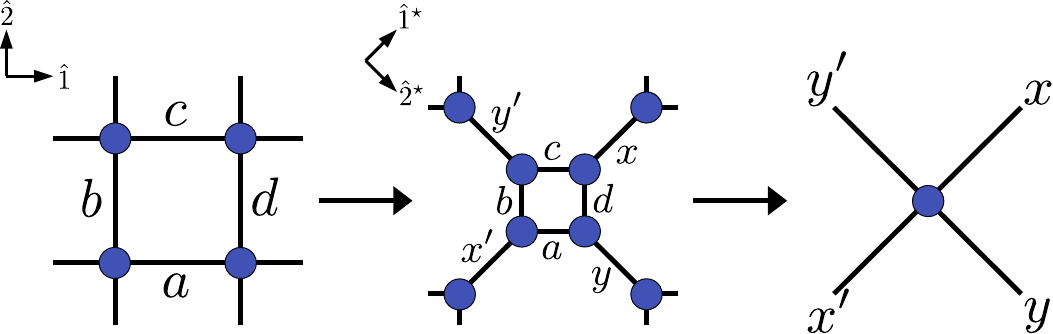}
  \caption{
    A coarse-graining step for a tensor network.
    Circles represent tensors, and closed indices should be contracted. From Ref.~\cite{Meurice:2020pxc}.
  }
  \label{trg}
\end{figure}

Tensor network formulations can also be constructed for Euclidean path integrals and 
in conjunction with the tensor renormalization group (TRG) have
been used to provide classical simulations of two dimensional models including both
gauge fields, scalars and fermions such as the non-abelian
Higgs, Schwinger and Gross-Neveu models provide classical simulations of lattice field theory models including
spin models, gauge theories, scalars and fermions,  the non-abelian Higgs, the Schwinger and Gross-Neveu models \cite{Meurice:2022xbk,Meurice:2020pxc}.
For more extensive sets of references and a road-map (the ``Kogut sequence"), Typically
the local Boltzmann factors are expanded on characters of the group and the
original fields integrated out exactly yielding a discrete tensor representation that can form
a starting point for quantum simulation. Furthermore, it can be shown that truncation of
these tensors to a finite number of representations does not break any
symmetries. These tensor networks can be coarse grained using renormalization group
ideas to allow computation of a variety of observable including
the free energy. Fig.~\ref{trg} shows an example of
this procedure for the two dimensional Ising model
Recently, improved 
renormalization group methods (loop-TNR and ATRG
algorithms---see Ref.~\cite{Meurice:2020pxc})
have been developed that are capable of yielding accurate results for three dimensional $U(1)$ gauge
theory and scalar theories in four dimensions \cite{Akiyama:2020ntf}. 
To handle fermions a Grassmann tensor
renormalization group has also been developed which can keep track of any signs arising
from the fermions \cite{Kadoh:2018hqq}. 

\subsection{Simple holographic lattice models}
\begin{figure}[htbp]
  \centering
  \includegraphics[width=0.9\hsize]{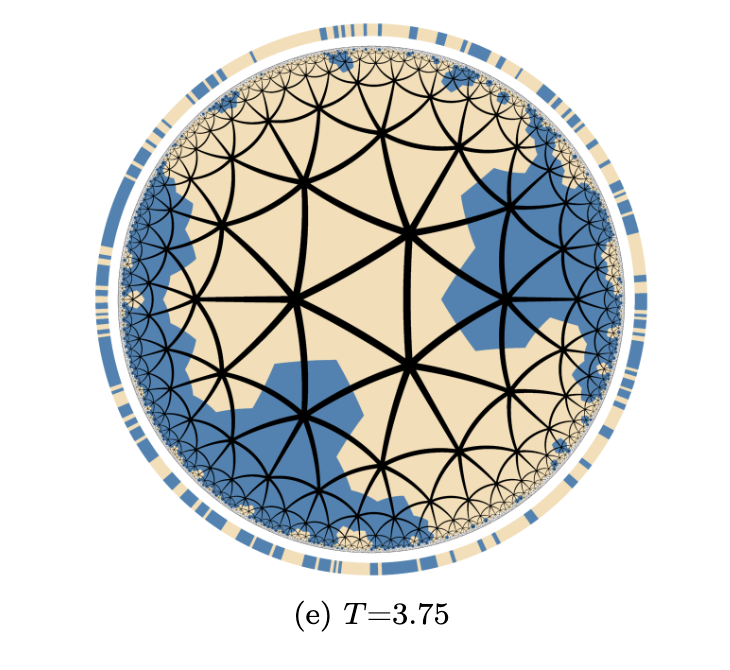}
  \caption{Sample configuration for Ising model on 2d hyperbolic disk
  }
  \label{isingpic}
\end{figure}

The original AdS/CFT correspondence provided strong evidence for a duality between
gravity in anti-deSitter space and strongly coupled quantum field theories living at the boundary
of that space \cite{Mald}. Subsequently ideas from quantum information science 
have served to strengthen this holographic connection between gravity and entangled quantum
systems---see a recent Snowmass review~\cite{Faulkner:2022mlp}. Typically this correspondence is studied in the limit where the bulk gravitational
fluctuations are vanishing. Quantum simulation can be used
in principle to probe this duality in more general regimes. A significant amount of
work has been done using classical simulation over the last decade to check
holography see Refs.~\cite{Cotler:2016fpe,Berkowitz:2016jlq,Hanada:2016zxj,Catterall:2020nmn,Catterall:2008yz,Catterall:2017lub,Hanada:2013rga}. However the next frontier in testing the holographic dictionary will likely come from quantum simulation.

Motivated by this a variety of simple scalar, Ising and fermion matter systems
have been simulated 
on tessellations of hyperbolic space (Wick rotated anti-deSitter space) in two and three
dimensions \cite{Asaduzzaman:2020hjl,Asaduzzaman:2021bcw,Brower:2020huh}. Results like the RT formula
for the entanglement entropy arises naturally via lattice strong coupling expansions. An example
Ising configuration generated by a Monte Carlo algorithm
is shown in fig.~\ref{isingpic}.
In spite of the fact that the discrete spaces are only invariant under a subgroup
of the continuum isometry group the boundary theories are seen to exhibit power law
correlations as expected for a conformal field theory. Furthermore it is possible to include
dynamical gravity effects by allowing the tessellation to fluctuate. Preliminary results
in this case suggest the existence of both conformal and non-conformal phases depending on the size of
the gravitational fluctuations \cite{Asaduzzaman:2021ufo}.

Extending this approach to a Hamiltonian formulation allows for quantum simulation and would
allow a connection to be made to tensor network models of quantum error correction \cite{Pastawski:2015qua}. The central idea of holographic error correction is that the quantum
state of qubits residing deep inside the bulk of such a tessellated hyperbolic space is robust to
errors in the quantum state of the boundary qubits.

\subsection{Future near-term goals}
Both the algorithms and hardware needed for quantum simulation
are progressing rapidly and so any projection for the next decade will necessarily
be somewhat uncertain. Nevertheless the successes that have been achieved so far
suggest some interesting goals may be achievable with the next five to ten years. These include:
\begin{itemize}
\item Simulation of quantum circuits
representing gauge theories coupled to Dirac fermions in (1+1) and 
(2+1) dimensions on small lattices with control over time-digitization errors. 
\item Classical tensor-network calculations of similar systems for validation purposes. Use of tensor networks for initial state preparation.
\item Classical and quantum simulation of gauge theories and fermions
formulated on discretizations of two and three dimensional hyperbolic space. 
Studies of quantum error correction on such networks.
\item{Identifying and implementing problems that can provide qualitative understanding of non-equilibrium dynamics of strong interactions, including mechanisms for equilibration and thermalization, starting from small prototype models and using potentially analog simulators with larger Hilbert spaces, as well as classical Hamiltonian-simulation methods such as tensor networks, see e.g., Refs.~\cite{Mueller:2021gxd,Zhou:2021kdl,deJong:2021wsd,pichler2016real,Zache:2018cqq,Desaules:2022kse}.}
\end{itemize}
Crucial to the long term success of these efforts will be the development of
reliable quantum error correction algorithms and hardware. This will in turn depend
partly on the fabrication of systems with many thousands of qubits.
Keeping gate fidelities and decoherence times acceptable at such scales will be an important mission of the quantum hardware community.

\section{Theory for Quantum Sensing of Fundamental Physics}

The emerging ability to manipulate and control the quantum state of a system, as well as the technological advances that enable it, opens a host of new opportunities to directly probe fundamental physics and to search for new physics. Particle theorists are playing a vital role in identifying the physics opportunities for novel sensing experiments and drawing new connections between technology and fundamental tests. The theory-experiment-technology connection in the broadly-defined area of quantum sensing is strong and mutually beneficial to all these communities. Researchers developing sensors and systems for ``practial'' quantum applications in the areas of superconducting circuits, quantum optics, single photon detection, atomic physics, or superconducting cavities have been made aware of the utility of their technology to searching for new physics by particle theorists. Similarly, HEP theorists have been informed by their experimental colleagues about state-of-the-art technology, triggering new ideas. This back-and-forth is particularly strong in universities and labs in the US.

In this section we will present some of the physics targets for quantum sensor searches and proceed to highlight some of the technologies that theorists, often in collaboration with experimentalists, have proposed to pursue these target. We conclude with a a discussion and outlook. Other Snowmass white papers that deal with sensing with a theory component include Refs.~\cite{Carney:2022rlu, Essig:2022yzw}.

\subsection{Fundamental physics targets}
Some of the physics questions on which sensing experiments aim to shed light are central to enhancing our understating of nature. These include:
\paragraph*{\bf Searching for new particles:} An understanding of which degrees of freedom  are included in the rule book of nature -- in other words, which particles exist -- has always been at the heart of fundamental physics. Cosmic rays (in the early 20th century) and collider experiments (in its later half and into the 21st) have pushed the frontier in search of heavier particles. New light degrees of freedom that are feebly interacting, of which neutrinos are an example, are well motivated, and can exist within a large parameter space~\cite{Essig:2013lka}. Axions~\cite{Peccei:1977ur}, axion-like particles, dark photons~\cite{Holdom:1985ag}, and milli-charge particles are among the best motivated. Proposed and ongoing sensing experiments, with strong theory involvement, are making headway in this search.
\paragraph*{\bf Dark matter:} Our and other galaxies were brought to form by the gravitational pull of dark matter which still dominates galactic mass. The beyond the standard model nature of dark matter motivates the search for new degrees of freedom, as in the previous paragraph, which also make up all or part of our dark matter halo. New sensing initiatives have originated from the theory community, going back to Ref.~\cite{Sikivie:1983ip}, but many new ideas have been put forth recently~(see Refs.~\cite{Adams:2022pbo, Antypas:2022asj, Berlin:2022hfx} for Snowmass summaries). Dark matter candidates that can be probed by quantum sensing setups are remarkably varied, from ultralight wavelike dark matter (masses of $10^{-21}$~eV to about 1 eV) of which axions~\cite{Peccei:1977ur}, dark photons~\cite{Holdom:1985ag} are the most well-known, to light WIMP-like particles (in the keV to GeV range), as well as Planck mass particles and above~\cite{Carney:2022gse}.
\paragraph*{\bf Gravitational waves:}
The discovery of gravitational waves (GWs) by LIGO~\cite{LIGOScientific:2016aoc} has profound implications, both in confirming the predictions of general relativity, and in opening a new window for exploring the Universe. Expanding the range of frequencies in which we can observe GWs is of vital importance, akin to the detection of light beyond the visible spectrum. Quantum technology is playing an important role in enabling and enhancing LIGO's reach. Furthermore, theorists have put forth ideas for detecting gravitational waves in a large range of frequencies~\cite{Berlin:2021txa, Domcke:2022rgu, Fedderke:2022kxq, Fedderke:2021kuy}, some of which are being actively pursued experimentally~\cite{Berlin:2022hfx}.
\paragraph*{\bf Tests of quantum mechanics:}
Quantum mechanics (QM) is an odd theory. Yet, it has withstood many experimental tests, and hence it is at the very least an excellent approximation of Nature. Despite its oddity, it has also proven theoretically very difficult to modify due to constraints imposed by unitarity and causality~\cite{Weinberg:1989us, Polchinski:1990py}. High energy theorists are well positioned to address this because any successful modification of QM would need to be consistent, or at least embedded within the standard model of particle physics, a theory which provides an excellent description of nature at high energies. A modification of quantum mechanics must find a home in quantum field theory (QFT). Indeed, recently a new nonlinear extension of quantum mechanics that is inherently unitary and causal has been proposed by HEP theorists by starting with a modification of QFT and working back to the consequences for QM~\cite{Kaplan:2021qpv}. This extension, and likely others that may follow, can be tested experimentally with dedicated quantum sensing  setups~\cite{Polkovnikov:2022nwk}. 
\paragraph*{\bf Tests of gravity in a quantum setting:}
Theories of gravity and quantum mechanics have yet to be unified to a single framework at arbitrarily short lengths scales. From a theoretical point of view, gravity and quantum theory are compatible within a long distance effective field theory. Still, given the difficulty in unifying them at short distance it is interesting to test gravity at laboratory length scales in situations in which things are manifestly quantum. Theorists have proposed such tests employing recent advances in atomic physics and opto-mechanical systems~\cite{Carney:2022dku}.

\subsection{Theory for sensing experiments}

In this subsection we demonstrate the role theorists are playing in the arena of sensing with quantum technology by providing explicit examples of experimental efforts which were spawned by theory-experiment collaboration as well as proposals for novel experiments. The takeaway from this list is that theory is contributing to the development of an innovative and varied research program in the area of quantum sensing. A similar point has been made more broadly in Ref.~\cite{Essig:2022yzw}. 
The list below is by no means complete (apologies for omissions), but the degree of its incompleteness serves to strengthen the point that the HEP theory community is making a strong and broad contribution in proposing innovative ways to search for new physics. 

\paragraph*{\bf Atom interferometers - MAGIS:} Atom interferometers operated in free fall provide an exciting opportunity to search for gravitational waves in the mid-band (around a Hz, between LIGO and LISA) as well as for several classes of wavelike dark matter~\cite{MAGIS-100:2021etm}. The flagship experiment in this class, MAGIS~100, a 100 meter network of interferometers is in design stages~\cite{MAGIS-100:2021etm} and will be at Fermilab in the coming years. The initial proposal to search for gravitational waves with atom interferometers was put forth by a collaboration of HEP theorists and atomic experimentalists~\cite{Dimopoulos:2006nk, Dimopoulos:2007cj, Graham:2012sy}.

\paragraph*{\bf Atomic clocks and other atomic experiments:}
Atomic physics provides ever increasing precision in time keeping and frequency determination. The comparison of two atomic clocks, or of an atomic clock to another high quality device can be sensitive to light bosonic dark matter, such as a dilaton, as proposed by theorists~\cite{Derevianko:2013oaa,Arvanitaki:2014faa, Stadnik:2014tta} (and reviewed in greater detail in Ref.~\cite{Antypas:2022asj}. 
Atomic clock technology is also of interest for future space missions to probe fundamental physics~\cite{Alonso:2022oot, Tsai:2021lly}. 
In addition, precision atomic spectroscopy and a systematic comparison of transitions among nuclear isotopes has been shown, again in HEP theory-experiment collaboration, to be sensitive to new long range forces~\cite{Berengut:2017zuo}.

\paragraph*{\bf Condensed matter systems for low recoil searches:} The program to detect dark matter directly in the lab has focused for several decades on dark matter masses above 10~GeV. The leading detectors in this range, liquid noble element experiments, do not have sensitivity to dark matter below this range because detection thresholds are above the energy deposited in dark matter collisions. A need for lower threshold detectors has thus arisen. Many ideas have been put forth and are pursued experimentally. Not surprisingly, many such efforts have been spawned by theorists or in a theory-experiment collaboration. 
The connection to quantum technology is simply that the systems that are needed to explore this region of parameter space are often used to detect single photons or single quanta. 
Examples include Skipper CCDs and the SENSEI experiment~\cite{Crisler:2018gci} (which has potential for growth~\cite{Aguilar-Arevalo:2022kqd}), proposals to use transition edge sensors~\cite{Hochberg:2015pha}, a proposal to search for dark matter with superconducting nano-wires~\cite{Hochberg:2021yud}, and magnetic bubble chambers~\cite{Bunting:2017net} (with prototype results in Ref.~\cite{Chen:2020jia}).

\paragraph*{\bf Cavity based Dark Matter searches:}
High quality radio frequency cavities are a key component of the axion haloscope originally proposed by Sikivie~\cite{Sikivie:1983ip} and are currently being used to search for dark matter axions in the 1-10~GHz mass range~\cite{Adams:2022pbo}. Quantum techniques, such as squeezing (demonstrated in the HAYSTAC experiment~\cite{HAYSTAC:2020kwv}) and photon counting~\cite{Dixit:2020ymh} holds promise to hasten the search in this range. A recent theoretical analysis showed that squeezing can be distributed over a coherent network of sensors to provide additional gain in the scan rate~\cite{Brady}.

Recent proposals by theorists and experimentalists have been made to extend cavity based searches to lower axion masses by leveraging the extremely high quality of superconducting cavities by considering the upconversion of one cavity mode to another~\cite{Berlin:2019ahk, Berlin:2020vrk,Lasenby:2019prg} with interesting potential coverage. R\&D is ongoing with theory input to pursue this goal~\cite{Berlin:2022hfx,Giaccone:2022pke}. 

\paragraph*{\bf LC resonator DM searches: }

Axion, dark photon, and milli-charged particles searches at lower dark matter masses can be achieved by other resonant systems such as LC circuits. The DM-Radio experiment~\cite{DMRadio:2022pkf} was originated by a theory-experiment collaboration and will search for faint electromagnetic signals inside a well shielded region employing a squid~\cite{Chaudhuri:2014dla}. The ABRACADABRA-10cm  experiment~\cite{Ouellet:2018beu}, which also originated in the theory community~\cite{Kahn:2016aff}, proposed a similar detection strategy employing a toroidal geometry. ABRACADABRA is now merged with DM-Radio~\cite{DMRadio:2022pkf}. Pushing limits on the QCD axion can be achieved with quantum limited sensing and a large scale effort. A modest modification to such setups involves driving electromagnetic fields in a nearby shielded environment; such a ``direct deflection" experiment was proposed by theorists and would be sensitive to the same class of milli-charged dark matter models targeted by low-threshold detectors~\cite{Berlin:2019uco}.

\paragraph*{\bf Searching for new particles with cavities:}
High quality cavities can be utilized to search for new degrees of freedom independently without assuming they make up the dark matter. Dark SRF (reviewed in Ref.~\cite{Berlin:2022hfx}) is an ongoing light-shining-through-wall (LSW) experiment, with theory-experiments collaboration, searching for dark photons as part of the SQMS center at Fermilab. An ultra high-Q superconducting RF cavity is excited to the brim with photons, and a feeble appearance signal is searched for in a nearby cavity which is in tune with the first. The Dark SRF setup is optimized to search for the longitudinal polarization of the dark photon, which leads to enhanced sensitivity, as pointed out in a theoretical paper~\cite{Graham_PRD_2014}. Theorists have also pointed out that the Dark SRF experiment will also be sensitive to very light milli-charged particles via Schwinger pair production~\cite{Berlin:2020pey}. The Dark SRF experiment will be pushed to the quantum regime in the coming years~\cite{Berlin:2022hfx}.

Several theoretical proposals have been made to utilize the high quality of SRF cavities to search for axions. These include two cavities and a conversion region with a static magnetic field~\cite{Janish_PRD_2019} in an LSW setup, a two-cavity LSW employing the up-conversion technique~\cite{Gao:2020anb}, or a single cavity in which two modes are populated, and a third quiet mode is used as a signal mode~\cite{Bogorad_PRL_2019}. The later can in principle also be sensitive the the standard model Euler-Heisenberg light-by-light interaction. The feasibility of these various schemes is being studied by experimentalists and theorists in SQMS~\cite{Berlin:2022hfx}.

\paragraph*{\bf DM Searches with 
Dielectric Stacks and Dish Antennas:}
HEP Theorists have pointed out that both axion and dark photon dark matter can convert to photons at the surface of a mirror, and that using the appropriate geometry, the signal can be focused~\cite{Horns:2012jf}. This broadband search approach will benefit from single photon detection at the frequency in question. This idea led to a plethora of experimental efforts, including new limits, such as Refs.~\cite{Knirck:2018ojz, Tomita:2020usq, Brun:2019kak, FUNKExperiment:2020ofv} (some of which with strong theory involvement) and planned experiments~\cite {BREAD:2021tpx}. 

A related detection strategy also proposed by theorists~\cite{Jaeckel:2013eha, Millar:2016cjp}, is to use the transition between two dielectric materials to generate photons from dark matter. A repeated transition, a dielectric layered  stack, can be used to enhance the rate. The MADMAX experiment, a collaboration including theorists, has grown from this and is already in the prototype phase~\cite{MADMAX:2019pub}. 

\paragraph*{\bf Optics based searches:} Quantum optics is a mature field and a strong driver of progress in QIS. Optics-based searches for new particles and/or dark matter present interesting opportunites around the~eV scale. The LAMPPOST experiment~\cite{Chiles:2021gxk}, an optics realization of the dielectric stack was first proposed in a theory paper~\cite{ Baryakhtar:2018doz}. A prototype has already set new limits on dark photon dark matter.

Optics is also a key tool to search for particles that are not dark matter. The ALPS experiment~\cite{Arias_2012}, also with strong theory involvement, is a LSW experiment consisting of two optical cavities on either side of a wall, one acting as emitter and the other as receiver. The parametric gain achieved by a receiver cavity was first pointed out in Ref.~\cite{Sikivie:2007qm} in a theory-experiment collaboration.

More recently, nonlinear optics was proposed as a novel tool to search for new particles by theorists and experimentalists. In a process dubbed dark SPDC, a pump photon down-converts to a signal photon and an associated dark particle, either a dark photon or an axion~\cite{Estrada:2020dpg}. A ``missing energy at the optics table''experiment may have parametric advantages, and very different challenges, compared to light shining through walls experiments. 

\paragraph*{\bf Spin precession experiments:}

Some of the most powerful constraints on new physics that violates CP come from constraints on the (static) electric dipole moments of fundamental particles~\cite{Alarcon:2022ero}. 
The QCD axion, if it is dark matter, will lead to an oscillation of the neutron electric dipole moment with a frequency set by the axion mass. Theorists and experimentalists have proposed to search for this effect by looking for the precession of the polarized nuclear spins in an electric field in a resonant NMR setup~\cite{Budker:2013hfa}. The CASPER experiment~\cite{JacksonKimball:2017elr} is pursuing this, as well as spin precession effects due to direct axion coupling to matter, known as CASPER-Wind~\cite{JacksonKimball:2017elr} (see also Ref.~\cite{Gao:2022nuq} for a proposal using superfluid helium-3 and precision readout using atomic clocks). Using a liquid co-magnetometer, a prototype has already set limits with theoretical involvement~\cite{Wu:2019exd}. Other spin precision experiments, including a spin polarized torsion pendulum, have been shown to have interesting reach in a theory-experiment collaboration in~\cite{Graham:2017ivz}.

As a notable example, quantum spin gyroscopes known as noble-alkali co-magnetometers can also been used to detect the coupling of a gradient of the axion field to nuclear spins~\cite{Safronova:2017xyt, Bloch:2019lcy}. In a small scale effort dubbed NASDUCK~\cite{Bloch:2021vnn}, initiated by theorists, new limits on such axion dark matter couplings have been set.

Similar NMR techniques can also be used to search for the spin dependent long range force which is mediated by the QCD axion. Again, in strong collaboration with theorists, the ARIADNE project aims to look for a force between a Helium-3 polarized sensor and a nearby tungsten target~\cite{ARIADNE:2017tdd}.

\paragraph*{\bf Single particle traps:}

One area of advance in the quantum regime is the ability to trap and manipulate the quantum state of a single charged particle, either an ion or electron. In the realm of fundamental physics, this allows one to study the properties of single particles. The classic example is the anomalous gyromagnetic moment of the electron, $(g-2)_e$, for which the theory-experiment comparison represents the most precise test of the standard model to date~\cite{Hanneke:2008tm}.

Recently, it was pointed out that a single particle can be a sensitive target for dark matter direct detection. Millicharged dark matter can cause energy to be transferred from the laboratory to the cold particle trap at a rate which is higher than observed~\cite{Budker:2021quh, Carney:2021irt}. In addition, dark photon dark matter can be resonantly absorbed by a trapped electron leading to quantum jumps in its axial energy level, as has been shown in a recent proof of concept experiment~\cite{Fan:2022uwu}.

\paragraph*{\bf Opto-mechanical sensors:}
The ability to bring macroscopic objects to their ground state and sense their motion at quantum-limited precision is opening new opportunities for sensing. An array of quantum-limited sensors was proposed by theorists to search for the gravitational interaction of very heavy dark matter~\cite{Carney:2019pza} and for $B-L$ vector boson dark matter~\cite{Carney:2019cio}. A theoretical analysis of going beyond the standard quantum limit was presented in Ref.~\cite{Ghosh:2019rsc}. An ambitious program, Windchime, was also proposed~\cite{Windchime:2022whs}. Theorists have also analyzed the reach of torsion balance experiments to some of the same targets in Ref.~\cite{Graham:2015ifn} and a dedicated search can make further progress.

\subsection{Outlook}

There are conclusions to draw from this long, yet partial, list of contributions from theorists to the search for new physics:
\begin{itemize}
    \item The search for new physics requires us to cast a wide net~\cite{hochberg2022new}. Theorists are contributing to the search, in particular searches for new particles, dark matter, and gravitational waves, in a central way.
    \item Though some of the contributions date back decades, many were made in the past decade and at an increasing rate. The field of novel ideas to strengthen our searches for new physics is vibrant and growing. 
    \item Though they may lead to small or large projects, these contributions are enabled by the vibrant HEP research environment in which theorists and experimentalists interact, learn and exchange ideas. Maintaining the funding for research in the US, particularly for theory, is vital in order to reap the benefits from new innovations. A shortage in funding for small scale efforts in this genre is a current bottleneck in expanding the search for new physics.
    \item Not all theory-experiment collaborations are within the scope of quantum science (see Ref.~\cite{Essig:2022yzw} for a discussion that is synergistic to this one, but with a broader scope). However, a large number of quantum and quantum-related technologies have been identified by theorists and experimentalists as new tools in the search for new physics. This is not a coincidence; quantum information science requires isolation and control of delicate systems as well as reading out the most feeble of signals. These challenges are shared by the endeavour to look for new physics and
    this justifies an investment by the HEP community in the research that connects quantum technology to fundamental physics. HEP theorists are well poised to make the needed connections between these fields. 
    \item Gravitational waves have, for a while, been removed from the scope of HEP. Regardless of how they are detected, GWs are of great interest to the HEP community. The emergence of new GW detection schemes serves to strengthen this point further. 
\end{itemize}


\section{Formal Aspects of Quantum Information and Fundamental Physics}

Let us now consider more broadly the theoretical underpinnings of the links between quantum information science (QIS) and fundamental physics.  
Within QIS itself the main directions can be categorized into broad themes of quantum information theory, quantum computing, and implementations.  This section focuses on the former set, where topics such as entropy, entanglement measures, quantum error correction, and quantum communication play a prominent role.  
These concepts have remarkably enriched theoretical physics, especially in the context of quantum field theories and quantum gravity.   

A particularly fruitful setting at the confluence of these ideas is the AdS/CFT correspondence \cite{Mald} (also known as gauge/gravity duality), which relates a higher-dimensional gravitational theory to a lower-dimensional non-gravitational one.  A cornerstone feature of these dualities is the specific type of rearrangement (or transmutation) of the degrees of freedom between the two sides, which allows one to extract genuinely quantum features on the CFT side from classical ones in the dual AdS, as well as strongly gravitational effects in AdS from non-gravitational ones in the CFT, with both pictures nevertheless retaining the requisite degree of locality.
Within the last decade or so, it has become evident that many key quantities of interest in QI theory actually have a rather simple description on the gravity side, belying the intricacy of the holographic mapping as well as the operational definition of the QI construct itself -- in some sense, the bulk geometry in the gravitational theory naturally `knows about' entanglement!  

This section will explain how this revelation allowed for a change in perspective which fueled the vibrant activity in the recent years, then briefly indicate some of the fascinating resulting developments, and end with a broad outlook of the big-picture questions that the community is now well poised to answer over the next decade.  Much of the narrative is adapted from the excellent and pedagogically written white paper \cite{Faulkner:2022mlp}. 
A related theme focused on quantum aspects of black holes and the emergence of spacetime is nicely summarized in the white paper \cite{Bousso:2022ntt}. Other white papers on related topics include \cite{Giddings:2022jda,Meurice:2022xbk,Blake:2022uyo}. Some useful pedagogical resources on the subject include Refs.~\cite{Nielsen:2010aa,Wilde:2013aa} for an introduction to quantum information in quantum mechanics, Refs.~\cite{Hagg:2012alg,Ohya:2004qnt} for algebraic treatment of QFTs, \cite{Calabrese:2009qy} for computation of entanglement in conformally invariant QFTs, and 
Refs.~\cite{Nishioka:2009un,VanRaamsdonk:2016exw,Rangamani:2016dms,Headrick:2019eth} for reviews of developments in the context of holography.
The referencing in this section will be kept minimal, restricted to other relevant white papers and to seminal works initiating a given subject or reviews thereof.  

\subsection{Change of perspective}

With the advent of QIS, the traditional and intuitively natural ways of organizing physics, e.g.\ by energy scale or interaction strength, by field content or observables of interest, or even by its computational tractability, has been amended by a wholly different perspective of focusing rather on the information content.   Such an approach allows for a description applicable to a bewildering variety of physical systems, unifying previously disparate avenues of exploration and simultaneously deepening our understanding.  

Of particular significance is the shift of focus towards genuinely quantum phenomena.  Although quantum field theory, providing a framework for explaining the fundamental laws of nature, is built on quantum mechanics as its central pillar, the full quantumness is seldom utilized in the traditional approach. 
Instead, one typically focuses on observables that can be computed using correlation functions of local operators, which can then be translated  into physical quantities such as scattering cross-sections and dynamical response functions. The Wilsonian effective field theory paradigm is also primarily geared towards identifying relevant operators and understanding their correlation functions below some cut-off scale. Furthermore, techniques such as path integrals, the renormalization group, effective theories, symmetries, and dualities are all primarily geared in textbook treatments towards computing such correlation functions. 

There is, however, more information to be mined by generalizing the framework to one which is more cognizant of the underlying quantum mechanical structure. In particular, such a perspective has to account for the fact that composite quantum mechanical systems exist in tensor product superpositions, which leads to the essential concept of entanglement. Focusing on these aspects can not only help us further elucidate the field theory framework, but it also appears better suited to addressing profound questions in the quantum gravitational setting.

\subsection{Recent developments}

The general themes that have been explored in this broad area include refinement of conventional field-theoretic tools to quantify spatially-ordered entanglement in QFTs, quantifying measures of entanglement in mixed states, and the revival of operator algebraic formulations of QFT. In a parallel development, the geometrization of information theoretic measures in the context of the holographic AdS/CFT correspondence has played an important role in furthering our understanding of the holographic dictionary. 
Additionally, progress has been made on questions relating to the complexity of state preparation, which is important for quantum simulations, and it has furthermore been argued that these ideas have a physics role to play in QFT and quantum gravity. 
Many of these developments have been accompanied by novel insights on the quantum information side, showing that the synergy between the fields benefits both sides.
Here we briefly mention the most important recent developments and open problems in these areas.

\paragraph*{\bf Characterizing quantum information:}
It has become increasingly evident that the structure of entanglement plays a key role in understanding many properties of physical systems.  Correspondingly a central goal is to define and quantify the amount of entanglement, with view towards understanding the organization of entanglement in any given state, in terms of how it is structured in space and between scales, and how it can be prepared.
This is by no means a simple task.  Already in the typical QIS setting involving finite systems (having Hilbert space of finite dimension which factorizes into Hilbert spaces of subsystems), there are many distinct entanglement measures.  They are typically based on entropic functions designed to be meaningful from the point of view of communication, or motivated from a resource theoretic perspective.  Examples include von Neumann entropy (which in the high energy community is referred to as entanglement entropy), relative entropy, entanglement of distillation, entanglement cost, entanglement of purification, etc..  

In the setting of interest to high energy physics, formulated in the language of a QFT and in particular gauge theory, things are far more complicated.  Not only is the Hilbert space infinite-dimensional, but it does not actually factorize \cite{Witten:2018zxz}.  Nevertheless, one can still define entanglement entropy and extract its regulator-independent universal features.  In the holographic context this takes a particularly nice form: the entanglement entropy of a subsystem defined by a spatial region in the CFT is given by the area of a certain (area-extremizing) surface in the bulk geometry associated to the specified region.  (This prescription is referred to as RT \cite{Ryu:2006bv} in the static context, HRT \cite{Hubeny:2007xt} in the general dynamical context, and QES \cite{Engelhardt:2014gca} when certain quantum corrections are included.)  The fact that entanglement entropy is computed by such a simple geometrical construct provided an early hint at the intriguing connection between entanglement and geometry which we are still trying to fathom.  The emerging lesson is that in some sense, entanglement builds spacetime \cite{VanRaamsdonk:2009ar} (sometimes referred to by the slogan ER=EPR \cite{Maldacena:2013xja}).  In the holographic context, the bulk geometry naturally implements a quantum error correcting encoding of the boundary physics \cite{Almheiri:2014lwa}.

Hence such geometrization of entanglement has been immensely useful not only at the practical level (of probing the behavior of entanglement in many systems of interest) but also at the conceptual level.  
The QI-based insights will in turn deepen our understanding of the physical frameworks.  For example, by connecting information quantities to basic data in QFT, we might expect to be able to put new constraints on the QFT theory space.

\paragraph*{\bf Symmetry, renormalization group flows, and phases:}

Quantum information ideas also interplay with some key non-perturbative structures in QFT.  For example, one can extract statistical properties of charge fluctuations in localized regions and study different charged sectors using symmetry resolved entanglement \cite{Goldstein:2017bua}, as well as formulate order parameters based on the relative entropy which probe superselection sectors of global charges in QFT \cite{haag2012local}.

Our understanding of effective field theories relies on the fact that dynamics below some cut-off scale is only sensitive to a handful of relevant parameters \cite{Polchinski:1983gv}. Since from a microscopic viewpoint, the effective description is attained by tracing out high energy degrees of freedom which is an intrinsically lossy process, it is natural to expect a measure of the number of degrees of freedom that is monotonic in energy scale, reaching a conformal fixed point in the IR.  Such functions (dubbed $c$, $F$, and $a$ in 2, 3, and 4 dimensions respectively) have been found \cite{Jafferis:2011zi,Komargodski:2011vj} starting with Zamolodchikov's c-theorem  \cite{Zamolodchikov:1986gt}.
While the conventional field theory methods relied on dimension-specific techniques, more recently entropic proofs based on the Markovian property of the CFT vacuum have been developed \cite{Casini:2017vbe}, which provide an interesting overarching principle.  One may then hope that analogous information theoretic perspective will provide similar monotone functions for higher-dimensional field theories (at least in five and six dimensions where superconformal critical points exist).

Entanglement and complexity also play an important role in the classification of IR fixed points of RG flows. In the context of many-body physics, both in condensed matter examples and engineered systems built from atoms, molecules, and photons,  
such fixed points describe distinct phases of matter. Classification of these phases is a major challenge in many-body physics, intimately tied to the classification of QFTs.  (See also the white papers \cite{Brauner:2022rvf,Cordova:2022ruw} for related reviews.)

\paragraph*{\bf Dynamics:}
The subject of dynamics may be viewed on several levels.  The most manifest one is simply understanding how specific systems evolve in time, and what universal features such time evolution exhibits.  A more primal level concerns the actual formulation of the quantities of interest in a generic time-dependent setting.  This is especially fruitful in the context of holography, where the notion of time differs between the two sides.  In particular, using general covariance of the bulk gravitational theory explicitly provides a useful guiding principle to developing the holographic dictionary by suggesting natural geometric quantities dual to specific QI or CFT quantities in the boundary theory.
At a deeper level, it is intriguing to observe that the formulation of the typical QI quantities explored in holography, defined at a single instant in time, does not manifest the dual bulk general covariance.  Understanding this symmetry in the boundary formulation has been relatively under-explored, and we may expect to see the insights gained from harnessing the full power of the gauge/gravity duality come to fruition in the next decade. 

Under time evolution, quantum information tends to spread across many degrees of freedom. It spreads both spatially, as signals propagate locally through the system, and internally, among the degrees of freedom at each location. Over the last decade we have come to better understand the remarkable fact that the spreading of quantum information obeys universal laws. In a wide variety of complex systems, including condensed matter, strongly interacting QFTs, and black holes, information spreads according to general principles and subject to fundamental bounds imposed by locality and by information-theoretic inequalities. These bounds play a key role in questions such as how locality emerges in quantum gravity and how to characterize the dynamics of strongly correlated materials.

Another important feature is scrambling, a dynamical process which effectively randomizes the quantum state, which is ubiquitous in chaotic quantum systems with many degrees of freedom, from the SYK model to black holes and large-$N$ CFTs \cite{Sekino:2008he}.  A far-reaching new perspective on scrambling, developed over the last decade, reformulates it in terms of out-of-time-order correlation functions (OTOCs) \cite{Shenker:2013pqa}, which measure the effect of small initial time perturbations on later time operators, and hence can be used to probe the onset of chaos.  This in turn carries connections to a wide range of topics such as emergence of bulk causality in AdS/CFT or bounds on CFTs from conformal bootstrap. 

Vacuum-subtracted energy density in quantum field theory can be negative due to quantum fluctuations. This negative energy can potentially give rise to acausal, or otherwise pathological, gravitational dynamics when coupling the QFT to gravity. For example, traversable wormholes might provide shortcuts between distant points. The Hawking black hole area theorem and Penrose singularity theorems rely on assumptions about non-existence of various forms of negative energy  \cite{Hawking:1973uf}. 
It is thus important to find general constraints on such negative energy. Surprisingly, these constraints have been shown to arise from quantum information considerations applied directly to the quantum field theory without gravity. 
For example, the positivity of relative entropy implies the Bekenstein bound \cite{Casini:2008cr}, the monotonicity of relative entropy gives the averaged null energy condition (ANEC) \cite{Faulkner:2016mzt}, the algebraic approach to quantum information provides a proof for the quantum null energy condition (QNEC \cite{Bousso:2015wca}), etc..
This story thus connects to the broader paradigm of gravity from quantum information. 

The interaction of a system with an external environment (whose unknown details are  traced over) can be described by a quantum channel acting on an open quantum system (naturally described by a density matrix).  Since such situations are ubiquitous in physics, it is of broad interest to determine the general rules to construct effective field theories for these open quantum systems.
One important problem is to ascertain sufficient conditions for a local effective field theory to emerge. For perturbative dynamics, this is difficult, since locality relies on the system losing memory of its interaction with the environment, 
whereas the relaxation timescale is long at weak coupling.  However holographic systems, where the dynamics is intrinsically strongly coupled, are fast scrambling and maximally ergodic in their dynamics, leading to a simple dynamics of probe effective field theories.  
Building on developments in real-time AdS/CFT 
\cite{Son:2002sd,Glorioso:2018mmw}, 
there has been renewed interest and progress in this subject. 

The concept of circuit complexity has also found its way into the holographic context over the last decade, catalyzed by the observation that its linear growth under time evolution mimics the growth of the Einstein-Rosen bridge in a two-sided black hole in the bulk.  Several specific geometrical duals of complexity have been proposed \cite{Stanford:2014jda,Brown:2015bva}
and substantiated by a tensor-network picture.  
A different set of ideas connecting holography to a notion of complexity is path-integral optimization, in which the bulk spacetime emerges from minimizing the number of operations required to prepare the state using a Euclidean path integral \cite{Caputa:2017urj}. 
While each of these proposals does seem to capture important qualitative features of holographic dualities, it remains to be seen how they are related to each other and in what regime (if any) they are correct. The idea that circuit complexity has a simple bulk dual in holographic theories has also led to a large amount of work attempting to define and quantify this notion in general field theories, including free ones, as well as in quantum mechanics.

One of the most notorious open problems in theoretical physics is the black hole information paradox, highlighting the clash between general relativity and quantum physics, specifically the thermality of Hawking radiation being incompatible with the unitarity of quantum evolution.  
This problem provides an excellent testing ground for putative formulations of quantum gravity, and over the last decade, many of the above-mentioned ideas have revitalized efforts in this direction, engendering significant progress and change of perspective.  This story is the focus of the white paper \cite{Bousso:2022ntt} and the earlier pedagogical review \cite{Almheiri:2020cfm}, explaining how using holographic entanglement (in particular the QES prescription) one can recover the Page curve for Hawking evaporation consistent with unitarity, and the key role played by spacetime wormholes. 
These can be studied in a controlled way in low dimensional settings such as the SYK model and its low-energy limit of JT gravity, though the story is much less clear for higher-dimensional gravity with propagating gravitational degrees of freedom.
A complementary perspective is offered in the white paper \cite{Giddings:2022jda} discussing quantum gravity more broadly. This work emphasizes that gravity requires modifications of the traditional notion of locality (which hinges on understanding the correct mathematical structure to define a subsystem in quantum gravity).  Further remarks on UV/IR mixing in quantum gravity and it possible signatures at low energies can be found in the white paper \cite{Berglund:2022qcc}.
A distinct viewpoint is presented by the approaches of the fuzzball and  microstate geometry programs, reviewed in the white paper \cite{Bena:2022ldq}, which give gravitational and quantum description of horizon-scale microstructure in terms of horizonless objects in string theory.  
The vibrant debate revolving around the black hole information paradox underscores the utility of exploring diverse approaches, in a healthy symbiosis though which we bootstrap our way towards more complete understanding.

\subsection{Outlook}

Despite the remarkable progress in our understanding of quantum information in fundamental physics, many important questions remain to be addressed.  Fortuitously there is high hope of progress on these over the next few years.

\paragraph*{\bf What is QFT?:} Notwithstanding its celebrated success as a framework for understanding the fundamental principles of nature, it is fair to say that we don't yet fully understand the nature of QFT \cite{Dedushenko:2022zwd}. 
In addition to the growing role of information-theoretic methods applied to QFT 
\cite{Faulkner:2022mlp}, the past decade has also seen new geometric approaches to computing observables such as  scattering amplitudes \cite{Bourjaily:2022bwx}, further development of  non-perturbative bootstrap techniques \cite{Poland:2022qrs,Hartman:2022zik}, a better understanding of symmetries and charges \cite{Cordova:2022ruw},  and ever deeper connections to many areas of mathematics \cite{Bah:2022xfv}. 
All of these developments have in different ways lent insights far transcending the standard textbook treatments of QFT. One natural question is: Which of these ingredients is a defining feature of the theory? What is the canonical toolkit of  a future quantum field theorist, as presented in a textbook a couple of decades hence? While these questions might be intangible at present, part of the progress will likely come from asking questions about the interconnections among these developments. Some of these are easier to fathom, being already somewhat developed, such as the connection between symmetry charges and operator algebraic formulations of QFT. On the other hand, while we have several different proofs of quantities that are RG monotones ($c$, $F$, and $a$ theorems), as yet we still lack a unified picture.  Nevertheless, understanding how the information theoretic data in QFTs relates to other observables, and developing further techniques to extract them, should provide valuable insight towards this goal.

\paragraph*{\bf Gauge invariance and information:}  In theories with gauge-invariant degrees of freedom, the decomposition of algebras comes with non-trivial centers and related superselection sectors. Several open questions remain in this context. What is the significance of algebraic centers for lattice gauge theories in the continuum limit?  
What is the role of topological entanglement in gapless theories? How do we describe gauge theory entanglement in the algebraic approach? What of interacting theories in the algebraic approach? Does the change in the nature of the von Neumann algebras, noted in the case of large $N$ confinement-deconfinement transition \cite{Witten:2021unn}, lead to new insights into the dynamics of confining gauge theories?

\paragraph*{\bf From fields to gravity and strings:} In the gravitational context, at the semiclassical level we now understand the relevance of the generalized entropy, an object that combines a Bekenstein-Hawking-like classical term with a quantum von Neumann entropy. It has played a crucial role both in the development of gravitational entropy bounds, and in the recent discussions of the black hole information paradox \cite{Bousso:2022ntt}.  An open problem is to better understand this quantity beyond the semiclassical regime. How does one pick subregions and define their algebras in this setting in a relational manner, maintaining diffeomorphism invariance? How does one  define these quantities in string theory? 
Can the information theory approach shed light on the question of which effective field theories can be consistently completed into a quantum theory of gravity, and which lie in the swampland? In the context of holography, it is interesting to ask about the information-theoretic nature of QFT wavefunctions. In particular, which aspects of their entanglement structure are central for the holographic entropy inequalities, and what does this tell us about quantum gravity?  Is there a principled way to discern which of the various conjectures relating to complexity are valid, and what  they are telling us about the nature of quantum gravitational wavefunctions?

\paragraph*{\bf Real-time dynamics and cosmology:}  QFT dynamics in cosmological spacetimes shares many characteristics of open quantum systems. To date, there isn't a clean description of the effective dynamics of quantum fields in an open system, and many questions remain to be addressed. Progress in these directions should help us better formulate the issues we need to confront in the cosmological context, such as the nature of observables or the temporal evolution of von Neumann entropy and other  information-theoretic quantities. It would also be interesting to tie these to the study of cosmological correlators, which has been tackled using bootstrap methods. 

\paragraph*{\bf Connections:}
While the list of interesting, important, and timely questions of course far exceeds the above sample, it is already evident that the most fruitful progress is rooted in their interconnection.  Physical systems which were traditionally far beyond the remit of HEP are gradually recognized to underlie deeper understanding.   The encouraging trend of diffusing topical boundaries is likely to accelerate further progress, which often comes from unexpected directions.  
The next decade might well be one of the most exciting ones in theoretical physics.

\begin{figure}{t}
    \centering
    \includegraphics[width=\columnwidth]{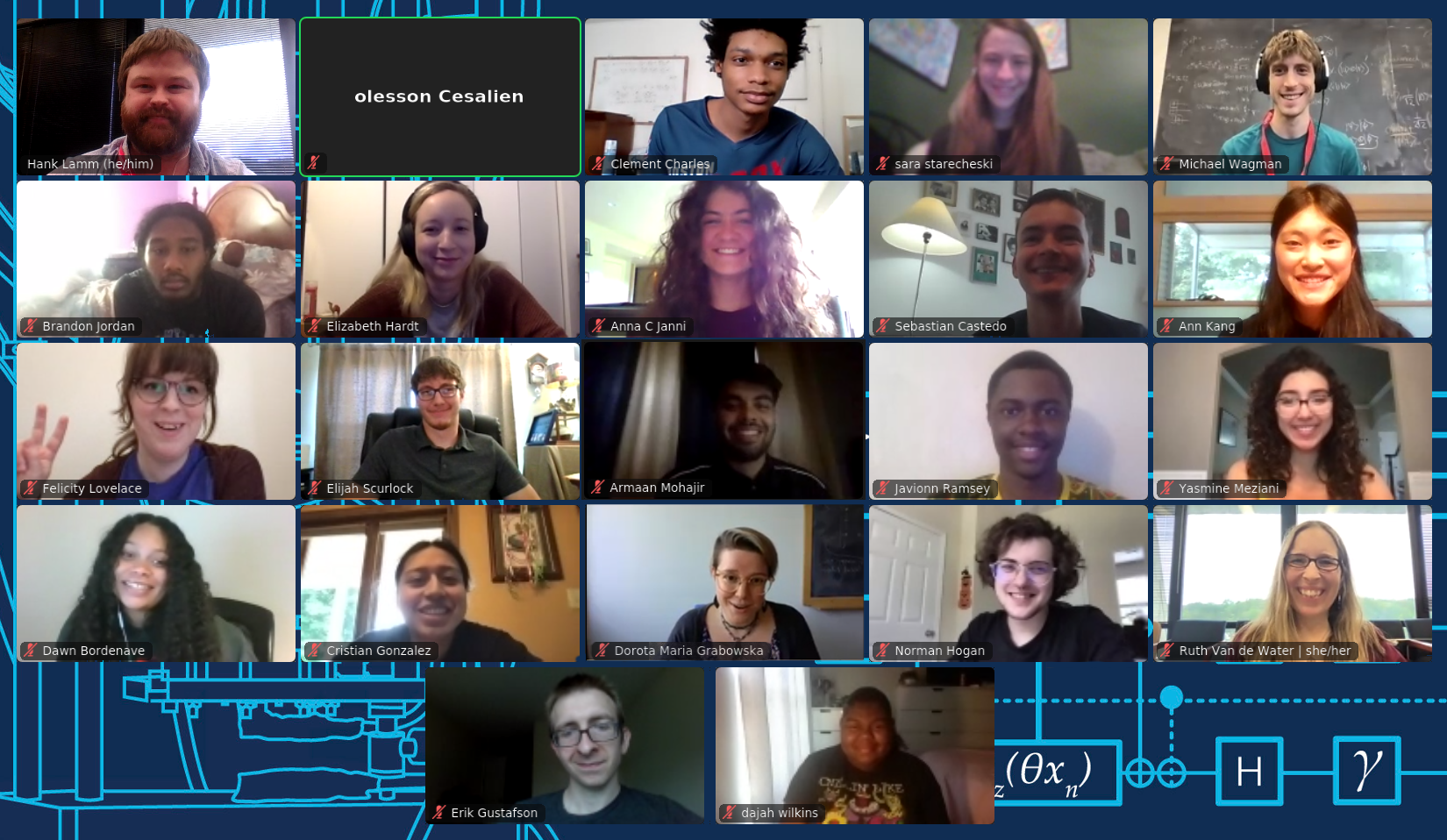}
    \caption{Students and organizers of the first Quantum Computing Internship for Physics Undergraduates Program (QCIPU) organized at Fermilab in the summer of 2021. }
    \label{fig:qcipu}
\end{figure}

\section{Quantum Workforce Development}

As quantum information science is becoming an industry, there is an anticipated need to develop a workforce that is educated in the physics, language, and tools of QIS~\cite{Hughes:2021lyh}. Emphasizing the need for diversity in the workforce while the field is in its infancy is important from the standpoint of equity, as well as to enrich the field with a plethora of viewpoints. In light of this theorists have taken initiatives that target students at a variety of levels. Examples include
\begin{itemize}
    \item References~\cite{Perry:2019bqg} is a textbook and/or course-module on quantum computing tailored to advanced high school students or physics undergraduates authored by a collaboration of HEP theorists and high school teachers brought together by a science teacher internship at Fermilab. The course has been successful in classroom setting, as described in~\cite{Hughes:2020ngh}. 
    \item The Quantum Computing Internship for Physics Undergraduates Program (QCIPU) is a summer school and internship program which provides training in QIS for undergraduate students, with an emphasis on physics sophomores and juniors from under-represented groups. The program
    is run and organized by Fermilab theorists, taking place virtually 
    (see Figure~\ref{fig:qcipu} and 
    \href{https://indico.fnal.gov/event/54760/}\href{https://indico.fnal.gov/event/54760/})
    in the past two summers, with plans for in-person programs in the future. Several students from the 2021 program have continued to work on research projects at Fermilab as part of the internship.
    \item At the graduate level and beyond, the SQMS center has organized two GGI summer schools, one (virtual) dedicated to Quantum sensing for fundamental physics, and a second (in person) on quantum simulation of QFT.
\end{itemize}
The continued success and further developments of programs such as QCIPU, especially as we leave the zoom era, will require increased community support, and devotion of effort and funding.

\section{Acknowledgements}

We thank the authors of snowmass white papers that were submitted to TF10. 
 RH is much obliged to Ying-Ying Li and Wanqiang Liu for useful discussions.
 Fermilab is operated by the Fermi Research Alliance, LLC under contract No. DE-AC02-07CH11359 with the United States Department of Energy.
The work of RH is also  supported by the U.S. Department of Energy, Office of Science, National Quantum Information Science Research Centers, Superconducting Quantum Materials and Systems Center (SQMS) under the contract No. DE-AC02-07CH11359.
VH has been supported by the U.S.\ Department of Energy grant DE-SC0009999. S.C has been
supported under U.S.\ Department of Energy grants DE-SC0009998 and DE-SC0019139. ZD is supported the U.S.\ Department of Energy grant DE-SC0021143, U.S.\ Department of Energy's Quantum Computing Application Teams program under fieldwork proposal number ERKJ347, and by the National Science Foundation QLCI grant OMA-2120757.

\bibliography{bibliography.bib}

\begin{thebibliography}{372}%
\makeatletter
\providecommand \@ifxundefined [1]{%
 \@ifx{#1\undefined}
}%
\providecommand \@ifnum [1]{%
 \ifnum #1\expandafter \@firstoftwo
 \else \expandafter \@secondoftwo
 \fi
}%
\providecommand \@ifx [1]{%
 \ifx #1\expandafter \@firstoftwo
 \else \expandafter \@secondoftwo
 \fi
}%
\providecommand \natexlab [1]{#1}%
\providecommand \enquote  [1]{``#1''}%
\providecommand \bibnamefont  [1]{#1}%
\providecommand \bibfnamefont [1]{#1}%
\providecommand \citenamefont [1]{#1}%
\providecommand \href@noop [0]{\@secondoftwo}%
\providecommand \href [0]{\begingroup \@sanitize@url \@href}%
\providecommand \@href[1]{\@@startlink{#1}\@@href}%
\providecommand \@@href[1]{\endgroup#1\@@endlink}%
\providecommand \@sanitize@url [0]{\catcode `\\12\catcode `\$12\catcode
  `\&12\catcode `\#12\catcode `\^12\catcode `\_12\catcode `\%12\relax}%
\providecommand \@@startlink[1]{}%
\providecommand \@@endlink[0]{}%
\providecommand \url  [0]{\begingroup\@sanitize@url \@url }%
\providecommand \@url [1]{\endgroup\@href {#1}{\urlprefix }}%
\providecommand \urlprefix  [0]{URL }%
\providecommand \Eprint [0]{\href }%
\providecommand \doibase [0]{http://dx.doi.org/}%
\providecommand \selectlanguage [0]{\@gobble}%
\providecommand \bibinfo  [0]{\@secondoftwo}%
\providecommand \bibfield  [0]{\@secondoftwo}%
\providecommand \translation [1]{[#1]}%
\providecommand \BibitemOpen [0]{}%
\providecommand \bibitemStop [0]{}%
\providecommand \bibitemNoStop [0]{.\EOS\space}%
\providecommand \EOS [0]{\spacefactor3000\relax}%
\providecommand \BibitemShut  [1]{\csname bibitem#1\endcsname}%
\let\auto@bib@innerbib\@empty
\bibitem [{\citenamefont {Feynman}(1982)}]{Feynman:1981tf}%
  \BibitemOpen
  \bibfield  {author} {\bibinfo {author} {\bibfnamefont {Richard~P.}\
  \bibnamefont {Feynman}},\ }\bibfield  {title} {\enquote {\bibinfo {title}
  {{Simulating physics with computers}},}\ }\href {\doibase 10.1007/BF02650179}
  {\bibfield  {journal} {\bibinfo  {journal} {Int. J. Theor. Phys.}\ }\textbf
  {\bibinfo {volume} {21}},\ \bibinfo {pages} {467--488} (\bibinfo {year}
  {1982})}\BibitemShut {NoStop}%
\bibitem [{\citenamefont {Jordan}\ \emph {et~al.}(2018)\citenamefont {Jordan},
  \citenamefont {Krovi}, \citenamefont {Lee},\ and\ \citenamefont
  {Preskill}}]{Jordan:2017lea}%
  \BibitemOpen
  \bibfield  {author} {\bibinfo {author} {\bibfnamefont {Stephen~P.}\
  \bibnamefont {Jordan}}, \bibinfo {author} {\bibfnamefont {Hari}\ \bibnamefont
  {Krovi}}, \bibinfo {author} {\bibfnamefont {Keith~S.M.}\ \bibnamefont {Lee}},
  \ and\ \bibinfo {author} {\bibfnamefont {John}\ \bibnamefont {Preskill}},\
  }\bibfield  {title} {\enquote {\bibinfo {title} {{BQP-completeness of
  Scattering in Scalar Quantum Field Theory}},}\ }\href {\doibase
  10.22331/q-2018-01-08-44} {\bibfield  {journal} {\bibinfo  {journal}
  {Quantum}\ }\textbf {\bibinfo {volume} {2}},\ \bibinfo {pages} {44} (\bibinfo
  {year} {2018})},\ \Eprint {http://arxiv.org/abs/1703.00454} {arXiv:1703.00454
  [quant-ph]} \BibitemShut {NoStop}%
\bibitem [{\citenamefont {Carena}\ \emph {et~al.}(2020)\citenamefont {Carena},
  \citenamefont {Lamm}, \citenamefont {Li}, \citenamefont {Lykken},
  \citenamefont {Wang},\ and\ \citenamefont {Yamauchi}}]{pqa_loi}%
  \BibitemOpen
  \bibfield  {author} {\bibinfo {author} {\bibfnamefont {Marcela}\ \bibnamefont
  {Carena}}, \bibinfo {author} {\bibfnamefont {Henry}\ \bibnamefont {Lamm}},
  \bibinfo {author} {\bibfnamefont {Ying-Ying}\ \bibnamefont {Li}}, \bibinfo
  {author} {\bibfnamefont {Joseph~D.}\ \bibnamefont {Lykken}}, \bibinfo
  {author} {\bibfnamefont {Lian-Tao}\ \bibnamefont {Wang}}, \ and\ \bibinfo
  {author} {\bibfnamefont {Yukari}\ \bibnamefont {Yamauchi}},\ }\bibfield
  {title} {\enquote {\bibinfo {title} {{Practical Quantum Advantages in High
  Energy Physics}},}\ }\href@noop {} {\bibfield  {journal} {\bibinfo  {journal}
  {\href{https://www.snowmass21.org/docs/files/summaries/TF/SNOWMASS21-TF10_TF0-CompF6_CompF0_Hank_Lamm-077.pdf}{Snowmass
  2021 LOI}}\ }\textbf {\bibinfo {volume} {TF10-077}} (\bibinfo {year}
  {2020})}\BibitemShut {NoStop}%
\bibitem [{\citenamefont {Bauer}\ \emph {et~al.}(2022)\citenamefont {Bauer},
  \citenamefont {Davoudi} \emph {et~al.}}]{Bauer:2022hpo}%
  \BibitemOpen
  \bibfield  {author} {\bibinfo {author} {\bibfnamefont {Christian~W.}\
  \bibnamefont {Bauer}}, \bibinfo {author} {\bibfnamefont {Zohreh}\
  \bibnamefont {Davoudi}},  \emph {et~al.},\ }\bibfield  {title} {\enquote
  {\bibinfo {title} {{Quantum Simulation for High Energy Physics}},}\
  }\href@noop {} {\  (\bibinfo {year} {2022})},\ \Eprint
  {http://arxiv.org/abs/2204.03381} {arXiv:2204.03381 [quant-ph]} \BibitemShut
  {NoStop}%
\bibitem [{\citenamefont {Jordan}\ \emph {et~al.}(2012)\citenamefont {Jordan},
  \citenamefont {Lee},\ and\ \citenamefont {Preskill}}]{Jordan:2011ne}%
  \BibitemOpen
  \bibfield  {author} {\bibinfo {author} {\bibfnamefont {Stephen~P.}\
  \bibnamefont {Jordan}}, \bibinfo {author} {\bibfnamefont {Keith S.~M.}\
  \bibnamefont {Lee}}, \ and\ \bibinfo {author} {\bibfnamefont {John}\
  \bibnamefont {Preskill}},\ }\bibfield  {title} {\enquote {\bibinfo {title}
  {{Quantum Algorithms for Quantum Field Theories}},}\ }\href {\doibase
  10.1126/science.1217069} {\bibfield  {journal} {\bibinfo  {journal}
  {Science}\ }\textbf {\bibinfo {volume} {336}},\ \bibinfo {pages} {1130--1133}
  (\bibinfo {year} {2012})},\ \Eprint {http://arxiv.org/abs/1111.3633}
  {arXiv:1111.3633 [quant-ph]} \BibitemShut {NoStop}%
\bibitem [{\citenamefont {Bauer}\ \emph
  {et~al.}(2021{\natexlab{a}})\citenamefont {Bauer}, \citenamefont {Freytsis},\
  and\ \citenamefont {Nachman}}]{Bauer:2021gup}%
  \BibitemOpen
  \bibfield  {author} {\bibinfo {author} {\bibfnamefont {Christian~W.}\
  \bibnamefont {Bauer}}, \bibinfo {author} {\bibfnamefont {Marat}\ \bibnamefont
  {Freytsis}}, \ and\ \bibinfo {author} {\bibfnamefont {Benjamin}\ \bibnamefont
  {Nachman}},\ }\bibfield  {title} {\enquote {\bibinfo {title} {{Simulating
  Collider Physics on Quantum Computers Using Effective Field Theories}},}\
  }\href {\doibase 10.1103/PhysRevLett.127.212001} {\bibfield  {journal}
  {\bibinfo  {journal} {Phys. Rev. Lett.}\ }\textbf {\bibinfo {volume} {127}},\
  \bibinfo {pages} {212001} (\bibinfo {year} {2021}{\natexlab{a}})},\ \Eprint
  {http://arxiv.org/abs/2102.05044} {arXiv:2102.05044 [hep-ph]} \BibitemShut
  {NoStop}%
\bibitem [{\citenamefont {Constantinou}\ \emph {et~al.}(2022)\citenamefont
  {Constantinou} \emph {et~al.}}]{Constantinou:2022yye}%
  \BibitemOpen
  \bibfield  {author} {\bibinfo {author} {\bibfnamefont {Martha}\ \bibnamefont
  {Constantinou}} \emph {et~al.},\ }\bibfield  {title} {\enquote {\bibinfo
  {title} {{Lattice QCD Calculations of Parton Physics}},}\ }\href@noop {} {\
  (\bibinfo {year} {2022})},\ \Eprint {http://arxiv.org/abs/2202.07193}
  {arXiv:2202.07193 [hep-lat]} \BibitemShut {NoStop}%
\bibitem [{\citenamefont {Lamm}\ \emph
  {et~al.}(2020{\natexlab{a}})\citenamefont {Lamm}, \citenamefont {Lawrence},\
  and\ \citenamefont {Yamauchi}}]{Lamm:2019uyc}%
  \BibitemOpen
  \bibfield  {author} {\bibinfo {author} {\bibfnamefont {Henry}\ \bibnamefont
  {Lamm}}, \bibinfo {author} {\bibfnamefont {Scott}\ \bibnamefont {Lawrence}},
  \ and\ \bibinfo {author} {\bibfnamefont {Yukari}\ \bibnamefont {Yamauchi}}
  (\bibinfo {collaboration} {NuQS}),\ }\bibfield  {title} {\enquote {\bibinfo
  {title} {{Parton physics on a quantum computer}},}\ }\href {\doibase
  10.1103/PhysRevResearch.2.013272} {\bibfield  {journal} {\bibinfo  {journal}
  {Phys. Rev. Res.}\ }\textbf {\bibinfo {volume} {2}},\ \bibinfo {pages}
  {013272} (\bibinfo {year} {2020}{\natexlab{a}})},\ \Eprint
  {http://arxiv.org/abs/1908.10439} {arXiv:1908.10439 [hep-lat]} \BibitemShut
  {NoStop}%
\bibitem [{\citenamefont {Echevarria}\ \emph {et~al.}(2021)\citenamefont
  {Echevarria}, \citenamefont {Egusquiza}, \citenamefont {Rico},\ and\
  \citenamefont {Schnell}}]{echevarria2021quantum}%
  \BibitemOpen
  \bibfield  {author} {\bibinfo {author} {\bibfnamefont {MG}~\bibnamefont
  {Echevarria}}, \bibinfo {author} {\bibfnamefont {IL}~\bibnamefont
  {Egusquiza}}, \bibinfo {author} {\bibfnamefont {E}~\bibnamefont {Rico}}, \
  and\ \bibinfo {author} {\bibfnamefont {G}~\bibnamefont {Schnell}},\
  }\bibfield  {title} {\enquote {\bibinfo {title} {Quantum simulation of
  light-front parton correlators},}\ }\href@noop {} {\bibfield  {journal}
  {\bibinfo  {journal} {Physical Review D}\ }\textbf {\bibinfo {volume}
  {104}},\ \bibinfo {pages} {014512} (\bibinfo {year} {2021})}\BibitemShut
  {NoStop}%
\bibitem [{\citenamefont {Li}\ \emph {et~al.}(2021)\citenamefont {Li},
  \citenamefont {Guo}, \citenamefont {Lai}, \citenamefont {Liu}, \citenamefont
  {Wang}, \citenamefont {Xing}, \citenamefont {Zhang},\ and\ \citenamefont
  {Zhu}}]{li2021partonic}%
  \BibitemOpen
  \bibfield  {author} {\bibinfo {author} {\bibfnamefont {Tianyin}\ \bibnamefont
  {Li}}, \bibinfo {author} {\bibfnamefont {Xingyu}\ \bibnamefont {Guo}},
  \bibinfo {author} {\bibfnamefont {Wai~Kin}\ \bibnamefont {Lai}}, \bibinfo
  {author} {\bibfnamefont {Xiaohui}\ \bibnamefont {Liu}}, \bibinfo {author}
  {\bibfnamefont {Enke}\ \bibnamefont {Wang}}, \bibinfo {author} {\bibfnamefont
  {Hongxi}\ \bibnamefont {Xing}}, \bibinfo {author} {\bibfnamefont {Dan-Bo}\
  \bibnamefont {Zhang}}, \ and\ \bibinfo {author} {\bibfnamefont {Shi-Liang}\
  \bibnamefont {Zhu}},\ }\bibfield  {title} {\enquote {\bibinfo {title}
  {Partonic structure by quantum computing},}\ }\href@noop {} {\bibfield
  {journal} {\bibinfo  {journal} {arXiv preprint arXiv:2106.03865}\ } (\bibinfo
  {year} {2021})}\BibitemShut {NoStop}%
\bibitem [{\citenamefont {Mueller}\ \emph {et~al.}(2020)\citenamefont
  {Mueller}, \citenamefont {Tarasov},\ and\ \citenamefont
  {Venugopalan}}]{mueller2020deeply}%
  \BibitemOpen
  \bibfield  {author} {\bibinfo {author} {\bibfnamefont {Niklas}\ \bibnamefont
  {Mueller}}, \bibinfo {author} {\bibfnamefont {Andrey}\ \bibnamefont
  {Tarasov}}, \ and\ \bibinfo {author} {\bibfnamefont {Raju}\ \bibnamefont
  {Venugopalan}},\ }\bibfield  {title} {\enquote {\bibinfo {title} {Deeply
  inelastic scattering structure functions on a hybrid quantum computer},}\
  }\href@noop {} {\bibfield  {journal} {\bibinfo  {journal} {Physical Review
  D}\ }\textbf {\bibinfo {volume} {102}},\ \bibinfo {pages} {016007} (\bibinfo
  {year} {2020})}\BibitemShut {NoStop}%
\bibitem [{\citenamefont {Qian}\ \emph {et~al.}(2021)\citenamefont {Qian},
  \citenamefont {Basili}, \citenamefont {Pal}, \citenamefont {Luecke},\ and\
  \citenamefont {Vary}}]{qian2021solving}%
  \BibitemOpen
  \bibfield  {author} {\bibinfo {author} {\bibfnamefont {Wenyang}\ \bibnamefont
  {Qian}}, \bibinfo {author} {\bibfnamefont {Robert}\ \bibnamefont {Basili}},
  \bibinfo {author} {\bibfnamefont {Soham}\ \bibnamefont {Pal}}, \bibinfo
  {author} {\bibfnamefont {Glenn}\ \bibnamefont {Luecke}}, \ and\ \bibinfo
  {author} {\bibfnamefont {James~P}\ \bibnamefont {Vary}},\ }\bibfield  {title}
  {\enquote {\bibinfo {title} {Solving hadron structures with variational
  quantum eigensolvers},}\ }\href@noop {} {\bibfield  {journal} {\bibinfo
  {journal} {arXiv preprint arXiv:2112.01927}\ } (\bibinfo {year}
  {2021})}\BibitemShut {NoStop}%
\bibitem [{\citenamefont {Bauer}\ and\ \citenamefont
  {Schwartz}(2007)}]{Bauer:2006mk}%
  \BibitemOpen
  \bibfield  {author} {\bibinfo {author} {\bibfnamefont {Christian~W.}\
  \bibnamefont {Bauer}}\ and\ \bibinfo {author} {\bibfnamefont {Matthew~D.}\
  \bibnamefont {Schwartz}},\ }\bibfield  {title} {\enquote {\bibinfo {title}
  {{Event Generation from Effective Field Theory}},}\ }\href {\doibase
  10.1103/PhysRevD.76.074004} {\bibfield  {journal} {\bibinfo  {journal} {Phys.
  Rev. D}\ }\textbf {\bibinfo {volume} {76}},\ \bibinfo {pages} {074004}
  (\bibinfo {year} {2007})},\ \Eprint {http://arxiv.org/abs/hep-ph/0607296}
  {arXiv:hep-ph/0607296} \BibitemShut {NoStop}%
\bibitem [{\citenamefont {Bauer}\ \emph
  {et~al.}(2021{\natexlab{b}})\citenamefont {Bauer}, \citenamefont {de~Jong},
  \citenamefont {Nachman},\ and\ \citenamefont {Provasoli}}]{Bauer:2019qxa}%
  \BibitemOpen
  \bibfield  {author} {\bibinfo {author} {\bibfnamefont {Christian~W.}\
  \bibnamefont {Bauer}}, \bibinfo {author} {\bibfnamefont {Wibe~A.}\
  \bibnamefont {de~Jong}}, \bibinfo {author} {\bibfnamefont {Benjamin}\
  \bibnamefont {Nachman}}, \ and\ \bibinfo {author} {\bibfnamefont {Davide}\
  \bibnamefont {Provasoli}},\ }\bibfield  {title} {\enquote {\bibinfo {title}
  {{Quantum Algorithm for High Energy Physics Simulations}},}\ }\href {\doibase
  10.1103/PhysRevLett.126.062001} {\bibfield  {journal} {\bibinfo  {journal}
  {Phys. Rev. Lett.}\ }\textbf {\bibinfo {volume} {126}},\ \bibinfo {pages}
  {062001} (\bibinfo {year} {2021}{\natexlab{b}})},\ \Eprint
  {http://arxiv.org/abs/1904.03196} {arXiv:1904.03196 [hep-ph]} \BibitemShut
  {NoStop}%
\bibitem [{\citenamefont {Williams}\ \emph {et~al.}(2021)\citenamefont
  {Williams}, \citenamefont {Malik}, \citenamefont {Spannowsky},\ and\
  \citenamefont {Bepari}}]{williams2021quantum}%
  \BibitemOpen
  \bibfield  {author} {\bibinfo {author} {\bibfnamefont {Simon}\ \bibnamefont
  {Williams}}, \bibinfo {author} {\bibfnamefont {Sarah}\ \bibnamefont {Malik}},
  \bibinfo {author} {\bibfnamefont {Michael}\ \bibnamefont {Spannowsky}}, \
  and\ \bibinfo {author} {\bibfnamefont {Khadeejah}\ \bibnamefont {Bepari}},\
  }\bibfield  {title} {\enquote {\bibinfo {title} {A quantum walk approach to
  simulating parton showers},}\ }\href@noop {} {\bibfield  {journal} {\bibinfo
  {journal} {arXiv preprint arXiv:2109.13975}\ } (\bibinfo {year}
  {2021})}\BibitemShut {NoStop}%
\bibitem [{\citenamefont {Gustafson}\ \emph {et~al.}(2022)\citenamefont
  {Gustafson}, \citenamefont {Prestel}, \citenamefont {Spannowsky},\ and\
  \citenamefont {Williams}}]{gustafson2022collider}%
  \BibitemOpen
  \bibfield  {author} {\bibinfo {author} {\bibfnamefont {G{\"o}sta}\
  \bibnamefont {Gustafson}}, \bibinfo {author} {\bibfnamefont {Stefan}\
  \bibnamefont {Prestel}}, \bibinfo {author} {\bibfnamefont {Michael}\
  \bibnamefont {Spannowsky}}, \ and\ \bibinfo {author} {\bibfnamefont {Simon}\
  \bibnamefont {Williams}},\ }\bibfield  {title} {\enquote {\bibinfo {title}
  {Collider events on a quantum computer},}\ }\href@noop {} {\bibfield
  {journal} {\bibinfo  {journal} {arXiv preprint arXiv:2207.10694}\ } (\bibinfo
  {year} {2022})}\BibitemShut {NoStop}%
\bibitem [{\citenamefont {Zohar}\ \emph {et~al.}(2016)\citenamefont {Zohar},
  \citenamefont {Cirac},\ and\ \citenamefont {Reznik}}]{Zohar:2015hwa}%
  \BibitemOpen
  \bibfield  {author} {\bibinfo {author} {\bibfnamefont {Erez}\ \bibnamefont
  {Zohar}}, \bibinfo {author} {\bibfnamefont {J.~Ignacio}\ \bibnamefont
  {Cirac}}, \ and\ \bibinfo {author} {\bibfnamefont {Benni}\ \bibnamefont
  {Reznik}},\ }\bibfield  {title} {\enquote {\bibinfo {title} {{Quantum
  Simulations of Lattice Gauge Theories using Ultracold Atoms in Optical
  Lattices}},}\ }\href {\doibase 10.1088/0034-4885/79/1/014401} {\bibfield
  {journal} {\bibinfo  {journal} {Rept. Prog. Phys.}\ }\textbf {\bibinfo
  {volume} {79}},\ \bibinfo {pages} {014401} (\bibinfo {year} {2016})},\
  \Eprint {http://arxiv.org/abs/1503.02312} {arXiv:1503.02312 [quant-ph]}
  \BibitemShut {NoStop}%
\bibitem [{\citenamefont {Bermudez}\ \emph {et~al.}(2010)\citenamefont
  {Bermudez}, \citenamefont {Mazza}, \citenamefont {Rizzi}, \citenamefont
  {Goldman}, \citenamefont {Lewenstein},\ and\ \citenamefont
  {Martin-Delgado}}]{Bermudez:2010da}%
  \BibitemOpen
  \bibfield  {author} {\bibinfo {author} {\bibfnamefont {A.}~\bibnamefont
  {Bermudez}}, \bibinfo {author} {\bibfnamefont {L.}~\bibnamefont {Mazza}},
  \bibinfo {author} {\bibfnamefont {M.}~\bibnamefont {Rizzi}}, \bibinfo
  {author} {\bibfnamefont {N.}~\bibnamefont {Goldman}}, \bibinfo {author}
  {\bibfnamefont {M.}~\bibnamefont {Lewenstein}}, \ and\ \bibinfo {author}
  {\bibfnamefont {M.~A.}\ \bibnamefont {Martin-Delgado}},\ }\bibfield  {title}
  {\enquote {\bibinfo {title} {{Wilson Fermions and Axion Electrodynamics in
  Optical Lattices}},}\ }\href {\doibase 10.1103/PhysRevLett.105.190404}
  {\bibfield  {journal} {\bibinfo  {journal} {Phys. Rev. Lett.}\ }\textbf
  {\bibinfo {volume} {105}},\ \bibinfo {pages} {190404} (\bibinfo {year}
  {2010})},\ \Eprint {http://arxiv.org/abs/1004.5101} {arXiv:1004.5101
  [cond-mat.quant-gas]} \BibitemShut {NoStop}%
\bibitem [{\citenamefont {Laflamme}\ \emph {et~al.}(2016)\citenamefont
  {Laflamme}, \citenamefont {Evans}, \citenamefont {Dalmonte}, \citenamefont
  {Gerber}, \citenamefont {Mej\'\i{}a-D\'\i{}az}, \citenamefont {Bietenholz},
  \citenamefont {Wiese},\ and\ \citenamefont {Zoller}}]{Laflamme:2015wma}%
  \BibitemOpen
  \bibfield  {author} {\bibinfo {author} {\bibfnamefont {C.}~\bibnamefont
  {Laflamme}}, \bibinfo {author} {\bibfnamefont {W.}~\bibnamefont {Evans}},
  \bibinfo {author} {\bibfnamefont {M.}~\bibnamefont {Dalmonte}}, \bibinfo
  {author} {\bibfnamefont {U.}~\bibnamefont {Gerber}}, \bibinfo {author}
  {\bibfnamefont {H.}~\bibnamefont {Mej\'\i{}a-D\'\i{}az}}, \bibinfo {author}
  {\bibfnamefont {W.}~\bibnamefont {Bietenholz}}, \bibinfo {author}
  {\bibfnamefont {U.~J.}\ \bibnamefont {Wiese}}, \ and\ \bibinfo {author}
  {\bibfnamefont {P.}~\bibnamefont {Zoller}},\ }\bibfield  {title} {\enquote
  {\bibinfo {title} {{$\mathbb{C}$P(N\ensuremath{-}1) quantum field theories
  with alkaline-earth atoms in optical lattices}},}\ }\href {\doibase
  10.1016/j.aop.2016.03.012} {\bibfield  {journal} {\bibinfo  {journal} {Annals
  Phys.}\ }\textbf {\bibinfo {volume} {370}},\ \bibinfo {pages} {117--127}
  (\bibinfo {year} {2016})},\ \Eprint {http://arxiv.org/abs/1507.06788}
  {arXiv:1507.06788 [quant-ph]} \BibitemShut {NoStop}%
\bibitem [{\citenamefont {Meurice}(2011)}]{Meurice:2011aji}%
  \BibitemOpen
  \bibfield  {author} {\bibinfo {author} {\bibfnamefont {Y.}~\bibnamefont
  {Meurice}},\ }\bibfield  {title} {\enquote {\bibinfo {title} {{QCD
  calculations with optical lattices?}}}\ }\href {\doibase 10.22323/1.139.0040}
  {\bibfield  {journal} {\bibinfo  {journal} {PoS}\ }\textbf {\bibinfo {volume}
  {LATTICE2011}},\ \bibinfo {pages} {040} (\bibinfo {year} {2011})},\ \Eprint
  {http://arxiv.org/abs/1202.1605} {arXiv:1202.1605 [hep-lat]} \BibitemShut
  {NoStop}%
\bibitem [{\citenamefont {Bazavov}\ \emph {et~al.}(2015)\citenamefont
  {Bazavov}, \citenamefont {Meurice}, \citenamefont {Tsai}, \citenamefont
  {Unmuth-Yockey},\ and\ \citenamefont {Zhang}}]{Bazavov:2015kka}%
  \BibitemOpen
  \bibfield  {author} {\bibinfo {author} {\bibfnamefont {Alexei}\ \bibnamefont
  {Bazavov}}, \bibinfo {author} {\bibfnamefont {Yannick}\ \bibnamefont
  {Meurice}}, \bibinfo {author} {\bibfnamefont {Shan-Wen}\ \bibnamefont
  {Tsai}}, \bibinfo {author} {\bibfnamefont {Judah}\ \bibnamefont
  {Unmuth-Yockey}}, \ and\ \bibinfo {author} {\bibfnamefont {Jin}\ \bibnamefont
  {Zhang}},\ }\bibfield  {title} {\enquote {\bibinfo {title} {{Gauge-invariant
  implementation of the Abelian Higgs model on optical lattices}},}\ }\href
  {\doibase 10.1103/PhysRevD.92.076003} {\bibfield  {journal} {\bibinfo
  {journal} {Phys. Rev. D}\ }\textbf {\bibinfo {volume} {92}},\ \bibinfo
  {pages} {076003} (\bibinfo {year} {2015})},\ \Eprint
  {http://arxiv.org/abs/1503.08354} {arXiv:1503.08354 [hep-lat]} \BibitemShut
  {NoStop}%
\bibitem [{\citenamefont {Tagliacozzo}\ \emph
  {et~al.}(2013{\natexlab{a}})\citenamefont {Tagliacozzo}, \citenamefont
  {Celi}, \citenamefont {Orland},\ and\ \citenamefont
  {Lewenstein}}]{Tagliacozzo:2012df}%
  \BibitemOpen
  \bibfield  {author} {\bibinfo {author} {\bibfnamefont {L.}~\bibnamefont
  {Tagliacozzo}}, \bibinfo {author} {\bibfnamefont {A.}~\bibnamefont {Celi}},
  \bibinfo {author} {\bibfnamefont {P.}~\bibnamefont {Orland}}, \ and\ \bibinfo
  {author} {\bibfnamefont {M.}~\bibnamefont {Lewenstein}},\ }\bibfield  {title}
  {\enquote {\bibinfo {title} {{Simulations of non-Abelian gauge theories with
  optical lattices}},}\ }\href {\doibase 10.1038/ncomms3615} {\bibfield
  {journal} {\bibinfo  {journal} {Nature Commun.}\ }\textbf {\bibinfo {volume}
  {4}},\ \bibinfo {pages} {2615} (\bibinfo {year} {2013}{\natexlab{a}})},\
  \Eprint {http://arxiv.org/abs/1211.2704} {arXiv:1211.2704
  [cond-mat.quant-gas]} \BibitemShut {NoStop}%
\bibitem [{\citenamefont {Davoudi}\ \emph
  {et~al.}(2020{\natexlab{a}})\citenamefont {Davoudi}, \citenamefont {Hafezi},
  \citenamefont {Monroe}, \citenamefont {Pagano}, \citenamefont {Seif},\ and\
  \citenamefont {Shaw}}]{Davoudi:2019bhy}%
  \BibitemOpen
  \bibfield  {author} {\bibinfo {author} {\bibfnamefont {Zohreh}\ \bibnamefont
  {Davoudi}}, \bibinfo {author} {\bibfnamefont {Mohammad}\ \bibnamefont
  {Hafezi}}, \bibinfo {author} {\bibfnamefont {Christopher}\ \bibnamefont
  {Monroe}}, \bibinfo {author} {\bibfnamefont {Guido}\ \bibnamefont {Pagano}},
  \bibinfo {author} {\bibfnamefont {Alireza}\ \bibnamefont {Seif}}, \ and\
  \bibinfo {author} {\bibfnamefont {Andrew}\ \bibnamefont {Shaw}},\ }\bibfield
  {title} {\enquote {\bibinfo {title} {{Towards analog quantum simulations of
  lattice gauge theories with trapped ions}},}\ }\href {\doibase
  10.1103/PhysRevResearch.2.023015} {\bibfield  {journal} {\bibinfo  {journal}
  {Phys. Rev. Res.}\ }\textbf {\bibinfo {volume} {2}},\ \bibinfo {pages}
  {023015} (\bibinfo {year} {2020}{\natexlab{a}})},\ \Eprint
  {http://arxiv.org/abs/1908.03210} {arXiv:1908.03210 [quant-ph]} \BibitemShut
  {NoStop}%
\bibitem [{\citenamefont {Monroe}\ \emph {et~al.}(2021)\citenamefont {Monroe}
  \emph {et~al.}}]{Monroe:2019asq}%
  \BibitemOpen
  \bibfield  {author} {\bibinfo {author} {\bibfnamefont {C.}~\bibnamefont
  {Monroe}} \emph {et~al.},\ }\bibfield  {title} {\enquote {\bibinfo {title}
  {{Programmable quantum simulations of spin systems with trapped ions}},}\
  }\href {\doibase 10.1103/RevModPhys.93.025001} {\bibfield  {journal}
  {\bibinfo  {journal} {Rev. Mod. Phys.}\ }\textbf {\bibinfo {volume} {93}},\
  \bibinfo {pages} {025001} (\bibinfo {year} {2021})},\ \Eprint
  {http://arxiv.org/abs/1912.07845} {arXiv:1912.07845 [quant-ph]} \BibitemShut
  {NoStop}%
\bibitem [{\citenamefont {Nguyen}\ \emph {et~al.}(2022)\citenamefont {Nguyen},
  \citenamefont {Tran}, \citenamefont {Zhu}, \citenamefont {Green},
  \citenamefont {Alderete}, \citenamefont {Davoudi},\ and\ \citenamefont
  {Linke}}]{Nguyen:2021hyk}%
  \BibitemOpen
  \bibfield  {author} {\bibinfo {author} {\bibfnamefont {Nhung~H.}\
  \bibnamefont {Nguyen}}, \bibinfo {author} {\bibfnamefont {Minh~C.}\
  \bibnamefont {Tran}}, \bibinfo {author} {\bibfnamefont {Yingyue}\
  \bibnamefont {Zhu}}, \bibinfo {author} {\bibfnamefont {Alaina~M.}\
  \bibnamefont {Green}}, \bibinfo {author} {\bibfnamefont {C.~Huerta}\
  \bibnamefont {Alderete}}, \bibinfo {author} {\bibfnamefont {Zohreh}\
  \bibnamefont {Davoudi}}, \ and\ \bibinfo {author} {\bibfnamefont
  {Norbert~M.}\ \bibnamefont {Linke}},\ }\bibfield  {title} {\enquote {\bibinfo
  {title} {{Digital Quantum Simulation of the Schwinger Model and Symmetry
  Protection with Trapped Ions}},}\ }\href {\doibase
  10.1103/PRXQuantum.3.020324} {\bibfield  {journal} {\bibinfo  {journal} {PRX
  Quantum}\ }\textbf {\bibinfo {volume} {3}},\ \bibinfo {pages} {020324}
  (\bibinfo {year} {2022})},\ \Eprint {http://arxiv.org/abs/2112.14262}
  {arXiv:2112.14262 [quant-ph]} \BibitemShut {NoStop}%
\bibitem [{\citenamefont {Brooks}(2022)}]{Brooksetal}%
  \BibitemOpen
  \bibfield  {author} {\bibinfo {author} {\bibfnamefont {S~et~al.}\
  \bibnamefont {Brooks}},\ }\bibfield  {title} {\enquote {\bibinfo {title}
  {{Ion Coulomb crystals in storage rings for quantum information science}},}\
  }\href@noop {} {\  (\bibinfo {year} {2022})},\ \Eprint
  {http://arxiv.org/abs/2203.06809} {2203.06809 [physics.acc-ph]} \BibitemShut
  {NoStop}%
\bibitem [{\citenamefont {Martinez}\ \emph
  {et~al.}(2016{\natexlab{a}})\citenamefont {Martinez} \emph
  {et~al.}}]{Martinez:2016yna}%
  \BibitemOpen
  \bibfield  {author} {\bibinfo {author} {\bibfnamefont {E.~A.}\ \bibnamefont
  {Martinez}} \emph {et~al.},\ }\bibfield  {title} {\enquote {\bibinfo {title}
  {{Real-time dynamics of lattice gauge theories with a few-qubit quantum
  computer}},}\ }\href {\doibase 10.1038/nature18318} {\bibfield  {journal}
  {\bibinfo  {journal} {Nature}\ }\textbf {\bibinfo {volume} {534}},\ \bibinfo
  {pages} {516--519} (\bibinfo {year} {2016}{\natexlab{a}})},\ \Eprint
  {http://arxiv.org/abs/1605.04570} {arXiv:1605.04570 [quant-ph]} \BibitemShut
  {NoStop}%
\bibitem [{\citenamefont {Lamm}\ \emph {et~al.}(2019)\citenamefont {Lamm},
  \citenamefont {Lawrence},\ and\ \citenamefont {Yamauchi}}]{Lamm:2019bik}%
  \BibitemOpen
  \bibfield  {author} {\bibinfo {author} {\bibfnamefont {Henry}\ \bibnamefont
  {Lamm}}, \bibinfo {author} {\bibfnamefont {Scott}\ \bibnamefont {Lawrence}},
  \ and\ \bibinfo {author} {\bibfnamefont {Yukari}\ \bibnamefont {Yamauchi}}
  (\bibinfo {collaboration} {NuQS}),\ }\bibfield  {title} {\enquote {\bibinfo
  {title} {{General Methods for Digital Quantum Simulation of Gauge
  Theories}},}\ }\href {\doibase 10.1103/PhysRevD.100.034518} {\bibfield
  {journal} {\bibinfo  {journal} {Phys. Rev. D}\ }\textbf {\bibinfo {volume}
  {100}},\ \bibinfo {pages} {034518} (\bibinfo {year} {2019})},\ \Eprint
  {http://arxiv.org/abs/1903.08807} {arXiv:1903.08807 [hep-lat]} \BibitemShut
  {NoStop}%
\bibitem [{\citenamefont {Raychowdhury}\ and\ \citenamefont
  {Stryker}(2020{\natexlab{a}})}]{Raychowdhury:2018osk}%
  \BibitemOpen
  \bibfield  {author} {\bibinfo {author} {\bibfnamefont {Indrakshi}\
  \bibnamefont {Raychowdhury}}\ and\ \bibinfo {author} {\bibfnamefont
  {Jesse~R.}\ \bibnamefont {Stryker}},\ }\bibfield  {title} {\enquote {\bibinfo
  {title} {{Solving Gauss's Law on Digital Quantum Computers with
  Loop-String-Hadron Digitization}},}\ }\href {\doibase
  10.1103/PhysRevResearch.2.033039} {\bibfield  {journal} {\bibinfo  {journal}
  {Phys. Rev. Res.}\ }\textbf {\bibinfo {volume} {2}},\ \bibinfo {pages}
  {033039} (\bibinfo {year} {2020}{\natexlab{a}})},\ \Eprint
  {http://arxiv.org/abs/1812.07554} {arXiv:1812.07554 [hep-lat]} \BibitemShut
  {NoStop}%
\bibitem [{\citenamefont {Gustafson}\ \emph
  {et~al.}(2019{\natexlab{a}})\citenamefont {Gustafson}, \citenamefont
  {Meurice},\ and\ \citenamefont {Unmuth-Yockey}}]{PhysRevD.99.094503}%
  \BibitemOpen
  \bibfield  {author} {\bibinfo {author} {\bibfnamefont {Erik}\ \bibnamefont
  {Gustafson}}, \bibinfo {author} {\bibfnamefont {Y.}~\bibnamefont {Meurice}},
  \ and\ \bibinfo {author} {\bibfnamefont {Judah}\ \bibnamefont
  {Unmuth-Yockey}},\ }\bibfield  {title} {\enquote {\bibinfo {title} {Quantum
  simulation of scattering in the quantum ising model},}\ }\href {\doibase
  10.1103/PhysRevD.99.094503} {\bibfield  {journal} {\bibinfo  {journal} {Phys.
  Rev. D}\ }\textbf {\bibinfo {volume} {99}},\ \bibinfo {pages} {094503}
  (\bibinfo {year} {2019}{\natexlab{a}})}\BibitemShut {NoStop}%
\bibitem [{\citenamefont {Alam}\ \emph {et~al.}(2022)\citenamefont {Alam} \emph
  {et~al.}}]{Alam:2022crs}%
  \BibitemOpen
  \bibfield  {author} {\bibinfo {author} {\bibfnamefont {M.~Sohaib}\
  \bibnamefont {Alam}} \emph {et~al.},\ }\bibfield  {title} {\enquote {\bibinfo
  {title} {{Quantum computing hardware for HEP algorithms and sensing}},}\ }in\
  \href@noop {} {\emph {\bibinfo {booktitle} {{2022 Snowmass Summer Study}}}}\
  (\bibinfo {year} {2022})\ \Eprint {http://arxiv.org/abs/2204.08605}
  {arXiv:2204.08605 [quant-ph]} \BibitemShut {NoStop}%
\bibitem [{\citenamefont {Kurkcuoglu}\ \emph {et~al.}(2021)\citenamefont
  {Kurkcuoglu}, \citenamefont {Alam}, \citenamefont {Job}, \citenamefont {Li},
  \citenamefont {Macridin}, \citenamefont {Perdue},\ and\ \citenamefont
  {Providence}}]{Kurkcuoglu:2021dnw}%
  \BibitemOpen
  \bibfield  {author} {\bibinfo {author} {\bibfnamefont {Doga~Murat}\
  \bibnamefont {Kurkcuoglu}}, \bibinfo {author} {\bibfnamefont {M.~Sohaib}\
  \bibnamefont {Alam}}, \bibinfo {author} {\bibfnamefont {Joshua~Adam}\
  \bibnamefont {Job}}, \bibinfo {author} {\bibfnamefont {Andy C.~Y.}\
  \bibnamefont {Li}}, \bibinfo {author} {\bibfnamefont {Alexandru}\
  \bibnamefont {Macridin}}, \bibinfo {author} {\bibfnamefont {Gabriel~N.}\
  \bibnamefont {Perdue}}, \ and\ \bibinfo {author} {\bibfnamefont {Stephen}\
  \bibnamefont {Providence}},\ }\bibfield  {title} {\enquote {\bibinfo {title}
  {{Quantum simulation of $\phi^4$ theories in qudit systems}},}\ }\href@noop
  {} {\  (\bibinfo {year} {2021})},\ \Eprint {http://arxiv.org/abs/2108.13357}
  {arXiv:2108.13357 [quant-ph]} \BibitemShut {NoStop}%
\bibitem [{\citenamefont {{Keesling}}\ \emph {et~al.}(2019)\citenamefont
  {{Keesling}}, \citenamefont {{Omran}}, \citenamefont {{Levine}},
  \citenamefont {{Bernien}}, \citenamefont {{Pichler}}, \citenamefont {{Choi}},
  \citenamefont {{Samajdar}}, \citenamefont {{Schwartz}}, \citenamefont
  {{Silvi}}, \citenamefont {{Sachdev}}, \citenamefont {{Zoller}}, \citenamefont
  {{Endres}}, \citenamefont {{Greiner}}, \citenamefont {{Vuleti{\'c}}},
  \citenamefont {{}},\ and\ \citenamefont {{Lukin}}}]{keesling}%
  \BibitemOpen
  \bibfield  {author} {\bibinfo {author} {\bibfnamefont {Alexander}\
  \bibnamefont {{Keesling}}}, \bibinfo {author} {\bibfnamefont {Ahmed}\
  \bibnamefont {{Omran}}}, \bibinfo {author} {\bibfnamefont {Harry}\
  \bibnamefont {{Levine}}}, \bibinfo {author} {\bibfnamefont {Hannes}\
  \bibnamefont {{Bernien}}}, \bibinfo {author} {\bibfnamefont {Hannes}\
  \bibnamefont {{Pichler}}}, \bibinfo {author} {\bibfnamefont {Soonwon}\
  \bibnamefont {{Choi}}}, \bibinfo {author} {\bibfnamefont {Rhine}\
  \bibnamefont {{Samajdar}}}, \bibinfo {author} {\bibfnamefont {Sylvain}\
  \bibnamefont {{Schwartz}}}, \bibinfo {author} {\bibfnamefont {Pietro}\
  \bibnamefont {{Silvi}}}, \bibinfo {author} {\bibfnamefont {Subir}\
  \bibnamefont {{Sachdev}}}, \bibinfo {author} {\bibfnamefont {Peter}\
  \bibnamefont {{Zoller}}}, \bibinfo {author} {\bibfnamefont {Manuel}\
  \bibnamefont {{Endres}}}, \bibinfo {author} {\bibfnamefont {Markus}\
  \bibnamefont {{Greiner}}}, \bibinfo {author} {\bibnamefont {{Vuleti{\'c}}}},
  \bibinfo {author} {\bibfnamefont {Vladan}\ \bibnamefont {{}}}, \ and\
  \bibinfo {author} {\bibfnamefont {Mikhail~D.}\ \bibnamefont {{Lukin}}},\
  }\bibfield  {title} {\enquote {\bibinfo {title} {{Quantum Kibble-Zurek
  mechanism and critical dynamics on a programmable Rydberg simulator}},}\
  }\href {\doibase 10.1038/s41586-019-1070-1} {\bibfield  {journal} {\bibinfo
  {journal} {\nat}\ }\textbf {\bibinfo {volume} {568}},\ \bibinfo {pages}
  {207--211} (\bibinfo {year} {2019})},\ \Eprint
  {http://arxiv.org/abs/1809.05540} {arXiv:1809.05540 [quant-ph]} \BibitemShut
  {NoStop}%
\bibitem [{\citenamefont {{Bernien}}\ \emph {et~al.}(2017)\citenamefont
  {{Bernien}}, \citenamefont {{Schwartz}}, \citenamefont {{Keesling}},
  \citenamefont {{Levine}}, \citenamefont {{Omran}}, \citenamefont {{Pichler}},
  \citenamefont {{Choi}}, \citenamefont {{Zibrov}}, \citenamefont {{Endres}},
  \citenamefont {{Greiner}}, \citenamefont {{Vuleti{\'c}}},\ and\ \citenamefont
  {{Lukin}}}]{51atom}%
  \BibitemOpen
  \bibfield  {author} {\bibinfo {author} {\bibfnamefont {Hannes}\ \bibnamefont
  {{Bernien}}}, \bibinfo {author} {\bibfnamefont {Sylvain}\ \bibnamefont
  {{Schwartz}}}, \bibinfo {author} {\bibfnamefont {Alexander}\ \bibnamefont
  {{Keesling}}}, \bibinfo {author} {\bibfnamefont {Harry}\ \bibnamefont
  {{Levine}}}, \bibinfo {author} {\bibfnamefont {Ahmed}\ \bibnamefont
  {{Omran}}}, \bibinfo {author} {\bibfnamefont {Hannes}\ \bibnamefont
  {{Pichler}}}, \bibinfo {author} {\bibfnamefont {Soonwon}\ \bibnamefont
  {{Choi}}}, \bibinfo {author} {\bibfnamefont {Alexander~S.}\ \bibnamefont
  {{Zibrov}}}, \bibinfo {author} {\bibfnamefont {Manuel}\ \bibnamefont
  {{Endres}}}, \bibinfo {author} {\bibfnamefont {Markus}\ \bibnamefont
  {{Greiner}}}, \bibinfo {author} {\bibfnamefont {Vladan}\ \bibnamefont
  {{Vuleti{\'c}}}}, \ and\ \bibinfo {author} {\bibfnamefont {Mikhail~D.}\
  \bibnamefont {{Lukin}}},\ }\bibfield  {title} {\enquote {\bibinfo {title}
  {{Probing many-body dynamics on a 51-atom quantum simulator}},}\ }\href
  {\doibase 10.1038/nature24622} {\bibfield  {journal} {\bibinfo  {journal}
  {\nat}\ }\textbf {\bibinfo {volume} {551}},\ \bibinfo {pages} {579--584}
  (\bibinfo {year} {2017})},\ \Eprint {http://arxiv.org/abs/1707.04344}
  {arXiv:1707.04344 [quant-ph]} \BibitemShut {NoStop}%
\bibitem [{\citenamefont {{Notarnicola}}\ \emph {et~al.}(2020)\citenamefont
  {{Notarnicola}}, \citenamefont {{Collura}},\ and\ \citenamefont
  {{Montangero}}}]{qedryd}%
  \BibitemOpen
  \bibfield  {author} {\bibinfo {author} {\bibfnamefont {Simone}\ \bibnamefont
  {{Notarnicola}}}, \bibinfo {author} {\bibfnamefont {Mario}\ \bibnamefont
  {{Collura}}}, \ and\ \bibinfo {author} {\bibfnamefont {Simone}\ \bibnamefont
  {{Montangero}}},\ }\bibfield  {title} {\enquote {\bibinfo {title}
  {{Real-time-dynamics quantum simulation of (1 +1 )-dimensional lattice QED
  with Rydberg atoms}},}\ }\href {\doibase 10.1103/PhysRevResearch.2.013288}
  {\bibfield  {journal} {\bibinfo  {journal} {Physical Review Research}\
  }\textbf {\bibinfo {volume} {2}},\ \bibinfo {eid} {013288} (\bibinfo {year}
  {2020})},\ \Eprint {http://arxiv.org/abs/1907.12579} {arXiv:1907.12579
  [cond-mat.quant-gas]} \BibitemShut {NoStop}%
\bibitem [{\citenamefont {{Celi}}\ \emph {et~al.}(2020)\citenamefont {{Celi}},
  \citenamefont {{Vermersch}}, \citenamefont {{Viyuela}}, \citenamefont
  {{Pichler}}, \citenamefont {{Lukin}},\ and\ \citenamefont {{Zoller}}}]{celi}%
  \BibitemOpen
  \bibfield  {author} {\bibinfo {author} {\bibfnamefont {Alessio}\ \bibnamefont
  {{Celi}}}, \bibinfo {author} {\bibfnamefont {Beno{\^\i}t}\ \bibnamefont
  {{Vermersch}}}, \bibinfo {author} {\bibfnamefont {Oscar}\ \bibnamefont
  {{Viyuela}}}, \bibinfo {author} {\bibfnamefont {Hannes}\ \bibnamefont
  {{Pichler}}}, \bibinfo {author} {\bibfnamefont {Mikhail~D.}\ \bibnamefont
  {{Lukin}}}, \ and\ \bibinfo {author} {\bibfnamefont {Peter}\ \bibnamefont
  {{Zoller}}},\ }\bibfield  {title} {\enquote {\bibinfo {title} {{Emerging
  Two-Dimensional Gauge Theories in Rydberg Configurable Arrays}},}\ }\href
  {\doibase 10.1103/PhysRevX.10.021057} {\bibfield  {journal} {\bibinfo
  {journal} {Physical Review X}\ }\textbf {\bibinfo {volume} {10}},\ \bibinfo
  {eid} {021057} (\bibinfo {year} {2020})},\ \Eprint
  {http://arxiv.org/abs/1907.03311} {arXiv:1907.03311 [quant-ph]} \BibitemShut
  {NoStop}%
\bibitem [{\citenamefont {Surace}\ \emph {et~al.}(2020)\citenamefont {Surace},
  \citenamefont {Mazza}, \citenamefont {Giudici}, \citenamefont {Lerose},
  \citenamefont {Gambassi},\ and\ \citenamefont
  {Dalmonte}}]{surace2020lattice}%
  \BibitemOpen
  \bibfield  {author} {\bibinfo {author} {\bibfnamefont {Federica~M}\
  \bibnamefont {Surace}}, \bibinfo {author} {\bibfnamefont {Paolo~P}\
  \bibnamefont {Mazza}}, \bibinfo {author} {\bibfnamefont {Giuliano}\
  \bibnamefont {Giudici}}, \bibinfo {author} {\bibfnamefont {Alessio}\
  \bibnamefont {Lerose}}, \bibinfo {author} {\bibfnamefont {Andrea}\
  \bibnamefont {Gambassi}}, \ and\ \bibinfo {author} {\bibfnamefont {Marcello}\
  \bibnamefont {Dalmonte}},\ }\bibfield  {title} {\enquote {\bibinfo {title}
  {Lattice gauge theories and string dynamics in rydberg atom quantum
  simulators},}\ }\href@noop {} {\bibfield  {journal} {\bibinfo  {journal}
  {Physical Review X}\ }\textbf {\bibinfo {volume} {10}},\ \bibinfo {pages}
  {021041} (\bibinfo {year} {2020})}\BibitemShut {NoStop}%
\bibitem [{\citenamefont {Meurice}(2021{\natexlab{a}})}]{Meurice:2021pvj}%
  \BibitemOpen
  \bibfield  {author} {\bibinfo {author} {\bibfnamefont {Yannick}\ \bibnamefont
  {Meurice}},\ }\bibfield  {title} {\enquote {\bibinfo {title} {{Theoretical
  methods to design and test quantum simulators for the compact Abelian Higgs
  model}},}\ }\href {\doibase 10.1103/PhysRevD.104.094513} {\bibfield
  {journal} {\bibinfo  {journal} {Phys. Rev. D}\ }\textbf {\bibinfo {volume}
  {104}},\ \bibinfo {pages} {094513} (\bibinfo {year} {2021}{\natexlab{a}})},\
  \Eprint {http://arxiv.org/abs/2107.11366} {arXiv:2107.11366 [quant-ph]}
  \BibitemShut {NoStop}%
\bibitem [{\citenamefont {et~al.}(2007)}]{Bauer_Overview}%
  \BibitemOpen
  \bibfield  {author} {\bibinfo {author} {\bibfnamefont {P.~Bauer}\
  \bibnamefont {et~al.}},\ }\bibfield  {title} {\enquote {\bibinfo {title} {A
  review of the q-drop phenomenon in high gradient srf cavities.}}\ }in\
  \href@noop {} {\emph {\bibinfo {booktitle} {Fermilab technical report}}}\
  (\bibinfo {year} {2007})\BibitemShut {NoStop}%
\bibitem [{\citenamefont {Kogut}\ and\ \citenamefont
  {Susskind}(1975)}]{PhysRevD.11.395}%
  \BibitemOpen
  \bibfield  {author} {\bibinfo {author} {\bibfnamefont {John}\ \bibnamefont
  {Kogut}}\ and\ \bibinfo {author} {\bibfnamefont {Leonard}\ \bibnamefont
  {Susskind}},\ }\bibfield  {title} {\enquote {\bibinfo {title} {Hamiltonian
  formulation of wilson's lattice gauge theories},}\ }\href {\doibase
  10.1103/PhysRevD.11.395} {\bibfield  {journal} {\bibinfo  {journal} {Phys.
  Rev. D}\ }\textbf {\bibinfo {volume} {11}},\ \bibinfo {pages} {395--408}
  (\bibinfo {year} {1975})}\BibitemShut {NoStop}%
\bibitem [{\citenamefont {Liu}\ \emph {et~al.}(2013)\citenamefont {Liu},
  \citenamefont {Meurice}, \citenamefont {Qin}, \citenamefont {Unmuth-Yockey},
  \citenamefont {Xiang}, \citenamefont {Xie}, \citenamefont {Yu},\ and\
  \citenamefont {Zou}}]{Liu:2013nsa}%
  \BibitemOpen
  \bibfield  {author} {\bibinfo {author} {\bibfnamefont {Yuzhi}\ \bibnamefont
  {Liu}}, \bibinfo {author} {\bibfnamefont {Y.}~\bibnamefont {Meurice}},
  \bibinfo {author} {\bibfnamefont {M.~P.}\ \bibnamefont {Qin}}, \bibinfo
  {author} {\bibfnamefont {J.}~\bibnamefont {Unmuth-Yockey}}, \bibinfo {author}
  {\bibfnamefont {T.}~\bibnamefont {Xiang}}, \bibinfo {author} {\bibfnamefont
  {Z.~Y.}\ \bibnamefont {Xie}}, \bibinfo {author} {\bibfnamefont {J.~F.}\
  \bibnamefont {Yu}}, \ and\ \bibinfo {author} {\bibfnamefont {Haiyuan}\
  \bibnamefont {Zou}},\ }\bibfield  {title} {\enquote {\bibinfo {title} {{Exact
  Blocking Formulas for Spin and Gauge Models}},}\ }\href {\doibase
  10.1103/PhysRevD.88.056005} {\bibfield  {journal} {\bibinfo  {journal} {Phys.
  Rev. D}\ }\textbf {\bibinfo {volume} {88}},\ \bibinfo {pages} {056005}
  (\bibinfo {year} {2013})},\ \Eprint {http://arxiv.org/abs/1307.6543}
  {arXiv:1307.6543 [hep-lat]} \BibitemShut {NoStop}%
\bibitem [{\citenamefont {Meurice}\ \emph
  {et~al.}(2022{\natexlab{a}})\citenamefont {Meurice}, \citenamefont {Osborn},
  \citenamefont {Sakai}, \citenamefont {Unmuth-Yockey}, \citenamefont
  {Catterall},\ and\ \citenamefont {Somma}}]{Meurice:2022xbk}%
  \BibitemOpen
  \bibfield  {author} {\bibinfo {author} {\bibfnamefont {Yannick}\ \bibnamefont
  {Meurice}}, \bibinfo {author} {\bibfnamefont {James~C.}\ \bibnamefont
  {Osborn}}, \bibinfo {author} {\bibfnamefont {Ryo}\ \bibnamefont {Sakai}},
  \bibinfo {author} {\bibfnamefont {Judah}\ \bibnamefont {Unmuth-Yockey}},
  \bibinfo {author} {\bibfnamefont {Simon}\ \bibnamefont {Catterall}}, \ and\
  \bibinfo {author} {\bibfnamefont {Rolando~D.}\ \bibnamefont {Somma}},\
  }\bibfield  {title} {\enquote {\bibinfo {title} {{Tensor networks for High
  Energy Physics: contribution to Snowmass 2021}},}\ \ }(\bibinfo {year}
  {2022})\ \Eprint {http://arxiv.org/abs/2203.04902} {arXiv:2203.04902
  [hep-lat]} \BibitemShut {NoStop}%
\bibitem [{\citenamefont {Meurice}\ \emph
  {et~al.}(2022{\natexlab{b}})\citenamefont {Meurice}, \citenamefont {Sakai},\
  and\ \citenamefont {Unmuth-Yockey}}]{Meurice:2020pxc}%
  \BibitemOpen
  \bibfield  {author} {\bibinfo {author} {\bibfnamefont {Yannick}\ \bibnamefont
  {Meurice}}, \bibinfo {author} {\bibfnamefont {Ryo}\ \bibnamefont {Sakai}}, \
  and\ \bibinfo {author} {\bibfnamefont {Judah}\ \bibnamefont
  {Unmuth-Yockey}},\ }\bibfield  {title} {\enquote {\bibinfo {title} {{Tensor
  lattice field theory for renormalization and quantum computing}},}\ }\href
  {\doibase 10.1103/RevModPhys.94.025005} {\bibfield  {journal} {\bibinfo
  {journal} {Rev. Mod. Phys.}\ }\textbf {\bibinfo {volume} {94}},\ \bibinfo
  {pages} {025005} (\bibinfo {year} {2022}{\natexlab{b}})},\ \Eprint
  {http://arxiv.org/abs/2010.06539} {arXiv:2010.06539 [hep-lat]} \BibitemShut
  {NoStop}%
\bibitem [{\citenamefont {Mathur}\ \emph {et~al.}(2010)\citenamefont {Mathur},
  \citenamefont {Raychowdhury},\ and\ \citenamefont {Anishetty}}]{mathur2010n}%
  \BibitemOpen
  \bibfield  {author} {\bibinfo {author} {\bibfnamefont {Manu}\ \bibnamefont
  {Mathur}}, \bibinfo {author} {\bibfnamefont {Indrakshi}\ \bibnamefont
  {Raychowdhury}}, \ and\ \bibinfo {author} {\bibfnamefont {Ramesh}\
  \bibnamefont {Anishetty}},\ }\bibfield  {title} {\enquote {\bibinfo {title}
  {Su (n) irreducible schwinger bosons},}\ }\href@noop {} {\bibfield  {journal}
  {\bibinfo  {journal} {Journal of mathematical physics}\ }\textbf {\bibinfo
  {volume} {51}},\ \bibinfo {pages} {093504} (\bibinfo {year}
  {2010})}\BibitemShut {NoStop}%
\bibitem [{\citenamefont {Wiese}(2013)}]{Wiese:2013uua}%
  \BibitemOpen
  \bibfield  {author} {\bibinfo {author} {\bibfnamefont {Uwe-Jens}\
  \bibnamefont {Wiese}},\ }\bibfield  {title} {\enquote {\bibinfo {title}
  {{Ultracold Quantum Gases and Lattice Systems: Quantum Simulation of Lattice
  Gauge Theories}},}\ }\href {\doibase 10.1002/andp.201300104} {\bibfield
  {journal} {\bibinfo  {journal} {Annalen Phys.}\ }\textbf {\bibinfo {volume}
  {525}},\ \bibinfo {pages} {777--796} (\bibinfo {year} {2013})},\ \Eprint
  {http://arxiv.org/abs/1305.1602} {arXiv:1305.1602 [quant-ph]} \BibitemShut
  {NoStop}%
\bibitem [{\citenamefont {Wiese}(2021)}]{Wiese:2021djl}%
  \BibitemOpen
  \bibfield  {author} {\bibinfo {author} {\bibfnamefont {Uwe-Jens}\
  \bibnamefont {Wiese}},\ }\bibfield  {title} {\enquote {\bibinfo {title}
  {{From quantum link models to D-theory: a resource efficient framework for
  the quantum simulation and computation of gauge theories}},}\ }\href
  {\doibase 10.1098/rsta.2021.0068} {\bibfield  {journal} {\bibinfo  {journal}
  {Phil. Trans. A. Math. Phys. Eng. Sci.}\ }\textbf {\bibinfo {volume} {380}},\
  \bibinfo {pages} {20210068} (\bibinfo {year} {2021})},\ \Eprint
  {http://arxiv.org/abs/2107.09335} {arXiv:2107.09335 [hep-lat]} \BibitemShut
  {NoStop}%
\bibitem [{\citenamefont {Bhattacharya}\ \emph {et~al.}(2020)\citenamefont
  {Bhattacharya}, \citenamefont {Buser}, \citenamefont {Chandrasekharan},
  \citenamefont {Gupta},\ and\ \citenamefont {Singh}}]{Bhattacharya:2020gpm}%
  \BibitemOpen
  \bibfield  {author} {\bibinfo {author} {\bibfnamefont {Tanmoy}\ \bibnamefont
  {Bhattacharya}}, \bibinfo {author} {\bibfnamefont {Alexander~J.}\
  \bibnamefont {Buser}}, \bibinfo {author} {\bibfnamefont {Shailesh}\
  \bibnamefont {Chandrasekharan}}, \bibinfo {author} {\bibfnamefont {Rajan}\
  \bibnamefont {Gupta}}, \ and\ \bibinfo {author} {\bibfnamefont {Hersh}\
  \bibnamefont {Singh}},\ }\href@noop {} {\enquote {\bibinfo {title} {{Qubit
  regularization of asymptotic freedom}},}\ } (\bibinfo {year} {2020}),\
  \Eprint {http://arxiv.org/abs/2012.02153} {arXiv:2012.02153 [hep-lat]}
  \BibitemShut {NoStop}%
\bibitem [{\citenamefont {Shaw}\ \emph {et~al.}(2020)\citenamefont {Shaw},
  \citenamefont {Lougovski}, \citenamefont {Stryker},\ and\ \citenamefont
  {Wiebe}}]{shaw2020quantum}%
  \BibitemOpen
  \bibfield  {author} {\bibinfo {author} {\bibfnamefont {Alexander~F}\
  \bibnamefont {Shaw}}, \bibinfo {author} {\bibfnamefont {Pavel}\ \bibnamefont
  {Lougovski}}, \bibinfo {author} {\bibfnamefont {Jesse~R}\ \bibnamefont
  {Stryker}}, \ and\ \bibinfo {author} {\bibfnamefont {Nathan}\ \bibnamefont
  {Wiebe}},\ }\bibfield  {title} {\enquote {\bibinfo {title} {Quantum
  algorithms for simulating the lattice schwinger model},}\ }\href@noop {}
  {\bibfield  {journal} {\bibinfo  {journal} {Quantum}\ }\textbf {\bibinfo
  {volume} {4}},\ \bibinfo {pages} {306} (\bibinfo {year} {2020})}\BibitemShut
  {NoStop}%
\bibitem [{\citenamefont {Ciavarella}\ \emph
  {et~al.}(2021{\natexlab{a}})\citenamefont {Ciavarella}, \citenamefont
  {Klco},\ and\ \citenamefont {Savage}}]{ciavarella2021trailhead}%
  \BibitemOpen
  \bibfield  {author} {\bibinfo {author} {\bibfnamefont {Anthony}\ \bibnamefont
  {Ciavarella}}, \bibinfo {author} {\bibfnamefont {Natalie}\ \bibnamefont
  {Klco}}, \ and\ \bibinfo {author} {\bibfnamefont {Martin~J}\ \bibnamefont
  {Savage}},\ }\bibfield  {title} {\enquote {\bibinfo {title} {Trailhead for
  quantum simulation of su (3) yang-mills lattice gauge theory in the local
  multiplet basis},}\ }\href@noop {} {\bibfield  {journal} {\bibinfo  {journal}
  {Physical Review D}\ }\textbf {\bibinfo {volume} {103}},\ \bibinfo {pages}
  {094501} (\bibinfo {year} {2021}{\natexlab{a}})}\BibitemShut {NoStop}%
\bibitem [{\citenamefont {Paulson}\ \emph {et~al.}(2021)\citenamefont
  {Paulson}, \citenamefont {Dellantonio}, \citenamefont {Haase}, \citenamefont
  {Celi}, \citenamefont {Kan}, \citenamefont {Jena}, \citenamefont {Kokail},
  \citenamefont {Van~Bijnen}, \citenamefont {Jansen}, \citenamefont {Zoller}
  \emph {et~al.}}]{paulson2021simulating}%
  \BibitemOpen
  \bibfield  {author} {\bibinfo {author} {\bibfnamefont {Danny}\ \bibnamefont
  {Paulson}}, \bibinfo {author} {\bibfnamefont {Luca}\ \bibnamefont
  {Dellantonio}}, \bibinfo {author} {\bibfnamefont {Jan~F}\ \bibnamefont
  {Haase}}, \bibinfo {author} {\bibfnamefont {Alessio}\ \bibnamefont {Celi}},
  \bibinfo {author} {\bibfnamefont {Angus}\ \bibnamefont {Kan}}, \bibinfo
  {author} {\bibfnamefont {Andrew}\ \bibnamefont {Jena}}, \bibinfo {author}
  {\bibfnamefont {Christian}\ \bibnamefont {Kokail}}, \bibinfo {author}
  {\bibfnamefont {Rick}\ \bibnamefont {Van~Bijnen}}, \bibinfo {author}
  {\bibfnamefont {Karl}\ \bibnamefont {Jansen}}, \bibinfo {author}
  {\bibfnamefont {Peter}\ \bibnamefont {Zoller}},  \emph {et~al.},\ }\bibfield
  {title} {\enquote {\bibinfo {title} {Simulating 2d effects in lattice gauge
  theories on a quantum computer},}\ }\href@noop {} {\bibfield  {journal}
  {\bibinfo  {journal} {PRX Quantum}\ }\textbf {\bibinfo {volume} {2}},\
  \bibinfo {pages} {030334} (\bibinfo {year} {2021})}\BibitemShut {NoStop}%
\bibitem [{\citenamefont {Kan}\ and\ \citenamefont
  {Nam}(2021)}]{kan2021lattice}%
  \BibitemOpen
  \bibfield  {author} {\bibinfo {author} {\bibfnamefont {Angus}\ \bibnamefont
  {Kan}}\ and\ \bibinfo {author} {\bibfnamefont {Yunseong}\ \bibnamefont
  {Nam}},\ }\bibfield  {title} {\enquote {\bibinfo {title} {Lattice quantum
  chromodynamics and electrodynamics on a universal quantum computer},}\
  }\href@noop {} {\bibfield  {journal} {\bibinfo  {journal} {arXiv preprint
  arXiv:2107.12769}\ } (\bibinfo {year} {2021})}\BibitemShut {NoStop}%
\bibitem [{\citenamefont {Maldacena}(1998)}]{Mald}%
  \BibitemOpen
  \bibfield  {author} {\bibinfo {author} {\bibfnamefont {Juan~M.}\ \bibnamefont
  {Maldacena}},\ }\bibfield  {title} {\enquote {\bibinfo {title} {The large n
  limit of superconformal field theories and supergravity},}\ }\href@noop {}
  {\bibfield  {journal} {\bibinfo  {journal} {Adv. Theor. Math. Phys.}\
  }\textbf {\bibinfo {volume} {2}},\ \bibinfo {pages} {231--252} (\bibinfo
  {year} {1998})},\ \Eprint {http://arxiv.org/abs/hep-th/9711200}
  {hep-th/9711200} \BibitemShut {NoStop}%
\bibitem [{\citenamefont {Ryu}\ and\ \citenamefont
  {Takayanagi}(2006{\natexlab{a}})}]{ryu2006holographic}%
  \BibitemOpen
  \bibfield  {author} {\bibinfo {author} {\bibfnamefont {Shinsei}\ \bibnamefont
  {Ryu}}\ and\ \bibinfo {author} {\bibfnamefont {Tadashi}\ \bibnamefont
  {Takayanagi}},\ }\bibfield  {title} {\enquote {\bibinfo {title} {Holographic
  derivation of entanglement entropy from the anti--de sitter space/conformal
  field theory correspondence},}\ }\href@noop {} {\bibfield  {journal}
  {\bibinfo  {journal} {Physical review letters}\ }\textbf {\bibinfo {volume}
  {96}},\ \bibinfo {pages} {181602} (\bibinfo {year}
  {2006}{\natexlab{a}})}\BibitemShut {NoStop}%
\bibitem [{\citenamefont {Van~Raamsdonk}(2010)}]{VanRaamsdonk:2010pw}%
  \BibitemOpen
  \bibfield  {author} {\bibinfo {author} {\bibfnamefont {Mark}\ \bibnamefont
  {Van~Raamsdonk}},\ }\bibfield  {title} {\enquote {\bibinfo {title} {{Building
  up spacetime with quantum entanglement}},}\ }\href {\doibase
  10.1142/S0218271810018529} {\bibfield  {journal} {\bibinfo  {journal} {Gen.
  Rel. Grav.}\ }\textbf {\bibinfo {volume} {42}},\ \bibinfo {pages}
  {2323--2329} (\bibinfo {year} {2010})},\ \Eprint
  {http://arxiv.org/abs/1005.3035} {arXiv:1005.3035 [hep-th]} \BibitemShut
  {NoStop}%
\bibitem [{\citenamefont {Faulkner}\ \emph {et~al.}(2022)\citenamefont
  {Faulkner}, \citenamefont {Hartman}, \citenamefont {Headrick}, \citenamefont
  {Rangamani},\ and\ \citenamefont {Swingle}}]{Faulkner:2022mlp}%
  \BibitemOpen
  \bibfield  {author} {\bibinfo {author} {\bibfnamefont {Thomas}\ \bibnamefont
  {Faulkner}}, \bibinfo {author} {\bibfnamefont {Thomas}\ \bibnamefont
  {Hartman}}, \bibinfo {author} {\bibfnamefont {Matthew}\ \bibnamefont
  {Headrick}}, \bibinfo {author} {\bibfnamefont {Mukund}\ \bibnamefont
  {Rangamani}}, \ and\ \bibinfo {author} {\bibfnamefont {Brian}\ \bibnamefont
  {Swingle}},\ }\bibfield  {title} {\enquote {\bibinfo {title} {{Snowmass white
  paper: Quantum information in quantum field theory and quantum gravity}},}\
  }in\ \href@noop {} {\emph {\bibinfo {booktitle} {{2022 Snowmass Summer
  Study}}}}\ (\bibinfo {year} {2022})\ \Eprint
  {http://arxiv.org/abs/2203.07117} {arXiv:2203.07117 [hep-th]} \BibitemShut
  {NoStop}%
\bibitem [{\citenamefont {Asaduzzaman}\ \emph {et~al.}(2020)\citenamefont
  {Asaduzzaman}, \citenamefont {Catterall}, \citenamefont {Hubisz},
  \citenamefont {Nelson},\ and\ \citenamefont
  {Unmuth-Yockey}}]{Asaduzzaman:2020hjl}%
  \BibitemOpen
  \bibfield  {author} {\bibinfo {author} {\bibfnamefont {Muhammad}\
  \bibnamefont {Asaduzzaman}}, \bibinfo {author} {\bibfnamefont {Simon}\
  \bibnamefont {Catterall}}, \bibinfo {author} {\bibfnamefont {Jay}\
  \bibnamefont {Hubisz}}, \bibinfo {author} {\bibfnamefont {Roice}\
  \bibnamefont {Nelson}}, \ and\ \bibinfo {author} {\bibfnamefont {Judah}\
  \bibnamefont {Unmuth-Yockey}},\ }\bibfield  {title} {\enquote {\bibinfo
  {title} {{Holography on tessellations of hyperbolic space}},}\ }\href
  {\doibase 10.1103/PhysRevD.102.034511} {\bibfield  {journal} {\bibinfo
  {journal} {Phys. Rev. D}\ }\textbf {\bibinfo {volume} {102}},\ \bibinfo
  {pages} {034511} (\bibinfo {year} {2020})},\ \Eprint
  {http://arxiv.org/abs/2005.12726} {arXiv:2005.12726 [hep-lat]} \BibitemShut
  {NoStop}%
\bibitem [{\citenamefont {Asaduzzaman}\ \emph {et~al.}(2021)\citenamefont
  {Asaduzzaman}, \citenamefont {Catterall}, \citenamefont {Hubisz},
  \citenamefont {Nelson},\ and\ \citenamefont
  {Unmuth-Yockey}}]{Asaduzzaman:2021bcw}%
  \BibitemOpen
  \bibfield  {author} {\bibinfo {author} {\bibfnamefont {Muhammad}\
  \bibnamefont {Asaduzzaman}}, \bibinfo {author} {\bibfnamefont {Simon}\
  \bibnamefont {Catterall}}, \bibinfo {author} {\bibfnamefont {Jay}\
  \bibnamefont {Hubisz}}, \bibinfo {author} {\bibfnamefont {Roice}\
  \bibnamefont {Nelson}}, \ and\ \bibinfo {author} {\bibfnamefont {Judah}\
  \bibnamefont {Unmuth-Yockey}},\ }\bibfield  {title} {\enquote {\bibinfo
  {title} {{Holography for Ising spins on the hyperbolic plane}},}\ }\href@noop
  {} {\  (\bibinfo {year} {2021})},\ \Eprint {http://arxiv.org/abs/2112.00184}
  {arXiv:2112.00184 [hep-lat]} \BibitemShut {NoStop}%
\bibitem [{\citenamefont {Evenbly}\ and\ \citenamefont
  {Vidal}(2011)}]{Evenbly}%
  \BibitemOpen
  \bibfield  {author} {\bibinfo {author} {\bibfnamefont {G}~\bibnamefont
  {Evenbly}}\ and\ \bibinfo {author} {\bibfnamefont {G}~\bibnamefont {Vidal}},\
  }\bibfield  {title} {\enquote {\bibinfo {title} {Tensor network states and
  geometry},}\ }\href {\doibase 10.1007/s10955-011-0237-4} {\bibfield
  {journal} {\bibinfo  {journal} {J. Stat. Phys}\ }\textbf {\bibinfo {volume}
  {145}} (\bibinfo {year} {2011}),\ 10.1007/s10955-011-0237-4},\ \Eprint
  {http://arxiv.org/abs/1108.1082} {1108.1082} \BibitemShut {NoStop}%
\bibitem [{\citenamefont {Pastawski}\ \emph {et~al.}(2015)\citenamefont
  {Pastawski}, \citenamefont {Yoshida}, \citenamefont {Harlow},\ and\
  \citenamefont {Preskill}}]{Pastawski:2015qua}%
  \BibitemOpen
  \bibfield  {author} {\bibinfo {author} {\bibfnamefont {Fernando}\
  \bibnamefont {Pastawski}}, \bibinfo {author} {\bibfnamefont {Beni}\
  \bibnamefont {Yoshida}}, \bibinfo {author} {\bibfnamefont {Daniel}\
  \bibnamefont {Harlow}}, \ and\ \bibinfo {author} {\bibfnamefont {John}\
  \bibnamefont {Preskill}},\ }\bibfield  {title} {\enquote {\bibinfo {title}
  {{Holographic quantum error-correcting codes: Toy models for the
  bulk/boundary correspondence}},}\ }\href {\doibase 10.1007/JHEP06(2015)149}
  {\bibfield  {journal} {\bibinfo  {journal} {JHEP}\ }\textbf {\bibinfo
  {volume} {06}},\ \bibinfo {pages} {149} (\bibinfo {year} {2015})},\ \Eprint
  {http://arxiv.org/abs/1503.06237} {arXiv:1503.06237 [hep-th]} \BibitemShut
  {NoStop}%
\bibitem [{\citenamefont {Kogut}(1983)}]{RevModPhys.55.775}%
  \BibitemOpen
  \bibfield  {author} {\bibinfo {author} {\bibfnamefont {John~B.}\ \bibnamefont
  {Kogut}},\ }\bibfield  {title} {\enquote {\bibinfo {title} {The lattice gauge
  theory approach to quantum chromodynamics},}\ }\href {\doibase
  10.1103/RevModPhys.55.775} {\bibfield  {journal} {\bibinfo  {journal} {Rev.
  Mod. Phys.}\ }\textbf {\bibinfo {volume} {55}},\ \bibinfo {pages} {775--836}
  (\bibinfo {year} {1983})}\BibitemShut {NoStop}%
\bibitem [{\citenamefont {Kogut}(1979)}]{RevModPhys.51.659}%
  \BibitemOpen
  \bibfield  {author} {\bibinfo {author} {\bibfnamefont {John~B.}\ \bibnamefont
  {Kogut}},\ }\bibfield  {title} {\enquote {\bibinfo {title} {An introduction
  to lattice gauge theory and spin systems},}\ }\href {\doibase
  10.1103/RevModPhys.51.659} {\bibfield  {journal} {\bibinfo  {journal} {Rev.
  Mod. Phys.}\ }\textbf {\bibinfo {volume} {51}},\ \bibinfo {pages} {659--713}
  (\bibinfo {year} {1979})}\BibitemShut {NoStop}%
\bibitem [{\citenamefont {Carena}\ \emph {et~al.}(2021)\citenamefont {Carena},
  \citenamefont {Lamm}, \citenamefont {Li},\ and\ \citenamefont
  {Liu}}]{PhysRevD.104.094519}%
  \BibitemOpen
  \bibfield  {author} {\bibinfo {author} {\bibfnamefont {Marcela}\ \bibnamefont
  {Carena}}, \bibinfo {author} {\bibfnamefont {Henry}\ \bibnamefont {Lamm}},
  \bibinfo {author} {\bibfnamefont {Ying-Ying}\ \bibnamefont {Li}}, \ and\
  \bibinfo {author} {\bibfnamefont {Wanqiang}\ \bibnamefont {Liu}},\ }\bibfield
   {title} {\enquote {\bibinfo {title} {Lattice renormalization of quantum
  simulations},}\ }\href {\doibase 10.1103/PhysRevD.104.094519} {\bibfield
  {journal} {\bibinfo  {journal} {Phys. Rev. D}\ }\textbf {\bibinfo {volume}
  {104}},\ \bibinfo {pages} {094519} (\bibinfo {year} {2021})}\BibitemShut
  {NoStop}%
\bibitem [{\citenamefont {Brice\~no}\ \emph {et~al.}(2021)\citenamefont
  {Brice\~no}, \citenamefont {Guerrero}, \citenamefont {Hansen},\ and\
  \citenamefont {Sturzu}}]{PhysRevD.103.014506}%
  \BibitemOpen
  \bibfield  {author} {\bibinfo {author} {\bibfnamefont {Ra\'ul~A.}\
  \bibnamefont {Brice\~no}}, \bibinfo {author} {\bibfnamefont {Juan~V.}\
  \bibnamefont {Guerrero}}, \bibinfo {author} {\bibfnamefont {Maxwell~T.}\
  \bibnamefont {Hansen}}, \ and\ \bibinfo {author} {\bibfnamefont
  {Alexandru~M.}\ \bibnamefont {Sturzu}},\ }\bibfield  {title} {\enquote
  {\bibinfo {title} {Role of boundary conditions in quantum computations of
  scattering observables},}\ }\href {\doibase 10.1103/PhysRevD.103.014506}
  {\bibfield  {journal} {\bibinfo  {journal} {Phys. Rev. D}\ }\textbf {\bibinfo
  {volume} {103}},\ \bibinfo {pages} {014506} (\bibinfo {year}
  {2021})}\BibitemShut {NoStop}%
\bibitem [{\citenamefont {Luo}\ \emph {et~al.}(1999)\citenamefont {Luo},
  \citenamefont {Guo}, \citenamefont {Kroger},\ and\ \citenamefont
  {Schutte}}]{Luo:1998dx}%
  \BibitemOpen
  \bibfield  {author} {\bibinfo {author} {\bibfnamefont {Xiang-Qian}\
  \bibnamefont {Luo}}, \bibinfo {author} {\bibfnamefont {Shuo-Hong}\
  \bibnamefont {Guo}}, \bibinfo {author} {\bibfnamefont {Helmut}\ \bibnamefont
  {Kroger}}, \ and\ \bibinfo {author} {\bibfnamefont {Dieter}\ \bibnamefont
  {Schutte}},\ }\bibfield  {title} {\enquote {\bibinfo {title} {{Improved
  lattice gauge field Hamiltonian}},}\ }\href {\doibase
  10.1103/PhysRevD.59.034503} {\bibfield  {journal} {\bibinfo  {journal} {Phys.
  Rev. D}\ }\textbf {\bibinfo {volume} {59}},\ \bibinfo {pages} {034503}
  (\bibinfo {year} {1999})},\ \Eprint {http://arxiv.org/abs/hep-lat/9804029}
  {arXiv:hep-lat/9804029} \BibitemShut {NoStop}%
\bibitem [{\citenamefont {Carlsson}\ and\ \citenamefont
  {McKellar}(2001)}]{Carlsson:2001wp}%
  \BibitemOpen
  \bibfield  {author} {\bibinfo {author} {\bibfnamefont {Jesse}\ \bibnamefont
  {Carlsson}}\ and\ \bibinfo {author} {\bibfnamefont {Bruce~H.J.}\ \bibnamefont
  {McKellar}},\ }\bibfield  {title} {\enquote {\bibinfo {title} {{Direct
  improvement of Hamiltonian lattice gauge theory}},}\ }\href {\doibase
  10.1103/PhysRevD.64.094503} {\bibfield  {journal} {\bibinfo  {journal} {Phys.
  Rev. D}\ }\textbf {\bibinfo {volume} {64}},\ \bibinfo {pages} {094503}
  (\bibinfo {year} {2001})},\ \Eprint {http://arxiv.org/abs/hep-lat/0105018}
  {arXiv:hep-lat/0105018} \BibitemShut {NoStop}%
\bibitem [{\citenamefont {Spitz}\ and\ \citenamefont
  {Berges}(2019)}]{Spitz:2018eps}%
  \BibitemOpen
  \bibfield  {author} {\bibinfo {author} {\bibfnamefont {Daniel}\ \bibnamefont
  {Spitz}}\ and\ \bibinfo {author} {\bibfnamefont {J\"urgen}\ \bibnamefont
  {Berges}},\ }\bibfield  {title} {\enquote {\bibinfo {title} {{Schwinger pair
  production and string breaking in non-Abelian gauge theory from real-time
  lattice improved Hamiltonians}},}\ }\href {\doibase
  10.1103/PhysRevD.99.036020} {\bibfield  {journal} {\bibinfo  {journal} {Phys.
  Rev. D}\ }\textbf {\bibinfo {volume} {99}},\ \bibinfo {pages} {036020}
  (\bibinfo {year} {2019})},\ \Eprint {http://arxiv.org/abs/1812.05835}
  {arXiv:1812.05835 [hep-ph]} \BibitemShut {NoStop}%
\bibitem [{\citenamefont {Carena}\ \emph {et~al.}(2022)\citenamefont {Carena},
  \citenamefont {Lamm}, \citenamefont {Li},\ and\ \citenamefont
  {Liu}}]{Carena:2022kpg}%
  \BibitemOpen
  \bibfield  {author} {\bibinfo {author} {\bibfnamefont {Marcela}\ \bibnamefont
  {Carena}}, \bibinfo {author} {\bibfnamefont {Henry}\ \bibnamefont {Lamm}},
  \bibinfo {author} {\bibfnamefont {Ying-Ying}\ \bibnamefont {Li}}, \ and\
  \bibinfo {author} {\bibfnamefont {Wanqiang}\ \bibnamefont {Liu}},\ }\bibfield
   {title} {\enquote {\bibinfo {title} {{Improved Hamiltonians for Quantum
  Simulations}},}\ }\href@noop {} {\  (\bibinfo {year} {2022})},\ \Eprint
  {http://arxiv.org/abs/2203.02823} {arXiv:2203.02823 [hep-lat]} \BibitemShut
  {NoStop}%
\bibitem [{\citenamefont {Luscher}(1986)}]{Luscher:1986pf}%
  \BibitemOpen
  \bibfield  {author} {\bibinfo {author} {\bibfnamefont {M.}~\bibnamefont
  {Luscher}},\ }\bibfield  {title} {\enquote {\bibinfo {title} {{Volume
  Dependence of the Energy Spectrum in Massive Quantum Field Theories. 2.
  Scattering States}},}\ }\href {\doibase 10.1007/BF01211097} {\bibfield
  {journal} {\bibinfo  {journal} {Commun. Math. Phys.}\ }\textbf {\bibinfo
  {volume} {105}},\ \bibinfo {pages} {153--188} (\bibinfo {year}
  {1986})}\BibitemShut {NoStop}%
\bibitem [{\citenamefont {Briceno}\ \emph {et~al.}(2018)\citenamefont
  {Briceno}, \citenamefont {Dudek},\ and\ \citenamefont
  {Young}}]{Briceno:2017max}%
  \BibitemOpen
  \bibfield  {author} {\bibinfo {author} {\bibfnamefont {Raul~A.}\ \bibnamefont
  {Briceno}}, \bibinfo {author} {\bibfnamefont {Jozef~J.}\ \bibnamefont
  {Dudek}}, \ and\ \bibinfo {author} {\bibfnamefont {Ross~D.}\ \bibnamefont
  {Young}},\ }\bibfield  {title} {\enquote {\bibinfo {title} {{Scattering
  processes and resonances from lattice QCD}},}\ }\href {\doibase
  10.1103/RevModPhys.90.025001} {\bibfield  {journal} {\bibinfo  {journal}
  {Rev. Mod. Phys.}\ }\textbf {\bibinfo {volume} {90}},\ \bibinfo {pages}
  {025001} (\bibinfo {year} {2018})},\ \Eprint
  {http://arxiv.org/abs/1706.06223} {arXiv:1706.06223 [hep-lat]} \BibitemShut
  {NoStop}%
\bibitem [{\citenamefont {Davoudi}(2018)}]{Davoudi:2018wgb}%
  \BibitemOpen
  \bibfield  {author} {\bibinfo {author} {\bibfnamefont {Zohreh}\ \bibnamefont
  {Davoudi}},\ }\bibfield  {title} {\enquote {\bibinfo {title} {{The path from
  finite to infinite volume: Hadronic observables from lattice QCD}},}\ }\href
  {\doibase 10.22323/1.334.0014} {\bibfield  {journal} {\bibinfo  {journal}
  {PoS}\ }\textbf {\bibinfo {volume} {LATTICE2018}},\ \bibinfo {pages} {014}
  (\bibinfo {year} {2018})},\ \Eprint {http://arxiv.org/abs/1812.11899}
  {arXiv:1812.11899 [hep-lat]} \BibitemShut {NoStop}%
\bibitem [{\citenamefont {Bulava}\ \emph {et~al.}(2022)\citenamefont {Bulava}
  \emph {et~al.}}]{Bulava:2022ovd}%
  \BibitemOpen
  \bibfield  {author} {\bibinfo {author} {\bibfnamefont {John}\ \bibnamefont
  {Bulava}} \emph {et~al.},\ }\bibfield  {title} {\enquote {\bibinfo {title}
  {{Hadron Spectroscopy with Lattice QCD}},}\ }in\ \href@noop {} {\emph
  {\bibinfo {booktitle} {{2022 Snowmass Summer Study}}}}\ (\bibinfo {year}
  {2022})\ \Eprint {http://arxiv.org/abs/2203.03230} {arXiv:2203.03230
  [hep-lat]} \BibitemShut {NoStop}%
\bibitem [{\citenamefont {Jordan}\ and\ \citenamefont
  {Wigner}(1928)}]{Jordan:1928wi}%
  \BibitemOpen
  \bibfield  {author} {\bibinfo {author} {\bibfnamefont {Pascual}\ \bibnamefont
  {Jordan}}\ and\ \bibinfo {author} {\bibfnamefont {Eugene~P.}\ \bibnamefont
  {Wigner}},\ }\bibfield  {title} {\enquote {\bibinfo {title} {{About the Pauli
  exclusion principle}},}\ }\href {\doibase 10.1007/BF01331938} {\bibfield
  {journal} {\bibinfo  {journal} {Z. Phys.}\ }\textbf {\bibinfo {volume}
  {47}},\ \bibinfo {pages} {631--651} (\bibinfo {year} {1928})}\BibitemShut
  {NoStop}%
\bibitem [{\citenamefont {Bravyi}\ and\ \citenamefont
  {Kitaev}(2002)}]{Bravyi2002FermionicQC}%
  \BibitemOpen
  \bibfield  {author} {\bibinfo {author} {\bibfnamefont {Sergey}\ \bibnamefont
  {Bravyi}}\ and\ \bibinfo {author} {\bibfnamefont {Alexei~Y.}\ \bibnamefont
  {Kitaev}},\ }\bibfield  {title} {\enquote {\bibinfo {title} {Fermionic
  quantum computation},}\ }\href@noop {} {\bibfield  {journal} {\bibinfo
  {journal} {Annals of Physics}\ }\textbf {\bibinfo {volume} {298}},\ \bibinfo
  {pages} {210--226} (\bibinfo {year} {2002})}\BibitemShut {NoStop}%
\bibitem [{\citenamefont {Chen}\ and\ \citenamefont
  {Kapustin}(2019)}]{Chen:2018nog}%
  \BibitemOpen
  \bibfield  {author} {\bibinfo {author} {\bibfnamefont {Yu-An}\ \bibnamefont
  {Chen}}\ and\ \bibinfo {author} {\bibfnamefont {Anton}\ \bibnamefont
  {Kapustin}},\ }\bibfield  {title} {\enquote {\bibinfo {title} {{Bosonization
  in three spatial dimensions and a 2-form gauge theory}},}\ }\href {\doibase
  10.1103/PhysRevB.100.245127} {\bibfield  {journal} {\bibinfo  {journal}
  {Phys. Rev. B}\ }\textbf {\bibinfo {volume} {100}},\ \bibinfo {pages}
  {245127} (\bibinfo {year} {2019})},\ \Eprint
  {http://arxiv.org/abs/1807.07081} {arXiv:1807.07081 [cond-mat.str-el]}
  \BibitemShut {NoStop}%
\bibitem [{\citenamefont {Muschik}\ \emph {et~al.}(2017)\citenamefont
  {Muschik}, \citenamefont {Heyl}, \citenamefont {Martinez}, \citenamefont
  {Monz}, \citenamefont {Schindler}, \citenamefont {Vogell}, \citenamefont
  {Dalmonte}, \citenamefont {Hauke}, \citenamefont {Blatt},\ and\ \citenamefont
  {Zoller}}]{Muschik:2016tws}%
  \BibitemOpen
  \bibfield  {author} {\bibinfo {author} {\bibfnamefont {Christine}\
  \bibnamefont {Muschik}}, \bibinfo {author} {\bibfnamefont {Markus}\
  \bibnamefont {Heyl}}, \bibinfo {author} {\bibfnamefont {Esteban}\
  \bibnamefont {Martinez}}, \bibinfo {author} {\bibfnamefont {Thomas}\
  \bibnamefont {Monz}}, \bibinfo {author} {\bibfnamefont {Philipp}\
  \bibnamefont {Schindler}}, \bibinfo {author} {\bibfnamefont {Berit}\
  \bibnamefont {Vogell}}, \bibinfo {author} {\bibfnamefont {Marcello}\
  \bibnamefont {Dalmonte}}, \bibinfo {author} {\bibfnamefont {Philipp}\
  \bibnamefont {Hauke}}, \bibinfo {author} {\bibfnamefont {Rainer}\
  \bibnamefont {Blatt}}, \ and\ \bibinfo {author} {\bibfnamefont {Peter}\
  \bibnamefont {Zoller}},\ }\bibfield  {title} {\enquote {\bibinfo {title}
  {{U(1) Wilson lattice gauge theories in digital quantum simulators}},}\
  }\href {\doibase 10.1088/1367-2630/aa89ab} {\bibfield  {journal} {\bibinfo
  {journal} {New J. Phys.}\ }\textbf {\bibinfo {volume} {19}},\ \bibinfo
  {pages} {103020} (\bibinfo {year} {2017})},\ \Eprint
  {http://arxiv.org/abs/1612.08653} {arXiv:1612.08653 [quant-ph]} \BibitemShut
  {NoStop}%
\bibitem [{\citenamefont {Zohar}\ and\ \citenamefont
  {Cirac}(2018)}]{Zohar:2018cwb}%
  \BibitemOpen
  \bibfield  {author} {\bibinfo {author} {\bibfnamefont {Erez}\ \bibnamefont
  {Zohar}}\ and\ \bibinfo {author} {\bibfnamefont {J.~Ignacio}\ \bibnamefont
  {Cirac}},\ }\bibfield  {title} {\enquote {\bibinfo {title} {{Eliminating
  fermionic matter fields in lattice gauge theories}},}\ }\href {\doibase
  10.1103/PhysRevB.98.075119} {\bibfield  {journal} {\bibinfo  {journal} {Phys.
  Rev. B}\ }\textbf {\bibinfo {volume} {98}},\ \bibinfo {pages} {075119}
  (\bibinfo {year} {2018})},\ \Eprint {http://arxiv.org/abs/1805.05347}
  {arXiv:1805.05347 [quant-ph]} \BibitemShut {NoStop}%
\bibitem [{\citenamefont {Zohar}\ and\ \citenamefont
  {Cirac}(2019)}]{Zohar:2019ygc}%
  \BibitemOpen
  \bibfield  {author} {\bibinfo {author} {\bibfnamefont {Erez}\ \bibnamefont
  {Zohar}}\ and\ \bibinfo {author} {\bibfnamefont {J.~Ignacio}\ \bibnamefont
  {Cirac}},\ }\bibfield  {title} {\enquote {\bibinfo {title} {{Removing
  Staggered Fermionic Matter in $U(N)$ and $SU(N)$ Lattice Gauge Theories}},}\
  }\href {\doibase 10.1103/PhysRevD.99.114511} {\bibfield  {journal} {\bibinfo
  {journal} {Phys. Rev. D}\ }\textbf {\bibinfo {volume} {99}},\ \bibinfo
  {pages} {114511} (\bibinfo {year} {2019})},\ \Eprint
  {http://arxiv.org/abs/1905.00652} {arXiv:1905.00652 [quant-ph]} \BibitemShut
  {NoStop}%
\bibitem [{\citenamefont {Hackett}\ \emph {et~al.}(2019)\citenamefont
  {Hackett}, \citenamefont {Howe}, \citenamefont {Hughes}, \citenamefont {Jay},
  \citenamefont {Neil},\ and\ \citenamefont {Simone}}]{Hackett:2018cel}%
  \BibitemOpen
  \bibfield  {author} {\bibinfo {author} {\bibfnamefont {Daniel~C.}\
  \bibnamefont {Hackett}}, \bibinfo {author} {\bibfnamefont {Kiel}\
  \bibnamefont {Howe}}, \bibinfo {author} {\bibfnamefont {Ciaran}\ \bibnamefont
  {Hughes}}, \bibinfo {author} {\bibfnamefont {William}\ \bibnamefont {Jay}},
  \bibinfo {author} {\bibfnamefont {Ethan~T.}\ \bibnamefont {Neil}}, \ and\
  \bibinfo {author} {\bibfnamefont {James~N.}\ \bibnamefont {Simone}},\
  }\bibfield  {title} {\enquote {\bibinfo {title} {{Digitizing Gauge Fields:
  Lattice Monte Carlo Results for Future Quantum Computers}},}\ }\href
  {\doibase 10.1103/PhysRevA.99.062341} {\bibfield  {journal} {\bibinfo
  {journal} {Phys.\ Rev.\ A}\ }\textbf {\bibinfo {volume} {99}},\ \bibinfo
  {pages} {062341} (\bibinfo {year} {2019})},\ \Eprint
  {http://arxiv.org/abs/1811.03629} {arXiv:1811.03629 [quant-ph]} \BibitemShut
  {NoStop}%
\bibitem [{\citenamefont {Alexandru}\ \emph {et~al.}(2019)\citenamefont
  {Alexandru}, \citenamefont {Bedaque}, \citenamefont {Harmalkar},
  \citenamefont {Lamm}, \citenamefont {Lawrence},\ and\ \citenamefont
  {Warrington}}]{Alexandru:2019nsa}%
  \BibitemOpen
  \bibfield  {author} {\bibinfo {author} {\bibfnamefont {Andrei}\ \bibnamefont
  {Alexandru}}, \bibinfo {author} {\bibfnamefont {Paulo~F.}\ \bibnamefont
  {Bedaque}}, \bibinfo {author} {\bibfnamefont {Siddhartha}\ \bibnamefont
  {Harmalkar}}, \bibinfo {author} {\bibfnamefont {Henry}\ \bibnamefont {Lamm}},
  \bibinfo {author} {\bibfnamefont {Scott}\ \bibnamefont {Lawrence}}, \ and\
  \bibinfo {author} {\bibfnamefont {Neill~C.}\ \bibnamefont {Warrington}}
  (\bibinfo {collaboration} {NuQS}),\ }\bibfield  {title} {\enquote {\bibinfo
  {title} {Gluon field digitization for quantum computers},}\ }\href {\doibase
  10.1103/PhysRevD.100.114501} {\bibfield  {journal} {\bibinfo  {journal}
  {Phys.Rev.D}\ }\textbf {\bibinfo {volume} {100}},\ \bibinfo {pages} {114501}
  (\bibinfo {year} {2019})},\ \Eprint {http://arxiv.org/abs/1906.11213}
  {arXiv:1906.11213 [hep-lat]} \BibitemShut {NoStop}%
\bibitem [{\citenamefont {Singh}\ and\ \citenamefont
  {Chandrasekharan}(2019)}]{Singh:2019uwd}%
  \BibitemOpen
  \bibfield  {author} {\bibinfo {author} {\bibfnamefont {Hersh}\ \bibnamefont
  {Singh}}\ and\ \bibinfo {author} {\bibfnamefont {Shailesh}\ \bibnamefont
  {Chandrasekharan}},\ }\bibfield  {title} {\enquote {\bibinfo {title} {{Qubit
  regularization of the $O(3)$ sigma model}},}\ }\href {\doibase
  10.1103/PhysRevD.100.054505} {\bibfield  {journal} {\bibinfo  {journal}
  {Phys. Rev. D}\ }\textbf {\bibinfo {volume} {100}},\ \bibinfo {pages}
  {054505} (\bibinfo {year} {2019})},\ \Eprint
  {http://arxiv.org/abs/1905.13204} {arXiv:1905.13204 [hep-lat]} \BibitemShut
  {NoStop}%
\bibitem [{\citenamefont {Singh}(2019)}]{Singh:2019jog}%
  \BibitemOpen
  \bibfield  {author} {\bibinfo {author} {\bibfnamefont {Hersh}\ \bibnamefont
  {Singh}},\ }\href@noop {} {\enquote {\bibinfo {title} {{Qubit $O(N)$
  nonlinear sigma models}},}\ } (\bibinfo {year} {2019}),\ \Eprint
  {http://arxiv.org/abs/1911.12353} {arXiv:1911.12353 [hep-lat]} \BibitemShut
  {NoStop}%
\bibitem [{\citenamefont {Davoudi}\ \emph
  {et~al.}(2020{\natexlab{b}})\citenamefont {Davoudi}, \citenamefont
  {Raychowdhury},\ and\ \citenamefont {Shaw}}]{Davoudi:2020yln}%
  \BibitemOpen
  \bibfield  {author} {\bibinfo {author} {\bibfnamefont {Zohreh}\ \bibnamefont
  {Davoudi}}, \bibinfo {author} {\bibfnamefont {Indrakshi}\ \bibnamefont
  {Raychowdhury}}, \ and\ \bibinfo {author} {\bibfnamefont {Andrew}\
  \bibnamefont {Shaw}},\ }\href@noop {} {\enquote {\bibinfo {title} {{Search
  for Efficient Formulations for Hamiltonian Simulation of non-Abelian Lattice
  Gauge Theories}},}\ } (\bibinfo {year} {2020}{\natexlab{b}}),\ \Eprint
  {http://arxiv.org/abs/2009.11802} {arXiv:2009.11802 [hep-lat]} \BibitemShut
  {NoStop}%
\bibitem [{\citenamefont {Alexandru}\ \emph {et~al.}(2022)\citenamefont
  {Alexandru}, \citenamefont {Bedaque}, \citenamefont {Carosso},\ and\
  \citenamefont {Sheng}}]{alexandru2022universality}%
  \BibitemOpen
  \bibfield  {author} {\bibinfo {author} {\bibfnamefont {Andrei}\ \bibnamefont
  {Alexandru}}, \bibinfo {author} {\bibfnamefont {Paulo~F}\ \bibnamefont
  {Bedaque}}, \bibinfo {author} {\bibfnamefont {Andrea}\ \bibnamefont
  {Carosso}}, \ and\ \bibinfo {author} {\bibfnamefont {Andy}\ \bibnamefont
  {Sheng}},\ }\bibfield  {title} {\enquote {\bibinfo {title} {Universality of a
  truncated sigma-model},}\ }\href@noop {} {\bibfield  {journal} {\bibinfo
  {journal} {Physics Letters B}\ ,\ \bibinfo {pages} {137230}} (\bibinfo {year}
  {2022})}\BibitemShut {NoStop}%
\bibitem [{\citenamefont {Barata}\ \emph {et~al.}(2020)\citenamefont {Barata},
  \citenamefont {Mueller}, \citenamefont {Tarasov},\ and\ \citenamefont
  {Venugopalan}}]{Barata:2020jtq}%
  \BibitemOpen
  \bibfield  {author} {\bibinfo {author} {\bibfnamefont {Jo\~ao}\ \bibnamefont
  {Barata}}, \bibinfo {author} {\bibfnamefont {Niklas}\ \bibnamefont
  {Mueller}}, \bibinfo {author} {\bibfnamefont {Andrey}\ \bibnamefont
  {Tarasov}}, \ and\ \bibinfo {author} {\bibfnamefont {Raju}\ \bibnamefont
  {Venugopalan}},\ }\href@noop {} {\enquote {\bibinfo {title} {{Single-particle
  digitization strategy for quantum computation of a $\phi^4$ scalar field
  theory}},}\ } (\bibinfo {year} {2020}),\ \Eprint
  {http://arxiv.org/abs/2012.00020} {arXiv:2012.00020 [hep-th]} \BibitemShut
  {NoStop}%
\bibitem [{\citenamefont {Kreshchuk}\ \emph
  {et~al.}(2020{\natexlab{a}})\citenamefont {Kreshchuk}, \citenamefont {Jia},
  \citenamefont {Kirby}, \citenamefont {Goldstein}, \citenamefont {Vary},\ and\
  \citenamefont {Love}}]{Kreshchuk:2020kcz}%
  \BibitemOpen
  \bibfield  {author} {\bibinfo {author} {\bibfnamefont {Michael}\ \bibnamefont
  {Kreshchuk}}, \bibinfo {author} {\bibfnamefont {Shaoyang}\ \bibnamefont
  {Jia}}, \bibinfo {author} {\bibfnamefont {William~M.}\ \bibnamefont {Kirby}},
  \bibinfo {author} {\bibfnamefont {Gary}\ \bibnamefont {Goldstein}}, \bibinfo
  {author} {\bibfnamefont {James~P.}\ \bibnamefont {Vary}}, \ and\ \bibinfo
  {author} {\bibfnamefont {Peter~J.}\ \bibnamefont {Love}},\ }\href@noop {}
  {\enquote {\bibinfo {title} {{Light-Front Field Theory on Current Quantum
  Computers}},}\ } (\bibinfo {year} {2020}{\natexlab{a}}),\ \Eprint
  {http://arxiv.org/abs/2009.07885} {arXiv:2009.07885 [quant-ph]} \BibitemShut
  {NoStop}%
\bibitem [{\citenamefont {Ji}\ \emph {et~al.}(2020)\citenamefont {Ji},
  \citenamefont {Lamm},\ and\ \citenamefont {Zhu}}]{Ji:2020kjk}%
  \BibitemOpen
  \bibfield  {author} {\bibinfo {author} {\bibfnamefont {Yao}\ \bibnamefont
  {Ji}}, \bibinfo {author} {\bibfnamefont {Henry}\ \bibnamefont {Lamm}}, \ and\
  \bibinfo {author} {\bibfnamefont {Shuchen}\ \bibnamefont {Zhu}} (\bibinfo
  {collaboration} {NuQS}),\ }\bibfield  {title} {\enquote {\bibinfo {title}
  {{Gluon Field Digitization via Group Space Decimation for Quantum
  Computers}},}\ }\href {\doibase 10.1103/PhysRevD.102.114513} {\bibfield
  {journal} {\bibinfo  {journal} {Phys. Rev. D}\ }\textbf {\bibinfo {volume}
  {102}},\ \bibinfo {pages} {114513} (\bibinfo {year} {2020})},\ \Eprint
  {http://arxiv.org/abs/2005.14221} {arXiv:2005.14221 [hep-lat]} \BibitemShut
  {NoStop}%
\bibitem [{\citenamefont {Zohar}\ \emph
  {et~al.}(2013{\natexlab{a}})\citenamefont {Zohar}, \citenamefont {Cirac},\
  and\ \citenamefont {Reznik}}]{Zohar:2013zla}%
  \BibitemOpen
  \bibfield  {author} {\bibinfo {author} {\bibfnamefont {Erez}\ \bibnamefont
  {Zohar}}, \bibinfo {author} {\bibfnamefont {J.~Ignacio}\ \bibnamefont
  {Cirac}}, \ and\ \bibinfo {author} {\bibfnamefont {Benni}\ \bibnamefont
  {Reznik}},\ }\bibfield  {title} {\enquote {\bibinfo {title} {{Quantum
  simulations of gauge theories with ultracold atoms: local gauge invariance
  from angular momentum conservation}},}\ }\href {\doibase
  10.1103/PhysRevA.88.023617} {\bibfield  {journal} {\bibinfo  {journal} {Phys.
  Rev.}\ }\textbf {\bibinfo {volume} {A88}},\ \bibinfo {pages} {023617}
  (\bibinfo {year} {2013}{\natexlab{a}})},\ \Eprint
  {http://arxiv.org/abs/1303.5040} {arXiv:1303.5040 [quant-ph]} \BibitemShut
  {NoStop}%
\bibitem [{\citenamefont {Hasenfratz}\ and\ \citenamefont
  {Niedermayer}(2001{\natexlab{a}})}]{Hasenfratz:2001iz}%
  \BibitemOpen
  \bibfield  {author} {\bibinfo {author} {\bibfnamefont {Peter}\ \bibnamefont
  {Hasenfratz}}\ and\ \bibinfo {author} {\bibfnamefont {Ferenc}\ \bibnamefont
  {Niedermayer}},\ }\bibfield  {title} {\enquote {\bibinfo {title} {{Asymptotic
  freedom with discrete spin variables?}}}\ }\bibfield  {booktitle} {\emph
  {\bibinfo {booktitle} {{Proceedings, 2001 Europhysics Conference on High
  Energy Physics (EPS-HEP 2001): Budapest, Hungary, July 12-18, 2001}}},\
  }\href {\doibase 10.22323/1.007.0229} {\bibfield  {journal} {\bibinfo
  {journal} {PoS}\ }\textbf {\bibinfo {volume} {HEP2001}},\ \bibinfo {pages}
  {229} (\bibinfo {year} {2001}{\natexlab{a}})},\ \Eprint
  {http://arxiv.org/abs/hep-lat/0112003} {arXiv:hep-lat/0112003 [hep-lat]}
  \BibitemShut {NoStop}%
\bibitem [{\citenamefont {Caracciolo}\ \emph
  {et~al.}(2001{\natexlab{a}})\citenamefont {Caracciolo}, \citenamefont
  {Montanari},\ and\ \citenamefont {Pelissetto}}]{Caracciolo:2001jd}%
  \BibitemOpen
  \bibfield  {author} {\bibinfo {author} {\bibfnamefont {Sergio}\ \bibnamefont
  {Caracciolo}}, \bibinfo {author} {\bibfnamefont {Andrea}\ \bibnamefont
  {Montanari}}, \ and\ \bibinfo {author} {\bibfnamefont {Andrea}\ \bibnamefont
  {Pelissetto}},\ }\bibfield  {title} {\enquote {\bibinfo {title}
  {{Asymptotically free models and discrete nonAbelian groups}},}\ }\href
  {\doibase 10.1016/S0370-2693(01)00674-8} {\bibfield  {journal} {\bibinfo
  {journal} {Phys. Lett.}\ }\textbf {\bibinfo {volume} {B513}},\ \bibinfo
  {pages} {223--231} (\bibinfo {year} {2001}{\natexlab{a}})},\ \Eprint
  {http://arxiv.org/abs/hep-lat/0103017} {arXiv:hep-lat/0103017 [hep-lat]}
  \BibitemShut {NoStop}%
\bibitem [{\citenamefont {Hasenfratz}\ and\ \citenamefont
  {Niedermayer}(2001{\natexlab{b}})}]{Hasenfratz:2000hd}%
  \BibitemOpen
  \bibfield  {author} {\bibinfo {author} {\bibfnamefont {Peter}\ \bibnamefont
  {Hasenfratz}}\ and\ \bibinfo {author} {\bibfnamefont {Ferenc}\ \bibnamefont
  {Niedermayer}},\ }\bibfield  {title} {\enquote {\bibinfo {title}
  {{Asymptotically free theories based on discrete subgroups}},}\ }\bibfield
  {booktitle} {\emph {\bibinfo {booktitle} {{Lattice field theory. Proceedings,
  18th International Symposium, Lattice 2000, Bangalore, India, August 17-22,
  2000}}},\ }\href {\doibase 10.1016/S0920-5632(01)00870-2} {\bibfield
  {journal} {\bibinfo  {journal} {Nucl. Phys. Proc. Suppl.}\ }\textbf {\bibinfo
  {volume} {94}},\ \bibinfo {pages} {575--578} (\bibinfo {year}
  {2001}{\natexlab{b}})},\ \Eprint {http://arxiv.org/abs/hep-lat/0011056}
  {arXiv:hep-lat/0011056 [hep-lat]} \BibitemShut {NoStop}%
\bibitem [{\citenamefont {Patrascioiu}\ and\ \citenamefont
  {Seiler}(1998)}]{PhysRevE.57.111}%
  \BibitemOpen
  \bibfield  {author} {\bibinfo {author} {\bibfnamefont {Adrian}\ \bibnamefont
  {Patrascioiu}}\ and\ \bibinfo {author} {\bibfnamefont {Erhard}\ \bibnamefont
  {Seiler}},\ }\bibfield  {title} {\enquote {\bibinfo {title} {Continuum limit
  of two-dimensional spin models with continuous symmetry and conformal quantum
  field theory},}\ }\href {\doibase 10.1103/PhysRevE.57.111} {\bibfield
  {journal} {\bibinfo  {journal} {Phys. Rev. E}\ }\textbf {\bibinfo {volume}
  {57}},\ \bibinfo {pages} {111--119} (\bibinfo {year} {1998})}\BibitemShut
  {NoStop}%
\bibitem [{\citenamefont {Krcmar}\ \emph {et~al.}(2016)\citenamefont {Krcmar},
  \citenamefont {Gendiar},\ and\ \citenamefont {Nishino}}]{PhysRevE.94.022134}%
  \BibitemOpen
  \bibfield  {author} {\bibinfo {author} {\bibfnamefont {Roman}\ \bibnamefont
  {Krcmar}}, \bibinfo {author} {\bibfnamefont {Andrej}\ \bibnamefont
  {Gendiar}}, \ and\ \bibinfo {author} {\bibfnamefont {Tomotoshi}\ \bibnamefont
  {Nishino}},\ }\bibfield  {title} {\enquote {\bibinfo {title} {Phase diagram
  of a truncated tetrahedral model},}\ }\href {\doibase
  10.1103/PhysRevE.94.022134} {\bibfield  {journal} {\bibinfo  {journal} {Phys.
  Rev. E}\ }\textbf {\bibinfo {volume} {94}},\ \bibinfo {pages} {022134}
  (\bibinfo {year} {2016})}\BibitemShut {NoStop}%
\bibitem [{\citenamefont {Caracciolo}\ \emph
  {et~al.}(2001{\natexlab{b}})\citenamefont {Caracciolo}, \citenamefont
  {Montanari},\ and\ \citenamefont {Pelissetto}}]{article}%
  \BibitemOpen
  \bibfield  {author} {\bibinfo {author} {\bibfnamefont {Sergio}\ \bibnamefont
  {Caracciolo}}, \bibinfo {author} {\bibfnamefont {Andrea}\ \bibnamefont
  {Montanari}}, \ and\ \bibinfo {author} {\bibfnamefont {Andrea}\ \bibnamefont
  {Pelissetto}},\ }\bibfield  {title} {\enquote {\bibinfo {title}
  {Asymptotically free models and discrete non-abelian groups},}\ }\href
  {\doibase 10.1016/S0370-2693(01)00674-8} {\bibfield  {journal} {\bibinfo
  {journal} {Physics Letters B}\ }\textbf {\bibinfo {volume} {513}},\ \bibinfo
  {pages} {223--231} (\bibinfo {year} {2001}{\natexlab{b}})}\BibitemShut
  {NoStop}%
\bibitem [{\citenamefont {Wilson}\ \emph {et~al.}(1994)\citenamefont {Wilson},
  \citenamefont {Walhout}, \citenamefont {Harindranath}, \citenamefont {Zhang},
  \citenamefont {Perry},\ and\ \citenamefont {Glazek}}]{Wilson:1994fk}%
  \BibitemOpen
  \bibfield  {author} {\bibinfo {author} {\bibfnamefont {Kenneth~G.}\
  \bibnamefont {Wilson}}, \bibinfo {author} {\bibfnamefont {Timothy~S.}\
  \bibnamefont {Walhout}}, \bibinfo {author} {\bibfnamefont {Avaroth}\
  \bibnamefont {Harindranath}}, \bibinfo {author} {\bibfnamefont {Wei-Min}\
  \bibnamefont {Zhang}}, \bibinfo {author} {\bibfnamefont {Robert~J.}\
  \bibnamefont {Perry}}, \ and\ \bibinfo {author} {\bibfnamefont
  {Stanislaw~D.}\ \bibnamefont {Glazek}},\ }\bibfield  {title} {\enquote
  {\bibinfo {title} {{Nonperturbative QCD: A Weak coupling treatment on the
  light front}},}\ }\href {\doibase 10.1103/PhysRevD.49.6720} {\bibfield
  {journal} {\bibinfo  {journal} {Phys. Rev. D}\ }\textbf {\bibinfo {volume}
  {49}},\ \bibinfo {pages} {6720--6766} (\bibinfo {year} {1994})},\ \Eprint
  {http://arxiv.org/abs/hep-th/9401153} {arXiv:hep-th/9401153} \BibitemShut
  {NoStop}%
\bibitem [{\citenamefont {Gustafson}\ \emph {et~al.}(2020)\citenamefont
  {Gustafson}, \citenamefont {Kawai}, \citenamefont {Lamm}, \citenamefont
  {Raychowdhury}, \citenamefont {Singh}, \citenamefont {Stryker},\ and\
  \citenamefont {Unmuth-Yockey}}]{digi_loi}%
  \BibitemOpen
  \bibfield  {author} {\bibinfo {author} {\bibfnamefont {Erik}\ \bibnamefont
  {Gustafson}}, \bibinfo {author} {\bibfnamefont {Hiroki}\ \bibnamefont
  {Kawai}}, \bibinfo {author} {\bibfnamefont {Henry}\ \bibnamefont {Lamm}},
  \bibinfo {author} {\bibfnamefont {Indrakshi}\ \bibnamefont {Raychowdhury}},
  \bibinfo {author} {\bibfnamefont {Hersh}\ \bibnamefont {Singh}}, \bibinfo
  {author} {\bibfnamefont {Jesse}\ \bibnamefont {Stryker}}, \ and\ \bibinfo
  {author} {\bibfnamefont {Judah}\ \bibnamefont {Unmuth-Yockey}},\ }\bibfield
  {title} {\enquote {\bibinfo {title} {{Exploring Digitizations of Quantum
  Fields for Quantum Devices}},}\ }\href@noop {} {\bibfield  {journal}
  {\bibinfo  {journal}
  {\href{https://www.snowmass21.org/docs/files/summaries/TF/SNOWMASS21-TF10_TF0-CompF6_CompF0_Hank_Lamm-079.pdf}{Snowmass
  2021 LOI}}\ }\textbf {\bibinfo {volume} {TF10-97}} (\bibinfo {year}
  {2020})}\BibitemShut {NoStop}%
\bibitem [{\citenamefont {Zohar}\ \emph
  {et~al.}(2012{\natexlab{a}})\citenamefont {Zohar}, \citenamefont {Cirac},\
  and\ \citenamefont {Reznik}}]{Zohar:2012ay}%
  \BibitemOpen
  \bibfield  {author} {\bibinfo {author} {\bibfnamefont {Erez}\ \bibnamefont
  {Zohar}}, \bibinfo {author} {\bibfnamefont {J.~Ignacio}\ \bibnamefont
  {Cirac}}, \ and\ \bibinfo {author} {\bibfnamefont {Benni}\ \bibnamefont
  {Reznik}},\ }\bibfield  {title} {\enquote {\bibinfo {title} {{Simulating
  Compact Quantum Electrodynamics with ultracold atoms: Probing confinement and
  nonperturbative effects}},}\ }\href {\doibase 10.1103/PhysRevLett.109.125302}
  {\bibfield  {journal} {\bibinfo  {journal} {Phys. Rev. Lett.}\ }\textbf
  {\bibinfo {volume} {109}},\ \bibinfo {pages} {125302} (\bibinfo {year}
  {2012}{\natexlab{a}})},\ \Eprint {http://arxiv.org/abs/1204.6574}
  {arXiv:1204.6574 [quant-ph]} \BibitemShut {NoStop}%
\bibitem [{\citenamefont {Zohar}\ \emph
  {et~al.}(2013{\natexlab{b}})\citenamefont {Zohar}, \citenamefont {Cirac},\
  and\ \citenamefont {Reznik}}]{Zohar:2012xf}%
  \BibitemOpen
  \bibfield  {author} {\bibinfo {author} {\bibfnamefont {Erez}\ \bibnamefont
  {Zohar}}, \bibinfo {author} {\bibfnamefont {J.~Ignacio}\ \bibnamefont
  {Cirac}}, \ and\ \bibinfo {author} {\bibfnamefont {Benni}\ \bibnamefont
  {Reznik}},\ }\bibfield  {title} {\enquote {\bibinfo {title} {{Cold-Atom
  Quantum Simulator for SU(2) Yang-Mills Lattice Gauge Theory}},}\ }\href
  {\doibase 10.1103/PhysRevLett.110.125304} {\bibfield  {journal} {\bibinfo
  {journal} {Phys. Rev. Lett.}\ }\textbf {\bibinfo {volume} {110}},\ \bibinfo
  {pages} {125304} (\bibinfo {year} {2013}{\natexlab{b}})},\ \Eprint
  {http://arxiv.org/abs/1211.2241} {arXiv:1211.2241 [quant-ph]} \BibitemShut
  {NoStop}%
\bibitem [{\citenamefont {Zohar}\ and\ \citenamefont
  {Burrello}(2015)}]{Zohar:2014qma}%
  \BibitemOpen
  \bibfield  {author} {\bibinfo {author} {\bibfnamefont {Erez}\ \bibnamefont
  {Zohar}}\ and\ \bibinfo {author} {\bibfnamefont {Michele}\ \bibnamefont
  {Burrello}},\ }\bibfield  {title} {\enquote {\bibinfo {title} {{Formulation
  of lattice gauge theories for quantum simulations}},}\ }\href {\doibase
  10.1103/PhysRevD.91.054506} {\bibfield  {journal} {\bibinfo  {journal} {Phys.
  Rev.}\ }\textbf {\bibinfo {volume} {D91}},\ \bibinfo {pages} {054506}
  (\bibinfo {year} {2015})},\ \Eprint {http://arxiv.org/abs/1409.3085}
  {arXiv:1409.3085 [quant-ph]} \BibitemShut {NoStop}%
\bibitem [{\citenamefont {Zohar}\ \emph {et~al.}(2017)\citenamefont {Zohar},
  \citenamefont {Farace}, \citenamefont {Reznik},\ and\ \citenamefont
  {Cirac}}]{Zohar:2016iic}%
  \BibitemOpen
  \bibfield  {author} {\bibinfo {author} {\bibfnamefont {Erez}\ \bibnamefont
  {Zohar}}, \bibinfo {author} {\bibfnamefont {Alessandro}\ \bibnamefont
  {Farace}}, \bibinfo {author} {\bibfnamefont {Benni}\ \bibnamefont {Reznik}},
  \ and\ \bibinfo {author} {\bibfnamefont {J.~Ignacio}\ \bibnamefont {Cirac}},\
  }\bibfield  {title} {\enquote {\bibinfo {title} {{Digital lattice gauge
  theories}},}\ }\href {\doibase 10.1103/PhysRevA.95.023604} {\bibfield
  {journal} {\bibinfo  {journal} {Phys. Rev.}\ }\textbf {\bibinfo {volume}
  {A95}},\ \bibinfo {pages} {023604} (\bibinfo {year} {2017})},\ \Eprint
  {http://arxiv.org/abs/1607.08121} {arXiv:1607.08121 [quant-ph]} \BibitemShut
  {NoStop}%
\bibitem [{\citenamefont {Klco}\ \emph {et~al.}(2020)\citenamefont {Klco},
  \citenamefont {Stryker},\ and\ \citenamefont {Savage}}]{Klco:2019evd}%
  \BibitemOpen
  \bibfield  {author} {\bibinfo {author} {\bibfnamefont {Natalie}\ \bibnamefont
  {Klco}}, \bibinfo {author} {\bibfnamefont {Jesse~R.}\ \bibnamefont
  {Stryker}}, \ and\ \bibinfo {author} {\bibfnamefont {Martin~J.}\ \bibnamefont
  {Savage}},\ }\bibfield  {title} {\enquote {\bibinfo {title} {{SU(2)
  non-Abelian gauge field theory in one dimension on digital quantum
  computers}},}\ }\href {\doibase 10.1103/PhysRevD.101.074512} {\bibfield
  {journal} {\bibinfo  {journal} {Phys. Rev. D}\ }\textbf {\bibinfo {volume}
  {101}},\ \bibinfo {pages} {074512} (\bibinfo {year} {2020})},\ \Eprint
  {http://arxiv.org/abs/1908.06935} {arXiv:1908.06935 [quant-ph]} \BibitemShut
  {NoStop}%
\bibitem [{\citenamefont {Ciavarella}\ \emph
  {et~al.}(2021{\natexlab{b}})\citenamefont {Ciavarella}, \citenamefont
  {Klco},\ and\ \citenamefont {Savage}}]{Ciavarella:2021nmj}%
  \BibitemOpen
  \bibfield  {author} {\bibinfo {author} {\bibfnamefont {Anthony}\ \bibnamefont
  {Ciavarella}}, \bibinfo {author} {\bibfnamefont {Natalie}\ \bibnamefont
  {Klco}}, \ and\ \bibinfo {author} {\bibfnamefont {Martin~J.}\ \bibnamefont
  {Savage}},\ }\href@noop {} {\enquote {\bibinfo {title} {{A Trailhead for
  Quantum Simulation of SU(3) Yang-Mills Lattice Gauge Theory in the Local
  Multiplet Basis}},}\ } (\bibinfo {year} {2021}{\natexlab{b}}),\ \Eprint
  {http://arxiv.org/abs/2101.10227} {arXiv:2101.10227 [quant-ph]} \BibitemShut
  {NoStop}%
\bibitem [{\citenamefont {Bender}\ \emph
  {et~al.}(2018{\natexlab{a}})\citenamefont {Bender}, \citenamefont {Zohar},
  \citenamefont {Farace},\ and\ \citenamefont {Cirac}}]{Bender:2018rdp}%
  \BibitemOpen
  \bibfield  {author} {\bibinfo {author} {\bibfnamefont {Julian}\ \bibnamefont
  {Bender}}, \bibinfo {author} {\bibfnamefont {Erez}\ \bibnamefont {Zohar}},
  \bibinfo {author} {\bibfnamefont {Alessandro}\ \bibnamefont {Farace}}, \ and\
  \bibinfo {author} {\bibfnamefont {J.~Ignacio}\ \bibnamefont {Cirac}},\
  }\bibfield  {title} {\enquote {\bibinfo {title} {{Digital quantum simulation
  of lattice gauge theories in three spatial dimensions}},}\ }\href {\doibase
  10.1088/1367-2630/aadb71} {\bibfield  {journal} {\bibinfo  {journal} {New J.
  Phys.}\ }\textbf {\bibinfo {volume} {20}},\ \bibinfo {pages} {093001}
  (\bibinfo {year} {2018}{\natexlab{a}})},\ \Eprint
  {http://arxiv.org/abs/1804.02082} {arXiv:1804.02082 [quant-ph]} \BibitemShut
  {NoStop}%
\bibitem [{\citenamefont {Liu}\ and\ \citenamefont {Xin}(2020)}]{Liu:2020eoa}%
  \BibitemOpen
  \bibfield  {author} {\bibinfo {author} {\bibfnamefont {Junyu}\ \bibnamefont
  {Liu}}\ and\ \bibinfo {author} {\bibfnamefont {Yuan}\ \bibnamefont {Xin}},\
  }\href@noop {} {\enquote {\bibinfo {title} {{Quantum simulation of quantum
  field theories as quantum chemistry}},}\ } (\bibinfo {year} {2020}),\ \Eprint
  {http://arxiv.org/abs/2004.13234} {arXiv:2004.13234 [hep-th]} \BibitemShut
  {NoStop}%
\bibitem [{\citenamefont {Yamamoto}(2021)}]{Yamamoto:2020eqi}%
  \BibitemOpen
  \bibfield  {author} {\bibinfo {author} {\bibfnamefont {Arata}\ \bibnamefont
  {Yamamoto}},\ }\bibfield  {title} {\enquote {\bibinfo {title} {{Real-time
  simulation of (2+1)-dimensional lattice gauge theory on qubits}},}\ }\href
  {\doibase 10.1093/ptep/ptaa171} {\bibfield  {journal} {\bibinfo  {journal}
  {PTEP}\ }\textbf {\bibinfo {volume} {2021}},\ \bibinfo {pages} {013B06}
  (\bibinfo {year} {2021})},\ \Eprint {http://arxiv.org/abs/2008.11395}
  {arXiv:2008.11395 [hep-lat]} \BibitemShut {NoStop}%
\bibitem [{\citenamefont {Haase}\ \emph {et~al.}(2020)\citenamefont {Haase},
  \citenamefont {Dellantonio}, \citenamefont {Celi}, \citenamefont {Paulson},
  \citenamefont {Kan}, \citenamefont {Jansen},\ and\ \citenamefont
  {Muschik}}]{Haase:2020kaj}%
  \BibitemOpen
  \bibfield  {author} {\bibinfo {author} {\bibfnamefont {Jan~F.}\ \bibnamefont
  {Haase}}, \bibinfo {author} {\bibfnamefont {Luca}\ \bibnamefont
  {Dellantonio}}, \bibinfo {author} {\bibfnamefont {Alessio}\ \bibnamefont
  {Celi}}, \bibinfo {author} {\bibfnamefont {Danny}\ \bibnamefont {Paulson}},
  \bibinfo {author} {\bibfnamefont {Angus}\ \bibnamefont {Kan}}, \bibinfo
  {author} {\bibfnamefont {Karl}\ \bibnamefont {Jansen}}, \ and\ \bibinfo
  {author} {\bibfnamefont {Christine~A.}\ \bibnamefont {Muschik}},\ }\href@noop
  {} {\enquote {\bibinfo {title} {{A resource efficient approach for quantum
  and classical simulations of gauge theories in particle physics}},}\ }
  (\bibinfo {year} {2020}),\ \Eprint {http://arxiv.org/abs/2006.14160}
  {arXiv:2006.14160 [quant-ph]} \BibitemShut {NoStop}%
\bibitem [{\citenamefont {Bazavov}\ \emph {et~al.}(2019)\citenamefont
  {Bazavov}, \citenamefont {Catterall}, \citenamefont {Jha},\ and\
  \citenamefont {Unmuth-Yockey}}]{PhysRevD.99.114507}%
  \BibitemOpen
  \bibfield  {author} {\bibinfo {author} {\bibfnamefont {Alexei}\ \bibnamefont
  {Bazavov}}, \bibinfo {author} {\bibfnamefont {Simon}\ \bibnamefont
  {Catterall}}, \bibinfo {author} {\bibfnamefont {Raghav~G.}\ \bibnamefont
  {Jha}}, \ and\ \bibinfo {author} {\bibfnamefont {Judah}\ \bibnamefont
  {Unmuth-Yockey}},\ }\bibfield  {title} {\enquote {\bibinfo {title} {Tensor
  renormalization group study of the non-abelian higgs model in two
  dimensions},}\ }\href {\doibase 10.1103/PhysRevD.99.114507} {\bibfield
  {journal} {\bibinfo  {journal} {Phys. Rev. D}\ }\textbf {\bibinfo {volume}
  {99}},\ \bibinfo {pages} {114507} (\bibinfo {year} {2019})}\BibitemShut
  {NoStop}%
\bibitem [{\citenamefont {Zhang}\ \emph {et~al.}(2018)\citenamefont {Zhang},
  \citenamefont {Unmuth-Yockey}, \citenamefont {Zeiher}, \citenamefont
  {Bazavov}, \citenamefont {Tsai},\ and\ \citenamefont
  {Meurice}}]{Zhang:2018ufj}%
  \BibitemOpen
  \bibfield  {author} {\bibinfo {author} {\bibfnamefont {Jin}\ \bibnamefont
  {Zhang}}, \bibinfo {author} {\bibfnamefont {J.}~\bibnamefont
  {Unmuth-Yockey}}, \bibinfo {author} {\bibfnamefont {J.}~\bibnamefont
  {Zeiher}}, \bibinfo {author} {\bibfnamefont {A.}~\bibnamefont {Bazavov}},
  \bibinfo {author} {\bibfnamefont {S.~W.}\ \bibnamefont {Tsai}}, \ and\
  \bibinfo {author} {\bibfnamefont {Y.}~\bibnamefont {Meurice}},\ }\bibfield
  {title} {\enquote {\bibinfo {title} {{Quantum simulation of the universal
  features of the Polyakov loop}},}\ }\href {\doibase
  10.1103/PhysRevLett.121.223201} {\bibfield  {journal} {\bibinfo  {journal}
  {Phys. Rev. Lett.}\ }\textbf {\bibinfo {volume} {121}},\ \bibinfo {pages}
  {223201} (\bibinfo {year} {2018})},\ \Eprint
  {http://arxiv.org/abs/1803.11166} {arXiv:1803.11166 [hep-lat]} \BibitemShut
  {NoStop}%
\bibitem [{\citenamefont {Unmuth-Yockey}\ \emph {et~al.}(2018)\citenamefont
  {Unmuth-Yockey}, \citenamefont {Zhang}, \citenamefont {Bazavov},
  \citenamefont {Meurice},\ and\ \citenamefont {Tsai}}]{Unmuth-Yockey:2018ugm}%
  \BibitemOpen
  \bibfield  {author} {\bibinfo {author} {\bibfnamefont {Judah}\ \bibnamefont
  {Unmuth-Yockey}}, \bibinfo {author} {\bibfnamefont {Jin}\ \bibnamefont
  {Zhang}}, \bibinfo {author} {\bibfnamefont {Alexei}\ \bibnamefont {Bazavov}},
  \bibinfo {author} {\bibfnamefont {Yannick}\ \bibnamefont {Meurice}}, \ and\
  \bibinfo {author} {\bibfnamefont {Shan-Wen}\ \bibnamefont {Tsai}},\
  }\bibfield  {title} {\enquote {\bibinfo {title} {{Universal features of the
  Abelian Polyakov loop in 1+1 dimensions}},}\ }\href {\doibase
  10.1103/PhysRevD.98.094511} {\bibfield  {journal} {\bibinfo  {journal} {Phys.
  Rev.}\ }\textbf {\bibinfo {volume} {D98}},\ \bibinfo {pages} {094511}
  (\bibinfo {year} {2018})},\ \Eprint {http://arxiv.org/abs/1807.09186}
  {arXiv:1807.09186 [hep-lat]} \BibitemShut {NoStop}%
\bibitem [{\citenamefont {Unmuth-Yockey}(2019)}]{Unmuth-Yockey:2018xak}%
  \BibitemOpen
  \bibfield  {author} {\bibinfo {author} {\bibfnamefont {Judah~F.}\
  \bibnamefont {Unmuth-Yockey}},\ }\bibfield  {title} {\enquote {\bibinfo
  {title} {{Gauge-invariant rotor Hamiltonian from dual variables of 3D $U(1)$
  gauge theory}},}\ }\href {\doibase 10.1103/PhysRevD.99.074502} {\bibfield
  {journal} {\bibinfo  {journal} {Phys.\ Rev.\ D}\ }\textbf {\bibinfo {volume}
  {99}},\ \bibinfo {pages} {074502} (\bibinfo {year} {2019})},\ \Eprint
  {http://arxiv.org/abs/1811.05884} {arXiv:1811.05884 [hep-lat]} \BibitemShut
  {NoStop}%
\bibitem [{\citenamefont {Kreshchuk}\ \emph
  {et~al.}(2020{\natexlab{b}})\citenamefont {Kreshchuk}, \citenamefont {Kirby},
  \citenamefont {Goldstein}, \citenamefont {Beauchemin},\ and\ \citenamefont
  {Love}}]{Kreshchuk:2020dla}%
  \BibitemOpen
  \bibfield  {author} {\bibinfo {author} {\bibfnamefont {Michael}\ \bibnamefont
  {Kreshchuk}}, \bibinfo {author} {\bibfnamefont {William~M.}\ \bibnamefont
  {Kirby}}, \bibinfo {author} {\bibfnamefont {Gary}\ \bibnamefont {Goldstein}},
  \bibinfo {author} {\bibfnamefont {Hugo}\ \bibnamefont {Beauchemin}}, \ and\
  \bibinfo {author} {\bibfnamefont {Peter~J.}\ \bibnamefont {Love}},\
  }\href@noop {} {\enquote {\bibinfo {title} {{Quantum Simulation of Quantum
  Field Theory in the Light-Front Formulation}},}\ } (\bibinfo {year}
  {2020}{\natexlab{b}}),\ \Eprint {http://arxiv.org/abs/2002.04016}
  {arXiv:2002.04016 [quant-ph]} \BibitemShut {NoStop}%
\bibitem [{\citenamefont {Kreshchuk}\ \emph
  {et~al.}(2020{\natexlab{c}})\citenamefont {Kreshchuk}, \citenamefont {Jia},
  \citenamefont {Kirby}, \citenamefont {Goldstein}, \citenamefont {Vary},\ and\
  \citenamefont {Love}}]{Kreshchuk:2020aiq}%
  \BibitemOpen
  \bibfield  {author} {\bibinfo {author} {\bibfnamefont {Michael}\ \bibnamefont
  {Kreshchuk}}, \bibinfo {author} {\bibfnamefont {Shaoyang}\ \bibnamefont
  {Jia}}, \bibinfo {author} {\bibfnamefont {William~M.}\ \bibnamefont {Kirby}},
  \bibinfo {author} {\bibfnamefont {Gary}\ \bibnamefont {Goldstein}}, \bibinfo
  {author} {\bibfnamefont {James~P.}\ \bibnamefont {Vary}}, \ and\ \bibinfo
  {author} {\bibfnamefont {Peter~J.}\ \bibnamefont {Love}},\ }\href@noop {}
  {\enquote {\bibinfo {title} {{Simulating Hadronic Physics on NISQ devices
  using Basis Light-Front Quantization}},}\ } (\bibinfo {year}
  {2020}{\natexlab{c}}),\ \Eprint {http://arxiv.org/abs/2011.13443}
  {arXiv:2011.13443 [quant-ph]} \BibitemShut {NoStop}%
\bibitem [{\citenamefont {Raychowdhury}\ and\ \citenamefont
  {Stryker}(2020{\natexlab{b}})}]{Raychowdhury:2019iki}%
  \BibitemOpen
  \bibfield  {author} {\bibinfo {author} {\bibfnamefont {Indrakshi}\
  \bibnamefont {Raychowdhury}}\ and\ \bibinfo {author} {\bibfnamefont
  {Jesse~R.}\ \bibnamefont {Stryker}},\ }\bibfield  {title} {\enquote {\bibinfo
  {title} {{Loop, String, and Hadron Dynamics in SU(2) Hamiltonian Lattice
  Gauge Theories}},}\ }\href {\doibase 10.1103/PhysRevD.101.114502} {\bibfield
  {journal} {\bibinfo  {journal} {Phys. Rev. D}\ }\textbf {\bibinfo {volume}
  {101}},\ \bibinfo {pages} {114502} (\bibinfo {year} {2020}{\natexlab{b}})},\
  \Eprint {http://arxiv.org/abs/1912.06133} {arXiv:1912.06133 [hep-lat]}
  \BibitemShut {NoStop}%
\bibitem [{\citenamefont {Wiese}(2014)}]{Wiese:2014rla}%
  \BibitemOpen
  \bibfield  {author} {\bibinfo {author} {\bibfnamefont {Uwe-Jens}\
  \bibnamefont {Wiese}},\ }\bibfield  {title} {\enquote {\bibinfo {title}
  {{Towards Quantum Simulating QCD}},}\ }\bibfield  {booktitle} {\emph
  {\bibinfo {booktitle} {{Proceedings, 24th International Conference on
  Ultra-Relativistic Nucleus-Nucleus Collisions (Quark Matter 2014): Darmstadt,
  Germany, May 19-24, 2014}}},\ }\href {\doibase
  10.1016/j.nuclphysa.2014.09.102} {\bibfield  {journal} {\bibinfo  {journal}
  {Nucl. Phys.}\ }\textbf {\bibinfo {volume} {A931}},\ \bibinfo {pages}
  {246--256} (\bibinfo {year} {2014})},\ \Eprint
  {http://arxiv.org/abs/1409.7414} {arXiv:1409.7414 [hep-th]} \BibitemShut
  {NoStop}%
\bibitem [{\citenamefont {Luo}\ \emph {et~al.}(2019)\citenamefont {Luo},
  \citenamefont {Shen}, \citenamefont {Highman}, \citenamefont {Clark},
  \citenamefont {DeMarco}, \citenamefont {El-Khadra},\ and\ \citenamefont
  {Gadway}}]{Luo:2019vmi}%
  \BibitemOpen
  \bibfield  {author} {\bibinfo {author} {\bibfnamefont {Di}~\bibnamefont
  {Luo}}, \bibinfo {author} {\bibfnamefont {Jiayu}\ \bibnamefont {Shen}},
  \bibinfo {author} {\bibfnamefont {Michael}\ \bibnamefont {Highman}}, \bibinfo
  {author} {\bibfnamefont {Bryan~K.}\ \bibnamefont {Clark}}, \bibinfo {author}
  {\bibfnamefont {Brian}\ \bibnamefont {DeMarco}}, \bibinfo {author}
  {\bibfnamefont {Aida~X.}\ \bibnamefont {El-Khadra}}, \ and\ \bibinfo {author}
  {\bibfnamefont {Bryce}\ \bibnamefont {Gadway}},\ }\href@noop {} {\enquote
  {\bibinfo {title} {{A Framework for Simulating Gauge Theories with Dipolar
  Spin Systems}},}\ } (\bibinfo {year} {2019}),\ \Eprint
  {http://arxiv.org/abs/1912.11488} {arXiv:1912.11488 [quant-ph]} \BibitemShut
  {NoStop}%
\bibitem [{\citenamefont {Brower}\ \emph {et~al.}(2019)\citenamefont {Brower},
  \citenamefont {Berenstein},\ and\ \citenamefont {Kawai}}]{Brower:2020huh}%
  \BibitemOpen
  \bibfield  {author} {\bibinfo {author} {\bibfnamefont {Richard~C.}\
  \bibnamefont {Brower}}, \bibinfo {author} {\bibfnamefont {David}\
  \bibnamefont {Berenstein}}, \ and\ \bibinfo {author} {\bibfnamefont {Hiroki}\
  \bibnamefont {Kawai}},\ }\bibfield  {title} {\enquote {\bibinfo {title}
  {{Lattice Gauge Theory for a Quantum Computer}},}\ }\href@noop {} {\bibfield
  {journal} {\bibinfo  {journal} {PoS}\ }\textbf {\bibinfo {volume}
  {LATTICE2019}},\ \bibinfo {pages} {112} (\bibinfo {year} {2019})},\ \Eprint
  {http://arxiv.org/abs/2002.10028} {arXiv:2002.10028 [hep-lat]} \BibitemShut
  {NoStop}%
\bibitem [{\citenamefont {Mathis}\ \emph {et~al.}(2020)\citenamefont {Mathis},
  \citenamefont {Mazzola},\ and\ \citenamefont {Tavernelli}}]{Mathis:2020fuo}%
  \BibitemOpen
  \bibfield  {author} {\bibinfo {author} {\bibfnamefont {Simon~V.}\
  \bibnamefont {Mathis}}, \bibinfo {author} {\bibfnamefont {Guglielmo}\
  \bibnamefont {Mazzola}}, \ and\ \bibinfo {author} {\bibfnamefont {Ivano}\
  \bibnamefont {Tavernelli}},\ }\bibfield  {title} {\enquote {\bibinfo {title}
  {{Toward scalable simulations of Lattice Gauge Theories on quantum
  computers}},}\ }\href {\doibase 10.1103/PhysRevD.102.094501} {\bibfield
  {journal} {\bibinfo  {journal} {Phys. Rev. D}\ }\textbf {\bibinfo {volume}
  {102}},\ \bibinfo {pages} {094501} (\bibinfo {year} {2020})},\ \Eprint
  {http://arxiv.org/abs/2005.10271} {arXiv:2005.10271 [quant-ph]} \BibitemShut
  {NoStop}%
\bibitem [{\citenamefont {Buser}\ \emph {et~al.}(2020)\citenamefont {Buser},
  \citenamefont {Bhattacharya}, \citenamefont {Cincio},\ and\ \citenamefont
  {Gupta}}]{Buser:2020uzs}%
  \BibitemOpen
  \bibfield  {author} {\bibinfo {author} {\bibfnamefont {Alexander~J.}\
  \bibnamefont {Buser}}, \bibinfo {author} {\bibfnamefont {Tanmoy}\
  \bibnamefont {Bhattacharya}}, \bibinfo {author} {\bibfnamefont {Lukasz}\
  \bibnamefont {Cincio}}, \ and\ \bibinfo {author} {\bibfnamefont {Rajan}\
  \bibnamefont {Gupta}},\ }\href@noop {} {\enquote {\bibinfo {title} {{Quantum
  simulation of the qubit-regularized O(3)-sigma model}},}\ } (\bibinfo {year}
  {2020}),\ \Eprint {http://arxiv.org/abs/2006.15746} {arXiv:2006.15746
  [quant-ph]} \BibitemShut {NoStop}%
\bibitem [{\citenamefont {Meurice}(2019)}]{Meurice:2019ddf}%
  \BibitemOpen
  \bibfield  {author} {\bibinfo {author} {\bibfnamefont {Yannick}\ \bibnamefont
  {Meurice}},\ }\bibfield  {title} {\enquote {\bibinfo {title} {{Examples of
  symmetry-preserving truncations in tensor field theory}},}\ }\href {\doibase
  10.1103/PhysRevD.100.014506} {\bibfield  {journal} {\bibinfo  {journal}
  {Phys. Rev. D}\ }\textbf {\bibinfo {volume} {100}},\ \bibinfo {pages}
  {014506} (\bibinfo {year} {2019})},\ \Eprint
  {http://arxiv.org/abs/1903.01918} {arXiv:1903.01918 [hep-lat]} \BibitemShut
  {NoStop}%
\bibitem [{\citenamefont {Meurice}(2020)}]{Meurice:2020gcd}%
  \BibitemOpen
  \bibfield  {author} {\bibinfo {author} {\bibfnamefont {Yannick}\ \bibnamefont
  {Meurice}},\ }\bibfield  {title} {\enquote {\bibinfo {title} {{Discrete
  aspects of continuous symmetries in the tensorial formulation of Abelian
  gauge theories}},}\ }\href {\doibase 10.1103/PhysRevD.102.014506} {\bibfield
  {journal} {\bibinfo  {journal} {Phys. Rev. D}\ }\textbf {\bibinfo {volume}
  {102}},\ \bibinfo {pages} {014506} (\bibinfo {year} {2020})},\ \Eprint
  {http://arxiv.org/abs/2003.10986} {arXiv:2003.10986 [hep-lat]} \BibitemShut
  {NoStop}%
\bibitem [{\citenamefont {Peruzzo}\ \emph {et~al.}(2014)\citenamefont
  {Peruzzo}, \citenamefont {McClean}, \citenamefont {Shadbolt}, \citenamefont
  {Yung}, \citenamefont {Zhou}, \citenamefont {Love}, \citenamefont
  {Aspuru-Guzik},\ and\ \citenamefont {O’brien}}]{peruzzo2014variational}%
  \BibitemOpen
  \bibfield  {author} {\bibinfo {author} {\bibfnamefont {Alberto}\ \bibnamefont
  {Peruzzo}}, \bibinfo {author} {\bibfnamefont {Jarrod}\ \bibnamefont
  {McClean}}, \bibinfo {author} {\bibfnamefont {Peter}\ \bibnamefont
  {Shadbolt}}, \bibinfo {author} {\bibfnamefont {Man-Hong}\ \bibnamefont
  {Yung}}, \bibinfo {author} {\bibfnamefont {Xiao-Qi}\ \bibnamefont {Zhou}},
  \bibinfo {author} {\bibfnamefont {Peter~J}\ \bibnamefont {Love}}, \bibinfo
  {author} {\bibfnamefont {Al{\'a}n}\ \bibnamefont {Aspuru-Guzik}}, \ and\
  \bibinfo {author} {\bibfnamefont {Jeremy~L}\ \bibnamefont {O’brien}},\
  }\bibfield  {title} {\enquote {\bibinfo {title} {A variational eigenvalue
  solver on a photonic quantum processor},}\ }\href@noop {} {\bibfield
  {journal} {\bibinfo  {journal} {Nature communications}\ }\textbf {\bibinfo
  {volume} {5}},\ \bibinfo {pages} {4213} (\bibinfo {year} {2014})}\BibitemShut
  {NoStop}%
\bibitem [{\citenamefont {Kokail}\ \emph {et~al.}(2018)\citenamefont {Kokail}
  \emph {et~al.}}]{Kokail:2018eiw}%
  \BibitemOpen
  \bibfield  {author} {\bibinfo {author} {\bibfnamefont {Christian}\
  \bibnamefont {Kokail}} \emph {et~al.},\ }\href@noop {} {\enquote {\bibinfo
  {title} {{Self-Verifying Variational Quantum Simulation of the Lattice
  Schwinger Model}},}\ } (\bibinfo {year} {2018}),\ \Eprint
  {http://arxiv.org/abs/1810.03421} {arXiv:1810.03421 [quant-ph]} \BibitemShut
  {NoStop}%
\bibitem [{\citenamefont {Abrams}\ and\ \citenamefont
  {Lloyd}(1999)}]{Abrams:1998pd}%
  \BibitemOpen
  \bibfield  {author} {\bibinfo {author} {\bibfnamefont {Daniel~S.}\
  \bibnamefont {Abrams}}\ and\ \bibinfo {author} {\bibfnamefont {Seth}\
  \bibnamefont {Lloyd}},\ }\bibfield  {title} {\enquote {\bibinfo {title} {{A
  Quantum algorithm providing exponential speed increase for finding
  eigenvalues and eigenvectors}},}\ }\href {\doibase
  10.1103/PhysRevLett.83.5162} {\bibfield  {journal} {\bibinfo  {journal}
  {Phys. Rev. Lett.}\ }\textbf {\bibinfo {volume} {83}},\ \bibinfo {pages}
  {5162--5165} (\bibinfo {year} {1999})},\ \Eprint
  {http://arxiv.org/abs/quant-ph/9807070} {arXiv:quant-ph/9807070 [quant-ph]}
  \BibitemShut {NoStop}%
\bibitem [{\citenamefont {Nielsen}\ and\ \citenamefont
  {Chuang}(2000)}]{nielsen2000quantum}%
  \BibitemOpen
  \bibfield  {author} {\bibinfo {author} {\bibfnamefont {Michael~A}\
  \bibnamefont {Nielsen}}\ and\ \bibinfo {author} {\bibfnamefont {Isaac~L}\
  \bibnamefont {Chuang}},\ }\href@noop {} {\enquote {\bibinfo {title} {Quantum
  computation and quantum information},}\ } (\bibinfo {year}
  {2000})\BibitemShut {NoStop}%
\bibitem [{\citenamefont {Wiebe}\ and\ \citenamefont
  {Granade}(2016)}]{PhysRevLett.117.010503}%
  \BibitemOpen
  \bibfield  {author} {\bibinfo {author} {\bibfnamefont {Nathan}\ \bibnamefont
  {Wiebe}}\ and\ \bibinfo {author} {\bibfnamefont {Chris}\ \bibnamefont
  {Granade}},\ }\bibfield  {title} {\enquote {\bibinfo {title} {Efficient
  bayesian phase estimation},}\ }\href {\doibase
  10.1103/PhysRevLett.117.010503} {\bibfield  {journal} {\bibinfo  {journal}
  {Phys. Rev. Lett.}\ }\textbf {\bibinfo {volume} {117}},\ \bibinfo {pages}
  {010503} (\bibinfo {year} {2016})}\BibitemShut {NoStop}%
\bibitem [{\citenamefont {Farhi}\ \emph {et~al.}(2000)\citenamefont {Farhi},
  \citenamefont {Goldstone}, \citenamefont {Gutmann},\ and\ \citenamefont
  {Sipser}}]{farhi2000quantum}%
  \BibitemOpen
  \bibfield  {author} {\bibinfo {author} {\bibfnamefont {Edward}\ \bibnamefont
  {Farhi}}, \bibinfo {author} {\bibfnamefont {Jeffrey}\ \bibnamefont
  {Goldstone}}, \bibinfo {author} {\bibfnamefont {Sam}\ \bibnamefont
  {Gutmann}}, \ and\ \bibinfo {author} {\bibfnamefont {Michael}\ \bibnamefont
  {Sipser}},\ }\bibfield  {title} {\enquote {\bibinfo {title} {Quantum
  computation by adiabatic evolution},}\ }\href@noop {} {\bibfield  {journal}
  {\bibinfo  {journal} {arXiv preprint quant-ph/0001106}\ } (\bibinfo {year}
  {2000})}\BibitemShut {NoStop}%
\bibitem [{\citenamefont {Farhi}\ \emph {et~al.}(2001)\citenamefont {Farhi},
  \citenamefont {Goldstone}, \citenamefont {Gutmann}, \citenamefont {Lapan},
  \citenamefont {Lundgren},\ and\ \citenamefont {Preda}}]{Farhi472}%
  \BibitemOpen
  \bibfield  {author} {\bibinfo {author} {\bibfnamefont {Edward}\ \bibnamefont
  {Farhi}}, \bibinfo {author} {\bibfnamefont {Jeffrey}\ \bibnamefont
  {Goldstone}}, \bibinfo {author} {\bibfnamefont {Sam}\ \bibnamefont
  {Gutmann}}, \bibinfo {author} {\bibfnamefont {Joshua}\ \bibnamefont {Lapan}},
  \bibinfo {author} {\bibfnamefont {Andrew}\ \bibnamefont {Lundgren}}, \ and\
  \bibinfo {author} {\bibfnamefont {Daniel}\ \bibnamefont {Preda}},\ }\bibfield
   {title} {\enquote {\bibinfo {title} {A quantum adiabatic evolution algorithm
  applied to random instances of an np-complete problem},}\ }\href {\doibase
  10.1126/science.1057726} {\bibfield  {journal} {\bibinfo  {journal}
  {Science}\ }\textbf {\bibinfo {volume} {292}},\ \bibinfo {pages} {472--475}
  (\bibinfo {year} {2001})}\BibitemShut {NoStop}%
\bibitem [{\citenamefont {Kaplan}\ \emph {et~al.}(2017)\citenamefont {Kaplan},
  \citenamefont {Klco},\ and\ \citenamefont {Roggero}}]{Kaplan:2017ccd}%
  \BibitemOpen
  \bibfield  {author} {\bibinfo {author} {\bibfnamefont {David~B.}\
  \bibnamefont {Kaplan}}, \bibinfo {author} {\bibfnamefont {Natalie}\
  \bibnamefont {Klco}}, \ and\ \bibinfo {author} {\bibfnamefont {Alessandro}\
  \bibnamefont {Roggero}},\ }\href@noop {} {\enquote {\bibinfo {title} {{Ground
  States via Spectral Combing on a Quantum Computer}},}\ } (\bibinfo {year}
  {2017}),\ \Eprint {http://arxiv.org/abs/1709.08250} {arXiv:1709.08250
  [quant-ph]} \BibitemShut {NoStop}%
\bibitem [{\citenamefont {Ciavarella}\ and\ \citenamefont
  {Chernyshev}(2022)}]{Ciavarella:2021lel}%
  \BibitemOpen
  \bibfield  {author} {\bibinfo {author} {\bibfnamefont {Anthony~N.}\
  \bibnamefont {Ciavarella}}\ and\ \bibinfo {author} {\bibfnamefont {Ivan~A.}\
  \bibnamefont {Chernyshev}},\ }\bibfield  {title} {\enquote {\bibinfo {title}
  {{Preparation of the SU(3) lattice Yang-Mills vacuum with variational quantum
  methods}},}\ }\href {\doibase 10.1103/PhysRevD.105.074504} {\bibfield
  {journal} {\bibinfo  {journal} {Phys. Rev. D}\ }\textbf {\bibinfo {volume}
  {105}},\ \bibinfo {pages} {074504} (\bibinfo {year} {2022})},\ \Eprint
  {http://arxiv.org/abs/2112.09083} {arXiv:2112.09083 [quant-ph]} \BibitemShut
  {NoStop}%
\bibitem [{\citenamefont {Farrell}\ \emph {et~al.}(2022)\citenamefont
  {Farrell}, \citenamefont {Chernyshev}, \citenamefont {Powell}, \citenamefont
  {Zemlevskiy}, \citenamefont {Illa},\ and\ \citenamefont
  {Savage}}]{Farrell:2022wyt}%
  \BibitemOpen
  \bibfield  {author} {\bibinfo {author} {\bibfnamefont {Roland~C.}\
  \bibnamefont {Farrell}}, \bibinfo {author} {\bibfnamefont {Ivan~A.}\
  \bibnamefont {Chernyshev}}, \bibinfo {author} {\bibfnamefont {Sarah J.~M.}\
  \bibnamefont {Powell}}, \bibinfo {author} {\bibfnamefont {Nikita~A.}\
  \bibnamefont {Zemlevskiy}}, \bibinfo {author} {\bibfnamefont {Marc}\
  \bibnamefont {Illa}}, \ and\ \bibinfo {author} {\bibfnamefont {Martin~J.}\
  \bibnamefont {Savage}},\ }\bibfield  {title} {\enquote {\bibinfo {title}
  {{Preparations for Quantum Simulations of Quantum Chromodynamics in 1+1
  Dimensions: (I) Axial Gauge}},}\ }\href@noop {} {\  (\bibinfo {year}
  {2022})},\ \Eprint {http://arxiv.org/abs/2207.01731} {arXiv:2207.01731
  [quant-ph]} \BibitemShut {NoStop}%
\bibitem [{\citenamefont {Atas}\ \emph {et~al.}(2021)\citenamefont {Atas},
  \citenamefont {Zhang}, \citenamefont {Lewis}, \citenamefont {Jahanpour},
  \citenamefont {Haase},\ and\ \citenamefont {Muschik}}]{Atas:2021ext}%
  \BibitemOpen
  \bibfield  {author} {\bibinfo {author} {\bibfnamefont {Yasar~Y.}\
  \bibnamefont {Atas}}, \bibinfo {author} {\bibfnamefont {Jinglei}\
  \bibnamefont {Zhang}}, \bibinfo {author} {\bibfnamefont {Randy}\ \bibnamefont
  {Lewis}}, \bibinfo {author} {\bibfnamefont {Amin}\ \bibnamefont {Jahanpour}},
  \bibinfo {author} {\bibfnamefont {Jan~F.}\ \bibnamefont {Haase}}, \ and\
  \bibinfo {author} {\bibfnamefont {Christine~A.}\ \bibnamefont {Muschik}},\
  }\bibfield  {title} {\enquote {\bibinfo {title} {{SU(2) hadrons on a quantum
  computer via a variational approach}},}\ }\href {\doibase
  10.1038/s41467-021-26825-4} {\bibfield  {journal} {\bibinfo  {journal}
  {Nature Commun.}\ }\textbf {\bibinfo {volume} {12}},\ \bibinfo {pages} {6499}
  (\bibinfo {year} {2021})},\ \Eprint {http://arxiv.org/abs/2102.08920}
  {arXiv:2102.08920 [quant-ph]} \BibitemShut {NoStop}%
\bibitem [{\citenamefont {Atas}\ \emph {et~al.}(2022)\citenamefont {Atas},
  \citenamefont {Haase}, \citenamefont {Zhang}, \citenamefont {Wei},
  \citenamefont {Pfaendler}, \citenamefont {Lewis},\ and\ \citenamefont
  {Muschik}}]{Atas:2022dqm}%
  \BibitemOpen
  \bibfield  {author} {\bibinfo {author} {\bibfnamefont {Yasar~Y.}\
  \bibnamefont {Atas}}, \bibinfo {author} {\bibfnamefont {Jan~F.}\ \bibnamefont
  {Haase}}, \bibinfo {author} {\bibfnamefont {Jinglei}\ \bibnamefont {Zhang}},
  \bibinfo {author} {\bibfnamefont {Victor}\ \bibnamefont {Wei}}, \bibinfo
  {author} {\bibfnamefont {Sieglinde M.~L.}\ \bibnamefont {Pfaendler}},
  \bibinfo {author} {\bibfnamefont {Randy}\ \bibnamefont {Lewis}}, \ and\
  \bibinfo {author} {\bibfnamefont {Christine~A.}\ \bibnamefont {Muschik}},\
  }\bibfield  {title} {\enquote {\bibinfo {title} {{Real-time evolution of
  SU(3) hadrons on a quantum computer}},}\ }\href@noop {} {\  (\bibinfo {year}
  {2022})},\ \Eprint {http://arxiv.org/abs/2207.03473} {arXiv:2207.03473
  [quant-ph]} \BibitemShut {NoStop}%
\bibitem [{\citenamefont {{Bilgin}}\ and\ \citenamefont
  {{Boixo}}(2010)}]{2010PhRvL.105q0405B}%
  \BibitemOpen
  \bibfield  {author} {\bibinfo {author} {\bibfnamefont {E.}~\bibnamefont
  {{Bilgin}}}\ and\ \bibinfo {author} {\bibfnamefont {S.}~\bibnamefont
  {{Boixo}}},\ }\bibfield  {title} {\enquote {\bibinfo {title} {{Preparing
  Thermal States of Quantum Systems by Dimension Reduction}},}\ }\href
  {\doibase 10.1103/PhysRevLett.105.170405} {\bibfield  {journal} {\bibinfo
  {journal} {Physical Review Letters}\ }\textbf {\bibinfo {volume} {105}},\
  \bibinfo {eid} {170405} (\bibinfo {year} {2010})},\ \Eprint
  {http://arxiv.org/abs/1008.4162} {arXiv:1008.4162 [quant-ph]} \BibitemShut
  {NoStop}%
\bibitem [{\citenamefont {Jordan}\ \emph
  {et~al.}(2014{\natexlab{a}})\citenamefont {Jordan}, \citenamefont {Lee},\
  and\ \citenamefont {Preskill}}]{Jordan:2011ci}%
  \BibitemOpen
  \bibfield  {author} {\bibinfo {author} {\bibfnamefont {Stephen~P.}\
  \bibnamefont {Jordan}}, \bibinfo {author} {\bibfnamefont {Keith S.~M.}\
  \bibnamefont {Lee}}, \ and\ \bibinfo {author} {\bibfnamefont {John}\
  \bibnamefont {Preskill}},\ }\bibfield  {title} {\enquote {\bibinfo {title}
  {{Quantum Computation of Scattering in Scalar Quantum Field Theories}},}\
  }\href@noop {} {\bibfield  {journal} {\bibinfo  {journal} {Quant. Inf.
  Comput.}\ }\textbf {\bibinfo {volume} {14}},\ \bibinfo {pages} {1014}
  (\bibinfo {year} {2014}{\natexlab{a}})},\ \Eprint
  {http://arxiv.org/abs/1112.4833} {arXiv:1112.4833 [hep-th]} \BibitemShut
  {NoStop}%
\bibitem [{\citenamefont {García-Álvarez}\ \emph {et~al.}(2015)\citenamefont
  {García-Álvarez}, \citenamefont {Casanova}, \citenamefont {Mezzacapo},
  \citenamefont {Egusquiza}, \citenamefont {Lamata}, \citenamefont {Romero},\
  and\ \citenamefont {Solano}}]{Garcia-Alvarez:2014uda}%
  \BibitemOpen
  \bibfield  {author} {\bibinfo {author} {\bibfnamefont {L.}~\bibnamefont
  {García-Álvarez}}, \bibinfo {author} {\bibfnamefont {J.}~\bibnamefont
  {Casanova}}, \bibinfo {author} {\bibfnamefont {A.}~\bibnamefont {Mezzacapo}},
  \bibinfo {author} {\bibfnamefont {I.~L.}\ \bibnamefont {Egusquiza}}, \bibinfo
  {author} {\bibfnamefont {L.}~\bibnamefont {Lamata}}, \bibinfo {author}
  {\bibfnamefont {G.}~\bibnamefont {Romero}}, \ and\ \bibinfo {author}
  {\bibfnamefont {E.}~\bibnamefont {Solano}},\ }\bibfield  {title} {\enquote
  {\bibinfo {title} {{Fermion-Fermion Scattering in Quantum Field Theory with
  Superconducting Circuits}},}\ }\href {\doibase
  10.1103/PhysRevLett.114.070502} {\bibfield  {journal} {\bibinfo  {journal}
  {Phys. Rev. Lett.}\ }\textbf {\bibinfo {volume} {114}},\ \bibinfo {pages}
  {070502} (\bibinfo {year} {2015})},\ \Eprint {http://arxiv.org/abs/1404.2868}
  {arXiv:1404.2868 [quant-ph]} \BibitemShut {NoStop}%
\bibitem [{\citenamefont {Jordan}\ \emph
  {et~al.}(2014{\natexlab{b}})\citenamefont {Jordan}, \citenamefont {Lee},\
  and\ \citenamefont {Preskill}}]{Jordan:2014tma}%
  \BibitemOpen
  \bibfield  {author} {\bibinfo {author} {\bibfnamefont {Stephen~P.}\
  \bibnamefont {Jordan}}, \bibinfo {author} {\bibfnamefont {Keith S.~M.}\
  \bibnamefont {Lee}}, \ and\ \bibinfo {author} {\bibfnamefont {John}\
  \bibnamefont {Preskill}},\ }\href@noop {} {\enquote {\bibinfo {title}
  {{Quantum Algorithms for Fermionic Quantum Field Theories}},}\ } (\bibinfo
  {year} {2014}{\natexlab{b}}),\ \Eprint {http://arxiv.org/abs/1404.7115}
  {arXiv:1404.7115 [hep-th]} \BibitemShut {NoStop}%
\bibitem [{\citenamefont {Hamed~Moosavian}\ and\ \citenamefont
  {Jordan}(2018)}]{Moosavian:2017tkv}%
  \BibitemOpen
  \bibfield  {author} {\bibinfo {author} {\bibfnamefont {Ali}\ \bibnamefont
  {Hamed~Moosavian}}\ and\ \bibinfo {author} {\bibfnamefont {Stephen}\
  \bibnamefont {Jordan}},\ }\bibfield  {title} {\enquote {\bibinfo {title}
  {{Faster Quantum Algorithm to simulate Fermionic Quantum Field Theory}},}\
  }\href {\doibase 10.1103/PhysRevA.98.012332} {\bibfield  {journal} {\bibinfo
  {journal} {Phys. Rev.}\ }\textbf {\bibinfo {volume} {A98}},\ \bibinfo {pages}
  {012332} (\bibinfo {year} {2018})},\ \Eprint
  {http://arxiv.org/abs/1711.04006} {arXiv:1711.04006 [quant-ph]} \BibitemShut
  {NoStop}%
\bibitem [{\citenamefont {Lamm}\ and\ \citenamefont
  {Lawrence}(2018)}]{Lamm:2018siq}%
  \BibitemOpen
  \bibfield  {author} {\bibinfo {author} {\bibfnamefont {Henry}\ \bibnamefont
  {Lamm}}\ and\ \bibinfo {author} {\bibfnamefont {Scott}\ \bibnamefont
  {Lawrence}},\ }\bibfield  {title} {\enquote {\bibinfo {title} {{Simulation of
  Nonequilibrium Dynamics on a Quantum Computer}},}\ }\href {\doibase
  10.1103/PhysRevLett.121.170501} {\bibfield  {journal} {\bibinfo  {journal}
  {Phys. Rev. Lett.}\ }\textbf {\bibinfo {volume} {121}},\ \bibinfo {pages}
  {170501} (\bibinfo {year} {2018})},\ \Eprint
  {http://arxiv.org/abs/1806.06649} {arXiv:1806.06649 [quant-ph]} \BibitemShut
  {NoStop}%
\bibitem [{\citenamefont {Gustafson}\ \emph
  {et~al.}(2019{\natexlab{b}})\citenamefont {Gustafson}, \citenamefont
  {Meurice},\ and\ \citenamefont {Unmuth-Yockey}}]{Gustafson:2019mpk}%
  \BibitemOpen
  \bibfield  {author} {\bibinfo {author} {\bibfnamefont {Erik}\ \bibnamefont
  {Gustafson}}, \bibinfo {author} {\bibfnamefont {Yannick}\ \bibnamefont
  {Meurice}}, \ and\ \bibinfo {author} {\bibfnamefont {Judah}\ \bibnamefont
  {Unmuth-Yockey}},\ }\href@noop {} {\enquote {\bibinfo {title} {{Quantum
  simulation of scattering in the quantum Ising model}},}\ } (\bibinfo {year}
  {2019}{\natexlab{b}}),\ \Eprint {http://arxiv.org/abs/1901.05944}
  {arXiv:1901.05944 [hep-lat]} \BibitemShut {NoStop}%
\bibitem [{\citenamefont {Gustafson}\ \emph
  {et~al.}(2021{\natexlab{a}})\citenamefont {Gustafson}, \citenamefont
  {Dreher}, \citenamefont {Hang},\ and\ \citenamefont
  {Meurice}}]{Gustafson:2019vsd}%
  \BibitemOpen
  \bibfield  {author} {\bibinfo {author} {\bibfnamefont {Erik}\ \bibnamefont
  {Gustafson}}, \bibinfo {author} {\bibfnamefont {Patrick}\ \bibnamefont
  {Dreher}}, \bibinfo {author} {\bibfnamefont {Zheyue}\ \bibnamefont {Hang}}, \
  and\ \bibinfo {author} {\bibfnamefont {Yannick}\ \bibnamefont {Meurice}},\
  }\bibfield  {title} {\enquote {\bibinfo {title} {{Benchmarking quantum
  computers for real-time evolution of a $(1+1)$ field theory with error
  mitigation}},}\ }\href@noop {} {\bibfield  {journal} {\bibinfo  {journal}
  {Quantum Sci. Technol.}\ }\textbf {\bibinfo {volume} {6}},\ \bibinfo {pages}
  {045020} (\bibinfo {year} {2021}{\natexlab{a}})},\ \Eprint
  {http://arxiv.org/abs/1910.09478} {arXiv:1910.09478 [hep-lat]} \BibitemShut
  {NoStop}%
\bibitem [{\citenamefont {Harmalkar}\ \emph {et~al.}(2020)\citenamefont
  {Harmalkar}, \citenamefont {Lamm},\ and\ \citenamefont
  {Lawrence}}]{Harmalkar:2020mpd}%
  \BibitemOpen
  \bibfield  {author} {\bibinfo {author} {\bibfnamefont {Siddhartha}\
  \bibnamefont {Harmalkar}}, \bibinfo {author} {\bibfnamefont {Henry}\
  \bibnamefont {Lamm}}, \ and\ \bibinfo {author} {\bibfnamefont {Scott}\
  \bibnamefont {Lawrence}} (\bibinfo {collaboration} {NuQS}),\ }\href@noop {}
  {\enquote {\bibinfo {title} {{Quantum Simulation of Field Theories Without
  State Preparation}},}\ } (\bibinfo {year} {2020}),\ \Eprint
  {http://arxiv.org/abs/2001.11490} {arXiv:2001.11490 [hep-lat]} \BibitemShut
  {NoStop}%
\bibitem [{\citenamefont {Gustafson}\ and\ \citenamefont
  {Lamm}(2021)}]{Gustafson:2020yfe}%
  \BibitemOpen
  \bibfield  {author} {\bibinfo {author} {\bibfnamefont {Erik~J.}\ \bibnamefont
  {Gustafson}}\ and\ \bibinfo {author} {\bibfnamefont {Henry}\ \bibnamefont
  {Lamm}},\ }\bibfield  {title} {\enquote {\bibinfo {title} {{Toward quantum
  simulations of $\mathbb{Z}_2$ gauge theory without state preparation}},}\
  }\href {\doibase 10.1103/PhysRevD.103.054507} {\bibfield  {journal} {\bibinfo
   {journal} {Phys. Rev. D}\ }\textbf {\bibinfo {volume} {103}},\ \bibinfo
  {pages} {054507} (\bibinfo {year} {2021})},\ \Eprint
  {http://arxiv.org/abs/2011.11677} {arXiv:2011.11677 [hep-lat]} \BibitemShut
  {NoStop}%
\bibitem [{\citenamefont {Klco}\ and\ \citenamefont
  {Savage}(2019)}]{Klco:2019xro}%
  \BibitemOpen
  \bibfield  {author} {\bibinfo {author} {\bibfnamefont {Natalie}\ \bibnamefont
  {Klco}}\ and\ \bibinfo {author} {\bibfnamefont {Martin~J.}\ \bibnamefont
  {Savage}},\ }\href@noop {} {\enquote {\bibinfo {title} {{Minimally-Entangled
  State Preparation of Localized Wavefunctions on Quantum Computers}},}\ }
  (\bibinfo {year} {2019}),\ \Eprint {http://arxiv.org/abs/1904.10440}
  {arXiv:1904.10440 [quant-ph]} \BibitemShut {NoStop}%
\bibitem [{\citenamefont {Riera}\ \emph {et~al.}(2012)\citenamefont {Riera},
  \citenamefont {Gogolin},\ and\ \citenamefont
  {Eisert}}]{PhysRevLett.108.080402}%
  \BibitemOpen
  \bibfield  {author} {\bibinfo {author} {\bibfnamefont {Arnau}\ \bibnamefont
  {Riera}}, \bibinfo {author} {\bibfnamefont {Christian}\ \bibnamefont
  {Gogolin}}, \ and\ \bibinfo {author} {\bibfnamefont {Jens}\ \bibnamefont
  {Eisert}},\ }\bibfield  {title} {\enquote {\bibinfo {title} {Thermalization
  in nature and on a quantum computer},}\ }\href {\doibase
  10.1103/PhysRevLett.108.080402} {\bibfield  {journal} {\bibinfo  {journal}
  {Phys. Rev. Lett.}\ }\textbf {\bibinfo {volume} {108}},\ \bibinfo {pages}
  {080402} (\bibinfo {year} {2012})}\BibitemShut {NoStop}%
\bibitem [{\citenamefont {Brand{\~a}o}\ and\ \citenamefont
  {Kastoryano}(2019)}]{brandao2019finite}%
  \BibitemOpen
  \bibfield  {author} {\bibinfo {author} {\bibfnamefont {Fernando~GSL}\
  \bibnamefont {Brand{\~a}o}}\ and\ \bibinfo {author} {\bibfnamefont
  {Michael~J}\ \bibnamefont {Kastoryano}},\ }\bibfield  {title} {\enquote
  {\bibinfo {title} {Finite correlation length implies efficient preparation of
  quantum thermal states},}\ }\href@noop {} {\bibfield  {journal} {\bibinfo
  {journal} {Communications in Mathematical Physics}\ }\textbf {\bibinfo
  {volume} {365}},\ \bibinfo {pages} {1--16} (\bibinfo {year}
  {2019})}\BibitemShut {NoStop}%
\bibitem [{\citenamefont {Clemente}\ \emph {et~al.}(2020)\citenamefont
  {Clemente} \emph {et~al.}}]{Clemente:2020lpr}%
  \BibitemOpen
  \bibfield  {author} {\bibinfo {author} {\bibfnamefont {Giuseppe}\
  \bibnamefont {Clemente}} \emph {et~al.} (\bibinfo {collaboration} {QuBiPF}),\
  }\bibfield  {title} {\enquote {\bibinfo {title} {{Quantum computation of
  thermal averages in the presence of a sign problem}},}\ }\href {\doibase
  10.1103/PhysRevD.101.074510} {\bibfield  {journal} {\bibinfo  {journal}
  {Phys. Rev. D}\ }\textbf {\bibinfo {volume} {101}},\ \bibinfo {pages}
  {074510} (\bibinfo {year} {2020})},\ \Eprint
  {http://arxiv.org/abs/2001.05328} {arXiv:2001.05328 [hep-lat]} \BibitemShut
  {NoStop}%
\bibitem [{\citenamefont {Motta}\ \emph {et~al.}(2020)\citenamefont {Motta},
  \citenamefont {Sun}, \citenamefont {Tan}, \citenamefont {O’Rourke},
  \citenamefont {Ye}, \citenamefont {Minnich}, \citenamefont {Brandao},\ and\
  \citenamefont {Chan}}]{motta2020determining}%
  \BibitemOpen
  \bibfield  {author} {\bibinfo {author} {\bibfnamefont {Mario}\ \bibnamefont
  {Motta}}, \bibinfo {author} {\bibfnamefont {Chong}\ \bibnamefont {Sun}},
  \bibinfo {author} {\bibfnamefont {Adrian~TK}\ \bibnamefont {Tan}}, \bibinfo
  {author} {\bibfnamefont {Matthew~J}\ \bibnamefont {O’Rourke}}, \bibinfo
  {author} {\bibfnamefont {Erika}\ \bibnamefont {Ye}}, \bibinfo {author}
  {\bibfnamefont {Austin~J}\ \bibnamefont {Minnich}}, \bibinfo {author}
  {\bibfnamefont {Fernando~GSL}\ \bibnamefont {Brandao}}, \ and\ \bibinfo
  {author} {\bibfnamefont {Garnet Kin-Lic}\ \bibnamefont {Chan}},\ }\bibfield
  {title} {\enquote {\bibinfo {title} {Determining eigenstates and thermal
  states on a quantum computer using quantum imaginary time evolution},}\
  }\href@noop {} {\bibfield  {journal} {\bibinfo  {journal} {Nature Physics}\
  }\textbf {\bibinfo {volume} {16}},\ \bibinfo {pages} {205--210} (\bibinfo
  {year} {2020})}\BibitemShut {NoStop}%
\bibitem [{\citenamefont {de~Jong}\ \emph {et~al.}(2021)\citenamefont
  {de~Jong}, \citenamefont {Lee}, \citenamefont {Mulligan}, \citenamefont
  {P\l{}osko\'n}, \citenamefont {Ringer},\ and\ \citenamefont
  {Yao}}]{deJong:2021wsd}%
  \BibitemOpen
  \bibfield  {author} {\bibinfo {author} {\bibfnamefont {Wibe~A.}\ \bibnamefont
  {de~Jong}}, \bibinfo {author} {\bibfnamefont {Kyle}\ \bibnamefont {Lee}},
  \bibinfo {author} {\bibfnamefont {James}\ \bibnamefont {Mulligan}}, \bibinfo
  {author} {\bibfnamefont {Mateusz}\ \bibnamefont {P\l{}osko\'n}}, \bibinfo
  {author} {\bibfnamefont {Felix}\ \bibnamefont {Ringer}}, \ and\ \bibinfo
  {author} {\bibfnamefont {Xiaojun}\ \bibnamefont {Yao}},\ }\href@noop {}
  {\enquote {\bibinfo {title} {{Quantum simulation of non-equilibrium dynamics
  and thermalization in the Schwinger model}},}\ } (\bibinfo {year} {2021}),\
  \Eprint {http://arxiv.org/abs/2106.08394} {arXiv:2106.08394 [quant-ph]}
  \BibitemShut {NoStop}%
\bibitem [{\citenamefont {Gustafson}\ \emph
  {et~al.}(2021{\natexlab{b}})\citenamefont {Gustafson}, \citenamefont {Zhu},
  \citenamefont {Dreher}, \citenamefont {Linke},\ and\ \citenamefont
  {Meurice}}]{Gustafson:2021imb}%
  \BibitemOpen
  \bibfield  {author} {\bibinfo {author} {\bibfnamefont {Erik}\ \bibnamefont
  {Gustafson}}, \bibinfo {author} {\bibfnamefont {Yingyue}\ \bibnamefont
  {Zhu}}, \bibinfo {author} {\bibfnamefont {Patrick}\ \bibnamefont {Dreher}},
  \bibinfo {author} {\bibfnamefont {Norbert~M.}\ \bibnamefont {Linke}}, \ and\
  \bibinfo {author} {\bibfnamefont {Yannick}\ \bibnamefont {Meurice}},\
  }\bibfield  {title} {\enquote {\bibinfo {title} {{Real-time quantum
  calculations of phase shifts using wave packet time delays}},}\ }\href
  {\doibase 10.1103/PhysRevD.104.054507} {\bibfield  {journal} {\bibinfo
  {journal} {Phys. Rev. D}\ }\textbf {\bibinfo {volume} {104}},\ \bibinfo
  {pages} {054507} (\bibinfo {year} {2021}{\natexlab{b}})},\ \Eprint
  {http://arxiv.org/abs/2103.06848} {arXiv:2103.06848 [hep-lat]} \BibitemShut
  {NoStop}%
\bibitem [{\citenamefont {Davoudi}\ \emph {et~al.}(2022)\citenamefont
  {Davoudi}, \citenamefont {Mueller},\ and\ \citenamefont
  {Powers}}]{Davoudi:2022uzo}%
  \BibitemOpen
  \bibfield  {author} {\bibinfo {author} {\bibfnamefont {Zohreh}\ \bibnamefont
  {Davoudi}}, \bibinfo {author} {\bibfnamefont {Niklas}\ \bibnamefont
  {Mueller}}, \ and\ \bibinfo {author} {\bibfnamefont {Connor}\ \bibnamefont
  {Powers}},\ }\bibfield  {title} {\enquote {\bibinfo {title} {{Toward Quantum
  Computing Phase Diagrams of Gauge Theories with Thermal Pure Quantum
  States}},}\ }\href@noop {} {\  (\bibinfo {year} {2022})},\ \Eprint
  {http://arxiv.org/abs/2208.13112} {arXiv:2208.13112 [hep-lat]} \BibitemShut
  {NoStop}%
\bibitem [{\citenamefont {Maiani}\ and\ \citenamefont
  {Testa}(1990)}]{Maiani:1990ca}%
  \BibitemOpen
  \bibfield  {author} {\bibinfo {author} {\bibfnamefont {L.}~\bibnamefont
  {Maiani}}\ and\ \bibinfo {author} {\bibfnamefont {M.}~\bibnamefont {Testa}},\
  }\bibfield  {title} {\enquote {\bibinfo {title} {{Final state interactions
  from Euclidean correlation functions}},}\ }\href {\doibase
  10.1016/0370-2693(90)90695-3} {\bibfield  {journal} {\bibinfo  {journal}
  {Phys. Lett. B}\ }\textbf {\bibinfo {volume} {245}},\ \bibinfo {pages}
  {585--590} (\bibinfo {year} {1990})}\BibitemShut {NoStop}%
\bibitem [{\citenamefont {Bruno}\ and\ \citenamefont
  {Hansen}(2020)}]{Bruno:2020kyl}%
  \BibitemOpen
  \bibfield  {author} {\bibinfo {author} {\bibfnamefont {Mattia}\ \bibnamefont
  {Bruno}}\ and\ \bibinfo {author} {\bibfnamefont {Maxwell~T.}\ \bibnamefont
  {Hansen}},\ }\href@noop {} {\enquote {\bibinfo {title} {{Variations on the
  Maiani-Testa approach and the inverse problem}},}\ } (\bibinfo {year}
  {2020}),\ \Eprint {http://arxiv.org/abs/2012.11488} {arXiv:2012.11488
  [hep-lat]} \BibitemShut {NoStop}%
\bibitem [{\citenamefont {Haah}\ \emph {et~al.}(2018)\citenamefont {Haah},
  \citenamefont {Hastings}, \citenamefont {Kothari},\ and\ \citenamefont
  {Low}}]{haah2018quantum}%
  \BibitemOpen
  \bibfield  {author} {\bibinfo {author} {\bibfnamefont {Jeongwan}\
  \bibnamefont {Haah}}, \bibinfo {author} {\bibfnamefont {Matthew~B}\
  \bibnamefont {Hastings}}, \bibinfo {author} {\bibfnamefont {Robin}\
  \bibnamefont {Kothari}}, \ and\ \bibinfo {author} {\bibfnamefont {Guang~Hao}\
  \bibnamefont {Low}},\ }\href@noop {} {\enquote {\bibinfo {title} {Quantum
  algorithm for simulating real time evolution of lattice hamiltonians},}\ }
  (\bibinfo {year} {2018})\BibitemShut {NoStop}%
\bibitem [{\citenamefont {Du}\ \emph {et~al.}(2020)\citenamefont {Du},
  \citenamefont {Vary}, \citenamefont {Zhao},\ and\ \citenamefont
  {Zuo}}]{Du:2020glq}%
  \BibitemOpen
  \bibfield  {author} {\bibinfo {author} {\bibfnamefont {Weijie}\ \bibnamefont
  {Du}}, \bibinfo {author} {\bibfnamefont {James~P.}\ \bibnamefont {Vary}},
  \bibinfo {author} {\bibfnamefont {Xingbo}\ \bibnamefont {Zhao}}, \ and\
  \bibinfo {author} {\bibfnamefont {Wei}\ \bibnamefont {Zuo}},\ }\href@noop {}
  {\enquote {\bibinfo {title} {{Quantum Simulation of Nuclear Inelastic
  Scattering}},}\ } (\bibinfo {year} {2020}),\ \Eprint
  {http://arxiv.org/abs/2006.01369} {arXiv:2006.01369 [nucl-th]} \BibitemShut
  {NoStop}%
\bibitem [{\citenamefont {Childs}\ \emph {et~al.}(2021)\citenamefont {Childs},
  \citenamefont {Su}, \citenamefont {Tran}, \citenamefont {Wiebe},\ and\
  \citenamefont {Zhu}}]{PhysRevX.11.011020}%
  \BibitemOpen
  \bibfield  {author} {\bibinfo {author} {\bibfnamefont {Andrew~M.}\
  \bibnamefont {Childs}}, \bibinfo {author} {\bibfnamefont {Yuan}\ \bibnamefont
  {Su}}, \bibinfo {author} {\bibfnamefont {Minh~C.}\ \bibnamefont {Tran}},
  \bibinfo {author} {\bibfnamefont {Nathan}\ \bibnamefont {Wiebe}}, \ and\
  \bibinfo {author} {\bibfnamefont {Shuchen}\ \bibnamefont {Zhu}},\ }\bibfield
  {title} {\enquote {\bibinfo {title} {Theory of trotter error with commutator
  scaling},}\ }\href {\doibase 10.1103/PhysRevX.11.011020} {\bibfield
  {journal} {\bibinfo  {journal} {Phys. Rev. X}\ }\textbf {\bibinfo {volume}
  {11}},\ \bibinfo {pages} {011020} (\bibinfo {year} {2021})}\BibitemShut
  {NoStop}%
\bibitem [{\citenamefont {Campbell}(2019)}]{PhysRevLett.123.070503}%
  \BibitemOpen
  \bibfield  {author} {\bibinfo {author} {\bibfnamefont {Earl}\ \bibnamefont
  {Campbell}},\ }\bibfield  {title} {\enquote {\bibinfo {title} {Random
  compiler for fast hamiltonian simulation},}\ }\href {\doibase
  10.1103/PhysRevLett.123.070503} {\bibfield  {journal} {\bibinfo  {journal}
  {Phys. Rev. Lett.}\ }\textbf {\bibinfo {volume} {123}},\ \bibinfo {pages}
  {070503} (\bibinfo {year} {2019})}\BibitemShut {NoStop}%
\bibitem [{\citenamefont {Cirstoiu}\ \emph {et~al.}(2020)\citenamefont
  {Cirstoiu}, \citenamefont {Holmes}, \citenamefont {Iosue}, \citenamefont
  {Cincio}, \citenamefont {Coles},\ and\ \citenamefont
  {Sornborger}}]{cirstoiu2020variational}%
  \BibitemOpen
  \bibfield  {author} {\bibinfo {author} {\bibfnamefont {Cristina}\
  \bibnamefont {Cirstoiu}}, \bibinfo {author} {\bibfnamefont {Zoe}\
  \bibnamefont {Holmes}}, \bibinfo {author} {\bibfnamefont {Joseph}\
  \bibnamefont {Iosue}}, \bibinfo {author} {\bibfnamefont {Lukasz}\
  \bibnamefont {Cincio}}, \bibinfo {author} {\bibfnamefont {Patrick~J}\
  \bibnamefont {Coles}}, \ and\ \bibinfo {author} {\bibfnamefont {Andrew}\
  \bibnamefont {Sornborger}},\ }\bibfield  {title} {\enquote {\bibinfo {title}
  {Variational fast forwarding for quantum simulation beyond the coherence
  time},}\ }\href@noop {} {\bibfield  {journal} {\bibinfo  {journal} {npj
  Quantum Information}\ }\textbf {\bibinfo {volume} {6}},\ \bibinfo {pages}
  {1--10} (\bibinfo {year} {2020})}\BibitemShut {NoStop}%
\bibitem [{\citenamefont {Gibbs}\ \emph {et~al.}(2021)\citenamefont {Gibbs},
  \citenamefont {Gili}, \citenamefont {Holmes}, \citenamefont {Commeau},
  \citenamefont {Arrasmith}, \citenamefont {Cincio}, \citenamefont {Coles},\
  and\ \citenamefont {Sornborger}}]{gibbs2021longtime}%
  \BibitemOpen
  \bibfield  {author} {\bibinfo {author} {\bibfnamefont {Joe}\ \bibnamefont
  {Gibbs}}, \bibinfo {author} {\bibfnamefont {Kaitlin}\ \bibnamefont {Gili}},
  \bibinfo {author} {\bibfnamefont {Zoë}\ \bibnamefont {Holmes}}, \bibinfo
  {author} {\bibfnamefont {Benjamin}\ \bibnamefont {Commeau}}, \bibinfo
  {author} {\bibfnamefont {Andrew}\ \bibnamefont {Arrasmith}}, \bibinfo
  {author} {\bibfnamefont {Lukasz}\ \bibnamefont {Cincio}}, \bibinfo {author}
  {\bibfnamefont {Patrick~J.}\ \bibnamefont {Coles}}, \ and\ \bibinfo {author}
  {\bibfnamefont {Andrew}\ \bibnamefont {Sornborger}},\ }\href@noop {}
  {\enquote {\bibinfo {title} {Long-time simulations with high fidelity on
  quantum hardware},}\ } (\bibinfo {year} {2021}),\ \Eprint
  {http://arxiv.org/abs/2102.04313} {arXiv:2102.04313 [quant-ph]} \BibitemShut
  {NoStop}%
\bibitem [{\citenamefont {Yao}\ \emph {et~al.}(2020)\citenamefont {Yao},
  \citenamefont {Gomes}, \citenamefont {Zhang}, \citenamefont {Iadecola},
  \citenamefont {Wang}, \citenamefont {Ho},\ and\ \citenamefont
  {Orth}}]{yao2020adaptive}%
  \BibitemOpen
  \bibfield  {author} {\bibinfo {author} {\bibfnamefont {Yong-Xin}\
  \bibnamefont {Yao}}, \bibinfo {author} {\bibfnamefont {Niladri}\ \bibnamefont
  {Gomes}}, \bibinfo {author} {\bibfnamefont {Feng}\ \bibnamefont {Zhang}},
  \bibinfo {author} {\bibfnamefont {Thomas}\ \bibnamefont {Iadecola}}, \bibinfo
  {author} {\bibfnamefont {Cai-Zhuang}\ \bibnamefont {Wang}}, \bibinfo {author}
  {\bibfnamefont {Kai-Ming}\ \bibnamefont {Ho}}, \ and\ \bibinfo {author}
  {\bibfnamefont {Peter~P}\ \bibnamefont {Orth}},\ }\bibfield  {title}
  {\enquote {\bibinfo {title} {Adaptive variational quantum dynamics
  simulations},}\ }\href@noop {} {\bibfield  {journal} {\bibinfo  {journal}
  {arXiv preprint arXiv:2011.00622}\ } (\bibinfo {year} {2020})}\BibitemShut
  {NoStop}%
\bibitem [{\citenamefont {Berry}\ \emph {et~al.}(2015)\citenamefont {Berry},
  \citenamefont {Childs}, \citenamefont {Cleve}, \citenamefont {Kothari},\ and\
  \citenamefont {Somma}}]{PhysRevLett.114.090502}%
  \BibitemOpen
  \bibfield  {author} {\bibinfo {author} {\bibfnamefont {Dominic~W.}\
  \bibnamefont {Berry}}, \bibinfo {author} {\bibfnamefont {Andrew~M.}\
  \bibnamefont {Childs}}, \bibinfo {author} {\bibfnamefont {Richard}\
  \bibnamefont {Cleve}}, \bibinfo {author} {\bibfnamefont {Robin}\ \bibnamefont
  {Kothari}}, \ and\ \bibinfo {author} {\bibfnamefont {Rolando~D.}\
  \bibnamefont {Somma}},\ }\bibfield  {title} {\enquote {\bibinfo {title}
  {Simulating hamiltonian dynamics with a truncated taylor series},}\ }\href
  {\doibase 10.1103/PhysRevLett.114.090502} {\bibfield  {journal} {\bibinfo
  {journal} {Phys. Rev. Lett.}\ }\textbf {\bibinfo {volume} {114}},\ \bibinfo
  {pages} {090502} (\bibinfo {year} {2015})}\BibitemShut {NoStop}%
\bibitem [{\citenamefont {Meurice}(2021{\natexlab{b}})}]{ymbook}%
  \BibitemOpen
  \bibfield  {author} {\bibinfo {author} {\bibfnamefont {Professor~Yannick}\
  \bibnamefont {Meurice}},\ }\href {\doibase 10.1088/978-0-7503-2187-7} {\emph
  {\bibinfo {title} {{Quantum Field Theory}}}}\ (\bibinfo  {publisher} {IOP},\
  \bibinfo {year} {2021})\BibitemShut {NoStop}%
\bibitem [{\citenamefont {Low}\ and\ \citenamefont
  {Chuang}(2019)}]{Low2019hamiltonian}%
  \BibitemOpen
  \bibfield  {author} {\bibinfo {author} {\bibfnamefont {Guang~Hao}\
  \bibnamefont {Low}}\ and\ \bibinfo {author} {\bibfnamefont {Isaac~L.}\
  \bibnamefont {Chuang}},\ }\bibfield  {title} {\enquote {\bibinfo {title}
  {Hamiltonian {S}imulation by {Q}ubitization},}\ }\href {\doibase
  10.22331/q-2019-07-12-163} {\bibfield  {journal} {\bibinfo  {journal}
  {{Quantum}}\ }\textbf {\bibinfo {volume} {3}},\ \bibinfo {pages} {163}
  (\bibinfo {year} {2019})}\BibitemShut {NoStop}%
\bibitem [{\citenamefont {Echevarria}\ \emph {et~al.}(2020)\citenamefont
  {Echevarria}, \citenamefont {Egusquiza}, \citenamefont {Rico},\ and\
  \citenamefont {Schnell}}]{Echevarria:2020wct}%
  \BibitemOpen
  \bibfield  {author} {\bibinfo {author} {\bibfnamefont {M.G.}\ \bibnamefont
  {Echevarria}}, \bibinfo {author} {\bibfnamefont {I.L.}\ \bibnamefont
  {Egusquiza}}, \bibinfo {author} {\bibfnamefont {E.}~\bibnamefont {Rico}}, \
  and\ \bibinfo {author} {\bibfnamefont {G.}~\bibnamefont {Schnell}},\
  }\href@noop {} {\enquote {\bibinfo {title} {{Quantum Simulation of
  Light-Front Parton Correlators}},}\ } (\bibinfo {year} {2020}),\ \Eprint
  {http://arxiv.org/abs/2011.01275} {arXiv:2011.01275 [quant-ph]} \BibitemShut
  {NoStop}%
\bibitem [{\citenamefont {Ciavarella}(2020)}]{Ciavarella:2020vqm}%
  \BibitemOpen
  \bibfield  {author} {\bibinfo {author} {\bibfnamefont {Anthony}\ \bibnamefont
  {Ciavarella}},\ }\bibfield  {title} {\enquote {\bibinfo {title} {{Algorithm
  for quantum computation of particle decays}},}\ }\href {\doibase
  10.1103/PhysRevD.102.094505} {\bibfield  {journal} {\bibinfo  {journal}
  {Phys. Rev. D}\ }\textbf {\bibinfo {volume} {102}},\ \bibinfo {pages}
  {094505} (\bibinfo {year} {2020})},\ \Eprint
  {http://arxiv.org/abs/2007.04447} {arXiv:2007.04447 [hep-th]} \BibitemShut
  {NoStop}%
\bibitem [{\citenamefont {Cohen}\ \emph {et~al.}(2021)\citenamefont {Cohen},
  \citenamefont {Lamm}, \citenamefont {Lawrence},\ and\ \citenamefont
  {Yamauchi}}]{Cohen:2021imf}%
  \BibitemOpen
  \bibfield  {author} {\bibinfo {author} {\bibfnamefont {Thomas~D.}\
  \bibnamefont {Cohen}}, \bibinfo {author} {\bibfnamefont {Henry}\ \bibnamefont
  {Lamm}}, \bibinfo {author} {\bibfnamefont {Scott}\ \bibnamefont {Lawrence}},
  \ and\ \bibinfo {author} {\bibfnamefont {Yukari}\ \bibnamefont {Yamauchi}},\
  }\href@noop {} {\enquote {\bibinfo {title} {{Quantum algorithms for transport
  coefficients in gauge theories}},}\ } (\bibinfo {year} {2021}),\ \Eprint
  {http://arxiv.org/abs/2104.02024} {arXiv:2104.02024 [hep-lat]} \BibitemShut
  {NoStop}%
\bibitem [{\citenamefont {Pedernales}\ \emph {et~al.}(2014)\citenamefont
  {Pedernales}, \citenamefont {Di~Candia}, \citenamefont {Egusquiza},
  \citenamefont {Casanova},\ and\ \citenamefont
  {Solano}}]{PhysRevLett.113.020505}%
  \BibitemOpen
  \bibfield  {author} {\bibinfo {author} {\bibfnamefont {J.~S.}\ \bibnamefont
  {Pedernales}}, \bibinfo {author} {\bibfnamefont {R.}~\bibnamefont
  {Di~Candia}}, \bibinfo {author} {\bibfnamefont {I.~L.}\ \bibnamefont
  {Egusquiza}}, \bibinfo {author} {\bibfnamefont {J.}~\bibnamefont {Casanova}},
  \ and\ \bibinfo {author} {\bibfnamefont {E.}~\bibnamefont {Solano}},\
  }\bibfield  {title} {\enquote {\bibinfo {title} {Efficient quantum algorithm
  for computing $n$-time correlation functions},}\ }\href {\doibase
  10.1103/PhysRevLett.113.020505} {\bibfield  {journal} {\bibinfo  {journal}
  {Phys. Rev. Lett.}\ }\textbf {\bibinfo {volume} {113}},\ \bibinfo {pages}
  {020505} (\bibinfo {year} {2014})}\BibitemShut {NoStop}%
\bibitem [{\citenamefont {Ortiz}\ \emph {et~al.}(2001)\citenamefont {Ortiz},
  \citenamefont {Gubernatis}, \citenamefont {Knill},\ and\ \citenamefont
  {Laflamme}}]{Ortiz:2000gc}%
  \BibitemOpen
  \bibfield  {author} {\bibinfo {author} {\bibfnamefont {G.}~\bibnamefont
  {Ortiz}}, \bibinfo {author} {\bibfnamefont {J.~E.}\ \bibnamefont
  {Gubernatis}}, \bibinfo {author} {\bibfnamefont {E.}~\bibnamefont {Knill}}, \
  and\ \bibinfo {author} {\bibfnamefont {R.}~\bibnamefont {Laflamme}},\
  }\bibfield  {title} {\enquote {\bibinfo {title} {{Quantum algorithms for
  fermionic simulations}},}\ }\href {\doibase 10.1103/PhysRevA.64.022319}
  {\bibfield  {journal} {\bibinfo  {journal} {Phys. Rev.}\ }\textbf {\bibinfo
  {volume} {A64}},\ \bibinfo {pages} {022319} (\bibinfo {year} {2001})},\
  \Eprint {http://arxiv.org/abs/cond-mat/0012334} {arXiv:cond-mat/0012334
  [cond-mat]} \BibitemShut {NoStop}%
\bibitem [{\citenamefont {Roggero}\ and\ \citenamefont
  {Carlson}(2018)}]{Roggero:2018hrn}%
  \BibitemOpen
  \bibfield  {author} {\bibinfo {author} {\bibfnamefont {Alessandro}\
  \bibnamefont {Roggero}}\ and\ \bibinfo {author} {\bibfnamefont {Joseph}\
  \bibnamefont {Carlson}},\ }\href@noop {} {\enquote {\bibinfo {title} {{Linear
  Response on a Quantum Computer}},}\ } (\bibinfo {year} {2018}),\ \Eprint
  {http://arxiv.org/abs/1804.01505} {arXiv:1804.01505 [quant-ph]} \BibitemShut
  {NoStop}%
\bibitem [{\citenamefont {Martinelli}\ \emph {et~al.}(1995)\citenamefont
  {Martinelli}, \citenamefont {Pittori}, \citenamefont {Sachrajda},
  \citenamefont {Testa},\ and\ \citenamefont {Vladikas}}]{Martinelli:1994ty}%
  \BibitemOpen
  \bibfield  {author} {\bibinfo {author} {\bibfnamefont {G.}~\bibnamefont
  {Martinelli}}, \bibinfo {author} {\bibfnamefont {C.}~\bibnamefont {Pittori}},
  \bibinfo {author} {\bibfnamefont {Christopher~T.}\ \bibnamefont {Sachrajda}},
  \bibinfo {author} {\bibfnamefont {M.}~\bibnamefont {Testa}}, \ and\ \bibinfo
  {author} {\bibfnamefont {A.}~\bibnamefont {Vladikas}},\ }\bibfield  {title}
  {\enquote {\bibinfo {title} {{A General method for nonperturbative
  renormalization of lattice operators}},}\ }\href {\doibase
  10.1016/0550-3213(95)00126-D} {\bibfield  {journal} {\bibinfo  {journal}
  {Nucl. Phys. B}\ }\textbf {\bibinfo {volume} {445}},\ \bibinfo {pages}
  {81--108} (\bibinfo {year} {1995})},\ \Eprint
  {http://arxiv.org/abs/hep-lat/9411010} {arXiv:hep-lat/9411010} \BibitemShut
  {NoStop}%
\bibitem [{\citenamefont {Aoki}\ \emph {et~al.}(2008)\citenamefont {Aoki} \emph
  {et~al.}}]{Aoki:2007xm}%
  \BibitemOpen
  \bibfield  {author} {\bibinfo {author} {\bibfnamefont {Y.}~\bibnamefont
  {Aoki}} \emph {et~al.},\ }\bibfield  {title} {\enquote {\bibinfo {title}
  {{Non-perturbative renormalization of quark bilinear operators and B(K) using
  domain wall fermions}},}\ }\href {\doibase 10.1103/PhysRevD.78.054510}
  {\bibfield  {journal} {\bibinfo  {journal} {Phys. Rev. D}\ }\textbf {\bibinfo
  {volume} {78}},\ \bibinfo {pages} {054510} (\bibinfo {year} {2008})},\
  \Eprint {http://arxiv.org/abs/0712.1061} {arXiv:0712.1061 [hep-lat]}
  \BibitemShut {NoStop}%
\bibitem [{\citenamefont {Sturm}\ \emph {et~al.}(2009)\citenamefont {Sturm},
  \citenamefont {Aoki}, \citenamefont {Christ}, \citenamefont {Izubuchi},
  \citenamefont {Sachrajda},\ and\ \citenamefont {Soni}}]{Sturm:2009kb}%
  \BibitemOpen
  \bibfield  {author} {\bibinfo {author} {\bibfnamefont {C.}~\bibnamefont
  {Sturm}}, \bibinfo {author} {\bibfnamefont {Y.}~\bibnamefont {Aoki}},
  \bibinfo {author} {\bibfnamefont {N.~H.}\ \bibnamefont {Christ}}, \bibinfo
  {author} {\bibfnamefont {T.}~\bibnamefont {Izubuchi}}, \bibinfo {author}
  {\bibfnamefont {C.~T.~C.}\ \bibnamefont {Sachrajda}}, \ and\ \bibinfo
  {author} {\bibfnamefont {A.}~\bibnamefont {Soni}},\ }\bibfield  {title}
  {\enquote {\bibinfo {title} {{Renormalization of quark bilinear operators in
  a momentum-subtraction scheme with a nonexceptional subtraction point}},}\
  }\href {\doibase 10.1103/PhysRevD.80.014501} {\bibfield  {journal} {\bibinfo
  {journal} {Phys. Rev. D}\ }\textbf {\bibinfo {volume} {80}},\ \bibinfo
  {pages} {014501} (\bibinfo {year} {2009})},\ \Eprint
  {http://arxiv.org/abs/0901.2599} {arXiv:0901.2599 [hep-ph]} \BibitemShut
  {NoStop}%
\bibitem [{\citenamefont {Stannigel}\ \emph {et~al.}(2014)\citenamefont
  {Stannigel}, \citenamefont {Hauke}, \citenamefont {Marcos}, \citenamefont
  {Hafezi}, \citenamefont {Diehl}, \citenamefont {Dalmonte},\ and\
  \citenamefont {Zoller}}]{Stannigel:2013zka}%
  \BibitemOpen
  \bibfield  {author} {\bibinfo {author} {\bibfnamefont {K.}~\bibnamefont
  {Stannigel}}, \bibinfo {author} {\bibfnamefont {P.}~\bibnamefont {Hauke}},
  \bibinfo {author} {\bibfnamefont {D.}~\bibnamefont {Marcos}}, \bibinfo
  {author} {\bibfnamefont {M.}~\bibnamefont {Hafezi}}, \bibinfo {author}
  {\bibfnamefont {S.}~\bibnamefont {Diehl}}, \bibinfo {author} {\bibfnamefont
  {M.}~\bibnamefont {Dalmonte}}, \ and\ \bibinfo {author} {\bibfnamefont
  {P.}~\bibnamefont {Zoller}},\ }\bibfield  {title} {\enquote {\bibinfo {title}
  {{Constrained dynamics via the Zeno effect in quantum simulation:
  Implementing non-Abelian lattice gauge theories with cold atoms}},}\ }\href
  {\doibase 10.1103/PhysRevLett.112.120406} {\bibfield  {journal} {\bibinfo
  {journal} {Phys. Rev. Lett.}\ }\textbf {\bibinfo {volume} {112}},\ \bibinfo
  {pages} {120406} (\bibinfo {year} {2014})},\ \Eprint
  {http://arxiv.org/abs/1308.0528} {arXiv:1308.0528 [quant-ph]} \BibitemShut
  {NoStop}%
\bibitem [{\citenamefont {Stryker}(2019)}]{Stryker:2018efp}%
  \BibitemOpen
  \bibfield  {author} {\bibinfo {author} {\bibfnamefont {Jesse~R.}\
  \bibnamefont {Stryker}},\ }\bibfield  {title} {\enquote {\bibinfo {title}
  {{Oracles for Gauss's law on digital quantum computers}},}\ }\href {\doibase
  10.1103/PhysRevA.99.042301} {\bibfield  {journal} {\bibinfo  {journal} {Phys.
  Rev.}\ }\textbf {\bibinfo {volume} {A99}},\ \bibinfo {pages} {042301}
  (\bibinfo {year} {2019})},\ \Eprint {http://arxiv.org/abs/1812.01617}
  {arXiv:1812.01617 [quant-ph]} \BibitemShut {NoStop}%
\bibitem [{\citenamefont {Halimeh}\ and\ \citenamefont
  {Hauke}(2019)}]{Halimeh:2019svu}%
  \BibitemOpen
  \bibfield  {author} {\bibinfo {author} {\bibfnamefont {Jad~C.}\ \bibnamefont
  {Halimeh}}\ and\ \bibinfo {author} {\bibfnamefont {Philipp}\ \bibnamefont
  {Hauke}},\ }\href@noop {} {\enquote {\bibinfo {title} {{Reliability of
  lattice gauge theories}},}\ } (\bibinfo {year} {2019}),\ \Eprint
  {http://arxiv.org/abs/2001.00024} {arXiv:2001.00024 [cond-mat.quant-gas]}
  \BibitemShut {NoStop}%
\bibitem [{\citenamefont {Lamm}\ \emph
  {et~al.}(2020{\natexlab{b}})\citenamefont {Lamm}, \citenamefont {Lawrence},\
  and\ \citenamefont {Yamauchi}}]{Lamm:2020jwv}%
  \BibitemOpen
  \bibfield  {author} {\bibinfo {author} {\bibfnamefont {Henry}\ \bibnamefont
  {Lamm}}, \bibinfo {author} {\bibfnamefont {Scott}\ \bibnamefont {Lawrence}},
  \ and\ \bibinfo {author} {\bibfnamefont {Yukari}\ \bibnamefont {Yamauchi}}
  (\bibinfo {collaboration} {NuQS}),\ }\href@noop {} {\enquote {\bibinfo
  {title} {{Suppressing Coherent Gauge Drift in Quantum Simulations}},}\ }
  (\bibinfo {year} {2020}{\natexlab{b}}),\ \Eprint
  {http://arxiv.org/abs/2005.12688} {arXiv:2005.12688 [quant-ph]} \BibitemShut
  {NoStop}%
\bibitem [{\citenamefont {Tran}\ \emph {et~al.}(2020)\citenamefont {Tran},
  \citenamefont {Su}, \citenamefont {Carney},\ and\ \citenamefont
  {Taylor}}]{Tran:2020azk}%
  \BibitemOpen
  \bibfield  {author} {\bibinfo {author} {\bibfnamefont {Minh~C.}\ \bibnamefont
  {Tran}}, \bibinfo {author} {\bibfnamefont {Yuan}\ \bibnamefont {Su}},
  \bibinfo {author} {\bibfnamefont {Daniel}\ \bibnamefont {Carney}}, \ and\
  \bibinfo {author} {\bibfnamefont {Jacob~M.}\ \bibnamefont {Taylor}},\
  }\href@noop {} {\enquote {\bibinfo {title} {{Faster Digital Quantum
  Simulation by Symmetry Protection}},}\ } (\bibinfo {year} {2020}),\ \Eprint
  {http://arxiv.org/abs/2006.16248} {arXiv:2006.16248 [quant-ph]} \BibitemShut
  {NoStop}%
\bibitem [{\citenamefont {Halimeh}\ \emph
  {et~al.}(2020{\natexlab{a}})\citenamefont {Halimeh}, \citenamefont {Lang},
  \citenamefont {Mildenberger}, \citenamefont {Jiang},\ and\ \citenamefont
  {Hauke}}]{Halimeh:2020ecg}%
  \BibitemOpen
  \bibfield  {author} {\bibinfo {author} {\bibfnamefont {Jad~C.}\ \bibnamefont
  {Halimeh}}, \bibinfo {author} {\bibfnamefont {Haifeng}\ \bibnamefont {Lang}},
  \bibinfo {author} {\bibfnamefont {Julius}\ \bibnamefont {Mildenberger}},
  \bibinfo {author} {\bibfnamefont {Zhang}\ \bibnamefont {Jiang}}, \ and\
  \bibinfo {author} {\bibfnamefont {Philipp}\ \bibnamefont {Hauke}},\
  }\href@noop {} {\enquote {\bibinfo {title} {{Gauge-Symmetry Protection Using
  Single-Body Terms}},}\ } (\bibinfo {year} {2020}{\natexlab{a}}),\ \Eprint
  {http://arxiv.org/abs/2007.00668} {arXiv:2007.00668 [quant-ph]} \BibitemShut
  {NoStop}%
\bibitem [{\citenamefont {Halimeh}\ and\ \citenamefont
  {Hauke}(2020)}]{Halimeh:2020kyu}%
  \BibitemOpen
  \bibfield  {author} {\bibinfo {author} {\bibfnamefont {Jad~C.}\ \bibnamefont
  {Halimeh}}\ and\ \bibinfo {author} {\bibfnamefont {Philipp}\ \bibnamefont
  {Hauke}},\ }\href@noop {} {\enquote {\bibinfo {title} {{Staircase
  prethermalization and constrained dynamics in lattice gauge theories}},}\ }
  (\bibinfo {year} {2020}),\ \Eprint {http://arxiv.org/abs/2004.07248}
  {arXiv:2004.07248 [cond-mat.quant-gas]} \BibitemShut {NoStop}%
\bibitem [{\citenamefont {Halimeh}\ \emph
  {et~al.}(2020{\natexlab{b}})\citenamefont {Halimeh}, \citenamefont {Kasper},\
  and\ \citenamefont {Hauke}}]{Halimeh:2020djb}%
  \BibitemOpen
  \bibfield  {author} {\bibinfo {author} {\bibfnamefont {Jad~C.}\ \bibnamefont
  {Halimeh}}, \bibinfo {author} {\bibfnamefont {Valentin}\ \bibnamefont
  {Kasper}}, \ and\ \bibinfo {author} {\bibfnamefont {Philipp}\ \bibnamefont
  {Hauke}},\ }\href@noop {} {\enquote {\bibinfo {title} {{Fate of Lattice Gauge
  Theories Under Decoherence}},}\ } (\bibinfo {year} {2020}{\natexlab{b}}),\
  \Eprint {http://arxiv.org/abs/2009.07848} {arXiv:2009.07848
  [cond-mat.quant-gas]} \BibitemShut {NoStop}%
\bibitem [{\citenamefont {Halimeh}\ \emph
  {et~al.}(2020{\natexlab{c}})\citenamefont {Halimeh}, \citenamefont {Ott},
  \citenamefont {McCulloch}, \citenamefont {Yang},\ and\ \citenamefont
  {Hauke}}]{Halimeh:2020xfd}%
  \BibitemOpen
  \bibfield  {author} {\bibinfo {author} {\bibfnamefont {Jad~C.}\ \bibnamefont
  {Halimeh}}, \bibinfo {author} {\bibfnamefont {Robert}\ \bibnamefont {Ott}},
  \bibinfo {author} {\bibfnamefont {Ian~P.}\ \bibnamefont {McCulloch}},
  \bibinfo {author} {\bibfnamefont {Bing}\ \bibnamefont {Yang}}, \ and\
  \bibinfo {author} {\bibfnamefont {Philipp}\ \bibnamefont {Hauke}},\
  }\bibfield  {title} {\enquote {\bibinfo {title} {{Robustness of
  gauge-invariant dynamics against defects in ultracold-atom gauge
  theories}},}\ }\href {\doibase 10.1103/PhysRevResearch.2.033361} {\bibfield
  {journal} {\bibinfo  {journal} {Phys. Rev. Res.}\ }\textbf {\bibinfo {volume}
  {2}},\ \bibinfo {pages} {033361} (\bibinfo {year} {2020}{\natexlab{c}})},\
  \Eprint {http://arxiv.org/abs/2005.10249} {arXiv:2005.10249
  [cond-mat.quant-gas]} \BibitemShut {NoStop}%
\bibitem [{\citenamefont {Van~Damme}\ \emph {et~al.}(2020)\citenamefont
  {Van~Damme}, \citenamefont {Halimeh},\ and\ \citenamefont
  {Hauke}}]{VanDamme:2020rur}%
  \BibitemOpen
  \bibfield  {author} {\bibinfo {author} {\bibfnamefont {Maarten}\ \bibnamefont
  {Van~Damme}}, \bibinfo {author} {\bibfnamefont {Jad~C.}\ \bibnamefont
  {Halimeh}}, \ and\ \bibinfo {author} {\bibfnamefont {Philipp}\ \bibnamefont
  {Hauke}},\ }\href@noop {} {\enquote {\bibinfo {title} {{Gauge-Symmetry
  Violation Quantum Phase Transition in Lattice Gauge Theories}},}\ } (\bibinfo
  {year} {2020}),\ \Eprint {http://arxiv.org/abs/2010.07338} {arXiv:2010.07338
  [cond-mat.quant-gas]} \BibitemShut {NoStop}%
\bibitem [{\citenamefont {Kasper}\ \emph {et~al.}(2020)\citenamefont {Kasper},
  \citenamefont {Zache}, \citenamefont {Jendrzejewski}, \citenamefont
  {Lewenstein},\ and\ \citenamefont {Zohar}}]{Kasper:2020owz}%
  \BibitemOpen
  \bibfield  {author} {\bibinfo {author} {\bibfnamefont {Valentin}\
  \bibnamefont {Kasper}}, \bibinfo {author} {\bibfnamefont {Torsten~V.}\
  \bibnamefont {Zache}}, \bibinfo {author} {\bibfnamefont {Fred}\ \bibnamefont
  {Jendrzejewski}}, \bibinfo {author} {\bibfnamefont {Maciej}\ \bibnamefont
  {Lewenstein}}, \ and\ \bibinfo {author} {\bibfnamefont {Erez}\ \bibnamefont
  {Zohar}},\ }\href@noop {} {\enquote {\bibinfo {title} {{Non-Abelian gauge
  invariance from dynamical decoupling}},}\ } (\bibinfo {year} {2020}),\
  \Eprint {http://arxiv.org/abs/2012.08620} {arXiv:2012.08620 [quant-ph]}
  \BibitemShut {NoStop}%
\bibitem [{\citenamefont {Halimeh}\ \emph {et~al.}(2021)\citenamefont
  {Halimeh}, \citenamefont {Lang},\ and\ \citenamefont
  {Hauke}}]{Halimeh:2021vzf}%
  \BibitemOpen
  \bibfield  {author} {\bibinfo {author} {\bibfnamefont {Jad~C.}\ \bibnamefont
  {Halimeh}}, \bibinfo {author} {\bibfnamefont {Haifeng}\ \bibnamefont {Lang}},
  \ and\ \bibinfo {author} {\bibfnamefont {Philipp}\ \bibnamefont {Hauke}},\
  }\href@noop {} {\enquote {\bibinfo {title} {{Gauge protection in non-Abelian
  lattice gauge theories}},}\ } (\bibinfo {year} {2021}),\ \Eprint
  {http://arxiv.org/abs/2106.09032} {arXiv:2106.09032 [cond-mat.quant-gas]}
  \BibitemShut {NoStop}%
\bibitem [{\citenamefont {Zohar}\ \emph
  {et~al.}(2012{\natexlab{b}})\citenamefont {Zohar}, \citenamefont {Cirac},\
  and\ \citenamefont {Reznik}}]{zohar2012simulating}%
  \BibitemOpen
  \bibfield  {author} {\bibinfo {author} {\bibfnamefont {Erez}\ \bibnamefont
  {Zohar}}, \bibinfo {author} {\bibfnamefont {J~Ignacio}\ \bibnamefont
  {Cirac}}, \ and\ \bibinfo {author} {\bibfnamefont {Benni}\ \bibnamefont
  {Reznik}},\ }\bibfield  {title} {\enquote {\bibinfo {title} {Simulating
  compact quantum electrodynamics with ultracold atoms: Probing confinement and
  nonperturbative effects},}\ }\href@noop {} {\bibfield  {journal} {\bibinfo
  {journal} {Physical review letters}\ }\textbf {\bibinfo {volume} {109}},\
  \bibinfo {pages} {125302} (\bibinfo {year} {2012}{\natexlab{b}})}\BibitemShut
  {NoStop}%
\bibitem [{\citenamefont {Zohar}\ and\ \citenamefont
  {Reznik}(2011)}]{zohar2011confinement}%
  \BibitemOpen
  \bibfield  {author} {\bibinfo {author} {\bibfnamefont {Erez}\ \bibnamefont
  {Zohar}}\ and\ \bibinfo {author} {\bibfnamefont {Benni}\ \bibnamefont
  {Reznik}},\ }\bibfield  {title} {\enquote {\bibinfo {title} {Confinement and
  lattice quantum-electrodynamic electric flux tubes simulated with ultracold
  atoms},}\ }\href@noop {} {\bibfield  {journal} {\bibinfo  {journal} {Physical
  review letters}\ }\textbf {\bibinfo {volume} {107}},\ \bibinfo {pages}
  {275301} (\bibinfo {year} {2011})}\BibitemShut {NoStop}%
\bibitem [{\citenamefont {Zohar}\ \emph
  {et~al.}(2013{\natexlab{c}})\citenamefont {Zohar}, \citenamefont {Cirac},\
  and\ \citenamefont {Reznik}}]{zohar2013simulating}%
  \BibitemOpen
  \bibfield  {author} {\bibinfo {author} {\bibfnamefont {Erez}\ \bibnamefont
  {Zohar}}, \bibinfo {author} {\bibfnamefont {J~Ignacio}\ \bibnamefont
  {Cirac}}, \ and\ \bibinfo {author} {\bibfnamefont {Benni}\ \bibnamefont
  {Reznik}},\ }\bibfield  {title} {\enquote {\bibinfo {title} {Simulating
  (2+1)-dimensional lattice qed with dynamical matter using ultracold atoms},}\
  }\href@noop {} {\bibfield  {journal} {\bibinfo  {journal} {Physical review
  letters}\ }\textbf {\bibinfo {volume} {110}},\ \bibinfo {pages} {055302}
  (\bibinfo {year} {2013}{\natexlab{c}})}\BibitemShut {NoStop}%
\bibitem [{\citenamefont {Tagliacozzo}\ \emph
  {et~al.}(2013{\natexlab{b}})\citenamefont {Tagliacozzo}, \citenamefont
  {Celi}, \citenamefont {Zamora},\ and\ \citenamefont
  {Lewenstein}}]{tagliacozzo2013optical}%
  \BibitemOpen
  \bibfield  {author} {\bibinfo {author} {\bibfnamefont {Luca}\ \bibnamefont
  {Tagliacozzo}}, \bibinfo {author} {\bibfnamefont {Alessio}\ \bibnamefont
  {Celi}}, \bibinfo {author} {\bibfnamefont {Alejandro}\ \bibnamefont
  {Zamora}}, \ and\ \bibinfo {author} {\bibfnamefont {Maciej}\ \bibnamefont
  {Lewenstein}},\ }\bibfield  {title} {\enquote {\bibinfo {title} {Optical
  abelian lattice gauge theories},}\ }\href@noop {} {\bibfield  {journal}
  {\bibinfo  {journal} {Annals of Physics}\ }\textbf {\bibinfo {volume}
  {330}},\ \bibinfo {pages} {160--191} (\bibinfo {year}
  {2013}{\natexlab{b}})}\BibitemShut {NoStop}%
\bibitem [{\citenamefont {Ott}\ \emph {et~al.}(2021)\citenamefont {Ott},
  \citenamefont {Zache}, \citenamefont {Jendrzejewski},\ and\ \citenamefont
  {Berges}}]{ott2021scalable}%
  \BibitemOpen
  \bibfield  {author} {\bibinfo {author} {\bibfnamefont {Robert}\ \bibnamefont
  {Ott}}, \bibinfo {author} {\bibfnamefont {Torsten~V}\ \bibnamefont {Zache}},
  \bibinfo {author} {\bibfnamefont {Fred}\ \bibnamefont {Jendrzejewski}}, \
  and\ \bibinfo {author} {\bibfnamefont {J{\"u}rgen}\ \bibnamefont {Berges}},\
  }\bibfield  {title} {\enquote {\bibinfo {title} {Scalable cold-atom quantum
  simulator for two-dimensional qed},}\ }\href@noop {} {\bibfield  {journal}
  {\bibinfo  {journal} {Physical Review Letters}\ }\textbf {\bibinfo {volume}
  {127}},\ \bibinfo {pages} {130504} (\bibinfo {year} {2021})}\BibitemShut
  {NoStop}%
\bibitem [{\citenamefont {Gonz{\'a}lez-Cuadra}\ \emph
  {et~al.}(2022)\citenamefont {Gonz{\'a}lez-Cuadra}, \citenamefont {Zache},
  \citenamefont {Carrasco}, \citenamefont {Kraus},\ and\ \citenamefont
  {Zoller}}]{gonzalez2022hardware}%
  \BibitemOpen
  \bibfield  {author} {\bibinfo {author} {\bibfnamefont {Daniel}\ \bibnamefont
  {Gonz{\'a}lez-Cuadra}}, \bibinfo {author} {\bibfnamefont {Torsten~V}\
  \bibnamefont {Zache}}, \bibinfo {author} {\bibfnamefont {Jose}\ \bibnamefont
  {Carrasco}}, \bibinfo {author} {\bibfnamefont {Barbara}\ \bibnamefont
  {Kraus}}, \ and\ \bibinfo {author} {\bibfnamefont {Peter}\ \bibnamefont
  {Zoller}},\ }\bibfield  {title} {\enquote {\bibinfo {title} {Hardware
  efficient quantum simulation of non-abelian gauge theories with qudits on
  rydberg platforms},}\ }\href@noop {} {\bibfield  {journal} {\bibinfo
  {journal} {arXiv preprint arXiv:2203.15541}\ } (\bibinfo {year}
  {2022})}\BibitemShut {NoStop}%
\bibitem [{\citenamefont {Martinez}\ \emph
  {et~al.}(2016{\natexlab{b}})\citenamefont {Martinez}, \citenamefont
  {Muschik}, \citenamefont {Schindler}, \citenamefont {Nigg}, \citenamefont
  {Erhard}, \citenamefont {Heyl}, \citenamefont {Hauke}, \citenamefont
  {Dalmonte}, \citenamefont {Monz}, \citenamefont {Zoller} \emph
  {et~al.}}]{martinez2016real}%
  \BibitemOpen
  \bibfield  {author} {\bibinfo {author} {\bibfnamefont {Esteban~A}\
  \bibnamefont {Martinez}}, \bibinfo {author} {\bibfnamefont {Christine~A}\
  \bibnamefont {Muschik}}, \bibinfo {author} {\bibfnamefont {Philipp}\
  \bibnamefont {Schindler}}, \bibinfo {author} {\bibfnamefont {Daniel}\
  \bibnamefont {Nigg}}, \bibinfo {author} {\bibfnamefont {Alexander}\
  \bibnamefont {Erhard}}, \bibinfo {author} {\bibfnamefont {Markus}\
  \bibnamefont {Heyl}}, \bibinfo {author} {\bibfnamefont {Philipp}\
  \bibnamefont {Hauke}}, \bibinfo {author} {\bibfnamefont {Marcello}\
  \bibnamefont {Dalmonte}}, \bibinfo {author} {\bibfnamefont {Thomas}\
  \bibnamefont {Monz}}, \bibinfo {author} {\bibfnamefont {Peter}\ \bibnamefont
  {Zoller}},  \emph {et~al.},\ }\bibfield  {title} {\enquote {\bibinfo {title}
  {Real-time dynamics of lattice gauge theories with a few-qubit quantum
  computer},}\ }\href@noop {} {\bibfield  {journal} {\bibinfo  {journal}
  {Nature}\ }\textbf {\bibinfo {volume} {534}},\ \bibinfo {pages} {516--519}
  (\bibinfo {year} {2016}{\natexlab{b}})}\BibitemShut {NoStop}%
\bibitem [{\citenamefont {Klco}\ \emph {et~al.}(2018)\citenamefont {Klco},
  \citenamefont {Dumitrescu}, \citenamefont {McCaskey}, \citenamefont {Morris},
  \citenamefont {Pooser}, \citenamefont {Sanz}, \citenamefont {Solano},
  \citenamefont {Lougovski},\ and\ \citenamefont {Savage}}]{Klco:2018kyo}%
  \BibitemOpen
  \bibfield  {author} {\bibinfo {author} {\bibfnamefont {N.}~\bibnamefont
  {Klco}}, \bibinfo {author} {\bibfnamefont {E.~F.}\ \bibnamefont
  {Dumitrescu}}, \bibinfo {author} {\bibfnamefont {A.~J.}\ \bibnamefont
  {McCaskey}}, \bibinfo {author} {\bibfnamefont {T.~D.}\ \bibnamefont
  {Morris}}, \bibinfo {author} {\bibfnamefont {R.~C.}\ \bibnamefont {Pooser}},
  \bibinfo {author} {\bibfnamefont {M.}~\bibnamefont {Sanz}}, \bibinfo {author}
  {\bibfnamefont {E.}~\bibnamefont {Solano}}, \bibinfo {author} {\bibfnamefont
  {P.}~\bibnamefont {Lougovski}}, \ and\ \bibinfo {author} {\bibfnamefont
  {M.~J.}\ \bibnamefont {Savage}},\ }\bibfield  {title} {\enquote {\bibinfo
  {title} {{Quantum-classical computation of Schwinger model dynamics using
  quantum computers}},}\ }\href {\doibase 10.1103/PhysRevA.98.032331}
  {\bibfield  {journal} {\bibinfo  {journal} {Phys. Rev. A}\ }\textbf {\bibinfo
  {volume} {98}},\ \bibinfo {pages} {032331} (\bibinfo {year} {2018})},\
  \Eprint {http://arxiv.org/abs/1803.03326} {arXiv:1803.03326 [quant-ph]}
  \BibitemShut {NoStop}%
\bibitem [{\citenamefont {Kokail}\ \emph {et~al.}(2019)\citenamefont {Kokail},
  \citenamefont {Maier}, \citenamefont {van Bijnen}, \citenamefont {Brydges},
  \citenamefont {Joshi}, \citenamefont {Jurcevic}, \citenamefont {Muschik},
  \citenamefont {Silvi}, \citenamefont {Blatt}, \citenamefont {Roos} \emph
  {et~al.}}]{kokail2019self}%
  \BibitemOpen
  \bibfield  {author} {\bibinfo {author} {\bibfnamefont {Christian}\
  \bibnamefont {Kokail}}, \bibinfo {author} {\bibfnamefont {Christine}\
  \bibnamefont {Maier}}, \bibinfo {author} {\bibfnamefont {Rick}\ \bibnamefont
  {van Bijnen}}, \bibinfo {author} {\bibfnamefont {Tiff}\ \bibnamefont
  {Brydges}}, \bibinfo {author} {\bibfnamefont {Manoj~K}\ \bibnamefont
  {Joshi}}, \bibinfo {author} {\bibfnamefont {Petar}\ \bibnamefont {Jurcevic}},
  \bibinfo {author} {\bibfnamefont {Christine~A}\ \bibnamefont {Muschik}},
  \bibinfo {author} {\bibfnamefont {Pietro}\ \bibnamefont {Silvi}}, \bibinfo
  {author} {\bibfnamefont {Rainer}\ \bibnamefont {Blatt}}, \bibinfo {author}
  {\bibfnamefont {Christian~F}\ \bibnamefont {Roos}},  \emph {et~al.},\
  }\bibfield  {title} {\enquote {\bibinfo {title} {Self-verifying variational
  quantum simulation of lattice models},}\ }\href@noop {} {\bibfield  {journal}
  {\bibinfo  {journal} {Nature}\ }\textbf {\bibinfo {volume} {569}},\ \bibinfo
  {pages} {355--360} (\bibinfo {year} {2019})}\BibitemShut {NoStop}%
\bibitem [{\citenamefont {Lu}\ \emph {et~al.}(2019)\citenamefont {Lu},
  \citenamefont {Klco}, \citenamefont {Lukens}, \citenamefont {Morris},
  \citenamefont {Bansal}, \citenamefont {Ekstr{\"o}m}, \citenamefont {Hagen},
  \citenamefont {Papenbrock}, \citenamefont {Weiner}, \citenamefont {Savage}
  \emph {et~al.}}]{lu2019simulations}%
  \BibitemOpen
  \bibfield  {author} {\bibinfo {author} {\bibfnamefont {Hsuan-Hao}\
  \bibnamefont {Lu}}, \bibinfo {author} {\bibfnamefont {Natalie}\ \bibnamefont
  {Klco}}, \bibinfo {author} {\bibfnamefont {Joseph~M}\ \bibnamefont {Lukens}},
  \bibinfo {author} {\bibfnamefont {Titus~D}\ \bibnamefont {Morris}}, \bibinfo
  {author} {\bibfnamefont {Aaina}\ \bibnamefont {Bansal}}, \bibinfo {author}
  {\bibfnamefont {Andreas}\ \bibnamefont {Ekstr{\"o}m}}, \bibinfo {author}
  {\bibfnamefont {Gaute}\ \bibnamefont {Hagen}}, \bibinfo {author}
  {\bibfnamefont {Thomas}\ \bibnamefont {Papenbrock}}, \bibinfo {author}
  {\bibfnamefont {Andrew~M}\ \bibnamefont {Weiner}}, \bibinfo {author}
  {\bibfnamefont {Martin~J}\ \bibnamefont {Savage}},  \emph {et~al.},\
  }\bibfield  {title} {\enquote {\bibinfo {title} {Simulations of subatomic
  many-body physics on a quantum frequency processor},}\ }\href@noop {}
  {\bibfield  {journal} {\bibinfo  {journal} {Physical Review A}\ }\textbf
  {\bibinfo {volume} {100}},\ \bibinfo {pages} {012320} (\bibinfo {year}
  {2019})}\BibitemShut {NoStop}%
\bibitem [{\citenamefont {Alam}\ \emph {et~al.}(2021)\citenamefont {Alam},
  \citenamefont {Hadfield}, \citenamefont {Lamm},\ and\ \citenamefont
  {Li}}]{alam2021quantum}%
  \BibitemOpen
  \bibfield  {author} {\bibinfo {author} {\bibfnamefont {M~Sohaib}\
  \bibnamefont {Alam}}, \bibinfo {author} {\bibfnamefont {Stuart}\ \bibnamefont
  {Hadfield}}, \bibinfo {author} {\bibfnamefont {Henry}\ \bibnamefont {Lamm}},
  \ and\ \bibinfo {author} {\bibfnamefont {Andy~CY}\ \bibnamefont {Li}},\
  }\bibfield  {title} {\enquote {\bibinfo {title} {Quantum simulation of
  dihedral gauge theories},}\ }\href@noop {} {\bibfield  {journal} {\bibinfo
  {journal} {arXiv preprint arXiv:2108.13305}\ } (\bibinfo {year}
  {2021})}\BibitemShut {NoStop}%
\bibitem [{\citenamefont {Xu}\ and\ \citenamefont {Xue}(2021)}]{xu20213+}%
  \BibitemOpen
  \bibfield  {author} {\bibinfo {author} {\bibfnamefont {Bin}\ \bibnamefont
  {Xu}}\ and\ \bibinfo {author} {\bibfnamefont {Wei}\ \bibnamefont {Xue}},\
  }\bibfield  {title} {\enquote {\bibinfo {title} {{3+1 Dimension Schwinger
  Pair Production with Quantum Computers}},}\ }\href@noop {} {\bibfield
  {journal} {\bibinfo  {journal} {arXiv preprint arXiv:2112.06863}\ } (\bibinfo
  {year} {2021})}\BibitemShut {NoStop}%
\bibitem [{\citenamefont {Mildenberger}\ \emph {et~al.}(2022)\citenamefont
  {Mildenberger}, \citenamefont {Mruczkiewicz}, \citenamefont {Halimeh},
  \citenamefont {Jiang},\ and\ \citenamefont
  {Hauke}}]{mildenberger2022probing}%
  \BibitemOpen
  \bibfield  {author} {\bibinfo {author} {\bibfnamefont {Julius}\ \bibnamefont
  {Mildenberger}}, \bibinfo {author} {\bibfnamefont {Wojciech}\ \bibnamefont
  {Mruczkiewicz}}, \bibinfo {author} {\bibfnamefont {Jad~C}\ \bibnamefont
  {Halimeh}}, \bibinfo {author} {\bibfnamefont {Zhang}\ \bibnamefont {Jiang}},
  \ and\ \bibinfo {author} {\bibfnamefont {Philipp}\ \bibnamefont {Hauke}},\
  }\bibfield  {title} {\enquote {\bibinfo {title} {{Probing confinement in a Z2
  lattice gauge theory on a quantum computer}},}\ }\href@noop {} {\bibfield
  {journal} {\bibinfo  {journal} {arXiv preprint arXiv:2203.08905}\ } (\bibinfo
  {year} {2022})}\BibitemShut {NoStop}%
\bibitem [{\citenamefont {Schweizer}\ \emph {et~al.}(2019)\citenamefont
  {Schweizer}, \citenamefont {Grusdt}, \citenamefont {Berngruber},
  \citenamefont {Barbiero}, \citenamefont {Demler}, \citenamefont {Goldman},
  \citenamefont {Bloch},\ and\ \citenamefont
  {Aidelsburger}}]{schweizer2019floquet}%
  \BibitemOpen
  \bibfield  {author} {\bibinfo {author} {\bibfnamefont {Christian}\
  \bibnamefont {Schweizer}}, \bibinfo {author} {\bibfnamefont {Fabian}\
  \bibnamefont {Grusdt}}, \bibinfo {author} {\bibfnamefont {Moritz}\
  \bibnamefont {Berngruber}}, \bibinfo {author} {\bibfnamefont {Luca}\
  \bibnamefont {Barbiero}}, \bibinfo {author} {\bibfnamefont {Eugene}\
  \bibnamefont {Demler}}, \bibinfo {author} {\bibfnamefont {Nathan}\
  \bibnamefont {Goldman}}, \bibinfo {author} {\bibfnamefont {Immanuel}\
  \bibnamefont {Bloch}}, \ and\ \bibinfo {author} {\bibfnamefont {Monika}\
  \bibnamefont {Aidelsburger}},\ }\bibfield  {title} {\enquote {\bibinfo
  {title} {{Floquet approach to Z@ lattice gauge theories with ultracold atoms
  in optical lattices}},}\ }\href@noop {} {\bibfield  {journal} {\bibinfo
  {journal} {Nature Physics}\ }\textbf {\bibinfo {volume} {15}},\ \bibinfo
  {pages} {1168--1173} (\bibinfo {year} {2019})}\BibitemShut {NoStop}%
\bibitem [{\citenamefont {G{\"o}rg}\ \emph {et~al.}(2019)\citenamefont
  {G{\"o}rg}, \citenamefont {Sandholzer}, \citenamefont {Minguzzi},
  \citenamefont {Desbuquois}, \citenamefont {Messer},\ and\ \citenamefont
  {Esslinger}}]{gorg2019realization}%
  \BibitemOpen
  \bibfield  {author} {\bibinfo {author} {\bibfnamefont {Frederik}\
  \bibnamefont {G{\"o}rg}}, \bibinfo {author} {\bibfnamefont {Kilian}\
  \bibnamefont {Sandholzer}}, \bibinfo {author} {\bibfnamefont {Joaqu{\'\i}n}\
  \bibnamefont {Minguzzi}}, \bibinfo {author} {\bibfnamefont {R{\'e}mi}\
  \bibnamefont {Desbuquois}}, \bibinfo {author} {\bibfnamefont {Michael}\
  \bibnamefont {Messer}}, \ and\ \bibinfo {author} {\bibfnamefont {Tilman}\
  \bibnamefont {Esslinger}},\ }\bibfield  {title} {\enquote {\bibinfo {title}
  {Realization of density-dependent peierls phases to engineer quantized gauge
  fields coupled to ultracold matter},}\ }\href@noop {} {\bibfield  {journal}
  {\bibinfo  {journal} {Nature Physics}\ }\textbf {\bibinfo {volume} {15}},\
  \bibinfo {pages} {1161--1167} (\bibinfo {year} {2019})}\BibitemShut {NoStop}%
\bibitem [{\citenamefont {Mil}\ \emph {et~al.}(2019)\citenamefont {Mil},
  \citenamefont {Zache}, \citenamefont {Hegde}, \citenamefont {Xia},
  \citenamefont {Bhatt}, \citenamefont {Oberthaler}, \citenamefont {Hauke},
  \citenamefont {Berges},\ and\ \citenamefont
  {Jendrzejewski}}]{mil2019realizing}%
  \BibitemOpen
  \bibfield  {author} {\bibinfo {author} {\bibfnamefont {Alexander}\
  \bibnamefont {Mil}}, \bibinfo {author} {\bibfnamefont {Torsten~V}\
  \bibnamefont {Zache}}, \bibinfo {author} {\bibfnamefont {Apoorva}\
  \bibnamefont {Hegde}}, \bibinfo {author} {\bibfnamefont {Andy}\ \bibnamefont
  {Xia}}, \bibinfo {author} {\bibfnamefont {Rohit~P}\ \bibnamefont {Bhatt}},
  \bibinfo {author} {\bibfnamefont {Markus~K}\ \bibnamefont {Oberthaler}},
  \bibinfo {author} {\bibfnamefont {Philipp}\ \bibnamefont {Hauke}}, \bibinfo
  {author} {\bibfnamefont {J{\"u}rgen}\ \bibnamefont {Berges}}, \ and\ \bibinfo
  {author} {\bibfnamefont {Fred}\ \bibnamefont {Jendrzejewski}},\ }\bibfield
  {title} {\enquote {\bibinfo {title} {{Realizing a scalable building block of
  a U(1) gauge theory with cold atomic mixtures}},}\ }\href@noop {} {\bibfield
  {journal} {\bibinfo  {journal} {arXiv preprint arXiv:1909.07641}\ } (\bibinfo
  {year} {2019})}\BibitemShut {NoStop}%
\bibitem [{\citenamefont {Yang}\ \emph {et~al.}(2020)\citenamefont {Yang},
  \citenamefont {Sun}, \citenamefont {Ott}, \citenamefont {Wang}, \citenamefont
  {Zache}, \citenamefont {Halimeh}, \citenamefont {Yuan}, \citenamefont
  {Hauke},\ and\ \citenamefont {Pan}}]{yang2020observation}%
  \BibitemOpen
  \bibfield  {author} {\bibinfo {author} {\bibfnamefont {Bing}\ \bibnamefont
  {Yang}}, \bibinfo {author} {\bibfnamefont {Hui}\ \bibnamefont {Sun}},
  \bibinfo {author} {\bibfnamefont {Robert}\ \bibnamefont {Ott}}, \bibinfo
  {author} {\bibfnamefont {Han-Yi}\ \bibnamefont {Wang}}, \bibinfo {author}
  {\bibfnamefont {Torsten~V}\ \bibnamefont {Zache}}, \bibinfo {author}
  {\bibfnamefont {Jad~C}\ \bibnamefont {Halimeh}}, \bibinfo {author}
  {\bibfnamefont {Zhen-Sheng}\ \bibnamefont {Yuan}}, \bibinfo {author}
  {\bibfnamefont {Philipp}\ \bibnamefont {Hauke}}, \ and\ \bibinfo {author}
  {\bibfnamefont {Jian-Wei}\ \bibnamefont {Pan}},\ }\bibfield  {title}
  {\enquote {\bibinfo {title} {Observation of gauge invariance in a 71-site
  bose--hubbard quantum simulator},}\ }\href@noop {} {\bibfield  {journal}
  {\bibinfo  {journal} {Nature}\ }\textbf {\bibinfo {volume} {587}},\ \bibinfo
  {pages} {392--396} (\bibinfo {year} {2020})}\BibitemShut {NoStop}%
\bibitem [{\citenamefont {Zhou}\ \emph {et~al.}(2022)\citenamefont {Zhou},
  \citenamefont {Su}, \citenamefont {Halimeh}, \citenamefont {Ott},
  \citenamefont {Sun}, \citenamefont {Hauke}, \citenamefont {Yang},
  \citenamefont {Yuan}, \citenamefont {Berges},\ and\ \citenamefont
  {Pan}}]{Zhou:2021kdl}%
  \BibitemOpen
  \bibfield  {author} {\bibinfo {author} {\bibfnamefont {Zhao-Yu}\ \bibnamefont
  {Zhou}}, \bibinfo {author} {\bibfnamefont {Guo-Xian}\ \bibnamefont {Su}},
  \bibinfo {author} {\bibfnamefont {Jad~C.}\ \bibnamefont {Halimeh}}, \bibinfo
  {author} {\bibfnamefont {Robert}\ \bibnamefont {Ott}}, \bibinfo {author}
  {\bibfnamefont {Hui}\ \bibnamefont {Sun}}, \bibinfo {author} {\bibfnamefont
  {Philipp}\ \bibnamefont {Hauke}}, \bibinfo {author} {\bibfnamefont {Bing}\
  \bibnamefont {Yang}}, \bibinfo {author} {\bibfnamefont {Zhen-Sheng}\
  \bibnamefont {Yuan}}, \bibinfo {author} {\bibfnamefont {J\"urgen}\
  \bibnamefont {Berges}}, \ and\ \bibinfo {author} {\bibfnamefont {Jian-Wei}\
  \bibnamefont {Pan}},\ }\bibfield  {title} {\enquote {\bibinfo {title}
  {{Thermalization dynamics of a gauge theory on a quantum simulator}},}\
  }\href {\doibase 10.1126/science.abl6277} {\bibfield  {journal} {\bibinfo
  {journal} {Science}\ }\textbf {\bibinfo {volume} {377}},\ \bibinfo {pages}
  {abl6277} (\bibinfo {year} {2022})},\ \Eprint
  {http://arxiv.org/abs/2107.13563} {arXiv:2107.13563 [cond-mat.quant-gas]}
  \BibitemShut {NoStop}%
\bibitem [{\citenamefont {Hauke}\ \emph {et~al.}(2013)\citenamefont {Hauke},
  \citenamefont {Marcos}, \citenamefont {Dalmonte},\ and\ \citenamefont
  {Zoller}}]{Hauke:2013jga}%
  \BibitemOpen
  \bibfield  {author} {\bibinfo {author} {\bibfnamefont {Philipp}\ \bibnamefont
  {Hauke}}, \bibinfo {author} {\bibfnamefont {David}\ \bibnamefont {Marcos}},
  \bibinfo {author} {\bibfnamefont {Marcello}\ \bibnamefont {Dalmonte}}, \ and\
  \bibinfo {author} {\bibfnamefont {Peter}\ \bibnamefont {Zoller}},\ }\bibfield
   {title} {\enquote {\bibinfo {title} {{Quantum simulation of a lattice
  Schwinger model in a chain of trapped ions}},}\ }\href {\doibase
  10.1103/PhysRevX.3.041018} {\bibfield  {journal} {\bibinfo  {journal} {Phys.
  Rev.}\ }\textbf {\bibinfo {volume} {X3}},\ \bibinfo {pages} {041018}
  (\bibinfo {year} {2013})},\ \Eprint {http://arxiv.org/abs/1306.2162}
  {arXiv:1306.2162 [cond-mat.quant-gas]} \BibitemShut {NoStop}%
\bibitem [{\citenamefont {Macridin}\ \emph {et~al.}(2018)\citenamefont
  {Macridin}, \citenamefont {Spentzouris}, \citenamefont {Amundson},\ and\
  \citenamefont {Harnik}}]{Macridin:2018gdw}%
  \BibitemOpen
  \bibfield  {author} {\bibinfo {author} {\bibfnamefont {Alexandru}\
  \bibnamefont {Macridin}}, \bibinfo {author} {\bibfnamefont {Panagiotis}\
  \bibnamefont {Spentzouris}}, \bibinfo {author} {\bibfnamefont {James}\
  \bibnamefont {Amundson}}, \ and\ \bibinfo {author} {\bibfnamefont {Roni}\
  \bibnamefont {Harnik}},\ }\bibfield  {title} {\enquote {\bibinfo {title}
  {{Electron-Phonon Systems on a Universal Quantum Computer}},}\ }\href
  {\doibase 10.1103/PhysRevLett.121.110504} {\bibfield  {journal} {\bibinfo
  {journal} {Phys. Rev. Lett.}\ }\textbf {\bibinfo {volume} {121}},\ \bibinfo
  {pages} {110504} (\bibinfo {year} {2018})},\ \Eprint
  {http://arxiv.org/abs/1802.07347} {arXiv:1802.07347 [quant-ph]} \BibitemShut
  {NoStop}%
\bibitem [{\citenamefont {Davoudi}\ \emph {et~al.}(2021)\citenamefont
  {Davoudi}, \citenamefont {Linke},\ and\ \citenamefont
  {Pagano}}]{davoudi2021towards}%
  \BibitemOpen
  \bibfield  {author} {\bibinfo {author} {\bibfnamefont {Zohreh}\ \bibnamefont
  {Davoudi}}, \bibinfo {author} {\bibfnamefont {Norbert~M.}\ \bibnamefont
  {Linke}}, \ and\ \bibinfo {author} {\bibfnamefont {Guido}\ \bibnamefont
  {Pagano}},\ }\bibfield  {title} {\enquote {\bibinfo {title} {Toward
  simulating quantum field theories with controlled phonon-ion dynamics: A
  hybrid analog-digital approach},}\ }\href {\doibase
  10.1103/PhysRevResearch.3.043072} {\bibfield  {journal} {\bibinfo  {journal}
  {Phys. Rev. Research}\ }\textbf {\bibinfo {volume} {3}},\ \bibinfo {pages}
  {043072} (\bibinfo {year} {2021})}\BibitemShut {NoStop}%
\bibitem [{\citenamefont {Casanova}\ \emph {et~al.}(2011)\citenamefont
  {Casanova}, \citenamefont {Lamata}, \citenamefont {Egusquiza}, \citenamefont
  {Gerritsma}, \citenamefont {Roos}, \citenamefont {Garc{\'\i}a-Ripoll},\ and\
  \citenamefont {Solano}}]{casanova2011quantum}%
  \BibitemOpen
  \bibfield  {author} {\bibinfo {author} {\bibfnamefont {Jorge}\ \bibnamefont
  {Casanova}}, \bibinfo {author} {\bibfnamefont {Lucas}\ \bibnamefont
  {Lamata}}, \bibinfo {author} {\bibfnamefont {IL}~\bibnamefont {Egusquiza}},
  \bibinfo {author} {\bibfnamefont {Rene}\ \bibnamefont {Gerritsma}}, \bibinfo
  {author} {\bibfnamefont {Christian~F}\ \bibnamefont {Roos}}, \bibinfo
  {author} {\bibfnamefont {Juan~Jos{\'e}}\ \bibnamefont {Garc{\'\i}a-Ripoll}},
  \ and\ \bibinfo {author} {\bibfnamefont {Enrique}\ \bibnamefont {Solano}},\
  }\bibfield  {title} {\enquote {\bibinfo {title} {Quantum simulation of
  quantum field theories in trapped ions},}\ }\href@noop {} {\bibfield
  {journal} {\bibinfo  {journal} {Physical review letters}\ }\textbf {\bibinfo
  {volume} {107}},\ \bibinfo {pages} {260501} (\bibinfo {year}
  {2011})}\BibitemShut {NoStop}%
\bibitem [{\citenamefont {Banerjee}\ \emph {et~al.}(2013)\citenamefont
  {Banerjee}, \citenamefont {Bögli}, \citenamefont {Dalmonte}, \citenamefont
  {Rico}, \citenamefont {Stebler}, \citenamefont {Wiese},\ and\ \citenamefont
  {Zoller}}]{Banerjee:2012xg}%
  \BibitemOpen
  \bibfield  {author} {\bibinfo {author} {\bibfnamefont {D.}~\bibnamefont
  {Banerjee}}, \bibinfo {author} {\bibfnamefont {M.}~\bibnamefont {Bögli}},
  \bibinfo {author} {\bibfnamefont {M.}~\bibnamefont {Dalmonte}}, \bibinfo
  {author} {\bibfnamefont {E.}~\bibnamefont {Rico}}, \bibinfo {author}
  {\bibfnamefont {P.}~\bibnamefont {Stebler}}, \bibinfo {author} {\bibfnamefont
  {U.~J.}\ \bibnamefont {Wiese}}, \ and\ \bibinfo {author} {\bibfnamefont
  {P.}~\bibnamefont {Zoller}},\ }\bibfield  {title} {\enquote {\bibinfo {title}
  {{Atomic Quantum Simulation of U(N) and SU(N) Non-Abelian Lattice Gauge
  Theories}},}\ }\href {\doibase 10.1103/PhysRevLett.110.125303} {\bibfield
  {journal} {\bibinfo  {journal} {Phys. Rev. Lett.}\ }\textbf {\bibinfo
  {volume} {110}},\ \bibinfo {pages} {125303} (\bibinfo {year} {2013})},\
  \Eprint {http://arxiv.org/abs/1211.2242} {arXiv:1211.2242
  [cond-mat.quant-gas]} \BibitemShut {NoStop}%
\bibitem [{\citenamefont {Dasgupta}\ and\ \citenamefont
  {Raychowdhury}(2022)}]{dasgupta2022cold}%
  \BibitemOpen
  \bibfield  {author} {\bibinfo {author} {\bibfnamefont {Raka}\ \bibnamefont
  {Dasgupta}}\ and\ \bibinfo {author} {\bibfnamefont {Indrakshi}\ \bibnamefont
  {Raychowdhury}},\ }\bibfield  {title} {\enquote {\bibinfo {title} {Cold-atom
  quantum simulator for string and hadron dynamics in non-abelian lattice gauge
  theory},}\ }\href@noop {} {\bibfield  {journal} {\bibinfo  {journal}
  {Physical Review A}\ }\textbf {\bibinfo {volume} {105}},\ \bibinfo {pages}
  {023322} (\bibinfo {year} {2022})}\BibitemShut {NoStop}%
\bibitem [{\citenamefont {Zohar}\ \emph
  {et~al.}(2013{\natexlab{d}})\citenamefont {Zohar}, \citenamefont {Cirac},\
  and\ \citenamefont {Reznik}}]{zohar2013quantum}%
  \BibitemOpen
  \bibfield  {author} {\bibinfo {author} {\bibfnamefont {Erez}\ \bibnamefont
  {Zohar}}, \bibinfo {author} {\bibfnamefont {J~Ignacio}\ \bibnamefont
  {Cirac}}, \ and\ \bibinfo {author} {\bibfnamefont {Benni}\ \bibnamefont
  {Reznik}},\ }\bibfield  {title} {\enquote {\bibinfo {title} {Quantum
  simulations of gauge theories with ultracold atoms: Local gauge invariance
  from angular-momentum conservation},}\ }\href@noop {} {\bibfield  {journal}
  {\bibinfo  {journal} {Physical Review A}\ }\textbf {\bibinfo {volume} {88}},\
  \bibinfo {pages} {023617} (\bibinfo {year} {2013}{\natexlab{d}})}\BibitemShut
  {NoStop}%
\bibitem [{\citenamefont {Bender}\ \emph
  {et~al.}(2018{\natexlab{b}})\citenamefont {Bender}, \citenamefont {Zohar},
  \citenamefont {Farace},\ and\ \citenamefont {Cirac}}]{bender2018digital}%
  \BibitemOpen
  \bibfield  {author} {\bibinfo {author} {\bibfnamefont {Julian}\ \bibnamefont
  {Bender}}, \bibinfo {author} {\bibfnamefont {Erez}\ \bibnamefont {Zohar}},
  \bibinfo {author} {\bibfnamefont {Alessandro}\ \bibnamefont {Farace}}, \ and\
  \bibinfo {author} {\bibfnamefont {J~Ignacio}\ \bibnamefont {Cirac}},\
  }\bibfield  {title} {\enquote {\bibinfo {title} {Digital quantum simulation
  of lattice gauge theories in three spatial dimensions},}\ }\href@noop {}
  {\bibfield  {journal} {\bibinfo  {journal} {New Journal of Physics}\ }\textbf
  {\bibinfo {volume} {20}},\ \bibinfo {pages} {093001} (\bibinfo {year}
  {2018}{\natexlab{b}})}\BibitemShut {NoStop}%
\bibitem [{\citenamefont {White}(1992)}]{PhysRevLett.69.2863}%
  \BibitemOpen
  \bibfield  {author} {\bibinfo {author} {\bibfnamefont {Steven~R.}\
  \bibnamefont {White}},\ }\bibfield  {title} {\enquote {\bibinfo {title}
  {Density matrix formulation for quantum renormalization groups},}\ }\href
  {\doibase 10.1103/PhysRevLett.69.2863} {\bibfield  {journal} {\bibinfo
  {journal} {Phys. Rev. Lett.}\ }\textbf {\bibinfo {volume} {69}},\ \bibinfo
  {pages} {2863--2866} (\bibinfo {year} {1992})}\BibitemShut {NoStop}%
\bibitem [{\citenamefont {Ba\~nuls}\ \emph {et~al.}(2020)\citenamefont
  {Ba\~nuls} \emph {et~al.}}]{Banuls:2019bmf}%
  \BibitemOpen
  \bibfield  {author} {\bibinfo {author} {\bibfnamefont {M.~C.}\ \bibnamefont
  {Ba\~nuls}} \emph {et~al.},\ }\bibfield  {title} {\enquote {\bibinfo {title}
  {{Simulating Lattice Gauge Theories within Quantum Technologies}},}\ }\href
  {\doibase 10.1140/epjd/e2020-100571-8} {\bibfield  {journal} {\bibinfo
  {journal} {Eur. Phys. J. D}\ }\textbf {\bibinfo {volume} {74}},\ \bibinfo
  {pages} {165} (\bibinfo {year} {2020})},\ \Eprint
  {http://arxiv.org/abs/1911.00003} {arXiv:1911.00003 [quant-ph]} \BibitemShut
  {NoStop}%
\bibitem [{\citenamefont {Cirac}\ \emph {et~al.}(2021)\citenamefont {Cirac},
  \citenamefont {Perez-Garcia}, \citenamefont {Schuch},\ and\ \citenamefont
  {Verstraete}}]{Cirac:2020obd}%
  \BibitemOpen
  \bibfield  {author} {\bibinfo {author} {\bibfnamefont {J.~Ignacio}\
  \bibnamefont {Cirac}}, \bibinfo {author} {\bibfnamefont {David}\ \bibnamefont
  {Perez-Garcia}}, \bibinfo {author} {\bibfnamefont {Norbert}\ \bibnamefont
  {Schuch}}, \ and\ \bibinfo {author} {\bibfnamefont {Frank}\ \bibnamefont
  {Verstraete}},\ }\bibfield  {title} {\enquote {\bibinfo {title} {{Matrix
  product states and projected entangled pair states: Concepts, symmetries,
  theorems}},}\ }\href {\doibase 10.1103/RevModPhys.93.045003} {\bibfield
  {journal} {\bibinfo  {journal} {Rev. Mod. Phys.}\ }\textbf {\bibinfo {volume}
  {93}},\ \bibinfo {pages} {045003} (\bibinfo {year} {2021})},\ \Eprint
  {http://arxiv.org/abs/2011.12127} {arXiv:2011.12127 [quant-ph]} \BibitemShut
  {NoStop}%
\bibitem [{\citenamefont {Ba\~nuls}\ \emph {et~al.}(2014)\citenamefont
  {Ba\~nuls}, \citenamefont {Cichy}, \citenamefont {Cirac}, \citenamefont
  {Jansen},\ and\ \citenamefont {Saito}}]{mcbpos2013}%
  \BibitemOpen
  \bibfield  {author} {\bibinfo {author} {\bibfnamefont {Mari~Carmen}\
  \bibnamefont {Ba\~nuls}}, \bibinfo {author} {\bibfnamefont {Krzysztof}\
  \bibnamefont {Cichy}}, \bibinfo {author} {\bibfnamefont {J.~Ignacio}\
  \bibnamefont {Cirac}}, \bibinfo {author} {\bibfnamefont {Karl}\ \bibnamefont
  {Jansen}}, \ and\ \bibinfo {author} {\bibfnamefont {Hana}\ \bibnamefont
  {Saito}},\ }\bibfield  {title} {\enquote {\bibinfo {title} {{Matrix Product
  States for Lattice Field Theories}},}\ }\href {\doibase 10.22323/1.187.0332}
  {\bibfield  {journal} {\bibinfo  {journal} {PoS}\ }\textbf {\bibinfo {volume}
  {LATTICE2013}},\ \bibinfo {pages} {332} (\bibinfo {year} {2014})},\ \Eprint
  {http://arxiv.org/abs/1310.4118} {arXiv:1310.4118 [hep-lat]} \BibitemShut
  {NoStop}%
\bibitem [{\citenamefont {Meurice}\ \emph
  {et~al.}(2014{\natexlab{a}})\citenamefont {Meurice}, \citenamefont
  {Denbleyker}, \citenamefont {Liu}, \citenamefont {Xiang}, \citenamefont
  {Xie}, \citenamefont {Yu}, \citenamefont {Unmuth-Yockey},\ and\ \citenamefont
  {Zou}}]{ympos2013}%
  \BibitemOpen
  \bibfield  {author} {\bibinfo {author} {\bibfnamefont {Yannick}\ \bibnamefont
  {Meurice}}, \bibinfo {author} {\bibfnamefont {Alan}\ \bibnamefont
  {Denbleyker}}, \bibinfo {author} {\bibfnamefont {Yuzhi}\ \bibnamefont {Liu}},
  \bibinfo {author} {\bibfnamefont {Tao}\ \bibnamefont {Xiang}}, \bibinfo
  {author} {\bibfnamefont {Zhiyuan}\ \bibnamefont {Xie}}, \bibinfo {author}
  {\bibfnamefont {Ji-Feng}\ \bibnamefont {Yu}}, \bibinfo {author}
  {\bibfnamefont {Judah}\ \bibnamefont {Unmuth-Yockey}}, \ and\ \bibinfo
  {author} {\bibfnamefont {Haiyuan}\ \bibnamefont {Zou}},\ }\bibfield  {title}
  {\enquote {\bibinfo {title} {{Comparing Tensor Renormalization Group and
  Monte Carlo calculations for spin and gauge models}},}\ }\href {\doibase
  10.22323/1.187.0329} {\bibfield  {journal} {\bibinfo  {journal} {PoS}\
  }\textbf {\bibinfo {volume} {LATTICE2013}},\ \bibinfo {pages} {329} (\bibinfo
  {year} {2014}{\natexlab{a}})},\ \Eprint {http://arxiv.org/abs/1311.4826}
  {arXiv:1311.4826 [hep-lat]} \BibitemShut {NoStop}%
\bibitem [{\citenamefont {Van~Acoleyen}\ \emph {et~al.}(2014)\citenamefont
  {Van~Acoleyen}, \citenamefont {Buyens}, \citenamefont {Haegeman},\ and\
  \citenamefont {Verstraete}}]{kvapos2014}%
  \BibitemOpen
  \bibfield  {author} {\bibinfo {author} {\bibfnamefont {Karel}\ \bibnamefont
  {Van~Acoleyen}}, \bibinfo {author} {\bibfnamefont {Boye}\ \bibnamefont
  {Buyens}}, \bibinfo {author} {\bibfnamefont {Jutho}\ \bibnamefont
  {Haegeman}}, \ and\ \bibinfo {author} {\bibfnamefont {Frank}\ \bibnamefont
  {Verstraete}},\ }\bibfield  {title} {\enquote {\bibinfo {title} {{Matrix
  product states for Hamiltonian lattice gauge theories}},}\ }\href {\doibase
  10.22323/1.214.0308} {\bibfield  {journal} {\bibinfo  {journal} {PoS}\
  }\textbf {\bibinfo {volume} {LATTICE2014}},\ \bibinfo {pages} {308} (\bibinfo
  {year} {2014})},\ \Eprint {http://arxiv.org/abs/1411.0020} {arXiv:1411.0020
  [hep-lat]} \BibitemShut {NoStop}%
\bibitem [{\citenamefont {Shimizu}\ and\ \citenamefont
  {Kuramashi}(2014)}]{yspos2014}%
  \BibitemOpen
  \bibfield  {author} {\bibinfo {author} {\bibfnamefont {Yuya}\ \bibnamefont
  {Shimizu}}\ and\ \bibinfo {author} {\bibfnamefont {Yoshinobu}\ \bibnamefont
  {Kuramashi}},\ }\bibfield  {title} {\enquote {\bibinfo {title} {{Grassmann
  Tensor Renormalization Group Study of Lattice QED with Theta Term in Two
  Dimensions}},}\ }\href {\doibase 10.22323/1.214.0303} {\bibfield  {journal}
  {\bibinfo  {journal} {PoS}\ }\textbf {\bibinfo {volume} {LATTICE2014}},\
  \bibinfo {pages} {303} (\bibinfo {year} {2014})}\BibitemShut {NoStop}%
\bibitem [{\citenamefont {Saito}\ \emph {et~al.}(2014)\citenamefont {Saito},
  \citenamefont {Ba\~nuls}, \citenamefont {Cichy}, \citenamefont {Cirac},\ and\
  \citenamefont {Jansen}}]{mcbpos2014}%
  \BibitemOpen
  \bibfield  {author} {\bibinfo {author} {\bibfnamefont {Hana}\ \bibnamefont
  {Saito}}, \bibinfo {author} {\bibfnamefont {Mari~Carmen}\ \bibnamefont
  {Ba\~nuls}}, \bibinfo {author} {\bibfnamefont {Krzysztof}\ \bibnamefont
  {Cichy}}, \bibinfo {author} {\bibfnamefont {J.~Ignacio}\ \bibnamefont
  {Cirac}}, \ and\ \bibinfo {author} {\bibfnamefont {Karl}\ \bibnamefont
  {Jansen}},\ }\bibfield  {title} {\enquote {\bibinfo {title} {{The temperature
  dependence of the chiral condensate in the Schwinger model with Matrix
  Product States}},}\ }\href {\doibase 10.22323/1.214.0302} {\bibfield
  {journal} {\bibinfo  {journal} {PoS}\ }\textbf {\bibinfo {volume}
  {LATTICE2014}},\ \bibinfo {pages} {302} (\bibinfo {year} {2014})},\ \Eprint
  {http://arxiv.org/abs/1412.0596} {arXiv:1412.0596 [hep-lat]} \BibitemShut
  {NoStop}%
\bibitem [{\citenamefont {Unmuth-Yockey}\ \emph {et~al.}(2014)\citenamefont
  {Unmuth-Yockey}, \citenamefont {Meurice}, \citenamefont {Osborn},\ and\
  \citenamefont {Zou}}]{juypos2014}%
  \BibitemOpen
  \bibfield  {author} {\bibinfo {author} {\bibfnamefont {Judah~F}\ \bibnamefont
  {Unmuth-Yockey}}, \bibinfo {author} {\bibfnamefont {Yannick}\ \bibnamefont
  {Meurice}}, \bibinfo {author} {\bibfnamefont {James}\ \bibnamefont {Osborn}},
  \ and\ \bibinfo {author} {\bibfnamefont {Haiyuan}\ \bibnamefont {Zou}},\
  }\bibfield  {title} {\enquote {\bibinfo {title} {{Tensor renormalization
  group study of the 2d O(3) model}},}\ }\href {\doibase 10.22323/1.214.0325}
  {\bibfield  {journal} {\bibinfo  {journal} {PoS}\ }\textbf {\bibinfo {volume}
  {LATTICE2014}},\ \bibinfo {pages} {325} (\bibinfo {year} {2014})},\ \Eprint
  {http://arxiv.org/abs/1411.4213} {arXiv:1411.4213 [hep-lat]} \BibitemShut
  {NoStop}%
\bibitem [{\citenamefont {Meurice}\ \emph
  {et~al.}(2014{\natexlab{b}})\citenamefont {Meurice}, \citenamefont {Liu},
  \citenamefont {Unmuth-Yockey}, \citenamefont {Yang},\ and\ \citenamefont
  {Zou}}]{ympos2014}%
  \BibitemOpen
  \bibfield  {author} {\bibinfo {author} {\bibfnamefont {Yannick}\ \bibnamefont
  {Meurice}}, \bibinfo {author} {\bibfnamefont {Yuzhi}\ \bibnamefont {Liu}},
  \bibinfo {author} {\bibfnamefont {Judah}\ \bibnamefont {Unmuth-Yockey}},
  \bibinfo {author} {\bibfnamefont {Li-Ping}\ \bibnamefont {Yang}}, \ and\
  \bibinfo {author} {\bibfnamefont {Haiyuan}\ \bibnamefont {Zou}},\ }\bibfield
  {title} {\enquote {\bibinfo {title} {{Sampling versus Blocking}},}\ }\href
  {\doibase 10.22323/1.214.0319} {\bibfield  {journal} {\bibinfo  {journal}
  {PoS}\ }\textbf {\bibinfo {volume} {LATTICE2014}},\ \bibinfo {pages} {319}
  (\bibinfo {year} {2014}{\natexlab{b}})},\ \Eprint
  {http://arxiv.org/abs/1411.3392} {arXiv:1411.3392 [hep-lat]} \BibitemShut
  {NoStop}%
\bibitem [{\citenamefont {Ba\~nuls}\ \emph {et~al.}(2018)\citenamefont
  {Ba\~nuls}, \citenamefont {Cichy}, \citenamefont {Cirac}, \citenamefont
  {Jansen},\ and\ \citenamefont {K\"uhn}}]{mcbpos2018}%
  \BibitemOpen
  \bibfield  {author} {\bibinfo {author} {\bibfnamefont {Mari~Carmen}\
  \bibnamefont {Ba\~nuls}}, \bibinfo {author} {\bibfnamefont {Krzysztof}\
  \bibnamefont {Cichy}}, \bibinfo {author} {\bibfnamefont {J.~Ignacio}\
  \bibnamefont {Cirac}}, \bibinfo {author} {\bibfnamefont {Karl}\ \bibnamefont
  {Jansen}}, \ and\ \bibinfo {author} {\bibfnamefont {Stefan}\ \bibnamefont
  {K\"uhn}},\ }\bibfield  {title} {\enquote {\bibinfo {title} {{Tensor Networks
  and their use for Lattice Gauge Theories}},}\ }\href {\doibase
  10.22323/1.334.0022} {\bibfield  {journal} {\bibinfo  {journal} {PoS}\
  }\textbf {\bibinfo {volume} {LATTICE2018}},\ \bibinfo {pages} {022} (\bibinfo
  {year} {2018})},\ \Eprint {http://arxiv.org/abs/1810.12838} {arXiv:1810.12838
  [hep-lat]} \BibitemShut {NoStop}%
\bibitem [{\citenamefont {Ba\~nuls}\ and\ \citenamefont
  {Cichy}(2020)}]{Banuls:2019rao}%
  \BibitemOpen
  \bibfield  {author} {\bibinfo {author} {\bibfnamefont {Mari~Carmen}\
  \bibnamefont {Ba\~nuls}}\ and\ \bibinfo {author} {\bibfnamefont {Krzysztof}\
  \bibnamefont {Cichy}},\ }\bibfield  {title} {\enquote {\bibinfo {title}
  {{Review on Novel Methods for Lattice Gauge Theories}},}\ }\href {\doibase
  10.1088/1361-6633/ab6311} {\bibfield  {journal} {\bibinfo  {journal} {Rept.
  Prog. Phys.}\ }\textbf {\bibinfo {volume} {83}},\ \bibinfo {pages} {024401}
  (\bibinfo {year} {2020})},\ \Eprint {http://arxiv.org/abs/1910.00257}
  {arXiv:1910.00257 [hep-lat]} \BibitemShut {NoStop}%
\bibitem [{\citenamefont {Akiyama}\ \emph {et~al.}(2020)\citenamefont
  {Akiyama}, \citenamefont {Kadoh}, \citenamefont {Kuramashi}, \citenamefont
  {Yamashita},\ and\ \citenamefont {Yoshimura}}]{Akiyama:2020ntf}%
  \BibitemOpen
  \bibfield  {author} {\bibinfo {author} {\bibfnamefont {Shinichiro}\
  \bibnamefont {Akiyama}}, \bibinfo {author} {\bibfnamefont {Daisuke}\
  \bibnamefont {Kadoh}}, \bibinfo {author} {\bibfnamefont {Yoshinobu}\
  \bibnamefont {Kuramashi}}, \bibinfo {author} {\bibfnamefont {Takumi}\
  \bibnamefont {Yamashita}}, \ and\ \bibinfo {author} {\bibfnamefont {Yusuke}\
  \bibnamefont {Yoshimura}},\ }\bibfield  {title} {\enquote {\bibinfo {title}
  {{Tensor renormalization group approach to four-dimensional complex $\phi^4$
  theory at finite density}},}\ }\href {\doibase 10.1007/JHEP09(2020)177}
  {\bibfield  {journal} {\bibinfo  {journal} {JHEP}\ }\textbf {\bibinfo
  {volume} {09}},\ \bibinfo {pages} {177} (\bibinfo {year} {2020})},\ \Eprint
  {http://arxiv.org/abs/2005.04645} {arXiv:2005.04645 [hep-lat]} \BibitemShut
  {NoStop}%
\bibitem [{\citenamefont {Kadoh}\ \emph {et~al.}(2018)\citenamefont {Kadoh},
  \citenamefont {Kuramashi}, \citenamefont {Nakamura}, \citenamefont {Sakai},
  \citenamefont {Takeda},\ and\ \citenamefont {Yoshimura}}]{Kadoh:2018hqq}%
  \BibitemOpen
  \bibfield  {author} {\bibinfo {author} {\bibfnamefont {Daisuke}\ \bibnamefont
  {Kadoh}}, \bibinfo {author} {\bibfnamefont {Yoshinobu}\ \bibnamefont
  {Kuramashi}}, \bibinfo {author} {\bibfnamefont {Yoshifumi}\ \bibnamefont
  {Nakamura}}, \bibinfo {author} {\bibfnamefont {Ryo}\ \bibnamefont {Sakai}},
  \bibinfo {author} {\bibfnamefont {Shinji}\ \bibnamefont {Takeda}}, \ and\
  \bibinfo {author} {\bibfnamefont {Yusuke}\ \bibnamefont {Yoshimura}},\
  }\bibfield  {title} {\enquote {\bibinfo {title} {{Tensor network formulation
  for two-dimensional lattice $ \mathcal{N} $ = 1 Wess-Zumino model}},}\ }\href
  {\doibase 10.1007/JHEP03(2018)141} {\bibfield  {journal} {\bibinfo  {journal}
  {JHEP}\ }\textbf {\bibinfo {volume} {03}},\ \bibinfo {pages} {141} (\bibinfo
  {year} {2018})},\ \Eprint {http://arxiv.org/abs/1801.04183} {arXiv:1801.04183
  [hep-lat]} \BibitemShut {NoStop}%
\bibitem [{\citenamefont {Cotler}\ \emph {et~al.}(2017)\citenamefont {Cotler},
  \citenamefont {Gur-Ari}, \citenamefont {Hanada}, \citenamefont {Polchinski},
  \citenamefont {Saad}, \citenamefont {Shenker}, \citenamefont {Stanford},
  \citenamefont {Streicher},\ and\ \citenamefont {Tezuka}}]{Cotler:2016fpe}%
  \BibitemOpen
  \bibfield  {author} {\bibinfo {author} {\bibfnamefont {Jordan~S.}\
  \bibnamefont {Cotler}}, \bibinfo {author} {\bibfnamefont {Guy}\ \bibnamefont
  {Gur-Ari}}, \bibinfo {author} {\bibfnamefont {Masanori}\ \bibnamefont
  {Hanada}}, \bibinfo {author} {\bibfnamefont {Joseph}\ \bibnamefont
  {Polchinski}}, \bibinfo {author} {\bibfnamefont {Phil}\ \bibnamefont {Saad}},
  \bibinfo {author} {\bibfnamefont {Stephen~H.}\ \bibnamefont {Shenker}},
  \bibinfo {author} {\bibfnamefont {Douglas}\ \bibnamefont {Stanford}},
  \bibinfo {author} {\bibfnamefont {Alexandre}\ \bibnamefont {Streicher}}, \
  and\ \bibinfo {author} {\bibfnamefont {Masaki}\ \bibnamefont {Tezuka}},\
  }\bibfield  {title} {\enquote {\bibinfo {title} {{Black Holes and Random
  Matrices}},}\ }\href {\doibase 10.1007/JHEP05(2017)118} {\bibfield  {journal}
  {\bibinfo  {journal} {JHEP}\ }\textbf {\bibinfo {volume} {05}},\ \bibinfo
  {pages} {118} (\bibinfo {year} {2017})},\ \bibinfo {note} {[Erratum: JHEP 09,
  002 (2018)]},\ \Eprint {http://arxiv.org/abs/1611.04650} {arXiv:1611.04650
  [hep-th]} \BibitemShut {NoStop}%
\bibitem [{\citenamefont {Berkowitz}\ \emph {et~al.}(2016)\citenamefont
  {Berkowitz}, \citenamefont {Rinaldi}, \citenamefont {Hanada}, \citenamefont
  {Ishiki}, \citenamefont {Shimasaki},\ and\ \citenamefont
  {Vranas}}]{Berkowitz:2016jlq}%
  \BibitemOpen
  \bibfield  {author} {\bibinfo {author} {\bibfnamefont {Evan}\ \bibnamefont
  {Berkowitz}}, \bibinfo {author} {\bibfnamefont {Enrico}\ \bibnamefont
  {Rinaldi}}, \bibinfo {author} {\bibfnamefont {Masanori}\ \bibnamefont
  {Hanada}}, \bibinfo {author} {\bibfnamefont {Goro}\ \bibnamefont {Ishiki}},
  \bibinfo {author} {\bibfnamefont {Shinji}\ \bibnamefont {Shimasaki}}, \ and\
  \bibinfo {author} {\bibfnamefont {Pavlos}\ \bibnamefont {Vranas}},\
  }\bibfield  {title} {\enquote {\bibinfo {title} {{Precision lattice test of
  the gauge/gravity duality at large-$N$}},}\ }\href {\doibase
  10.1103/PhysRevD.94.094501} {\bibfield  {journal} {\bibinfo  {journal} {Phys.
  Rev. D}\ }\textbf {\bibinfo {volume} {94}},\ \bibinfo {pages} {094501}
  (\bibinfo {year} {2016})},\ \Eprint {http://arxiv.org/abs/1606.04951}
  {arXiv:1606.04951 [hep-lat]} \BibitemShut {NoStop}%
\bibitem [{\citenamefont {Hanada}\ \emph {et~al.}(2016)\citenamefont {Hanada},
  \citenamefont {Hyakutake}, \citenamefont {Ishiki},\ and\ \citenamefont
  {Nishimura}}]{Hanada:2016zxj}%
  \BibitemOpen
  \bibfield  {author} {\bibinfo {author} {\bibfnamefont {Masanori}\
  \bibnamefont {Hanada}}, \bibinfo {author} {\bibfnamefont {Yoshifumi}\
  \bibnamefont {Hyakutake}}, \bibinfo {author} {\bibfnamefont {Goro}\
  \bibnamefont {Ishiki}}, \ and\ \bibinfo {author} {\bibfnamefont {Jun}\
  \bibnamefont {Nishimura}},\ }\bibfield  {title} {\enquote {\bibinfo {title}
  {{Numerical tests of the gauge/gravity duality conjecture for D0-branes at
  finite temperature and finite N}},}\ }\href {\doibase
  10.1103/PhysRevD.94.086010} {\bibfield  {journal} {\bibinfo  {journal} {Phys.
  Rev. D}\ }\textbf {\bibinfo {volume} {94}},\ \bibinfo {pages} {086010}
  (\bibinfo {year} {2016})},\ \Eprint {http://arxiv.org/abs/1603.00538}
  {arXiv:1603.00538 [hep-th]} \BibitemShut {NoStop}%
\bibitem [{\citenamefont {Catterall}\ \emph {et~al.}(2020)\citenamefont
  {Catterall}, \citenamefont {Giedt}, \citenamefont {Jha}, \citenamefont
  {Schaich},\ and\ \citenamefont {Wiseman}}]{Catterall:2020nmn}%
  \BibitemOpen
  \bibfield  {author} {\bibinfo {author} {\bibfnamefont {Simon}\ \bibnamefont
  {Catterall}}, \bibinfo {author} {\bibfnamefont {Joel}\ \bibnamefont {Giedt}},
  \bibinfo {author} {\bibfnamefont {Raghav~G.}\ \bibnamefont {Jha}}, \bibinfo
  {author} {\bibfnamefont {David}\ \bibnamefont {Schaich}}, \ and\ \bibinfo
  {author} {\bibfnamefont {Toby}\ \bibnamefont {Wiseman}},\ }\bibfield  {title}
  {\enquote {\bibinfo {title} {{Three-dimensional super-Yang--Mills theory on
  the lattice and dual black branes}},}\ }\href {\doibase
  10.1103/PhysRevD.102.106009} {\bibfield  {journal} {\bibinfo  {journal}
  {Phys. Rev. D}\ }\textbf {\bibinfo {volume} {102}},\ \bibinfo {pages}
  {106009} (\bibinfo {year} {2020})},\ \Eprint
  {http://arxiv.org/abs/2010.00026} {arXiv:2010.00026 [hep-th]} \BibitemShut
  {NoStop}%
\bibitem [{\citenamefont {Catterall}\ and\ \citenamefont
  {Wiseman}(2008)}]{Catterall:2008yz}%
  \BibitemOpen
  \bibfield  {author} {\bibinfo {author} {\bibfnamefont {Simon}\ \bibnamefont
  {Catterall}}\ and\ \bibinfo {author} {\bibfnamefont {Toby}\ \bibnamefont
  {Wiseman}},\ }\bibfield  {title} {\enquote {\bibinfo {title} {{Black hole
  thermodynamics from simulations of lattice Yang-Mills theory}},}\ }\href
  {\doibase 10.1103/PhysRevD.78.041502} {\bibfield  {journal} {\bibinfo
  {journal} {Phys. Rev. D}\ }\textbf {\bibinfo {volume} {78}},\ \bibinfo
  {pages} {041502} (\bibinfo {year} {2008})},\ \Eprint
  {http://arxiv.org/abs/0803.4273} {arXiv:0803.4273 [hep-th]} \BibitemShut
  {NoStop}%
\bibitem [{\citenamefont {Catterall}\ \emph {et~al.}(2018)\citenamefont
  {Catterall}, \citenamefont {Jha}, \citenamefont {Schaich},\ and\
  \citenamefont {Wiseman}}]{Catterall:2017lub}%
  \BibitemOpen
  \bibfield  {author} {\bibinfo {author} {\bibfnamefont {Simon}\ \bibnamefont
  {Catterall}}, \bibinfo {author} {\bibfnamefont {Raghav~G.}\ \bibnamefont
  {Jha}}, \bibinfo {author} {\bibfnamefont {David}\ \bibnamefont {Schaich}}, \
  and\ \bibinfo {author} {\bibfnamefont {Toby}\ \bibnamefont {Wiseman}},\
  }\bibfield  {title} {\enquote {\bibinfo {title} {{Testing holography using
  lattice super-Yang-Mills theory on a 2-torus}},}\ }\href {\doibase
  10.1103/PhysRevD.97.086020} {\bibfield  {journal} {\bibinfo  {journal} {Phys.
  Rev. D}\ }\textbf {\bibinfo {volume} {97}},\ \bibinfo {pages} {086020}
  (\bibinfo {year} {2018})},\ \Eprint {http://arxiv.org/abs/1709.07025}
  {arXiv:1709.07025 [hep-th]} \BibitemShut {NoStop}%
\bibitem [{\citenamefont {Hanada}\ \emph {et~al.}(2014)\citenamefont {Hanada},
  \citenamefont {Hyakutake}, \citenamefont {Ishiki},\ and\ \citenamefont
  {Nishimura}}]{Hanada:2013rga}%
  \BibitemOpen
  \bibfield  {author} {\bibinfo {author} {\bibfnamefont {Masanori}\
  \bibnamefont {Hanada}}, \bibinfo {author} {\bibfnamefont {Yoshifumi}\
  \bibnamefont {Hyakutake}}, \bibinfo {author} {\bibfnamefont {Goro}\
  \bibnamefont {Ishiki}}, \ and\ \bibinfo {author} {\bibfnamefont {Jun}\
  \bibnamefont {Nishimura}},\ }\bibfield  {title} {\enquote {\bibinfo {title}
  {{Holographic description of quantum black hole on a computer}},}\ }\href
  {\doibase 10.1126/science.1250122} {\bibfield  {journal} {\bibinfo  {journal}
  {Science}\ }\textbf {\bibinfo {volume} {344}},\ \bibinfo {pages} {882--885}
  (\bibinfo {year} {2014})},\ \Eprint {http://arxiv.org/abs/1311.5607}
  {arXiv:1311.5607 [hep-th]} \BibitemShut {NoStop}%
\bibitem [{\citenamefont {Asaduzzaman}\ and\ \citenamefont
  {Catterall}(2021)}]{Asaduzzaman:2021ufo}%
  \BibitemOpen
  \bibfield  {author} {\bibinfo {author} {\bibfnamefont {Muhammad}\
  \bibnamefont {Asaduzzaman}}\ and\ \bibinfo {author} {\bibfnamefont {Simon}\
  \bibnamefont {Catterall}},\ }\bibfield  {title} {\enquote {\bibinfo {title}
  {{Scalar fields on fluctuating hyperbolic geometries}},}\ }in\ \href@noop {}
  {\emph {\bibinfo {booktitle} {{38th International Symposium on Lattice Field
  Theory}}}}\ (\bibinfo {year} {2021})\ \Eprint
  {http://arxiv.org/abs/2112.00927} {arXiv:2112.00927 [hep-lat]} \BibitemShut
  {NoStop}%
\bibitem [{\citenamefont {Mueller}\ \emph {et~al.}(2022)\citenamefont
  {Mueller}, \citenamefont {Zache},\ and\ \citenamefont
  {Ott}}]{Mueller:2021gxd}%
  \BibitemOpen
  \bibfield  {author} {\bibinfo {author} {\bibfnamefont {Niklas}\ \bibnamefont
  {Mueller}}, \bibinfo {author} {\bibfnamefont {Torsten~V.}\ \bibnamefont
  {Zache}}, \ and\ \bibinfo {author} {\bibfnamefont {Robert}\ \bibnamefont
  {Ott}},\ }\bibfield  {title} {\enquote {\bibinfo {title} {{Thermalization of
  Gauge Theories from their Entanglement Spectrum}},}\ }\href {\doibase
  10.1103/PhysRevLett.129.011601} {\bibfield  {journal} {\bibinfo  {journal}
  {Phys. Rev. Lett.}\ }\textbf {\bibinfo {volume} {129}},\ \bibinfo {pages}
  {011601} (\bibinfo {year} {2022})},\ \Eprint
  {http://arxiv.org/abs/2107.11416} {arXiv:2107.11416 [quant-ph]} \BibitemShut
  {NoStop}%
\bibitem [{\citenamefont {Pichler}\ \emph {et~al.}(2016)\citenamefont
  {Pichler}, \citenamefont {Dalmonte}, \citenamefont {Rico}, \citenamefont
  {Zoller},\ and\ \citenamefont {Montangero}}]{pichler2016real}%
  \BibitemOpen
  \bibfield  {author} {\bibinfo {author} {\bibfnamefont {Thomas}\ \bibnamefont
  {Pichler}}, \bibinfo {author} {\bibfnamefont {Marcello}\ \bibnamefont
  {Dalmonte}}, \bibinfo {author} {\bibfnamefont {Enrique}\ \bibnamefont
  {Rico}}, \bibinfo {author} {\bibfnamefont {Peter}\ \bibnamefont {Zoller}}, \
  and\ \bibinfo {author} {\bibfnamefont {Simone}\ \bibnamefont {Montangero}},\
  }\bibfield  {title} {\enquote {\bibinfo {title} {Real-time dynamics in u (1)
  lattice gauge theories with tensor networks},}\ }\href@noop {} {\bibfield
  {journal} {\bibinfo  {journal} {Physical Review X}\ }\textbf {\bibinfo
  {volume} {6}},\ \bibinfo {pages} {011023} (\bibinfo {year}
  {2016})}\BibitemShut {NoStop}%
\bibitem [{\citenamefont {Zache}\ \emph {et~al.}(2019)\citenamefont {Zache},
  \citenamefont {Mueller}, \citenamefont {Schneider}, \citenamefont
  {Jendrzejewski}, \citenamefont {Berges},\ and\ \citenamefont
  {Hauke}}]{Zache:2018cqq}%
  \BibitemOpen
  \bibfield  {author} {\bibinfo {author} {\bibfnamefont {T.~V.}\ \bibnamefont
  {Zache}}, \bibinfo {author} {\bibfnamefont {N.}~\bibnamefont {Mueller}},
  \bibinfo {author} {\bibfnamefont {J.~T.}\ \bibnamefont {Schneider}}, \bibinfo
  {author} {\bibfnamefont {F.}~\bibnamefont {Jendrzejewski}}, \bibinfo {author}
  {\bibfnamefont {J.}~\bibnamefont {Berges}}, \ and\ \bibinfo {author}
  {\bibfnamefont {P.}~\bibnamefont {Hauke}},\ }\bibfield  {title} {\enquote
  {\bibinfo {title} {{Dynamical Topological Transitions in the Massive
  Schwinger Model with a $\theta$ Term}},}\ }\href {\doibase
  10.1103/PhysRevLett.122.050403} {\bibfield  {journal} {\bibinfo  {journal}
  {Phys. Rev. Lett.}\ }\textbf {\bibinfo {volume} {122}},\ \bibinfo {pages}
  {050403} (\bibinfo {year} {2019})},\ \Eprint
  {http://arxiv.org/abs/1808.07885} {arXiv:1808.07885 [quant-ph]} \BibitemShut
  {NoStop}%
\bibitem [{\citenamefont {Desaules}\ \emph {et~al.}(2022)\citenamefont
  {Desaules}, \citenamefont {Hudomal}, \citenamefont {Banerjee}, \citenamefont
  {Sen}, \citenamefont {Papi\'c},\ and\ \citenamefont
  {Halimeh}}]{Desaules:2022kse}%
  \BibitemOpen
  \bibfield  {author} {\bibinfo {author} {\bibfnamefont {Jean-Yves}\
  \bibnamefont {Desaules}}, \bibinfo {author} {\bibfnamefont {Ana}\
  \bibnamefont {Hudomal}}, \bibinfo {author} {\bibfnamefont {Debasish}\
  \bibnamefont {Banerjee}}, \bibinfo {author} {\bibfnamefont {Arnab}\
  \bibnamefont {Sen}}, \bibinfo {author} {\bibfnamefont {Zlatko}\ \bibnamefont
  {Papi\'c}}, \ and\ \bibinfo {author} {\bibfnamefont {Jad~C.}\ \bibnamefont
  {Halimeh}},\ }\bibfield  {title} {\enquote {\bibinfo {title} {{Prominent
  quantum many-body scars in a truncated Schwinger model}},}\ }\href@noop {} {\
   (\bibinfo {year} {2022})},\ \Eprint {http://arxiv.org/abs/2204.01745}
  {arXiv:2204.01745 [cond-mat.quant-gas]} \BibitemShut {NoStop}%
\bibitem [{\citenamefont {Carney}\ \emph
  {et~al.}(2022{\natexlab{a}})\citenamefont {Carney} \emph
  {et~al.}}]{Carney:2022rlu}%
  \BibitemOpen
  \bibfield  {author} {\bibinfo {author} {\bibfnamefont {Daniel}\ \bibnamefont
  {Carney}} \emph {et~al.},\ }\bibfield  {title} {\enquote {\bibinfo {title}
  {{Snowmass 2021: Quantum Sensors for HEP Science -- Interferometers,
  Mechanics, Traps, and Clocks}},}\ }in\ \href@noop {} {\emph {\bibinfo
  {booktitle} {{2022 Snowmass Summer Study}}}}\ (\bibinfo {year} {2022})\
  \Eprint {http://arxiv.org/abs/2203.07250} {arXiv:2203.07250 [quant-ph]}
  \BibitemShut {NoStop}%
\bibitem [{\citenamefont {Essig}\ \emph {et~al.}(2022)\citenamefont {Essig},
  \citenamefont {Kahn}, \citenamefont {Knapen}, \citenamefont {Ringwald},\ and\
  \citenamefont {Toro}}]{Essig:2022yzw}%
  \BibitemOpen
  \bibfield  {author} {\bibinfo {author} {\bibfnamefont {Rouven}\ \bibnamefont
  {Essig}}, \bibinfo {author} {\bibfnamefont {Yonatan}\ \bibnamefont {Kahn}},
  \bibinfo {author} {\bibfnamefont {Simon}\ \bibnamefont {Knapen}}, \bibinfo
  {author} {\bibfnamefont {Andreas}\ \bibnamefont {Ringwald}}, \ and\ \bibinfo
  {author} {\bibfnamefont {Natalia}\ \bibnamefont {Toro}},\ }\bibfield  {title}
  {\enquote {\bibinfo {title} {{Snowmass2021 Theory Frontier: Theory Meets the
  Lab}},}\ }in\ \href@noop {} {\emph {\bibinfo {booktitle} {{2022 Snowmass
  Summer Study}}}}\ (\bibinfo {year} {2022})\ \Eprint
  {http://arxiv.org/abs/2203.10089} {arXiv:2203.10089 [hep-ph]} \BibitemShut
  {NoStop}%
\bibitem [{\citenamefont {Essig}\ \emph {et~al.}(2013)\citenamefont {Essig}
  \emph {et~al.}}]{Essig:2013lka}%
  \BibitemOpen
  \bibfield  {author} {\bibinfo {author} {\bibfnamefont {Rouven}\ \bibnamefont
  {Essig}} \emph {et~al.},\ }\bibfield  {title} {\enquote {\bibinfo {title}
  {{Working Group Report: New Light Weakly Coupled Particles}},}\ }in\
  \href@noop {} {\emph {\bibinfo {booktitle} {{Community Summer Study 2013}:
  {Snowmass on the Mississippi}}}}\ (\bibinfo {year} {2013})\ \Eprint
  {http://arxiv.org/abs/1311.0029} {arXiv:1311.0029 [hep-ph]} \BibitemShut
  {NoStop}%
\bibitem [{\citenamefont {Peccei}\ and\ \citenamefont
  {Quinn}(1977)}]{Peccei:1977ur}%
  \BibitemOpen
  \bibfield  {author} {\bibinfo {author} {\bibfnamefont {R.~D.}\ \bibnamefont
  {Peccei}}\ and\ \bibinfo {author} {\bibfnamefont {Helen~R.}\ \bibnamefont
  {Quinn}},\ }\bibfield  {title} {\enquote {\bibinfo {title} {{Constraints
  Imposed by CP Conservation in the Presence of Instantons}},}\ }\href
  {\doibase 10.1103/PhysRevD.16.1791} {\bibfield  {journal} {\bibinfo
  {journal} {Phys. Rev. D}\ }\textbf {\bibinfo {volume} {16}},\ \bibinfo
  {pages} {1791--1797} (\bibinfo {year} {1977})}\BibitemShut {NoStop}%
\bibitem [{\citenamefont {Holdom}(1986)}]{Holdom:1985ag}%
  \BibitemOpen
  \bibfield  {author} {\bibinfo {author} {\bibfnamefont {Bob}\ \bibnamefont
  {Holdom}},\ }\bibfield  {title} {\enquote {\bibinfo {title} {{Two U(1)'s and
  Epsilon Charge Shifts}},}\ }\href {\doibase 10.1016/0370-2693(86)91377-8}
  {\bibfield  {journal} {\bibinfo  {journal} {Phys. Lett. B}\ }\textbf
  {\bibinfo {volume} {166}},\ \bibinfo {pages} {196--198} (\bibinfo {year}
  {1986})}\BibitemShut {NoStop}%
\bibitem [{\citenamefont {Sikivie}(1983)}]{Sikivie:1983ip}%
  \BibitemOpen
  \bibfield  {author} {\bibinfo {author} {\bibfnamefont {P.}~\bibnamefont
  {Sikivie}},\ }\bibfield  {title} {\enquote {\bibinfo {title} {{Experimental
  Tests of the Invisible Axion}},}\ }\href {\doibase
  10.1103/PhysRevLett.51.1415} {\bibfield  {journal} {\bibinfo  {journal}
  {Phys. Rev. Lett.}\ }\textbf {\bibinfo {volume} {51}},\ \bibinfo {pages}
  {1415--1417} (\bibinfo {year} {1983})},\ \bibinfo {note} {[Erratum:
  Phys.Rev.Lett. 52, 695 (1984)]}\BibitemShut {NoStop}%
\bibitem [{\citenamefont {Adams}\ \emph {et~al.}(2022)\citenamefont {Adams}
  \emph {et~al.}}]{Adams:2022pbo}%
  \BibitemOpen
  \bibfield  {author} {\bibinfo {author} {\bibfnamefont {C.~B.}\ \bibnamefont
  {Adams}} \emph {et~al.},\ }\bibfield  {title} {\enquote {\bibinfo {title}
  {{Axion Dark Matter}},}\ }in\ \href@noop {} {\emph {\bibinfo {booktitle}
  {{2022 Snowmass Summer Study}}}}\ (\bibinfo {year} {2022})\ \Eprint
  {http://arxiv.org/abs/2203.14923} {arXiv:2203.14923 [hep-ex]} \BibitemShut
  {NoStop}%
\bibitem [{\citenamefont {Antypas}\ \emph {et~al.}(2022)\citenamefont {Antypas}
  \emph {et~al.}}]{Antypas:2022asj}%
  \BibitemOpen
  \bibfield  {author} {\bibinfo {author} {\bibfnamefont {D.}~\bibnamefont
  {Antypas}} \emph {et~al.},\ }\bibfield  {title} {\enquote {\bibinfo {title}
  {{New Horizons: Scalar and Vector Ultralight Dark Matter}},}\ }\href@noop {}
  {\  (\bibinfo {year} {2022})},\ \Eprint {http://arxiv.org/abs/2203.14915}
  {arXiv:2203.14915 [hep-ex]} \BibitemShut {NoStop}%
\bibitem [{\citenamefont {Berlin}\ \emph {et~al.}(2022)\citenamefont {Berlin}
  \emph {et~al.}}]{Berlin:2022hfx}%
  \BibitemOpen
  \bibfield  {author} {\bibinfo {author} {\bibfnamefont {Asher}\ \bibnamefont
  {Berlin}} \emph {et~al.},\ }\bibfield  {title} {\enquote {\bibinfo {title}
  {{Searches for New Particles, Dark Matter, and Gravitational Waves with SRF
  Cavities}},}\ }\href@noop {} {\  (\bibinfo {year} {2022})},\ \Eprint
  {http://arxiv.org/abs/2203.12714} {arXiv:2203.12714 [hep-ph]} \BibitemShut
  {NoStop}%
\bibitem [{\citenamefont {Carney}\ \emph
  {et~al.}(2022{\natexlab{b}})\citenamefont {Carney} \emph
  {et~al.}}]{Carney:2022gse}%
  \BibitemOpen
  \bibfield  {author} {\bibinfo {author} {\bibfnamefont {Daniel}\ \bibnamefont
  {Carney}} \emph {et~al.},\ }\bibfield  {title} {\enquote {\bibinfo {title}
  {{Snowmass2021 Cosmic Frontier White Paper: Ultraheavy particle dark
  matter}},}\ }\href@noop {} {\  (\bibinfo {year} {2022}{\natexlab{b}})},\
  \Eprint {http://arxiv.org/abs/2203.06508} {arXiv:2203.06508 [hep-ph]}
  \BibitemShut {NoStop}%
\bibitem [{\citenamefont {Abbott}\ \emph {et~al.}(2016)\citenamefont {Abbott}
  \emph {et~al.}}]{LIGOScientific:2016aoc}%
  \BibitemOpen
  \bibfield  {author} {\bibinfo {author} {\bibfnamefont {B.~P.}\ \bibnamefont
  {Abbott}} \emph {et~al.} (\bibinfo {collaboration} {LIGO Scientific,
  Virgo}),\ }\bibfield  {title} {\enquote {\bibinfo {title} {{Observation of
  Gravitational Waves from a Binary Black Hole Merger}},}\ }\href {\doibase
  10.1103/PhysRevLett.116.061102} {\bibfield  {journal} {\bibinfo  {journal}
  {Phys. Rev. Lett.}\ }\textbf {\bibinfo {volume} {116}},\ \bibinfo {pages}
  {061102} (\bibinfo {year} {2016})},\ \Eprint
  {http://arxiv.org/abs/1602.03837} {arXiv:1602.03837 [gr-qc]} \BibitemShut
  {NoStop}%
\bibitem [{\citenamefont {Berlin}\ \emph
  {et~al.}(2021{\natexlab{a}})\citenamefont {Berlin}, \citenamefont {Blas},
  \citenamefont {Tito~D'Agnolo}, \citenamefont {Ellis}, \citenamefont {Harnik},
  \citenamefont {Kahn},\ and\ \citenamefont
  {Sch\"utte-Engel}}]{Berlin:2021txa}%
  \BibitemOpen
  \bibfield  {author} {\bibinfo {author} {\bibfnamefont {Asher}\ \bibnamefont
  {Berlin}}, \bibinfo {author} {\bibfnamefont {Diego}\ \bibnamefont {Blas}},
  \bibinfo {author} {\bibfnamefont {Raffaele}\ \bibnamefont {Tito~D'Agnolo}},
  \bibinfo {author} {\bibfnamefont {Sebastian A.~R.}\ \bibnamefont {Ellis}},
  \bibinfo {author} {\bibfnamefont {Roni}\ \bibnamefont {Harnik}}, \bibinfo
  {author} {\bibfnamefont {Yonatan}\ \bibnamefont {Kahn}}, \ and\ \bibinfo
  {author} {\bibfnamefont {Jan}\ \bibnamefont {Sch\"utte-Engel}},\ }\bibfield
  {title} {\enquote {\bibinfo {title} {{Detecting High-Frequency Gravitational
  Waves with Microwave Cavities}},}\ }\href@noop {} {\  (\bibinfo {year}
  {2021}{\natexlab{a}})},\ \Eprint {http://arxiv.org/abs/2112.11465}
  {arXiv:2112.11465 [hep-ph]} \BibitemShut {NoStop}%
\bibitem [{\citenamefont {Domcke}\ \emph {et~al.}(2022)\citenamefont {Domcke},
  \citenamefont {Garcia-Cely},\ and\ \citenamefont {Rodd}}]{Domcke:2022rgu}%
  \BibitemOpen
  \bibfield  {author} {\bibinfo {author} {\bibfnamefont {Valerie}\ \bibnamefont
  {Domcke}}, \bibinfo {author} {\bibfnamefont {Camilo}\ \bibnamefont
  {Garcia-Cely}}, \ and\ \bibinfo {author} {\bibfnamefont {Nicholas~L.}\
  \bibnamefont {Rodd}},\ }\bibfield  {title} {\enquote {\bibinfo {title} {{A
  novel search for high-frequency gravitational waves with low-mass axion
  haloscopes}},}\ }\href@noop {} {\  (\bibinfo {year} {2022})},\ \Eprint
  {http://arxiv.org/abs/2202.00695} {arXiv:2202.00695 [hep-ph]} \BibitemShut
  {NoStop}%
\bibitem [{\citenamefont {Fedderke}\ \emph {et~al.}(2022)\citenamefont
  {Fedderke}, \citenamefont {Graham}, \citenamefont {Macintosh},\ and\
  \citenamefont {Rajendran}}]{Fedderke:2022kxq}%
  \BibitemOpen
  \bibfield  {author} {\bibinfo {author} {\bibfnamefont {Michael~A.}\
  \bibnamefont {Fedderke}}, \bibinfo {author} {\bibfnamefont {Peter~W.}\
  \bibnamefont {Graham}}, \bibinfo {author} {\bibfnamefont {Bruce}\
  \bibnamefont {Macintosh}}, \ and\ \bibinfo {author} {\bibfnamefont {Surjeet}\
  \bibnamefont {Rajendran}},\ }\bibfield  {title} {\enquote {\bibinfo {title}
  {{Astrometric Gravitational-Wave Detection via Stellar Interferometry}},}\
  }\href@noop {} {\  (\bibinfo {year} {2022})},\ \Eprint
  {http://arxiv.org/abs/2204.07677} {arXiv:2204.07677 [astro-ph.IM]}
  \BibitemShut {NoStop}%
\bibitem [{\citenamefont {Fedderke}\ \emph {et~al.}(2021)\citenamefont
  {Fedderke}, \citenamefont {Graham},\ and\ \citenamefont
  {Rajendran}}]{Fedderke:2021kuy}%
  \BibitemOpen
  \bibfield  {author} {\bibinfo {author} {\bibfnamefont {Michael~A.}\
  \bibnamefont {Fedderke}}, \bibinfo {author} {\bibfnamefont {Peter~W.}\
  \bibnamefont {Graham}}, \ and\ \bibinfo {author} {\bibfnamefont {Surjeet}\
  \bibnamefont {Rajendran}},\ }\bibfield  {title} {\enquote {\bibinfo {title}
  {{Asteroids for $\mu$Hz gravitational-wave detection}},}\ }\href@noop {} {\
  (\bibinfo {year} {2021})},\ \Eprint {http://arxiv.org/abs/2112.11431}
  {arXiv:2112.11431 [gr-qc]} \BibitemShut {NoStop}%
\bibitem [{\citenamefont {Weinberg}(1989)}]{Weinberg:1989us}%
  \BibitemOpen
  \bibfield  {author} {\bibinfo {author} {\bibfnamefont {Steven}\ \bibnamefont
  {Weinberg}},\ }\bibfield  {title} {\enquote {\bibinfo {title} {{Testing
  Quantum Mechanics}},}\ }\href {\doibase 10.1016/0003-4916(89)90276-5}
  {\bibfield  {journal} {\bibinfo  {journal} {Annals Phys.}\ }\textbf {\bibinfo
  {volume} {194}},\ \bibinfo {pages} {336} (\bibinfo {year}
  {1989})}\BibitemShut {NoStop}%
\bibitem [{\citenamefont {Polchinski}(1991)}]{Polchinski:1990py}%
  \BibitemOpen
  \bibfield  {author} {\bibinfo {author} {\bibfnamefont {Joseph}\ \bibnamefont
  {Polchinski}},\ }\bibfield  {title} {\enquote {\bibinfo {title} {{Weinberg's
  nonlinear quantum mechanics and the EPR paradox}},}\ }\href {\doibase
  10.1103/PhysRevLett.66.397} {\bibfield  {journal} {\bibinfo  {journal} {Phys.
  Rev. Lett.}\ }\textbf {\bibinfo {volume} {66}},\ \bibinfo {pages} {397--400}
  (\bibinfo {year} {1991})}\BibitemShut {NoStop}%
\bibitem [{\citenamefont {Kaplan}\ and\ \citenamefont
  {Rajendran}(2022)}]{Kaplan:2021qpv}%
  \BibitemOpen
  \bibfield  {author} {\bibinfo {author} {\bibfnamefont {David~E.}\
  \bibnamefont {Kaplan}}\ and\ \bibinfo {author} {\bibfnamefont {Surjeet}\
  \bibnamefont {Rajendran}},\ }\bibfield  {title} {\enquote {\bibinfo {title}
  {{Causal framework for nonlinear quantum mechanics}},}\ }\href {\doibase
  10.1103/PhysRevD.105.055002} {\bibfield  {journal} {\bibinfo  {journal}
  {Phys. Rev. D}\ }\textbf {\bibinfo {volume} {105}},\ \bibinfo {pages}
  {055002} (\bibinfo {year} {2022})},\ \Eprint
  {http://arxiv.org/abs/2106.10576} {arXiv:2106.10576 [hep-th]} \BibitemShut
  {NoStop}%
\bibitem [{\citenamefont {Polkovnikov}\ \emph {et~al.}(2022)\citenamefont
  {Polkovnikov}, \citenamefont {Gramolin}, \citenamefont {Kaplan},
  \citenamefont {Rajendran},\ and\ \citenamefont
  {Sushkov}}]{Polkovnikov:2022nwk}%
  \BibitemOpen
  \bibfield  {author} {\bibinfo {author} {\bibfnamefont {Mark}\ \bibnamefont
  {Polkovnikov}}, \bibinfo {author} {\bibfnamefont {Alexander~V.}\ \bibnamefont
  {Gramolin}}, \bibinfo {author} {\bibfnamefont {David~E.}\ \bibnamefont
  {Kaplan}}, \bibinfo {author} {\bibfnamefont {Surjeet}\ \bibnamefont
  {Rajendran}}, \ and\ \bibinfo {author} {\bibfnamefont {Alexander~O.}\
  \bibnamefont {Sushkov}},\ }\bibfield  {title} {\enquote {\bibinfo {title}
  {{Experimental limit on non-linear state-dependent terms in quantum
  theory}},}\ }\href@noop {} {\  (\bibinfo {year} {2022})},\ \Eprint
  {http://arxiv.org/abs/2204.11875} {arXiv:2204.11875 [quant-ph]} \BibitemShut
  {NoStop}%
\bibitem [{\citenamefont {Carney}\ \emph
  {et~al.}(2022{\natexlab{c}})\citenamefont {Carney}, \citenamefont {Chen},
  \citenamefont {Geraci}, \citenamefont {M\"uller}, \citenamefont {Panda},
  \citenamefont {Stamp},\ and\ \citenamefont {Taylor}}]{Carney:2022dku}%
  \BibitemOpen
  \bibfield  {author} {\bibinfo {author} {\bibfnamefont {Daniel}\ \bibnamefont
  {Carney}}, \bibinfo {author} {\bibfnamefont {Yanbei}\ \bibnamefont {Chen}},
  \bibinfo {author} {\bibfnamefont {Andrew}\ \bibnamefont {Geraci}}, \bibinfo
  {author} {\bibfnamefont {Holger}\ \bibnamefont {M\"uller}}, \bibinfo {author}
  {\bibfnamefont {Cristian~D.}\ \bibnamefont {Panda}}, \bibinfo {author}
  {\bibfnamefont {Philip C.~E.}\ \bibnamefont {Stamp}}, \ and\ \bibinfo
  {author} {\bibfnamefont {Jacob~M.}\ \bibnamefont {Taylor}},\ }\bibfield
  {title} {\enquote {\bibinfo {title} {{Snowmass 2021 White Paper: Tabletop
  experiments for infrared quantum gravity}},}\ }in\ \href@noop {} {\emph
  {\bibinfo {booktitle} {{2022 Snowmass Summer Study}}}}\ (\bibinfo {year}
  {2022})\ \Eprint {http://arxiv.org/abs/2203.11846} {arXiv:2203.11846 [gr-qc]}
  \BibitemShut {NoStop}%
\bibitem [{\citenamefont {Abe}\ \emph {et~al.}(2021)\citenamefont {Abe} \emph
  {et~al.}}]{MAGIS-100:2021etm}%
  \BibitemOpen
  \bibfield  {author} {\bibinfo {author} {\bibfnamefont {Mahiro}\ \bibnamefont
  {Abe}} \emph {et~al.} (\bibinfo {collaboration} {MAGIS-100}),\ }\bibfield
  {title} {\enquote {\bibinfo {title} {{Matter-wave Atomic Gradiometer
  Interferometric Sensor (MAGIS-100)}},}\ }\href {\doibase
  10.1088/2058-9565/abf719} {\bibfield  {journal} {\bibinfo  {journal} {Quantum
  Sci. Technol.}\ }\textbf {\bibinfo {volume} {6}},\ \bibinfo {pages} {044003}
  (\bibinfo {year} {2021})},\ \Eprint {http://arxiv.org/abs/2104.02835}
  {arXiv:2104.02835 [physics.atom-ph]} \BibitemShut {NoStop}%
\bibitem [{\citenamefont {Dimopoulos}\ \emph {et~al.}(2007)\citenamefont
  {Dimopoulos}, \citenamefont {Graham}, \citenamefont {Hogan},\ and\
  \citenamefont {Kasevich}}]{Dimopoulos:2006nk}%
  \BibitemOpen
  \bibfield  {author} {\bibinfo {author} {\bibfnamefont {Savas}\ \bibnamefont
  {Dimopoulos}}, \bibinfo {author} {\bibfnamefont {Peter~W.}\ \bibnamefont
  {Graham}}, \bibinfo {author} {\bibfnamefont {Jason~M.}\ \bibnamefont
  {Hogan}}, \ and\ \bibinfo {author} {\bibfnamefont {Mark~A.}\ \bibnamefont
  {Kasevich}},\ }\bibfield  {title} {\enquote {\bibinfo {title} {{Testing
  general relativity with atom interferometry}},}\ }\href {\doibase
  10.1103/PhysRevLett.98.111102} {\bibfield  {journal} {\bibinfo  {journal}
  {Phys. Rev. Lett.}\ }\textbf {\bibinfo {volume} {98}},\ \bibinfo {pages}
  {111102} (\bibinfo {year} {2007})},\ \Eprint
  {http://arxiv.org/abs/gr-qc/0610047} {arXiv:gr-qc/0610047} \BibitemShut
  {NoStop}%
\bibitem [{\citenamefont {Dimopoulos}\ \emph {et~al.}(2009)\citenamefont
  {Dimopoulos}, \citenamefont {Graham}, \citenamefont {Hogan}, \citenamefont
  {Kasevich},\ and\ \citenamefont {Rajendran}}]{Dimopoulos:2007cj}%
  \BibitemOpen
  \bibfield  {author} {\bibinfo {author} {\bibfnamefont {Savas}\ \bibnamefont
  {Dimopoulos}}, \bibinfo {author} {\bibfnamefont {Peter~W.}\ \bibnamefont
  {Graham}}, \bibinfo {author} {\bibfnamefont {Jason~M.}\ \bibnamefont
  {Hogan}}, \bibinfo {author} {\bibfnamefont {Mark~A.}\ \bibnamefont
  {Kasevich}}, \ and\ \bibinfo {author} {\bibfnamefont {Surjeet}\ \bibnamefont
  {Rajendran}},\ }\bibfield  {title} {\enquote {\bibinfo {title}
  {{Gravitational Wave Detection with Atom Interferometry}},}\ }\href {\doibase
  10.1016/j.physletb.2009.06.011} {\bibfield  {journal} {\bibinfo  {journal}
  {Phys. Lett. B}\ }\textbf {\bibinfo {volume} {678}},\ \bibinfo {pages}
  {37--40} (\bibinfo {year} {2009})},\ \Eprint {http://arxiv.org/abs/0712.1250}
  {arXiv:0712.1250 [gr-qc]} \BibitemShut {NoStop}%
\bibitem [{\citenamefont {Graham}\ \emph {et~al.}(2013)\citenamefont {Graham},
  \citenamefont {Hogan}, \citenamefont {Kasevich},\ and\ \citenamefont
  {Rajendran}}]{Graham:2012sy}%
  \BibitemOpen
  \bibfield  {author} {\bibinfo {author} {\bibfnamefont {Peter~W.}\
  \bibnamefont {Graham}}, \bibinfo {author} {\bibfnamefont {Jason~M.}\
  \bibnamefont {Hogan}}, \bibinfo {author} {\bibfnamefont {Mark~A.}\
  \bibnamefont {Kasevich}}, \ and\ \bibinfo {author} {\bibfnamefont {Surjeet}\
  \bibnamefont {Rajendran}},\ }\bibfield  {title} {\enquote {\bibinfo {title}
  {{A New Method for Gravitational Wave Detection with Atomic Sensors}},}\
  }\href {\doibase 10.1103/PhysRevLett.110.171102} {\bibfield  {journal}
  {\bibinfo  {journal} {Phys. Rev. Lett.}\ }\textbf {\bibinfo {volume} {110}},\
  \bibinfo {pages} {171102} (\bibinfo {year} {2013})},\ \Eprint
  {http://arxiv.org/abs/1206.0818} {arXiv:1206.0818 [quant-ph]} \BibitemShut
  {NoStop}%
\bibitem [{\citenamefont {Derevianko}\ and\ \citenamefont
  {Pospelov}(2014)}]{Derevianko:2013oaa}%
  \BibitemOpen
  \bibfield  {author} {\bibinfo {author} {\bibfnamefont {A.}~\bibnamefont
  {Derevianko}}\ and\ \bibinfo {author} {\bibfnamefont {M.}~\bibnamefont
  {Pospelov}},\ }\bibfield  {title} {\enquote {\bibinfo {title} {{Hunting for
  topological dark matter with atomic clocks}},}\ }\href {\doibase
  10.1038/nphys3137} {\bibfield  {journal} {\bibinfo  {journal} {Nature Phys.}\
  }\textbf {\bibinfo {volume} {10}},\ \bibinfo {pages} {933} (\bibinfo {year}
  {2014})},\ \Eprint {http://arxiv.org/abs/1311.1244} {arXiv:1311.1244
  [physics.atom-ph]} \BibitemShut {NoStop}%
\bibitem [{\citenamefont {Arvanitaki}\ \emph {et~al.}(2015)\citenamefont
  {Arvanitaki}, \citenamefont {Huang},\ and\ \citenamefont
  {Van~Tilburg}}]{Arvanitaki:2014faa}%
  \BibitemOpen
  \bibfield  {author} {\bibinfo {author} {\bibfnamefont {Asimina}\ \bibnamefont
  {Arvanitaki}}, \bibinfo {author} {\bibfnamefont {Junwu}\ \bibnamefont
  {Huang}}, \ and\ \bibinfo {author} {\bibfnamefont {Ken}\ \bibnamefont
  {Van~Tilburg}},\ }\bibfield  {title} {\enquote {\bibinfo {title} {{Searching
  for dilaton dark matter with atomic clocks}},}\ }\href {\doibase
  10.1103/PhysRevD.91.015015} {\bibfield  {journal} {\bibinfo  {journal} {Phys.
  Rev. D}\ }\textbf {\bibinfo {volume} {91}},\ \bibinfo {pages} {015015}
  (\bibinfo {year} {2015})},\ \Eprint {http://arxiv.org/abs/1405.2925}
  {arXiv:1405.2925 [hep-ph]} \BibitemShut {NoStop}%
\bibitem [{\citenamefont {Stadnik}\ and\ \citenamefont
  {Flambaum}(2015)}]{Stadnik:2014tta}%
  \BibitemOpen
  \bibfield  {author} {\bibinfo {author} {\bibfnamefont {Y.~V.}\ \bibnamefont
  {Stadnik}}\ and\ \bibinfo {author} {\bibfnamefont {V.~V.}\ \bibnamefont
  {Flambaum}},\ }\bibfield  {title} {\enquote {\bibinfo {title} {{Searching for
  dark matter and variation of fundamental constants with laser and maser
  interferometry}},}\ }\href {\doibase 10.1103/PhysRevLett.114.161301}
  {\bibfield  {journal} {\bibinfo  {journal} {Phys. Rev. Lett.}\ }\textbf
  {\bibinfo {volume} {114}},\ \bibinfo {pages} {161301} (\bibinfo {year}
  {2015})},\ \Eprint {http://arxiv.org/abs/1412.7801} {arXiv:1412.7801
  [hep-ph]} \BibitemShut {NoStop}%
\bibitem [{\citenamefont {Alonso}\ \emph {et~al.}(2022)\citenamefont {Alonso}
  \emph {et~al.}}]{Alonso:2022oot}%
  \BibitemOpen
  \bibfield  {author} {\bibinfo {author} {\bibfnamefont {Ivan}\ \bibnamefont
  {Alonso}} \emph {et~al.},\ }\bibfield  {title} {\enquote {\bibinfo {title}
  {{Cold Atoms in Space: Community Workshop Summary and Proposed Road-Map}},}\
  \ }(\bibinfo {year} {2022})\ \Eprint {http://arxiv.org/abs/2201.07789}
  {arXiv:2201.07789 [astro-ph.IM]} \BibitemShut {NoStop}%
\bibitem [{\citenamefont {Tsai}\ \emph {et~al.}(2021)\citenamefont {Tsai},
  \citenamefont {Eby},\ and\ \citenamefont {Safronova}}]{Tsai:2021lly}%
  \BibitemOpen
  \bibfield  {author} {\bibinfo {author} {\bibfnamefont {Yu-Dai}\ \bibnamefont
  {Tsai}}, \bibinfo {author} {\bibfnamefont {Joshua}\ \bibnamefont {Eby}}, \
  and\ \bibinfo {author} {\bibfnamefont {Marianna~S.}\ \bibnamefont
  {Safronova}},\ }\bibfield  {title} {\enquote {\bibinfo {title} {{SpaceQ -
  Direct Detection of Ultralight Dark Matter with Space Quantum Sensors}},}\
  }\href@noop {} {\  (\bibinfo {year} {2021})},\ \Eprint
  {http://arxiv.org/abs/2112.07674} {arXiv:2112.07674 [hep-ph]} \BibitemShut
  {NoStop}%
\bibitem [{\citenamefont {Berengut}\ \emph {et~al.}(2018)\citenamefont
  {Berengut} \emph {et~al.}}]{Berengut:2017zuo}%
  \BibitemOpen
  \bibfield  {author} {\bibinfo {author} {\bibfnamefont {Julian~C.}\
  \bibnamefont {Berengut}} \emph {et~al.},\ }\bibfield  {title} {\enquote
  {\bibinfo {title} {{Probing New Long-Range Interactions by Isotope Shift
  Spectroscopy}},}\ }\href {\doibase 10.1103/PhysRevLett.120.091801} {\bibfield
   {journal} {\bibinfo  {journal} {Phys. Rev. Lett.}\ }\textbf {\bibinfo
  {volume} {120}},\ \bibinfo {pages} {091801} (\bibinfo {year} {2018})},\
  \Eprint {http://arxiv.org/abs/1704.05068} {arXiv:1704.05068 [hep-ph]}
  \BibitemShut {NoStop}%
\bibitem [{\citenamefont {Crisler}\ \emph {et~al.}(2018)\citenamefont
  {Crisler}, \citenamefont {Essig}, \citenamefont {Estrada}, \citenamefont
  {Fernandez}, \citenamefont {Tiffenberg}, \citenamefont {Sofo~haro},
  \citenamefont {Volansky},\ and\ \citenamefont {Yu}}]{Crisler:2018gci}%
  \BibitemOpen
  \bibfield  {author} {\bibinfo {author} {\bibfnamefont {Michael}\ \bibnamefont
  {Crisler}}, \bibinfo {author} {\bibfnamefont {Rouven}\ \bibnamefont {Essig}},
  \bibinfo {author} {\bibfnamefont {Juan}\ \bibnamefont {Estrada}}, \bibinfo
  {author} {\bibfnamefont {Guillermo}\ \bibnamefont {Fernandez}}, \bibinfo
  {author} {\bibfnamefont {Javier}\ \bibnamefont {Tiffenberg}}, \bibinfo
  {author} {\bibfnamefont {Miguel}\ \bibnamefont {Sofo~haro}}, \bibinfo
  {author} {\bibfnamefont {Tomer}\ \bibnamefont {Volansky}}, \ and\ \bibinfo
  {author} {\bibfnamefont {Tien-Tien}\ \bibnamefont {Yu}} (\bibinfo
  {collaboration} {SENSEI}),\ }\bibfield  {title} {\enquote {\bibinfo {title}
  {{SENSEI: First Direct-Detection Constraints on sub-GeV Dark Matter from a
  Surface Run}},}\ }\href {\doibase 10.1103/PhysRevLett.121.061803} {\bibfield
  {journal} {\bibinfo  {journal} {Phys. Rev. Lett.}\ }\textbf {\bibinfo
  {volume} {121}},\ \bibinfo {pages} {061803} (\bibinfo {year} {2018})},\
  \Eprint {http://arxiv.org/abs/1804.00088} {arXiv:1804.00088 [hep-ex]}
  \BibitemShut {NoStop}%
\bibitem [{\citenamefont {Aguilar-Arevalo}\ \emph {et~al.}(2022)\citenamefont
  {Aguilar-Arevalo} \emph {et~al.}}]{Aguilar-Arevalo:2022kqd}%
  \BibitemOpen
  \bibfield  {author} {\bibinfo {author} {\bibfnamefont {Alexis}\ \bibnamefont
  {Aguilar-Arevalo}} \emph {et~al.},\ }\bibfield  {title} {\enquote {\bibinfo
  {title} {{The Oscura Experiment}},}\ }\href@noop {} {\  (\bibinfo {year}
  {2022})},\ \Eprint {http://arxiv.org/abs/2202.10518} {arXiv:2202.10518
  [astro-ph.IM]} \BibitemShut {NoStop}%
\bibitem [{\citenamefont {Hochberg}\ \emph {et~al.}(2016)\citenamefont
  {Hochberg}, \citenamefont {Zhao},\ and\ \citenamefont
  {Zurek}}]{Hochberg:2015pha}%
  \BibitemOpen
  \bibfield  {author} {\bibinfo {author} {\bibfnamefont {Yonit}\ \bibnamefont
  {Hochberg}}, \bibinfo {author} {\bibfnamefont {Yue}\ \bibnamefont {Zhao}}, \
  and\ \bibinfo {author} {\bibfnamefont {Kathryn~M.}\ \bibnamefont {Zurek}},\
  }\bibfield  {title} {\enquote {\bibinfo {title} {{Superconducting Detectors
  for Superlight Dark Matter}},}\ }\href {\doibase
  10.1103/PhysRevLett.116.011301} {\bibfield  {journal} {\bibinfo  {journal}
  {Phys. Rev. Lett.}\ }\textbf {\bibinfo {volume} {116}},\ \bibinfo {pages}
  {011301} (\bibinfo {year} {2016})},\ \Eprint
  {http://arxiv.org/abs/1504.07237} {arXiv:1504.07237 [hep-ph]} \BibitemShut
  {NoStop}%
\bibitem [{\citenamefont {Hochberg}\ \emph {et~al.}(2021)\citenamefont
  {Hochberg}, \citenamefont {Lehmann}, \citenamefont {Charaev}, \citenamefont
  {Chiles}, \citenamefont {Colangelo}, \citenamefont {Nam},\ and\ \citenamefont
  {Berggren}}]{Hochberg:2021yud}%
  \BibitemOpen
  \bibfield  {author} {\bibinfo {author} {\bibfnamefont {Yonit}\ \bibnamefont
  {Hochberg}}, \bibinfo {author} {\bibfnamefont {Benjamin~V.}\ \bibnamefont
  {Lehmann}}, \bibinfo {author} {\bibfnamefont {Ilya}\ \bibnamefont {Charaev}},
  \bibinfo {author} {\bibfnamefont {Jeff}\ \bibnamefont {Chiles}}, \bibinfo
  {author} {\bibfnamefont {Marco}\ \bibnamefont {Colangelo}}, \bibinfo {author}
  {\bibfnamefont {Sae~Woo}\ \bibnamefont {Nam}}, \ and\ \bibinfo {author}
  {\bibfnamefont {Karl~K.}\ \bibnamefont {Berggren}},\ }\bibfield  {title}
  {\enquote {\bibinfo {title} {{New Constraints on Dark Matter from
  Superconducting Nanowires}},}\ }\href@noop {} {\  (\bibinfo {year} {2021})},\
  \Eprint {http://arxiv.org/abs/2110.01586} {arXiv:2110.01586 [hep-ph]}
  \BibitemShut {NoStop}%
\bibitem [{\citenamefont {Bunting}\ \emph {et~al.}(2017)\citenamefont
  {Bunting}, \citenamefont {Gratta}, \citenamefont {Melia},\ and\ \citenamefont
  {Rajendran}}]{Bunting:2017net}%
  \BibitemOpen
  \bibfield  {author} {\bibinfo {author} {\bibfnamefont {Philip~C.}\
  \bibnamefont {Bunting}}, \bibinfo {author} {\bibfnamefont {Giorgio}\
  \bibnamefont {Gratta}}, \bibinfo {author} {\bibfnamefont {Tom}\ \bibnamefont
  {Melia}}, \ and\ \bibinfo {author} {\bibfnamefont {Surjeet}\ \bibnamefont
  {Rajendran}},\ }\bibfield  {title} {\enquote {\bibinfo {title} {{Magnetic
  Bubble Chambers and Sub-GeV Dark Matter Direct Detection}},}\ }\href
  {\doibase 10.1103/PhysRevD.95.095001} {\bibfield  {journal} {\bibinfo
  {journal} {Phys. Rev. D}\ }\textbf {\bibinfo {volume} {95}},\ \bibinfo
  {pages} {095001} (\bibinfo {year} {2017})},\ \Eprint
  {http://arxiv.org/abs/1701.06566} {arXiv:1701.06566 [hep-ph]} \BibitemShut
  {NoStop}%
\bibitem [{\citenamefont {Chen}\ \emph {et~al.}(2020)\citenamefont {Chen},
  \citenamefont {Mahapatra}, \citenamefont {Agnolet}, \citenamefont {Nippe},
  \citenamefont {Lu}, \citenamefont {Bunting}, \citenamefont {Melia},
  \citenamefont {Rajendran}, \citenamefont {Gratta},\ and\ \citenamefont
  {Long}}]{Chen:2020jia}%
  \BibitemOpen
  \bibfield  {author} {\bibinfo {author} {\bibfnamefont {Hao}\ \bibnamefont
  {Chen}}, \bibinfo {author} {\bibfnamefont {Rupak}\ \bibnamefont {Mahapatra}},
  \bibinfo {author} {\bibfnamefont {Glenn}\ \bibnamefont {Agnolet}}, \bibinfo
  {author} {\bibfnamefont {Michael}\ \bibnamefont {Nippe}}, \bibinfo {author}
  {\bibfnamefont {Minjie}\ \bibnamefont {Lu}}, \bibinfo {author} {\bibfnamefont
  {Philip~C.}\ \bibnamefont {Bunting}}, \bibinfo {author} {\bibfnamefont {Tom}\
  \bibnamefont {Melia}}, \bibinfo {author} {\bibfnamefont {Surjeet}\
  \bibnamefont {Rajendran}}, \bibinfo {author} {\bibfnamefont {Giorgio}\
  \bibnamefont {Gratta}}, \ and\ \bibinfo {author} {\bibfnamefont {Jeffrey}\
  \bibnamefont {Long}},\ }\bibfield  {title} {\enquote {\bibinfo {title}
  {{Quantum Detection using Magnetic Avalanches in Single-Molecule Magnets}},}\
  }\href@noop {} {\  (\bibinfo {year} {2020})},\ \Eprint
  {http://arxiv.org/abs/2002.09409} {arXiv:2002.09409 [physics.ins-det]}
  \BibitemShut {NoStop}%
\bibitem [{\citenamefont {Backes}\ \emph {et~al.}(2021)\citenamefont {Backes}
  \emph {et~al.}}]{HAYSTAC:2020kwv}%
  \BibitemOpen
  \bibfield  {author} {\bibinfo {author} {\bibfnamefont {K.~M.}\ \bibnamefont
  {Backes}} \emph {et~al.} (\bibinfo {collaboration} {HAYSTAC}),\ }\bibfield
  {title} {\enquote {\bibinfo {title} {{A quantum-enhanced search for dark
  matter axions}},}\ }\href {\doibase 10.1038/s41586-021-03226-7} {\bibfield
  {journal} {\bibinfo  {journal} {Nature}\ }\textbf {\bibinfo {volume} {590}},\
  \bibinfo {pages} {238--242} (\bibinfo {year} {2021})},\ \Eprint
  {http://arxiv.org/abs/2008.01853} {arXiv:2008.01853 [quant-ph]} \BibitemShut
  {NoStop}%
\bibitem [{\citenamefont {Dixit}\ \emph {et~al.}(2021)\citenamefont {Dixit},
  \citenamefont {Chakram}, \citenamefont {He}, \citenamefont {Agrawal},
  \citenamefont {Naik}, \citenamefont {Schuster},\ and\ \citenamefont
  {Chou}}]{Dixit:2020ymh}%
  \BibitemOpen
  \bibfield  {author} {\bibinfo {author} {\bibfnamefont {Akash~V.}\
  \bibnamefont {Dixit}}, \bibinfo {author} {\bibfnamefont {Srivatsan}\
  \bibnamefont {Chakram}}, \bibinfo {author} {\bibfnamefont {Kevin}\
  \bibnamefont {He}}, \bibinfo {author} {\bibfnamefont {Ankur}\ \bibnamefont
  {Agrawal}}, \bibinfo {author} {\bibfnamefont {Ravi~K.}\ \bibnamefont {Naik}},
  \bibinfo {author} {\bibfnamefont {David~I.}\ \bibnamefont {Schuster}}, \ and\
  \bibinfo {author} {\bibfnamefont {Aaron}\ \bibnamefont {Chou}},\ }\bibfield
  {title} {\enquote {\bibinfo {title} {{Searching for Dark Matter with a
  Superconducting Qubit}},}\ }\href {\doibase 10.1103/PhysRevLett.126.141302}
  {\bibfield  {journal} {\bibinfo  {journal} {Phys. Rev. Lett.}\ }\textbf
  {\bibinfo {volume} {126}},\ \bibinfo {pages} {141302} (\bibinfo {year}
  {2021})},\ \Eprint {http://arxiv.org/abs/2008.12231} {arXiv:2008.12231
  [hep-ex]} \BibitemShut {NoStop}%
\bibitem [{\citenamefont {Brady}\ \emph {et~al.}(2022)\citenamefont {Brady},
  \citenamefont {Gao}, \citenamefont {Harnik}, \citenamefont {Liu},
  \citenamefont {Zhang},\ and\ \citenamefont {Zhuang}}]{Brady}%
  \BibitemOpen
  \bibfield  {author} {\bibinfo {author} {\bibfnamefont {A.}~\bibnamefont
  {Brady}}, \bibinfo {author} {\bibfnamefont {C.}~\bibnamefont {Gao}}, \bibinfo
  {author} {\bibfnamefont {R.}~\bibnamefont {Harnik}}, \bibinfo {author}
  {\bibfnamefont {Z.}~\bibnamefont {Liu}}, \bibinfo {author} {\bibfnamefont
  {Z.}~\bibnamefont {Zhang}}, \ and\ \bibinfo {author} {\bibfnamefont
  {Q.}~\bibnamefont {Zhuang}},\ }\href@noop {} {\enquote {\bibinfo {title}
  {{Entangled sensor-networks for dark-matter searches}},}\ } (\bibinfo {year}
  {2022}),\ \Eprint {http://arxiv.org/abs/2203.05375} {arXiv:2203.05375}
  \BibitemShut {NoStop}%
\bibitem [{\citenamefont {Berlin}\ \emph
  {et~al.}(2020{\natexlab{a}})\citenamefont {Berlin}, \citenamefont {D'Agnolo},
  \citenamefont {Ellis}, \citenamefont {Nantista}, \citenamefont {Neilson},
  \citenamefont {Schuster}, \citenamefont {Tantawi}, \citenamefont {Toro},\
  and\ \citenamefont {Zhou}}]{Berlin:2019ahk}%
  \BibitemOpen
  \bibfield  {author} {\bibinfo {author} {\bibfnamefont {Asher}\ \bibnamefont
  {Berlin}}, \bibinfo {author} {\bibfnamefont {Raffaele~Tito}\ \bibnamefont
  {D'Agnolo}}, \bibinfo {author} {\bibfnamefont {Sebastian A.~R.}\ \bibnamefont
  {Ellis}}, \bibinfo {author} {\bibfnamefont {Christopher}\ \bibnamefont
  {Nantista}}, \bibinfo {author} {\bibfnamefont {Jeffrey}\ \bibnamefont
  {Neilson}}, \bibinfo {author} {\bibfnamefont {Philip}\ \bibnamefont
  {Schuster}}, \bibinfo {author} {\bibfnamefont {Sami}\ \bibnamefont
  {Tantawi}}, \bibinfo {author} {\bibfnamefont {Natalia}\ \bibnamefont {Toro}},
  \ and\ \bibinfo {author} {\bibfnamefont {Kevin}\ \bibnamefont {Zhou}},\
  }\bibfield  {title} {\enquote {\bibinfo {title} {{Axion Dark Matter Detection
  by Superconducting Resonant Frequency Conversion}},}\ }\href {\doibase
  10.1007/JHEP07(2020)088} {\bibfield  {journal} {\bibinfo  {journal} {JHEP}\
  }\textbf {\bibinfo {volume} {07}},\ \bibinfo {pages} {088} (\bibinfo {year}
  {2020}{\natexlab{a}})},\ \Eprint {http://arxiv.org/abs/1912.11048}
  {arXiv:1912.11048 [hep-ph]} \BibitemShut {NoStop}%
\bibitem [{\citenamefont {Berlin}\ \emph
  {et~al.}(2021{\natexlab{b}})\citenamefont {Berlin}, \citenamefont {D'Agnolo},
  \citenamefont {Ellis},\ and\ \citenamefont {Zhou}}]{Berlin:2020vrk}%
  \BibitemOpen
  \bibfield  {author} {\bibinfo {author} {\bibfnamefont {Asher}\ \bibnamefont
  {Berlin}}, \bibinfo {author} {\bibfnamefont {Raffaele~Tito}\ \bibnamefont
  {D'Agnolo}}, \bibinfo {author} {\bibfnamefont {Sebastian A.~R.}\ \bibnamefont
  {Ellis}}, \ and\ \bibinfo {author} {\bibfnamefont {Kevin}\ \bibnamefont
  {Zhou}},\ }\bibfield  {title} {\enquote {\bibinfo {title} {{Heterodyne
  broadband detection of axion dark matter}},}\ }\href {\doibase
  10.1103/PhysRevD.104.L111701} {\bibfield  {journal} {\bibinfo  {journal}
  {Phys. Rev. D}\ }\textbf {\bibinfo {volume} {104}},\ \bibinfo {pages}
  {L111701} (\bibinfo {year} {2021}{\natexlab{b}})},\ \Eprint
  {http://arxiv.org/abs/2007.15656} {arXiv:2007.15656 [hep-ph]} \BibitemShut
  {NoStop}%
\bibitem [{\citenamefont {Lasenby}(2020)}]{Lasenby:2019prg}%
  \BibitemOpen
  \bibfield  {author} {\bibinfo {author} {\bibfnamefont {Robert}\ \bibnamefont
  {Lasenby}},\ }\bibfield  {title} {\enquote {\bibinfo {title} {{Microwave
  cavity searches for low-frequency axion dark matter}},}\ }\href {\doibase
  10.1103/PhysRevD.102.015008} {\bibfield  {journal} {\bibinfo  {journal}
  {Phys. Rev. D}\ }\textbf {\bibinfo {volume} {102}},\ \bibinfo {pages}
  {015008} (\bibinfo {year} {2020})},\ \Eprint
  {http://arxiv.org/abs/1912.11056} {arXiv:1912.11056 [hep-ph]} \BibitemShut
  {NoStop}%
\bibitem [{\citenamefont {Giaccone}\ \emph {et~al.}(2022)\citenamefont
  {Giaccone} \emph {et~al.}}]{Giaccone:2022pke}%
  \BibitemOpen
  \bibfield  {author} {\bibinfo {author} {\bibfnamefont {B.}~\bibnamefont
  {Giaccone}} \emph {et~al.},\ }\bibfield  {title} {\enquote {\bibinfo {title}
  {{Design of axion and axion dark matter searches based on ultra high Q SRF
  cavities}},}\ }\href@noop {} {\  (\bibinfo {year} {2022})},\ \Eprint
  {http://arxiv.org/abs/2207.11346} {arXiv:2207.11346 [hep-ex]} \BibitemShut
  {NoStop}%
\bibitem [{\citenamefont {Brouwer}\ \emph {et~al.}(2022)\citenamefont {Brouwer}
  \emph {et~al.}}]{DMRadio:2022pkf}%
  \BibitemOpen
  \bibfield  {author} {\bibinfo {author} {\bibfnamefont {L.}~\bibnamefont
  {Brouwer}} \emph {et~al.} (\bibinfo {collaboration} {DMRadio}),\ }\bibfield
  {title} {\enquote {\bibinfo {title} {{DMRadio-m$^3$: A Search for the QCD
  Axion Below $1\,\mu$eV}},}\ }\href@noop {} {\  (\bibinfo {year} {2022})},\
  \Eprint {http://arxiv.org/abs/2204.13781} {arXiv:2204.13781 [hep-ex]}
  \BibitemShut {NoStop}%
\bibitem [{\citenamefont {Chaudhuri}\ \emph {et~al.}(2015)\citenamefont
  {Chaudhuri}, \citenamefont {Graham}, \citenamefont {Irwin}, \citenamefont
  {Mardon}, \citenamefont {Rajendran},\ and\ \citenamefont
  {Zhao}}]{Chaudhuri:2014dla}%
  \BibitemOpen
  \bibfield  {author} {\bibinfo {author} {\bibfnamefont {Saptarshi}\
  \bibnamefont {Chaudhuri}}, \bibinfo {author} {\bibfnamefont {Peter~W.}\
  \bibnamefont {Graham}}, \bibinfo {author} {\bibfnamefont {Kent}\ \bibnamefont
  {Irwin}}, \bibinfo {author} {\bibfnamefont {Jeremy}\ \bibnamefont {Mardon}},
  \bibinfo {author} {\bibfnamefont {Surjeet}\ \bibnamefont {Rajendran}}, \ and\
  \bibinfo {author} {\bibfnamefont {Yue}\ \bibnamefont {Zhao}},\ }\bibfield
  {title} {\enquote {\bibinfo {title} {{Radio for hidden-photon dark matter
  detection}},}\ }\href {\doibase 10.1103/PhysRevD.92.075012} {\bibfield
  {journal} {\bibinfo  {journal} {Phys. Rev. D}\ }\textbf {\bibinfo {volume}
  {92}},\ \bibinfo {pages} {075012} (\bibinfo {year} {2015})},\ \Eprint
  {http://arxiv.org/abs/1411.7382} {arXiv:1411.7382 [hep-ph]} \BibitemShut
  {NoStop}%
\bibitem [{\citenamefont {Ouellet}\ \emph {et~al.}(2019)\citenamefont {Ouellet}
  \emph {et~al.}}]{Ouellet:2018beu}%
  \BibitemOpen
  \bibfield  {author} {\bibinfo {author} {\bibfnamefont {Jonathan~L.}\
  \bibnamefont {Ouellet}} \emph {et~al.},\ }\bibfield  {title} {\enquote
  {\bibinfo {title} {{First Results from ABRACADABRA-10 cm: A Search for
  Sub-$\mu$eV Axion Dark Matter}},}\ }\href {\doibase
  10.1103/PhysRevLett.122.121802} {\bibfield  {journal} {\bibinfo  {journal}
  {Phys. Rev. Lett.}\ }\textbf {\bibinfo {volume} {122}},\ \bibinfo {pages}
  {121802} (\bibinfo {year} {2019})},\ \Eprint
  {http://arxiv.org/abs/1810.12257} {arXiv:1810.12257 [hep-ex]} \BibitemShut
  {NoStop}%
\bibitem [{\citenamefont {Kahn}\ \emph {et~al.}(2016)\citenamefont {Kahn},
  \citenamefont {Safdi},\ and\ \citenamefont {Thaler}}]{Kahn:2016aff}%
  \BibitemOpen
  \bibfield  {author} {\bibinfo {author} {\bibfnamefont {Yonatan}\ \bibnamefont
  {Kahn}}, \bibinfo {author} {\bibfnamefont {Benjamin~R.}\ \bibnamefont
  {Safdi}}, \ and\ \bibinfo {author} {\bibfnamefont {Jesse}\ \bibnamefont
  {Thaler}},\ }\bibfield  {title} {\enquote {\bibinfo {title} {{Broadband and
  Resonant Approaches to Axion Dark Matter Detection}},}\ }\href {\doibase
  10.1103/PhysRevLett.117.141801} {\bibfield  {journal} {\bibinfo  {journal}
  {Phys. Rev. Lett.}\ }\textbf {\bibinfo {volume} {117}},\ \bibinfo {pages}
  {141801} (\bibinfo {year} {2016})},\ \Eprint
  {http://arxiv.org/abs/1602.01086} {arXiv:1602.01086 [hep-ph]} \BibitemShut
  {NoStop}%
\bibitem [{\citenamefont {Berlin}\ \emph
  {et~al.}(2020{\natexlab{b}})\citenamefont {Berlin}, \citenamefont {D'Agnolo},
  \citenamefont {Ellis}, \citenamefont {Schuster},\ and\ \citenamefont
  {Toro}}]{Berlin:2019uco}%
  \BibitemOpen
  \bibfield  {author} {\bibinfo {author} {\bibfnamefont {Asher}\ \bibnamefont
  {Berlin}}, \bibinfo {author} {\bibfnamefont {Raffaele~Tito}\ \bibnamefont
  {D'Agnolo}}, \bibinfo {author} {\bibfnamefont {Sebastian A.~R.}\ \bibnamefont
  {Ellis}}, \bibinfo {author} {\bibfnamefont {Philip}\ \bibnamefont
  {Schuster}}, \ and\ \bibinfo {author} {\bibfnamefont {Natalia}\ \bibnamefont
  {Toro}},\ }\bibfield  {title} {\enquote {\bibinfo {title} {{Directly
  Deflecting Particle Dark Matter}},}\ }\href {\doibase
  10.1103/PhysRevLett.124.011801} {\bibfield  {journal} {\bibinfo  {journal}
  {Phys. Rev. Lett.}\ }\textbf {\bibinfo {volume} {124}},\ \bibinfo {pages}
  {011801} (\bibinfo {year} {2020}{\natexlab{b}})},\ \Eprint
  {http://arxiv.org/abs/1908.06982} {arXiv:1908.06982 [hep-ph]} \BibitemShut
  {NoStop}%
\bibitem [{\citenamefont {Graham}\ \emph {et~al.}(2014)\citenamefont {Graham},
  \citenamefont {Mardon}, \citenamefont {Rajendran},\ and\ \citenamefont
  {Zhao}}]{Graham_PRD_2014}%
  \BibitemOpen
  \bibfield  {author} {\bibinfo {author} {\bibfnamefont {Peter~W.}\
  \bibnamefont {Graham}}, \bibinfo {author} {\bibfnamefont {Jeremy}\
  \bibnamefont {Mardon}}, \bibinfo {author} {\bibfnamefont {Surjeet}\
  \bibnamefont {Rajendran}}, \ and\ \bibinfo {author} {\bibfnamefont {Yue}\
  \bibnamefont {Zhao}},\ }\bibfield  {title} {\enquote {\bibinfo {title}
  {Parametrically enhanced hidden photon search},}\ }\href {\doibase
  10.1103/PhysRevD.90.075017} {\bibfield  {journal} {\bibinfo  {journal} {Phys.
  Rev. D}\ }\textbf {\bibinfo {volume} {90}},\ \bibinfo {pages} {075017}
  (\bibinfo {year} {2014})}\BibitemShut {NoStop}%
\bibitem [{\citenamefont {Berlin}\ and\ \citenamefont
  {Hook}(2020)}]{Berlin:2020pey}%
  \BibitemOpen
  \bibfield  {author} {\bibinfo {author} {\bibfnamefont {Asher}\ \bibnamefont
  {Berlin}}\ and\ \bibinfo {author} {\bibfnamefont {Anson}\ \bibnamefont
  {Hook}},\ }\bibfield  {title} {\enquote {\bibinfo {title} {{Searching for
  Millicharged Particles with Superconducting Radio-Frequency Cavities}},}\
  }\href {\doibase 10.1103/PhysRevD.102.035010} {\bibfield  {journal} {\bibinfo
   {journal} {Phys. Rev. D}\ }\textbf {\bibinfo {volume} {102}},\ \bibinfo
  {pages} {035010} (\bibinfo {year} {2020})},\ \Eprint
  {http://arxiv.org/abs/2001.02679} {arXiv:2001.02679 [hep-ph]} \BibitemShut
  {NoStop}%
\bibitem [{\citenamefont {Janish}\ \emph {et~al.}(2019)\citenamefont {Janish},
  \citenamefont {Narayan}, \citenamefont {Rajendran},\ and\ \citenamefont
  {Riggins}}]{Janish_PRD_2019}%
  \BibitemOpen
  \bibfield  {author} {\bibinfo {author} {\bibfnamefont {Ryan}\ \bibnamefont
  {Janish}}, \bibinfo {author} {\bibfnamefont {Vijay}\ \bibnamefont {Narayan}},
  \bibinfo {author} {\bibfnamefont {Surjeet}\ \bibnamefont {Rajendran}}, \ and\
  \bibinfo {author} {\bibfnamefont {Paul}\ \bibnamefont {Riggins}},\ }\bibfield
   {title} {\enquote {\bibinfo {title} {Axion production and detection with
  superconducting rf cavities},}\ }\href {\doibase 10.1103/PhysRevD.100.015036}
  {\bibfield  {journal} {\bibinfo  {journal} {Phys. Rev. D}\ }\textbf {\bibinfo
  {volume} {100}},\ \bibinfo {pages} {015036} (\bibinfo {year}
  {2019})}\BibitemShut {NoStop}%
\bibitem [{\citenamefont {Gao}\ and\ \citenamefont
  {Harnik}(2021)}]{Gao:2020anb}%
  \BibitemOpen
  \bibfield  {author} {\bibinfo {author} {\bibfnamefont {Christina}\
  \bibnamefont {Gao}}\ and\ \bibinfo {author} {\bibfnamefont {Roni}\
  \bibnamefont {Harnik}},\ }\bibfield  {title} {\enquote {\bibinfo {title}
  {{Axion searches with two superconducting radio-frequency cavities}},}\
  }\href {\doibase 10.1007/JHEP07(2021)053} {\bibfield  {journal} {\bibinfo
  {journal} {JHEP}\ }\textbf {\bibinfo {volume} {07}},\ \bibinfo {pages} {053}
  (\bibinfo {year} {2021})},\ \Eprint {http://arxiv.org/abs/2011.01350}
  {arXiv:2011.01350 [hep-ph]} \BibitemShut {NoStop}%
\bibitem [{\citenamefont {Bogorad}\ \emph {et~al.}(2019)\citenamefont
  {Bogorad}, \citenamefont {Hook}, \citenamefont {Kahn},\ and\ \citenamefont
  {Soreq}}]{Bogorad_PRL_2019}%
  \BibitemOpen
  \bibfield  {author} {\bibinfo {author} {\bibfnamefont {Zachary}\ \bibnamefont
  {Bogorad}}, \bibinfo {author} {\bibfnamefont {Anson}\ \bibnamefont {Hook}},
  \bibinfo {author} {\bibfnamefont {Yonatan}\ \bibnamefont {Kahn}}, \ and\
  \bibinfo {author} {\bibfnamefont {Yotam}\ \bibnamefont {Soreq}},\ }\bibfield
  {title} {\enquote {\bibinfo {title} {Probing axionlike particles and the
  axiverse with superconducting radio-frequency cavities},}\ }\href {\doibase
  10.1103/PhysRevLett.123.021801} {\bibfield  {journal} {\bibinfo  {journal}
  {Phys. Rev. Lett.}\ }\textbf {\bibinfo {volume} {123}},\ \bibinfo {pages}
  {021801} (\bibinfo {year} {2019})}\BibitemShut {NoStop}%
\bibitem [{\citenamefont {Horns}\ \emph {et~al.}(2013)\citenamefont {Horns},
  \citenamefont {Jaeckel}, \citenamefont {Lindner}, \citenamefont {Lobanov},
  \citenamefont {Redondo},\ and\ \citenamefont {Ringwald}}]{Horns:2012jf}%
  \BibitemOpen
  \bibfield  {author} {\bibinfo {author} {\bibfnamefont {Dieter}\ \bibnamefont
  {Horns}}, \bibinfo {author} {\bibfnamefont {Joerg}\ \bibnamefont {Jaeckel}},
  \bibinfo {author} {\bibfnamefont {Axel}\ \bibnamefont {Lindner}}, \bibinfo
  {author} {\bibfnamefont {Andrei}\ \bibnamefont {Lobanov}}, \bibinfo {author}
  {\bibfnamefont {Javier}\ \bibnamefont {Redondo}}, \ and\ \bibinfo {author}
  {\bibfnamefont {Andreas}\ \bibnamefont {Ringwald}},\ }\bibfield  {title}
  {\enquote {\bibinfo {title} {{Searching for WISPy Cold Dark Matter with a
  Dish Antenna}},}\ }\href {\doibase 10.1088/1475-7516/2013/04/016} {\bibfield
  {journal} {\bibinfo  {journal} {JCAP}\ }\textbf {\bibinfo {volume} {04}},\
  \bibinfo {pages} {016} (\bibinfo {year} {2013})},\ \Eprint
  {http://arxiv.org/abs/1212.2970} {arXiv:1212.2970 [hep-ph]} \BibitemShut
  {NoStop}%
\bibitem [{\citenamefont {Knirck}\ \emph {et~al.}(2018)\citenamefont {Knirck},
  \citenamefont {Yamazaki}, \citenamefont {Okesaku}, \citenamefont {Asai},
  \citenamefont {Idehara},\ and\ \citenamefont {Inada}}]{Knirck:2018ojz}%
  \BibitemOpen
  \bibfield  {author} {\bibinfo {author} {\bibfnamefont {Stefan}\ \bibnamefont
  {Knirck}}, \bibinfo {author} {\bibfnamefont {Takayuki}\ \bibnamefont
  {Yamazaki}}, \bibinfo {author} {\bibfnamefont {Yoshiki}\ \bibnamefont
  {Okesaku}}, \bibinfo {author} {\bibfnamefont {Shoji}\ \bibnamefont {Asai}},
  \bibinfo {author} {\bibfnamefont {Toshitaka}\ \bibnamefont {Idehara}}, \ and\
  \bibinfo {author} {\bibfnamefont {Toshiaki}\ \bibnamefont {Inada}},\
  }\bibfield  {title} {\enquote {\bibinfo {title} {{First results from a hidden
  photon dark matter search in the meV sector using a plane-parabolic mirror
  system}},}\ }\href {\doibase 10.1088/1475-7516/2018/11/031} {\bibfield
  {journal} {\bibinfo  {journal} {JCAP}\ }\textbf {\bibinfo {volume} {11}},\
  \bibinfo {pages} {031} (\bibinfo {year} {2018})},\ \Eprint
  {http://arxiv.org/abs/1806.05120} {arXiv:1806.05120 [hep-ex]} \BibitemShut
  {NoStop}%
\bibitem [{\citenamefont {Tomita}\ \emph {et~al.}(2020)\citenamefont {Tomita},
  \citenamefont {Oguri}, \citenamefont {Inoue}, \citenamefont {Minowa},
  \citenamefont {Nagasaki}, \citenamefont {Suzuki},\ and\ \citenamefont
  {Tajima}}]{Tomita:2020usq}%
  \BibitemOpen
  \bibfield  {author} {\bibinfo {author} {\bibfnamefont {Nozomu}\ \bibnamefont
  {Tomita}}, \bibinfo {author} {\bibfnamefont {Shugo}\ \bibnamefont {Oguri}},
  \bibinfo {author} {\bibfnamefont {Yoshizumi}\ \bibnamefont {Inoue}}, \bibinfo
  {author} {\bibfnamefont {Makoto}\ \bibnamefont {Minowa}}, \bibinfo {author}
  {\bibfnamefont {Taketo}\ \bibnamefont {Nagasaki}}, \bibinfo {author}
  {\bibfnamefont {Jun'ya}\ \bibnamefont {Suzuki}}, \ and\ \bibinfo {author}
  {\bibfnamefont {Osamu}\ \bibnamefont {Tajima}},\ }\bibfield  {title}
  {\enquote {\bibinfo {title} {{Search for hidden-photon cold dark matter using
  a K-band cryogenic receiver}},}\ }\href {\doibase
  10.1088/1475-7516/2020/09/012} {\bibfield  {journal} {\bibinfo  {journal}
  {JCAP}\ }\textbf {\bibinfo {volume} {09}},\ \bibinfo {pages} {012} (\bibinfo
  {year} {2020})},\ \Eprint {http://arxiv.org/abs/2006.02828} {arXiv:2006.02828
  [hep-ex]} \BibitemShut {NoStop}%
\bibitem [{\citenamefont {Brun}\ \emph
  {et~al.}(2019{\natexlab{a}})\citenamefont {Brun}, \citenamefont {Chevalier},\
  and\ \citenamefont {Flouzat}}]{Brun:2019kak}%
  \BibitemOpen
  \bibfield  {author} {\bibinfo {author} {\bibfnamefont {Pierre}\ \bibnamefont
  {Brun}}, \bibinfo {author} {\bibfnamefont {Laurent}\ \bibnamefont
  {Chevalier}}, \ and\ \bibinfo {author} {\bibfnamefont {Christophe}\
  \bibnamefont {Flouzat}},\ }\bibfield  {title} {\enquote {\bibinfo {title}
  {{Direct Searches for Hidden-Photon Dark Matter with the SHUKET
  Experiment}},}\ }\href {\doibase 10.1103/PhysRevLett.122.201801} {\bibfield
  {journal} {\bibinfo  {journal} {Phys. Rev. Lett.}\ }\textbf {\bibinfo
  {volume} {122}},\ \bibinfo {pages} {201801} (\bibinfo {year}
  {2019}{\natexlab{a}})},\ \Eprint {http://arxiv.org/abs/1905.05579}
  {arXiv:1905.05579 [hep-ex]} \BibitemShut {NoStop}%
\bibitem [{\citenamefont {Andrianavalomahefa}\ \emph
  {et~al.}(2020)\citenamefont {Andrianavalomahefa} \emph
  {et~al.}}]{FUNKExperiment:2020ofv}%
  \BibitemOpen
  \bibfield  {author} {\bibinfo {author} {\bibfnamefont {A.}~\bibnamefont
  {Andrianavalomahefa}} \emph {et~al.} (\bibinfo {collaboration} {FUNK
  Experiment}),\ }\bibfield  {title} {\enquote {\bibinfo {title} {{Limits from
  the Funk Experiment on the Mixing Strength of Hidden-Photon Dark Matter in
  the Visible and Near-Ultraviolet Wavelength Range}},}\ }\href {\doibase
  10.1103/PhysRevD.102.042001} {\bibfield  {journal} {\bibinfo  {journal}
  {Phys. Rev. D}\ }\textbf {\bibinfo {volume} {102}},\ \bibinfo {pages}
  {042001} (\bibinfo {year} {2020})},\ \Eprint
  {http://arxiv.org/abs/2003.13144} {arXiv:2003.13144 [astro-ph.CO]}
  \BibitemShut {NoStop}%
\bibitem [{\citenamefont {Liu}\ \emph {et~al.}(2022)\citenamefont {Liu} \emph
  {et~al.}}]{BREAD:2021tpx}%
  \BibitemOpen
  \bibfield  {author} {\bibinfo {author} {\bibfnamefont {Jesse}\ \bibnamefont
  {Liu}} \emph {et~al.} (\bibinfo {collaboration} {BREAD}),\ }\bibfield
  {title} {\enquote {\bibinfo {title} {{Broadband Solenoidal Haloscope for
  Terahertz Axion Detection}},}\ }\href {\doibase
  10.1103/PhysRevLett.128.131801} {\bibfield  {journal} {\bibinfo  {journal}
  {Phys. Rev. Lett.}\ }\textbf {\bibinfo {volume} {128}},\ \bibinfo {pages}
  {131801} (\bibinfo {year} {2022})},\ \Eprint
  {http://arxiv.org/abs/2111.12103} {arXiv:2111.12103 [physics.ins-det]}
  \BibitemShut {NoStop}%
\bibitem [{\citenamefont {Jaeckel}\ and\ \citenamefont
  {Redondo}(2013)}]{Jaeckel:2013eha}%
  \BibitemOpen
  \bibfield  {author} {\bibinfo {author} {\bibfnamefont {Joerg}\ \bibnamefont
  {Jaeckel}}\ and\ \bibinfo {author} {\bibfnamefont {Javier}\ \bibnamefont
  {Redondo}},\ }\bibfield  {title} {\enquote {\bibinfo {title} {{Resonant to
  broadband searches for cold dark matter consisting of weakly interacting slim
  particles}},}\ }\href {\doibase 10.1103/PhysRevD.88.115002} {\bibfield
  {journal} {\bibinfo  {journal} {Phys. Rev. D}\ }\textbf {\bibinfo {volume}
  {88}},\ \bibinfo {pages} {115002} (\bibinfo {year} {2013})},\ \Eprint
  {http://arxiv.org/abs/1308.1103} {arXiv:1308.1103 [hep-ph]} \BibitemShut
  {NoStop}%
\bibitem [{\citenamefont {Millar}\ \emph {et~al.}(2017)\citenamefont {Millar},
  \citenamefont {Raffelt}, \citenamefont {Redondo},\ and\ \citenamefont
  {Steffen}}]{Millar:2016cjp}%
  \BibitemOpen
  \bibfield  {author} {\bibinfo {author} {\bibfnamefont {Alexander~J.}\
  \bibnamefont {Millar}}, \bibinfo {author} {\bibfnamefont {Georg~G.}\
  \bibnamefont {Raffelt}}, \bibinfo {author} {\bibfnamefont {Javier}\
  \bibnamefont {Redondo}}, \ and\ \bibinfo {author} {\bibfnamefont {Frank~D.}\
  \bibnamefont {Steffen}},\ }\bibfield  {title} {\enquote {\bibinfo {title}
  {{Dielectric Haloscopes to Search for Axion Dark Matter: Theoretical
  Foundations}},}\ }\href {\doibase 10.1088/1475-7516/2017/01/061} {\bibfield
  {journal} {\bibinfo  {journal} {JCAP}\ }\textbf {\bibinfo {volume} {01}},\
  \bibinfo {pages} {061} (\bibinfo {year} {2017})},\ \Eprint
  {http://arxiv.org/abs/1612.07057} {arXiv:1612.07057 [hep-ph]} \BibitemShut
  {NoStop}%
\bibitem [{\citenamefont {Brun}\ \emph
  {et~al.}(2019{\natexlab{b}})\citenamefont {Brun} \emph
  {et~al.}}]{MADMAX:2019pub}%
  \BibitemOpen
  \bibfield  {author} {\bibinfo {author} {\bibfnamefont {P.}~\bibnamefont
  {Brun}} \emph {et~al.} (\bibinfo {collaboration} {MADMAX}),\ }\bibfield
  {title} {\enquote {\bibinfo {title} {{A new experimental approach to probe
  QCD axion dark matter in the mass range above 40 $\mu$eV}},}\ }\href
  {\doibase 10.1140/epjc/s10052-019-6683-x} {\bibfield  {journal} {\bibinfo
  {journal} {Eur. Phys. J. C}\ }\textbf {\bibinfo {volume} {79}},\ \bibinfo
  {pages} {186} (\bibinfo {year} {2019}{\natexlab{b}})},\ \Eprint
  {http://arxiv.org/abs/1901.07401} {arXiv:1901.07401 [physics.ins-det]}
  \BibitemShut {NoStop}%
\bibitem [{\citenamefont {Chiles}\ \emph {et~al.}(2021)\citenamefont {Chiles}
  \emph {et~al.}}]{Chiles:2021gxk}%
  \BibitemOpen
  \bibfield  {author} {\bibinfo {author} {\bibfnamefont {Jeff}\ \bibnamefont
  {Chiles}} \emph {et~al.},\ }\bibfield  {title} {\enquote {\bibinfo {title}
  {{First Constraints on Dark Photon Dark Matter with Superconducting Nanowire
  Detectors in an Optical Haloscope}},}\ }\href@noop {} {\  (\bibinfo {year}
  {2021})},\ \Eprint {http://arxiv.org/abs/2110.01582} {arXiv:2110.01582
  [hep-ex]} \BibitemShut {NoStop}%
\bibitem [{\citenamefont {Baryakhtar}\ \emph {et~al.}(2018)\citenamefont
  {Baryakhtar}, \citenamefont {Huang},\ and\ \citenamefont
  {Lasenby}}]{Baryakhtar:2018doz}%
  \BibitemOpen
  \bibfield  {author} {\bibinfo {author} {\bibfnamefont {Masha}\ \bibnamefont
  {Baryakhtar}}, \bibinfo {author} {\bibfnamefont {Junwu}\ \bibnamefont
  {Huang}}, \ and\ \bibinfo {author} {\bibfnamefont {Robert}\ \bibnamefont
  {Lasenby}},\ }\bibfield  {title} {\enquote {\bibinfo {title} {{Axion and
  hidden photon dark matter detection with multilayer optical haloscopes}},}\
  }\href {\doibase 10.1103/PhysRevD.98.035006} {\bibfield  {journal} {\bibinfo
  {journal} {Phys. Rev. D}\ }\textbf {\bibinfo {volume} {98}},\ \bibinfo
  {pages} {035006} (\bibinfo {year} {2018})},\ \Eprint
  {http://arxiv.org/abs/1803.11455} {arXiv:1803.11455 [hep-ph]} \BibitemShut
  {NoStop}%
\bibitem [{\citenamefont {Arias}\ \emph {et~al.}(2012)\citenamefont {Arias},
  \citenamefont {Cadamuro}, \citenamefont {Goodsell}, \citenamefont {Jaeckel},
  \citenamefont {Redondo},\ and\ \citenamefont {Ringwald}}]{Arias_2012}%
  \BibitemOpen
  \bibfield  {author} {\bibinfo {author} {\bibfnamefont {Paola}\ \bibnamefont
  {Arias}}, \bibinfo {author} {\bibfnamefont {Davide}\ \bibnamefont
  {Cadamuro}}, \bibinfo {author} {\bibfnamefont {Mark}\ \bibnamefont
  {Goodsell}}, \bibinfo {author} {\bibfnamefont {Joerg}\ \bibnamefont
  {Jaeckel}}, \bibinfo {author} {\bibfnamefont {Javier}\ \bibnamefont
  {Redondo}}, \ and\ \bibinfo {author} {\bibfnamefont {Andreas}\ \bibnamefont
  {Ringwald}},\ }\bibfield  {title} {\enquote {\bibinfo {title} {{WISPy} cold
  dark matter},}\ }\href {\doibase 10.1088/1475-7516/2012/06/013} {\bibfield
  {journal} {\bibinfo  {journal} {Journal of Cosmology and Astroparticle
  Physics}\ }\textbf {\bibinfo {volume} {2012}},\ \bibinfo {pages} {013--013}
  (\bibinfo {year} {2012})}\BibitemShut {NoStop}%
\bibitem [{\citenamefont {Sikivie}\ \emph {et~al.}(2007)\citenamefont
  {Sikivie}, \citenamefont {Tanner},\ and\ \citenamefont {van
  Bibber}}]{Sikivie:2007qm}%
  \BibitemOpen
  \bibfield  {author} {\bibinfo {author} {\bibfnamefont {P.}~\bibnamefont
  {Sikivie}}, \bibinfo {author} {\bibfnamefont {D.B.}\ \bibnamefont {Tanner}},
  \ and\ \bibinfo {author} {\bibfnamefont {Karl}\ \bibnamefont {van Bibber}},\
  }\bibfield  {title} {\enquote {\bibinfo {title} {{Resonantly enhanced
  axion-photon regeneration}},}\ }\href {\doibase
  10.1103/PhysRevLett.98.172002} {\bibfield  {journal} {\bibinfo  {journal}
  {Phys. Rev. Lett.}\ }\textbf {\bibinfo {volume} {98}},\ \bibinfo {pages}
  {172002} (\bibinfo {year} {2007})},\ \Eprint
  {http://arxiv.org/abs/hep-ph/0701198} {arXiv:hep-ph/0701198} \BibitemShut
  {NoStop}%
\bibitem [{\citenamefont {Estrada}\ \emph {et~al.}(2021)\citenamefont
  {Estrada}, \citenamefont {Harnik}, \citenamefont {Rodrigues},\ and\
  \citenamefont {Senger}}]{Estrada:2020dpg}%
  \BibitemOpen
  \bibfield  {author} {\bibinfo {author} {\bibfnamefont {Juan}\ \bibnamefont
  {Estrada}}, \bibinfo {author} {\bibfnamefont {Roni}\ \bibnamefont {Harnik}},
  \bibinfo {author} {\bibfnamefont {Dario}\ \bibnamefont {Rodrigues}}, \ and\
  \bibinfo {author} {\bibfnamefont {Matias}\ \bibnamefont {Senger}},\
  }\bibfield  {title} {\enquote {\bibinfo {title} {{Searching for Dark
  Particles with Quantum Optics}},}\ }\href {\doibase
  10.1103/PRXQuantum.2.030340} {\bibfield  {journal} {\bibinfo  {journal} {PRX
  Quantum}\ }\textbf {\bibinfo {volume} {2}},\ \bibinfo {pages} {030340}
  (\bibinfo {year} {2021})},\ \Eprint {http://arxiv.org/abs/2012.04707}
  {arXiv:2012.04707 [hep-ph]} \BibitemShut {NoStop}%
\bibitem [{\citenamefont {Alarcon}\ \emph {et~al.}(2022)\citenamefont {Alarcon}
  \emph {et~al.}}]{Alarcon:2022ero}%
  \BibitemOpen
  \bibfield  {author} {\bibinfo {author} {\bibfnamefont {Ricardo}\ \bibnamefont
  {Alarcon}} \emph {et~al.},\ }\bibfield  {title} {\enquote {\bibinfo {title}
  {{Electric dipole moments and the search for new physics}},}\ }in\ \href@noop
  {} {\emph {\bibinfo {booktitle} {{2022 Snowmass Summer Study}}}}\ (\bibinfo
  {year} {2022})\ \Eprint {http://arxiv.org/abs/2203.08103} {arXiv:2203.08103
  [hep-ph]} \BibitemShut {NoStop}%
\bibitem [{\citenamefont {Budker}\ \emph {et~al.}(2014)\citenamefont {Budker},
  \citenamefont {Graham}, \citenamefont {Ledbetter}, \citenamefont
  {Rajendran},\ and\ \citenamefont {Sushkov}}]{Budker:2013hfa}%
  \BibitemOpen
  \bibfield  {author} {\bibinfo {author} {\bibfnamefont {Dmitry}\ \bibnamefont
  {Budker}}, \bibinfo {author} {\bibfnamefont {Peter~W.}\ \bibnamefont
  {Graham}}, \bibinfo {author} {\bibfnamefont {Micah}\ \bibnamefont
  {Ledbetter}}, \bibinfo {author} {\bibfnamefont {Surjeet}\ \bibnamefont
  {Rajendran}}, \ and\ \bibinfo {author} {\bibfnamefont {Alex}\ \bibnamefont
  {Sushkov}},\ }\bibfield  {title} {\enquote {\bibinfo {title} {{Proposal for a
  Cosmic Axion Spin Precession Experiment (CASPEr)}},}\ }\href {\doibase
  10.1103/PhysRevX.4.021030} {\bibfield  {journal} {\bibinfo  {journal} {Phys.
  Rev. X}\ }\textbf {\bibinfo {volume} {4}},\ \bibinfo {pages} {021030}
  (\bibinfo {year} {2014})},\ \Eprint {http://arxiv.org/abs/1306.6089}
  {arXiv:1306.6089 [hep-ph]} \BibitemShut {NoStop}%
\bibitem [{\citenamefont {Jackson~Kimball}\ \emph {et~al.}(2020)\citenamefont
  {Jackson~Kimball} \emph {et~al.}}]{JacksonKimball:2017elr}%
  \BibitemOpen
  \bibfield  {author} {\bibinfo {author} {\bibfnamefont {Derek~F.}\
  \bibnamefont {Jackson~Kimball}} \emph {et~al.},\ }\bibfield  {title}
  {\enquote {\bibinfo {title} {{Overview of the Cosmic Axion Spin Precession
  Experiment (CASPEr)}},}\ }\href {\doibase 10.1007/978-3-030-43761-9_13}
  {\bibfield  {journal} {\bibinfo  {journal} {Springer Proc. Phys.}\ }\textbf
  {\bibinfo {volume} {245}},\ \bibinfo {pages} {105--121} (\bibinfo {year}
  {2020})},\ \Eprint {http://arxiv.org/abs/1711.08999} {arXiv:1711.08999
  [physics.ins-det]} \BibitemShut {NoStop}%
\bibitem [{\citenamefont {Gao}\ \emph {et~al.}(2022)\citenamefont {Gao},
  \citenamefont {Halperin}, \citenamefont {Kahn}, \citenamefont {Nguyen},
  \citenamefont {Sch\"utte-Engel},\ and\ \citenamefont {Scott}}]{Gao:2022nuq}%
  \BibitemOpen
  \bibfield  {author} {\bibinfo {author} {\bibfnamefont {Christina}\
  \bibnamefont {Gao}}, \bibinfo {author} {\bibfnamefont {William}\ \bibnamefont
  {Halperin}}, \bibinfo {author} {\bibfnamefont {Yonatan}\ \bibnamefont
  {Kahn}}, \bibinfo {author} {\bibfnamefont {Man}\ \bibnamefont {Nguyen}},
  \bibinfo {author} {\bibfnamefont {Jan}\ \bibnamefont {Sch\"utte-Engel}}, \
  and\ \bibinfo {author} {\bibfnamefont {John~William}\ \bibnamefont {Scott}},\
  }\bibfield  {title} {\enquote {\bibinfo {title} {{Axion wind detection with
  the homogeneous precession domain of superfluid helium-3}},}\ }\href@noop {}
  {\  (\bibinfo {year} {2022})},\ \Eprint {http://arxiv.org/abs/2208.14454}
  {arXiv:2208.14454 [hep-ph]} \BibitemShut {NoStop}%
\bibitem [{\citenamefont {Wu}\ \emph {et~al.}(2019)\citenamefont {Wu} \emph
  {et~al.}}]{Wu:2019exd}%
  \BibitemOpen
  \bibfield  {author} {\bibinfo {author} {\bibfnamefont {Teng}\ \bibnamefont
  {Wu}} \emph {et~al.},\ }\bibfield  {title} {\enquote {\bibinfo {title}
  {{Search for Axionlike Dark Matter with a Liquid-State Nuclear Spin
  Comagnetometer}},}\ }\href {\doibase 10.1103/PhysRevLett.122.191302}
  {\bibfield  {journal} {\bibinfo  {journal} {Phys. Rev. Lett.}\ }\textbf
  {\bibinfo {volume} {122}},\ \bibinfo {pages} {191302} (\bibinfo {year}
  {2019})},\ \Eprint {http://arxiv.org/abs/1901.10843} {arXiv:1901.10843
  [hep-ex]} \BibitemShut {NoStop}%
\bibitem [{\citenamefont {Graham}\ \emph {et~al.}(2018)\citenamefont {Graham},
  \citenamefont {Kaplan}, \citenamefont {Mardon}, \citenamefont {Rajendran},
  \citenamefont {Terrano}, \citenamefont {Trahms},\ and\ \citenamefont
  {Wilkason}}]{Graham:2017ivz}%
  \BibitemOpen
  \bibfield  {author} {\bibinfo {author} {\bibfnamefont {Peter~W.}\
  \bibnamefont {Graham}}, \bibinfo {author} {\bibfnamefont {David~E.}\
  \bibnamefont {Kaplan}}, \bibinfo {author} {\bibfnamefont {Jeremy}\
  \bibnamefont {Mardon}}, \bibinfo {author} {\bibfnamefont {Surjeet}\
  \bibnamefont {Rajendran}}, \bibinfo {author} {\bibfnamefont {William~A.}\
  \bibnamefont {Terrano}}, \bibinfo {author} {\bibfnamefont {Lutz}\
  \bibnamefont {Trahms}}, \ and\ \bibinfo {author} {\bibfnamefont {Thomas}\
  \bibnamefont {Wilkason}},\ }\bibfield  {title} {\enquote {\bibinfo {title}
  {{Spin Precession Experiments for Light Axionic Dark Matter}},}\ }\href
  {\doibase 10.1103/PhysRevD.97.055006} {\bibfield  {journal} {\bibinfo
  {journal} {Phys. Rev. D}\ }\textbf {\bibinfo {volume} {97}},\ \bibinfo
  {pages} {055006} (\bibinfo {year} {2018})},\ \Eprint
  {http://arxiv.org/abs/1709.07852} {arXiv:1709.07852 [hep-ph]} \BibitemShut
  {NoStop}%
\bibitem [{\citenamefont {Safronova}\ \emph {et~al.}(2018)\citenamefont
  {Safronova}, \citenamefont {Budker}, \citenamefont {DeMille}, \citenamefont
  {Kimball}, \citenamefont {Derevianko},\ and\ \citenamefont
  {Clark}}]{Safronova:2017xyt}%
  \BibitemOpen
  \bibfield  {author} {\bibinfo {author} {\bibfnamefont {M.~S.}\ \bibnamefont
  {Safronova}}, \bibinfo {author} {\bibfnamefont {D.}~\bibnamefont {Budker}},
  \bibinfo {author} {\bibfnamefont {D.}~\bibnamefont {DeMille}}, \bibinfo
  {author} {\bibfnamefont {Derek F.~Jackson}\ \bibnamefont {Kimball}}, \bibinfo
  {author} {\bibfnamefont {A.}~\bibnamefont {Derevianko}}, \ and\ \bibinfo
  {author} {\bibfnamefont {C.~W.}\ \bibnamefont {Clark}},\ }\bibfield  {title}
  {\enquote {\bibinfo {title} {{Search for New Physics with Atoms and
  Molecules}},}\ }\href {\doibase 10.1103/RevModPhys.90.025008} {\bibfield
  {journal} {\bibinfo  {journal} {Rev. Mod. Phys.}\ }\textbf {\bibinfo {volume}
  {90}},\ \bibinfo {pages} {025008} (\bibinfo {year} {2018})},\ \Eprint
  {http://arxiv.org/abs/1710.01833} {arXiv:1710.01833 [physics.atom-ph]}
  \BibitemShut {NoStop}%
\bibitem [{\citenamefont {Bloch}\ \emph {et~al.}(2020)\citenamefont {Bloch},
  \citenamefont {Hochberg}, \citenamefont {Kuflik},\ and\ \citenamefont
  {Volansky}}]{Bloch:2019lcy}%
  \BibitemOpen
  \bibfield  {author} {\bibinfo {author} {\bibfnamefont {Itay~M.}\ \bibnamefont
  {Bloch}}, \bibinfo {author} {\bibfnamefont {Yonit}\ \bibnamefont {Hochberg}},
  \bibinfo {author} {\bibfnamefont {Eric}\ \bibnamefont {Kuflik}}, \ and\
  \bibinfo {author} {\bibfnamefont {Tomer}\ \bibnamefont {Volansky}},\
  }\bibfield  {title} {\enquote {\bibinfo {title} {{Axion-like Relics: New
  Constraints from Old Comagnetometer Data}},}\ }\href {\doibase
  10.1007/JHEP01(2020)167} {\bibfield  {journal} {\bibinfo  {journal} {JHEP}\
  }\textbf {\bibinfo {volume} {01}},\ \bibinfo {pages} {167} (\bibinfo {year}
  {2020})},\ \Eprint {http://arxiv.org/abs/1907.03767} {arXiv:1907.03767
  [hep-ph]} \BibitemShut {NoStop}%
\bibitem [{\citenamefont {Bloch}\ \emph {et~al.}(2022)\citenamefont {Bloch},
  \citenamefont {Ronen}, \citenamefont {Shaham}, \citenamefont {Katz},
  \citenamefont {Volansky},\ and\ \citenamefont {Katz}}]{Bloch:2021vnn}%
  \BibitemOpen
  \bibfield  {author} {\bibinfo {author} {\bibfnamefont {Itay~M.}\ \bibnamefont
  {Bloch}}, \bibinfo {author} {\bibfnamefont {Gil}\ \bibnamefont {Ronen}},
  \bibinfo {author} {\bibfnamefont {Roy}\ \bibnamefont {Shaham}}, \bibinfo
  {author} {\bibfnamefont {Ori}\ \bibnamefont {Katz}}, \bibinfo {author}
  {\bibfnamefont {Tomer}\ \bibnamefont {Volansky}}, \ and\ \bibinfo {author}
  {\bibfnamefont {Or}~\bibnamefont {Katz}} (\bibinfo {collaboration}
  {NASDUCK}),\ }\bibfield  {title} {\enquote {\bibinfo {title} {{New
  constraints on axion-like dark matter using a Floquet quantum detector}},}\
  }\href {\doibase 10.1126/sciadv.abl8919} {\bibfield  {journal} {\bibinfo
  {journal} {Sci. Adv.}\ }\textbf {\bibinfo {volume} {8}},\ \bibinfo {pages}
  {abl8919} (\bibinfo {year} {2022})},\ \Eprint
  {http://arxiv.org/abs/2105.04603} {arXiv:2105.04603 [hep-ph]} \BibitemShut
  {NoStop}%
\bibitem [{\citenamefont {Geraci}\ \emph {et~al.}(2018)\citenamefont {Geraci}
  \emph {et~al.}}]{ARIADNE:2017tdd}%
  \BibitemOpen
  \bibfield  {author} {\bibinfo {author} {\bibfnamefont {A.~A.}\ \bibnamefont
  {Geraci}} \emph {et~al.} (\bibinfo {collaboration} {ARIADNE}),\ }\bibfield
  {title} {\enquote {\bibinfo {title} {{Progress on the ARIADNE axion
  experiment}},}\ }\href {\doibase 10.1007/978-3-319-92726-8_18} {\bibfield
  {journal} {\bibinfo  {journal} {Springer Proc. Phys.}\ }\textbf {\bibinfo
  {volume} {211}},\ \bibinfo {pages} {151--161} (\bibinfo {year} {2018})},\
  \Eprint {http://arxiv.org/abs/1710.05413} {arXiv:1710.05413 [astro-ph.IM]}
  \BibitemShut {NoStop}%
\bibitem [{\citenamefont {Hanneke}\ \emph {et~al.}(2008)\citenamefont
  {Hanneke}, \citenamefont {Fogwell},\ and\ \citenamefont
  {Gabrielse}}]{Hanneke:2008tm}%
  \BibitemOpen
  \bibfield  {author} {\bibinfo {author} {\bibfnamefont {D.}~\bibnamefont
  {Hanneke}}, \bibinfo {author} {\bibfnamefont {S.}~\bibnamefont {Fogwell}}, \
  and\ \bibinfo {author} {\bibfnamefont {G.}~\bibnamefont {Gabrielse}},\
  }\bibfield  {title} {\enquote {\bibinfo {title} {{New Measurement of the
  Electron Magnetic Moment and the Fine Structure Constant}},}\ }\href
  {\doibase 10.1103/PhysRevLett.100.120801} {\bibfield  {journal} {\bibinfo
  {journal} {Phys. Rev. Lett.}\ }\textbf {\bibinfo {volume} {100}},\ \bibinfo
  {pages} {120801} (\bibinfo {year} {2008})},\ \Eprint
  {http://arxiv.org/abs/0801.1134} {arXiv:0801.1134 [physics.atom-ph]}
  \BibitemShut {NoStop}%
\bibitem [{\citenamefont {Budker}\ \emph {et~al.}(2022)\citenamefont {Budker},
  \citenamefont {Graham}, \citenamefont {Ramani}, \citenamefont
  {Schmidt-Kaler}, \citenamefont {Smorra},\ and\ \citenamefont
  {Ulmer}}]{Budker:2021quh}%
  \BibitemOpen
  \bibfield  {author} {\bibinfo {author} {\bibfnamefont {Dmitry}\ \bibnamefont
  {Budker}}, \bibinfo {author} {\bibfnamefont {Peter~W.}\ \bibnamefont
  {Graham}}, \bibinfo {author} {\bibfnamefont {Harikrishnan}\ \bibnamefont
  {Ramani}}, \bibinfo {author} {\bibfnamefont {Ferdinand}\ \bibnamefont
  {Schmidt-Kaler}}, \bibinfo {author} {\bibfnamefont {Christian}\ \bibnamefont
  {Smorra}}, \ and\ \bibinfo {author} {\bibfnamefont {Stefan}\ \bibnamefont
  {Ulmer}},\ }\bibfield  {title} {\enquote {\bibinfo {title} {{Millicharged
  Dark Matter Detection with Ion Traps}},}\ }\href {\doibase
  10.1103/PRXQuantum.3.010330} {\bibfield  {journal} {\bibinfo  {journal} {PRX
  Quantum}\ }\textbf {\bibinfo {volume} {3}},\ \bibinfo {pages} {010330}
  (\bibinfo {year} {2022})},\ \Eprint {http://arxiv.org/abs/2108.05283}
  {arXiv:2108.05283 [hep-ph]} \BibitemShut {NoStop}%
\bibitem [{\citenamefont {Carney}\ \emph
  {et~al.}(2021{\natexlab{a}})\citenamefont {Carney}, \citenamefont
  {H\"affner}, \citenamefont {Moore},\ and\ \citenamefont
  {Taylor}}]{Carney:2021irt}%
  \BibitemOpen
  \bibfield  {author} {\bibinfo {author} {\bibfnamefont {Daniel}\ \bibnamefont
  {Carney}}, \bibinfo {author} {\bibfnamefont {Hartmut}\ \bibnamefont
  {H\"affner}}, \bibinfo {author} {\bibfnamefont {David~C.}\ \bibnamefont
  {Moore}}, \ and\ \bibinfo {author} {\bibfnamefont {Jacob~M.}\ \bibnamefont
  {Taylor}},\ }\bibfield  {title} {\enquote {\bibinfo {title} {{Trapped
  Electrons and Ions as Particle Detectors}},}\ }\href {\doibase
  10.1103/PhysRevLett.127.061804} {\bibfield  {journal} {\bibinfo  {journal}
  {Phys. Rev. Lett.}\ }\textbf {\bibinfo {volume} {127}},\ \bibinfo {pages}
  {061804} (\bibinfo {year} {2021}{\natexlab{a}})},\ \Eprint
  {http://arxiv.org/abs/2104.05737} {arXiv:2104.05737 [quant-ph]} \BibitemShut
  {NoStop}%
\bibitem [{\citenamefont {Fan}\ \emph {et~al.}(2022)\citenamefont {Fan},
  \citenamefont {Gabrielse}, \citenamefont {Graham}, \citenamefont {Harnik},
  \citenamefont {Myers}, \citenamefont {Ramani}, \citenamefont {Sukra},
  \citenamefont {Wong},\ and\ \citenamefont {Xiao}}]{Fan:2022uwu}%
  \BibitemOpen
  \bibfield  {author} {\bibinfo {author} {\bibfnamefont {Xing}\ \bibnamefont
  {Fan}}, \bibinfo {author} {\bibfnamefont {Gerald}\ \bibnamefont {Gabrielse}},
  \bibinfo {author} {\bibfnamefont {Peter~W.}\ \bibnamefont {Graham}}, \bibinfo
  {author} {\bibfnamefont {Roni}\ \bibnamefont {Harnik}}, \bibinfo {author}
  {\bibfnamefont {Thomas~G.}\ \bibnamefont {Myers}}, \bibinfo {author}
  {\bibfnamefont {Harikrishnan}\ \bibnamefont {Ramani}}, \bibinfo {author}
  {\bibfnamefont {Benedict A.~D.}\ \bibnamefont {Sukra}}, \bibinfo {author}
  {\bibfnamefont {Samuel S.~Y.}\ \bibnamefont {Wong}}, \ and\ \bibinfo {author}
  {\bibfnamefont {Yawen}\ \bibnamefont {Xiao}},\ }\bibfield  {title} {\enquote
  {\bibinfo {title} {{One-Electron Quantum Cyclotron as a Milli-eV Dark-Photon
  Detector}},}\ }\href@noop {} {\  (\bibinfo {year} {2022})},\ \Eprint
  {http://arxiv.org/abs/2208.06519} {arXiv:2208.06519 [hep-ex]} \BibitemShut
  {NoStop}%
\bibitem [{\citenamefont {Carney}\ \emph {et~al.}(2020)\citenamefont {Carney},
  \citenamefont {Ghosh}, \citenamefont {Krnjaic},\ and\ \citenamefont
  {Taylor}}]{Carney:2019pza}%
  \BibitemOpen
  \bibfield  {author} {\bibinfo {author} {\bibfnamefont {Daniel}\ \bibnamefont
  {Carney}}, \bibinfo {author} {\bibfnamefont {Sohitri}\ \bibnamefont {Ghosh}},
  \bibinfo {author} {\bibfnamefont {Gordan}\ \bibnamefont {Krnjaic}}, \ and\
  \bibinfo {author} {\bibfnamefont {Jacob~M.}\ \bibnamefont {Taylor}},\
  }\bibfield  {title} {\enquote {\bibinfo {title} {{Proposal for gravitational
  direct detection of dark matter}},}\ }\href {\doibase
  10.1103/PhysRevD.102.072003} {\bibfield  {journal} {\bibinfo  {journal}
  {Phys. Rev. D}\ }\textbf {\bibinfo {volume} {102}},\ \bibinfo {pages}
  {072003} (\bibinfo {year} {2020})},\ \Eprint
  {http://arxiv.org/abs/1903.00492} {arXiv:1903.00492 [hep-ph]} \BibitemShut
  {NoStop}%
\bibitem [{\citenamefont {Carney}\ \emph
  {et~al.}(2021{\natexlab{b}})\citenamefont {Carney}, \citenamefont {Hook},
  \citenamefont {Liu}, \citenamefont {Taylor},\ and\ \citenamefont
  {Zhao}}]{Carney:2019cio}%
  \BibitemOpen
  \bibfield  {author} {\bibinfo {author} {\bibfnamefont {Daniel}\ \bibnamefont
  {Carney}}, \bibinfo {author} {\bibfnamefont {Anson}\ \bibnamefont {Hook}},
  \bibinfo {author} {\bibfnamefont {Zhen}\ \bibnamefont {Liu}}, \bibinfo
  {author} {\bibfnamefont {Jacob~M.}\ \bibnamefont {Taylor}}, \ and\ \bibinfo
  {author} {\bibfnamefont {Yue}\ \bibnamefont {Zhao}},\ }\bibfield  {title}
  {\enquote {\bibinfo {title} {{Ultralight dark matter detection with
  mechanical quantum sensors}},}\ }\href {\doibase 10.1088/1367-2630/abd9e7}
  {\bibfield  {journal} {\bibinfo  {journal} {New J. Phys.}\ }\textbf {\bibinfo
  {volume} {23}},\ \bibinfo {pages} {023041} (\bibinfo {year}
  {2021}{\natexlab{b}})},\ \Eprint {http://arxiv.org/abs/1908.04797}
  {arXiv:1908.04797 [hep-ph]} \BibitemShut {NoStop}%
\bibitem [{\citenamefont {Ghosh}\ \emph {et~al.}(2020)\citenamefont {Ghosh},
  \citenamefont {Carney}, \citenamefont {Shawhan},\ and\ \citenamefont
  {Taylor}}]{Ghosh:2019rsc}%
  \BibitemOpen
  \bibfield  {author} {\bibinfo {author} {\bibfnamefont {Sohitri}\ \bibnamefont
  {Ghosh}}, \bibinfo {author} {\bibfnamefont {Daniel}\ \bibnamefont {Carney}},
  \bibinfo {author} {\bibfnamefont {Peter}\ \bibnamefont {Shawhan}}, \ and\
  \bibinfo {author} {\bibfnamefont {Jacob~M.}\ \bibnamefont {Taylor}},\
  }\bibfield  {title} {\enquote {\bibinfo {title} {{Backaction-evading impulse
  measurement with mechanical quantum sensors}},}\ }\href {\doibase
  10.1103/PhysRevA.102.023525} {\bibfield  {journal} {\bibinfo  {journal}
  {Phys. Rev. A}\ }\textbf {\bibinfo {volume} {102}},\ \bibinfo {pages}
  {023525} (\bibinfo {year} {2020})},\ \Eprint
  {http://arxiv.org/abs/1910.11892} {arXiv:1910.11892 [quant-ph]} \BibitemShut
  {NoStop}%
\bibitem [{\citenamefont {Attanasio}\ \emph {et~al.}(2022)\citenamefont
  {Attanasio} \emph {et~al.}}]{Windchime:2022whs}%
  \BibitemOpen
  \bibfield  {author} {\bibinfo {author} {\bibfnamefont {Alaina}\ \bibnamefont
  {Attanasio}} \emph {et~al.} (\bibinfo {collaboration} {Windchime}),\
  }\bibfield  {title} {\enquote {\bibinfo {title} {{Snowmass 2021 White Paper:
  The Windchime Project}},}\ }in\ \href@noop {} {\emph {\bibinfo {booktitle}
  {{2022 Snowmass Summer Study}}}}\ (\bibinfo {year} {2022})\ \Eprint
  {http://arxiv.org/abs/2203.07242} {arXiv:2203.07242 [hep-ex]} \BibitemShut
  {NoStop}%
\bibitem [{\citenamefont {Graham}\ \emph {et~al.}(2016)\citenamefont {Graham},
  \citenamefont {Kaplan}, \citenamefont {Mardon}, \citenamefont {Rajendran},\
  and\ \citenamefont {Terrano}}]{Graham:2015ifn}%
  \BibitemOpen
  \bibfield  {author} {\bibinfo {author} {\bibfnamefont {Peter~W.}\
  \bibnamefont {Graham}}, \bibinfo {author} {\bibfnamefont {David~E.}\
  \bibnamefont {Kaplan}}, \bibinfo {author} {\bibfnamefont {Jeremy}\
  \bibnamefont {Mardon}}, \bibinfo {author} {\bibfnamefont {Surjeet}\
  \bibnamefont {Rajendran}}, \ and\ \bibinfo {author} {\bibfnamefont
  {William~A.}\ \bibnamefont {Terrano}},\ }\bibfield  {title} {\enquote
  {\bibinfo {title} {{Dark Matter Direct Detection with Accelerometers}},}\
  }\href {\doibase 10.1103/PhysRevD.93.075029} {\bibfield  {journal} {\bibinfo
  {journal} {Phys. Rev. D}\ }\textbf {\bibinfo {volume} {93}},\ \bibinfo
  {pages} {075029} (\bibinfo {year} {2016})},\ \Eprint
  {http://arxiv.org/abs/1512.06165} {arXiv:1512.06165 [hep-ph]} \BibitemShut
  {NoStop}%
\bibitem [{\citenamefont {Hochberg}\ \emph {et~al.}(2022)\citenamefont
  {Hochberg}, \citenamefont {Kahn}, \citenamefont {Leane}, \citenamefont
  {Rajendran}, \citenamefont {Van~Tilburg}, \citenamefont {Yu},\ and\
  \citenamefont {Zurek}}]{hochberg2022new}%
  \BibitemOpen
  \bibfield  {author} {\bibinfo {author} {\bibfnamefont {Yonit}\ \bibnamefont
  {Hochberg}}, \bibinfo {author} {\bibfnamefont {Yonatan~F}\ \bibnamefont
  {Kahn}}, \bibinfo {author} {\bibfnamefont {Rebecca~K}\ \bibnamefont {Leane}},
  \bibinfo {author} {\bibfnamefont {Surjeet}\ \bibnamefont {Rajendran}},
  \bibinfo {author} {\bibfnamefont {Ken}\ \bibnamefont {Van~Tilburg}}, \bibinfo
  {author} {\bibfnamefont {Tien-Tien}\ \bibnamefont {Yu}}, \ and\ \bibinfo
  {author} {\bibfnamefont {Kathryn~M}\ \bibnamefont {Zurek}},\ }\bibfield
  {title} {\enquote {\bibinfo {title} {New approaches to dark matter
  detection},}\ }\href@noop {} {\bibfield  {journal} {\bibinfo  {journal}
  {Nature Reviews Physics}\ ,\ \bibinfo {pages} {1--5}} (\bibinfo {year}
  {2022})}\BibitemShut {NoStop}%
\bibitem [{\citenamefont {Bousso}\ \emph {et~al.}(2022)\citenamefont {Bousso},
  \citenamefont {Dong}, \citenamefont {Engelhardt}, \citenamefont {Faulkner},
  \citenamefont {Hartman}, \citenamefont {Shenker},\ and\ \citenamefont
  {Stanford}}]{Bousso:2022ntt}%
  \BibitemOpen
  \bibfield  {author} {\bibinfo {author} {\bibfnamefont {Raphael}\ \bibnamefont
  {Bousso}}, \bibinfo {author} {\bibfnamefont {Xi}~\bibnamefont {Dong}},
  \bibinfo {author} {\bibfnamefont {Netta}\ \bibnamefont {Engelhardt}},
  \bibinfo {author} {\bibfnamefont {Thomas}\ \bibnamefont {Faulkner}}, \bibinfo
  {author} {\bibfnamefont {Thomas}\ \bibnamefont {Hartman}}, \bibinfo {author}
  {\bibfnamefont {Stephen~H.}\ \bibnamefont {Shenker}}, \ and\ \bibinfo
  {author} {\bibfnamefont {Douglas}\ \bibnamefont {Stanford}},\ }\bibfield
  {title} {\enquote {\bibinfo {title} {{Snowmass White Paper: Quantum Aspects
  of Black Holes and the Emergence of Spacetime}},}\ }\href@noop {} {\
  (\bibinfo {year} {2022})},\ \Eprint {http://arxiv.org/abs/2201.03096}
  {arXiv:2201.03096 [hep-th]} \BibitemShut {NoStop}%
\bibitem [{\citenamefont {Giddings}(2022)}]{Giddings:2022jda}%
  \BibitemOpen
  \bibfield  {author} {\bibinfo {author} {\bibfnamefont {Steven~B.}\
  \bibnamefont {Giddings}},\ }\bibfield  {title} {\enquote {\bibinfo {title}
  {{The deepest problem: some perspectives on quantum gravity}},}\ }\href@noop
  {} {\  (\bibinfo {year} {2022})},\ \Eprint {http://arxiv.org/abs/2202.08292}
  {arXiv:2202.08292 [hep-th]} \BibitemShut {NoStop}%
\bibitem [{\citenamefont {Blake}\ \emph {et~al.}(2022)\citenamefont {Blake},
  \citenamefont {Gu}, \citenamefont {Hartnoll}, \citenamefont {Liu},
  \citenamefont {Lucas}, \citenamefont {Rajagopal}, \citenamefont {Swingle},\
  and\ \citenamefont {Yoshida}}]{Blake:2022uyo}%
  \BibitemOpen
  \bibfield  {author} {\bibinfo {author} {\bibfnamefont {Mike}\ \bibnamefont
  {Blake}}, \bibinfo {author} {\bibfnamefont {Yingfei}\ \bibnamefont {Gu}},
  \bibinfo {author} {\bibfnamefont {Sean~A.}\ \bibnamefont {Hartnoll}},
  \bibinfo {author} {\bibfnamefont {Hong}\ \bibnamefont {Liu}}, \bibinfo
  {author} {\bibfnamefont {Andrew}\ \bibnamefont {Lucas}}, \bibinfo {author}
  {\bibfnamefont {Krishna}\ \bibnamefont {Rajagopal}}, \bibinfo {author}
  {\bibfnamefont {Brian}\ \bibnamefont {Swingle}}, \ and\ \bibinfo {author}
  {\bibfnamefont {Beni}\ \bibnamefont {Yoshida}},\ }\bibfield  {title}
  {\enquote {\bibinfo {title} {{Snowmass White Paper: New ideas for many-body
  quantum systems from string theory and black holes}},}\ }\href@noop {} {\
  (\bibinfo {year} {2022})},\ \Eprint {http://arxiv.org/abs/2203.04718}
  {arXiv:2203.04718 [hep-th]} \BibitemShut {NoStop}%
\bibitem [{\citenamefont {Nielsen}\ and\ \citenamefont
  {Chuang}(2002)}]{Nielsen:2010aa}%
  \BibitemOpen
  \bibfield  {author} {\bibinfo {author} {\bibfnamefont {Michael~A}\
  \bibnamefont {Nielsen}}\ and\ \bibinfo {author} {\bibfnamefont {Isaac}\
  \bibnamefont {Chuang}},\ }\href@noop {} {\emph {\bibinfo {title} {Quantum
  computation and quantum information}}}\ (\bibinfo  {publisher} {American
  Association of Physics Teachers},\ \bibinfo {year} {2002})\BibitemShut
  {NoStop}%
\bibitem [{\citenamefont {Wilde}(2013)}]{Wilde:2013aa}%
  \BibitemOpen
  \bibfield  {author} {\bibinfo {author} {\bibfnamefont {Mark~M}\ \bibnamefont
  {Wilde}},\ }\href@noop {} {\emph {\bibinfo {title} {Quantum information
  theory}}}\ (\bibinfo  {publisher} {Cambridge University Press},\ \bibinfo
  {year} {2013})\BibitemShut {NoStop}%
\bibitem [{\citenamefont {Haag}(2012{\natexlab{a}})}]{Hagg:2012alg}%
  \BibitemOpen
  \bibfield  {author} {\bibinfo {author} {\bibfnamefont {Rudolf}\ \bibnamefont
  {Haag}},\ }\href@noop {} {\emph {\bibinfo {title} {Local quantum physics:
  Fields, particles, algebras}}}\ (\bibinfo  {publisher} {Springer Science \&
  Business Media},\ \bibinfo {year} {2012})\BibitemShut {NoStop}%
\bibitem [{\citenamefont {Ohya}\ and\ \citenamefont
  {Petz}(2004)}]{Ohya:2004qnt}%
  \BibitemOpen
  \bibfield  {author} {\bibinfo {author} {\bibfnamefont {Masanori}\
  \bibnamefont {Ohya}}\ and\ \bibinfo {author} {\bibfnamefont {D{\'e}nes}\
  \bibnamefont {Petz}},\ }\href@noop {} {\emph {\bibinfo {title} {Quantum
  entropy and its use}}}\ (\bibinfo  {publisher} {Springer Science \& Business
  Media},\ \bibinfo {year} {2004})\BibitemShut {NoStop}%
\bibitem [{\citenamefont {Calabrese}\ and\ \citenamefont
  {Cardy}(2009)}]{Calabrese:2009qy}%
  \BibitemOpen
  \bibfield  {author} {\bibinfo {author} {\bibfnamefont {Pasquale}\
  \bibnamefont {Calabrese}}\ and\ \bibinfo {author} {\bibfnamefont {John}\
  \bibnamefont {Cardy}},\ }\bibfield  {title} {\enquote {\bibinfo {title}
  {{Entanglement entropy and conformal field theory}},}\ }\href {\doibase
  10.1088/1751-8113/42/50/504005} {\bibfield  {journal} {\bibinfo  {journal}
  {J. Phys. A}\ }\textbf {\bibinfo {volume} {42}},\ \bibinfo {pages} {504005}
  (\bibinfo {year} {2009})},\ \Eprint {http://arxiv.org/abs/0905.4013}
  {arXiv:0905.4013 [cond-mat.stat-mech]} \BibitemShut {NoStop}%
\bibitem [{\citenamefont {Nishioka}\ \emph {et~al.}(2009)\citenamefont
  {Nishioka}, \citenamefont {Ryu},\ and\ \citenamefont
  {Takayanagi}}]{Nishioka:2009un}%
  \BibitemOpen
  \bibfield  {author} {\bibinfo {author} {\bibfnamefont {Tatsuma}\ \bibnamefont
  {Nishioka}}, \bibinfo {author} {\bibfnamefont {Shinsei}\ \bibnamefont {Ryu}},
  \ and\ \bibinfo {author} {\bibfnamefont {Tadashi}\ \bibnamefont
  {Takayanagi}},\ }\bibfield  {title} {\enquote {\bibinfo {title} {{Holographic
  Entanglement Entropy: An Overview}},}\ }\href {\doibase
  10.1088/1751-8113/42/50/504008} {\bibfield  {journal} {\bibinfo  {journal}
  {J. Phys. A}\ }\textbf {\bibinfo {volume} {42}},\ \bibinfo {pages} {504008}
  (\bibinfo {year} {2009})},\ \Eprint {http://arxiv.org/abs/0905.0932}
  {arXiv:0905.0932 [hep-th]} \BibitemShut {NoStop}%
\bibitem [{\citenamefont {Van~Raamsdonk}(2017)}]{VanRaamsdonk:2016exw}%
  \BibitemOpen
  \bibfield  {author} {\bibinfo {author} {\bibfnamefont {Mark}\ \bibnamefont
  {Van~Raamsdonk}},\ }\bibfield  {title} {\enquote {\bibinfo {title} {{Lectures
  on Gravity and Entanglement}},}\ }in\ \href {\doibase
  10.1142/9789813149441_0005} {\emph {\bibinfo {booktitle} {{Theoretical
  Advanced Study Institute in Elementary Particle Physics}: {New Frontiers in
  Fields and Strings}}}}\ (\bibinfo {year} {2017})\ pp.\ \bibinfo {pages}
  {297--351},\ \Eprint {http://arxiv.org/abs/1609.00026} {arXiv:1609.00026
  [hep-th]} \BibitemShut {NoStop}%
\bibitem [{\citenamefont {Rangamani}\ and\ \citenamefont
  {Takayanagi}(2017)}]{Rangamani:2016dms}%
  \BibitemOpen
  \bibfield  {author} {\bibinfo {author} {\bibfnamefont {Mukund}\ \bibnamefont
  {Rangamani}}\ and\ \bibinfo {author} {\bibfnamefont {Tadashi}\ \bibnamefont
  {Takayanagi}},\ }\bibfield  {title} {\enquote {\bibinfo {title} {{Holographic
  Entanglement Entropy}},}\ }\href {\doibase 10.1007/978-3-319-52573-0} {\
  \textbf {\bibinfo {volume} {931}} (\bibinfo {year} {2017}),\
  10.1007/978-3-319-52573-0},\ \Eprint {http://arxiv.org/abs/1609.01287}
  {arXiv:1609.01287 [hep-th]} \BibitemShut {NoStop}%
\bibitem [{\citenamefont {Headrick}(2019)}]{Headrick:2019eth}%
  \BibitemOpen
  \bibfield  {author} {\bibinfo {author} {\bibfnamefont {Matthew}\ \bibnamefont
  {Headrick}},\ }\bibfield  {title} {\enquote {\bibinfo {title} {{Lectures on
  entanglement entropy in field theory and holography}},}\ }\href@noop {} {\
  (\bibinfo {year} {2019})},\ \Eprint {http://arxiv.org/abs/1907.08126}
  {arXiv:1907.08126 [hep-th]} \BibitemShut {NoStop}%
\bibitem [{\citenamefont {Witten}(2018)}]{Witten:2018zxz}%
  \BibitemOpen
  \bibfield  {author} {\bibinfo {author} {\bibfnamefont {Edward}\ \bibnamefont
  {Witten}},\ }\bibfield  {title} {\enquote {\bibinfo {title} {{APS Medal for
  Exceptional Achievement in Research: Invited article on entanglement
  properties of quantum field theory}},}\ }\href {\doibase
  10.1103/RevModPhys.90.045003} {\bibfield  {journal} {\bibinfo  {journal}
  {Rev. Mod. Phys.}\ }\textbf {\bibinfo {volume} {90}},\ \bibinfo {pages}
  {045003} (\bibinfo {year} {2018})},\ \Eprint
  {http://arxiv.org/abs/1803.04993} {arXiv:1803.04993 [hep-th]} \BibitemShut
  {NoStop}%
\bibitem [{\citenamefont {Ryu}\ and\ \citenamefont
  {Takayanagi}(2006{\natexlab{b}})}]{Ryu:2006bv}%
  \BibitemOpen
  \bibfield  {author} {\bibinfo {author} {\bibfnamefont {Shinsei}\ \bibnamefont
  {Ryu}}\ and\ \bibinfo {author} {\bibfnamefont {Tadashi}\ \bibnamefont
  {Takayanagi}},\ }\bibfield  {title} {\enquote {\bibinfo {title} {{Holographic
  derivation of entanglement entropy from AdS/CFT}},}\ }\href {\doibase
  10.1103/PhysRevLett.96.181602} {\bibfield  {journal} {\bibinfo  {journal}
  {Phys. Rev. Lett.}\ }\textbf {\bibinfo {volume} {96}},\ \bibinfo {pages}
  {181602} (\bibinfo {year} {2006}{\natexlab{b}})},\ \Eprint
  {http://arxiv.org/abs/hep-th/0603001} {arXiv:hep-th/0603001} \BibitemShut
  {NoStop}%
\bibitem [{\citenamefont {Hubeny}\ \emph {et~al.}(2007)\citenamefont {Hubeny},
  \citenamefont {Rangamani},\ and\ \citenamefont {Takayanagi}}]{Hubeny:2007xt}%
  \BibitemOpen
  \bibfield  {author} {\bibinfo {author} {\bibfnamefont {Veronika~E.}\
  \bibnamefont {Hubeny}}, \bibinfo {author} {\bibfnamefont {Mukund}\
  \bibnamefont {Rangamani}}, \ and\ \bibinfo {author} {\bibfnamefont {Tadashi}\
  \bibnamefont {Takayanagi}},\ }\bibfield  {title} {\enquote {\bibinfo {title}
  {{A Covariant holographic entanglement entropy proposal}},}\ }\href {\doibase
  10.1088/1126-6708/2007/07/062} {\bibfield  {journal} {\bibinfo  {journal}
  {JHEP}\ }\textbf {\bibinfo {volume} {07}},\ \bibinfo {pages} {062} (\bibinfo
  {year} {2007})},\ \Eprint {http://arxiv.org/abs/0705.0016} {arXiv:0705.0016
  [hep-th]} \BibitemShut {NoStop}%
\bibitem [{\citenamefont {Engelhardt}\ and\ \citenamefont
  {Wall}(2015)}]{Engelhardt:2014gca}%
  \BibitemOpen
  \bibfield  {author} {\bibinfo {author} {\bibfnamefont {Netta}\ \bibnamefont
  {Engelhardt}}\ and\ \bibinfo {author} {\bibfnamefont {Aron~C.}\ \bibnamefont
  {Wall}},\ }\bibfield  {title} {\enquote {\bibinfo {title} {{Quantum Extremal
  Surfaces: Holographic Entanglement Entropy beyond the Classical Regime}},}\
  }\href {\doibase 10.1007/JHEP01(2015)073} {\bibfield  {journal} {\bibinfo
  {journal} {JHEP}\ }\textbf {\bibinfo {volume} {01}},\ \bibinfo {pages} {073}
  (\bibinfo {year} {2015})},\ \Eprint {http://arxiv.org/abs/1408.3203}
  {arXiv:1408.3203 [hep-th]} \BibitemShut {NoStop}%
\bibitem [{\citenamefont {Van~Raamsdonk}(2009)}]{VanRaamsdonk:2009ar}%
  \BibitemOpen
  \bibfield  {author} {\bibinfo {author} {\bibfnamefont {Mark}\ \bibnamefont
  {Van~Raamsdonk}},\ }\bibfield  {title} {\enquote {\bibinfo {title} {{Comments
  on quantum gravity and entanglement}},}\ }\href@noop {} {\  (\bibinfo {year}
  {2009})},\ \Eprint {http://arxiv.org/abs/0907.2939} {arXiv:0907.2939
  [hep-th]} \BibitemShut {NoStop}%
\bibitem [{\citenamefont {Maldacena}\ and\ \citenamefont
  {Susskind}(2013)}]{Maldacena:2013xja}%
  \BibitemOpen
  \bibfield  {author} {\bibinfo {author} {\bibfnamefont {Juan}\ \bibnamefont
  {Maldacena}}\ and\ \bibinfo {author} {\bibfnamefont {Leonard}\ \bibnamefont
  {Susskind}},\ }\bibfield  {title} {\enquote {\bibinfo {title} {{Cool horizons
  for entangled black holes}},}\ }\href {\doibase 10.1002/prop.201300020}
  {\bibfield  {journal} {\bibinfo  {journal} {Fortsch. Phys.}\ }\textbf
  {\bibinfo {volume} {61}},\ \bibinfo {pages} {781--811} (\bibinfo {year}
  {2013})},\ \Eprint {http://arxiv.org/abs/1306.0533} {arXiv:1306.0533
  [hep-th]} \BibitemShut {NoStop}%
\bibitem [{\citenamefont {Almheiri}\ \emph {et~al.}(2015)\citenamefont
  {Almheiri}, \citenamefont {Dong},\ and\ \citenamefont
  {Harlow}}]{Almheiri:2014lwa}%
  \BibitemOpen
  \bibfield  {author} {\bibinfo {author} {\bibfnamefont {Ahmed}\ \bibnamefont
  {Almheiri}}, \bibinfo {author} {\bibfnamefont {Xi}~\bibnamefont {Dong}}, \
  and\ \bibinfo {author} {\bibfnamefont {Daniel}\ \bibnamefont {Harlow}},\
  }\bibfield  {title} {\enquote {\bibinfo {title} {{Bulk Locality and Quantum
  Error Correction in AdS/CFT}},}\ }\href {\doibase 10.1007/JHEP04(2015)163}
  {\bibfield  {journal} {\bibinfo  {journal} {JHEP}\ }\textbf {\bibinfo
  {volume} {04}},\ \bibinfo {pages} {163} (\bibinfo {year} {2015})},\ \Eprint
  {http://arxiv.org/abs/1411.7041} {arXiv:1411.7041 [hep-th]} \BibitemShut
  {NoStop}%
\bibitem [{\citenamefont {Goldstein}\ and\ \citenamefont
  {Sela}(2018)}]{Goldstein:2017bua}%
  \BibitemOpen
  \bibfield  {author} {\bibinfo {author} {\bibfnamefont {Moshe}\ \bibnamefont
  {Goldstein}}\ and\ \bibinfo {author} {\bibfnamefont {Eran}\ \bibnamefont
  {Sela}},\ }\bibfield  {title} {\enquote {\bibinfo {title} {{Symmetry-resolved
  entanglement in many-body systems}},}\ }\href {\doibase
  10.1103/PhysRevLett.120.200602} {\bibfield  {journal} {\bibinfo  {journal}
  {Phys. Rev. Lett.}\ }\textbf {\bibinfo {volume} {120}},\ \bibinfo {pages}
  {200602} (\bibinfo {year} {2018})},\ \Eprint
  {http://arxiv.org/abs/1711.09418} {arXiv:1711.09418 [cond-mat.stat-mech]}
  \BibitemShut {NoStop}%
\bibitem [{\citenamefont {Haag}(2012{\natexlab{b}})}]{haag2012local}%
  \BibitemOpen
  \bibfield  {author} {\bibinfo {author} {\bibfnamefont {Rudolf}\ \bibnamefont
  {Haag}},\ }\href@noop {} {\emph {\bibinfo {title} {Local quantum physics:
  Fields, particles, algebras}}}\ (\bibinfo  {publisher} {Springer Science \&
  Business Media},\ \bibinfo {year} {2012})\BibitemShut {NoStop}%
\bibitem [{\citenamefont {Polchinski}(1984)}]{Polchinski:1983gv}%
  \BibitemOpen
  \bibfield  {author} {\bibinfo {author} {\bibfnamefont {Joseph}\ \bibnamefont
  {Polchinski}},\ }\bibfield  {title} {\enquote {\bibinfo {title}
  {{Renormalization and Effective Lagrangians}},}\ }\href {\doibase
  10.1016/0550-3213(84)90287-6} {\bibfield  {journal} {\bibinfo  {journal}
  {Nucl. Phys. B}\ }\textbf {\bibinfo {volume} {231}},\ \bibinfo {pages}
  {269--295} (\bibinfo {year} {1984})}\BibitemShut {NoStop}%
\bibitem [{\citenamefont {Jafferis}\ \emph {et~al.}(2011)\citenamefont
  {Jafferis}, \citenamefont {Klebanov}, \citenamefont {Pufu},\ and\
  \citenamefont {Safdi}}]{Jafferis:2011zi}%
  \BibitemOpen
  \bibfield  {author} {\bibinfo {author} {\bibfnamefont {Daniel~L.}\
  \bibnamefont {Jafferis}}, \bibinfo {author} {\bibfnamefont {Igor~R.}\
  \bibnamefont {Klebanov}}, \bibinfo {author} {\bibfnamefont {Silviu~S.}\
  \bibnamefont {Pufu}}, \ and\ \bibinfo {author} {\bibfnamefont {Benjamin~R.}\
  \bibnamefont {Safdi}},\ }\bibfield  {title} {\enquote {\bibinfo {title}
  {{Towards the F-Theorem: N=2 Field Theories on the Three-Sphere}},}\ }\href
  {\doibase 10.1007/JHEP06(2011)102} {\bibfield  {journal} {\bibinfo  {journal}
  {JHEP}\ }\textbf {\bibinfo {volume} {06}},\ \bibinfo {pages} {102} (\bibinfo
  {year} {2011})},\ \Eprint {http://arxiv.org/abs/1103.1181} {arXiv:1103.1181
  [hep-th]} \BibitemShut {NoStop}%
\bibitem [{\citenamefont {Komargodski}\ and\ \citenamefont
  {Schwimmer}(2011)}]{Komargodski:2011vj}%
  \BibitemOpen
  \bibfield  {author} {\bibinfo {author} {\bibfnamefont {Zohar}\ \bibnamefont
  {Komargodski}}\ and\ \bibinfo {author} {\bibfnamefont {Adam}\ \bibnamefont
  {Schwimmer}},\ }\bibfield  {title} {\enquote {\bibinfo {title} {{On
  Renormalization Group Flows in Four Dimensions}},}\ }\href {\doibase
  10.1007/JHEP12(2011)099} {\bibfield  {journal} {\bibinfo  {journal} {JHEP}\
  }\textbf {\bibinfo {volume} {12}},\ \bibinfo {pages} {099} (\bibinfo {year}
  {2011})},\ \Eprint {http://arxiv.org/abs/1107.3987} {arXiv:1107.3987
  [hep-th]} \BibitemShut {NoStop}%
\bibitem [{\citenamefont {Zamolodchikov}(1986)}]{Zamolodchikov:1986gt}%
  \BibitemOpen
  \bibfield  {author} {\bibinfo {author} {\bibfnamefont {A.~B.}\ \bibnamefont
  {Zamolodchikov}},\ }\bibfield  {title} {\enquote {\bibinfo {title}
  {{Irreversibility of the Flux of the Renormalization Group in a 2D Field
  Theory}},}\ }\href@noop {} {\bibfield  {journal} {\bibinfo  {journal} {JETP
  Lett.}\ }\textbf {\bibinfo {volume} {43}},\ \bibinfo {pages} {730--732}
  (\bibinfo {year} {1986})}\BibitemShut {NoStop}%
\bibitem [{\citenamefont {Casini}\ \emph {et~al.}(2017)\citenamefont {Casini},
  \citenamefont {Test\'e},\ and\ \citenamefont {Torroba}}]{Casini:2017vbe}%
  \BibitemOpen
  \bibfield  {author} {\bibinfo {author} {\bibfnamefont {Horacio}\ \bibnamefont
  {Casini}}, \bibinfo {author} {\bibfnamefont {Eduardo}\ \bibnamefont
  {Test\'e}}, \ and\ \bibinfo {author} {\bibfnamefont {Gonzalo}\ \bibnamefont
  {Torroba}},\ }\bibfield  {title} {\enquote {\bibinfo {title} {{Markov
  Property of the Conformal Field Theory Vacuum and the a Theorem}},}\ }\href
  {\doibase 10.1103/PhysRevLett.118.261602} {\bibfield  {journal} {\bibinfo
  {journal} {Phys. Rev. Lett.}\ }\textbf {\bibinfo {volume} {118}},\ \bibinfo
  {pages} {261602} (\bibinfo {year} {2017})},\ \Eprint
  {http://arxiv.org/abs/1704.01870} {arXiv:1704.01870 [hep-th]} \BibitemShut
  {NoStop}%
\bibitem [{\citenamefont {Brauner}\ \emph {et~al.}(2022)\citenamefont
  {Brauner}, \citenamefont {Hartnoll}, \citenamefont {Kovtun}, \citenamefont
  {Liu}, \citenamefont {Mezei}, \citenamefont {Nicolis}, \citenamefont {Penco},
  \citenamefont {Shao},\ and\ \citenamefont {Son}}]{Brauner:2022rvf}%
  \BibitemOpen
  \bibfield  {author} {\bibinfo {author} {\bibfnamefont {Tomas}\ \bibnamefont
  {Brauner}}, \bibinfo {author} {\bibfnamefont {Sean~A.}\ \bibnamefont
  {Hartnoll}}, \bibinfo {author} {\bibfnamefont {Pavel}\ \bibnamefont
  {Kovtun}}, \bibinfo {author} {\bibfnamefont {Hong}\ \bibnamefont {Liu}},
  \bibinfo {author} {\bibfnamefont {M\'ark}\ \bibnamefont {Mezei}}, \bibinfo
  {author} {\bibfnamefont {Alberto}\ \bibnamefont {Nicolis}}, \bibinfo {author}
  {\bibfnamefont {Riccardo}\ \bibnamefont {Penco}}, \bibinfo {author}
  {\bibfnamefont {Shu-Heng}\ \bibnamefont {Shao}}, \ and\ \bibinfo {author}
  {\bibfnamefont {Dam~Thanh}\ \bibnamefont {Son}},\ }\bibfield  {title}
  {\enquote {\bibinfo {title} {{Snowmass White Paper: Effective Field Theories
  for Condensed Matter Systems}},}\ }in\ \href@noop {} {\emph {\bibinfo
  {booktitle} {{2022 Snowmass Summer Study}}}}\ (\bibinfo {year} {2022})\
  \Eprint {http://arxiv.org/abs/2203.10110} {arXiv:2203.10110 [hep-th]}
  \BibitemShut {NoStop}%
\bibitem [{\citenamefont {Cordova}\ \emph {et~al.}(2022)\citenamefont
  {Cordova}, \citenamefont {Dumitrescu}, \citenamefont {Intriligator},\ and\
  \citenamefont {Shao}}]{Cordova:2022ruw}%
  \BibitemOpen
  \bibfield  {author} {\bibinfo {author} {\bibfnamefont {Clay}\ \bibnamefont
  {Cordova}}, \bibinfo {author} {\bibfnamefont {Thomas~T.}\ \bibnamefont
  {Dumitrescu}}, \bibinfo {author} {\bibfnamefont {Kenneth}\ \bibnamefont
  {Intriligator}}, \ and\ \bibinfo {author} {\bibfnamefont {Shu-Heng}\
  \bibnamefont {Shao}},\ }\bibfield  {title} {\enquote {\bibinfo {title}
  {{Snowmass White Paper: Generalized Symmetries in Quantum Field Theory and
  Beyond}},}\ }in\ \href@noop {} {\emph {\bibinfo {booktitle} {{2022 Snowmass
  Summer Study}}}}\ (\bibinfo {year} {2022})\ \Eprint
  {http://arxiv.org/abs/2205.09545} {arXiv:2205.09545 [hep-th]} \BibitemShut
  {NoStop}%
\bibitem [{\citenamefont {Sekino}\ and\ \citenamefont
  {Susskind}(2008)}]{Sekino:2008he}%
  \BibitemOpen
  \bibfield  {author} {\bibinfo {author} {\bibfnamefont {Yasuhiro}\
  \bibnamefont {Sekino}}\ and\ \bibinfo {author} {\bibfnamefont {Leonard}\
  \bibnamefont {Susskind}},\ }\bibfield  {title} {\enquote {\bibinfo {title}
  {{Fast Scramblers}},}\ }\href {\doibase 10.1088/1126-6708/2008/10/065}
  {\bibfield  {journal} {\bibinfo  {journal} {JHEP}\ }\textbf {\bibinfo
  {volume} {10}},\ \bibinfo {pages} {065} (\bibinfo {year} {2008})},\ \Eprint
  {http://arxiv.org/abs/0808.2096} {arXiv:0808.2096 [hep-th]} \BibitemShut
  {NoStop}%
\bibitem [{\citenamefont {Shenker}\ and\ \citenamefont
  {Stanford}(2014)}]{Shenker:2013pqa}%
  \BibitemOpen
  \bibfield  {author} {\bibinfo {author} {\bibfnamefont {Stephen~H.}\
  \bibnamefont {Shenker}}\ and\ \bibinfo {author} {\bibfnamefont {Douglas}\
  \bibnamefont {Stanford}},\ }\bibfield  {title} {\enquote {\bibinfo {title}
  {{Black holes and the butterfly effect}},}\ }\href {\doibase
  10.1007/JHEP03(2014)067} {\bibfield  {journal} {\bibinfo  {journal} {JHEP}\
  }\textbf {\bibinfo {volume} {03}},\ \bibinfo {pages} {067} (\bibinfo {year}
  {2014})},\ \Eprint {http://arxiv.org/abs/1306.0622} {arXiv:1306.0622
  [hep-th]} \BibitemShut {NoStop}%
\bibitem [{\citenamefont {Hawking}\ and\ \citenamefont
  {Ellis}(2011)}]{Hawking:1973uf}%
  \BibitemOpen
  \bibfield  {author} {\bibinfo {author} {\bibfnamefont {S.~W.}\ \bibnamefont
  {Hawking}}\ and\ \bibinfo {author} {\bibfnamefont {G.~F.~R.}\ \bibnamefont
  {Ellis}},\ }\href {\doibase 10.1017/CBO9780511524646} {\emph {\bibinfo
  {title} {{The Large Scale Structure of Space-Time}}}},\ Cambridge Monographs
  on Mathematical Physics\ (\bibinfo  {publisher} {Cambridge University
  Press},\ \bibinfo {year} {2011})\BibitemShut {NoStop}%
\bibitem [{\citenamefont {Casini}(2008)}]{Casini:2008cr}%
  \BibitemOpen
  \bibfield  {author} {\bibinfo {author} {\bibfnamefont {H.}~\bibnamefont
  {Casini}},\ }\bibfield  {title} {\enquote {\bibinfo {title} {{Relative
  entropy and the Bekenstein bound}},}\ }\href {\doibase
  10.1088/0264-9381/25/20/205021} {\bibfield  {journal} {\bibinfo  {journal}
  {Class. Quant. Grav.}\ }\textbf {\bibinfo {volume} {25}},\ \bibinfo {pages}
  {205021} (\bibinfo {year} {2008})},\ \Eprint {http://arxiv.org/abs/0804.2182}
  {arXiv:0804.2182 [hep-th]} \BibitemShut {NoStop}%
\bibitem [{\citenamefont {Faulkner}\ \emph {et~al.}(2016)\citenamefont
  {Faulkner}, \citenamefont {Leigh}, \citenamefont {Parrikar},\ and\
  \citenamefont {Wang}}]{Faulkner:2016mzt}%
  \BibitemOpen
  \bibfield  {author} {\bibinfo {author} {\bibfnamefont {Thomas}\ \bibnamefont
  {Faulkner}}, \bibinfo {author} {\bibfnamefont {Robert~G.}\ \bibnamefont
  {Leigh}}, \bibinfo {author} {\bibfnamefont {Onkar}\ \bibnamefont {Parrikar}},
  \ and\ \bibinfo {author} {\bibfnamefont {Huajia}\ \bibnamefont {Wang}},\
  }\bibfield  {title} {\enquote {\bibinfo {title} {{Modular Hamiltonians for
  Deformed Half-Spaces and the Averaged Null Energy Condition}},}\ }\href
  {\doibase 10.1007/JHEP09(2016)038} {\bibfield  {journal} {\bibinfo  {journal}
  {JHEP}\ }\textbf {\bibinfo {volume} {09}},\ \bibinfo {pages} {038} (\bibinfo
  {year} {2016})},\ \Eprint {http://arxiv.org/abs/1605.08072} {arXiv:1605.08072
  [hep-th]} \BibitemShut {NoStop}%
\bibitem [{\citenamefont {Bousso}\ \emph {et~al.}(2016)\citenamefont {Bousso},
  \citenamefont {Fisher}, \citenamefont {Koeller}, \citenamefont
  {Leichenauer},\ and\ \citenamefont {Wall}}]{Bousso:2015wca}%
  \BibitemOpen
  \bibfield  {author} {\bibinfo {author} {\bibfnamefont {Raphael}\ \bibnamefont
  {Bousso}}, \bibinfo {author} {\bibfnamefont {Zachary}\ \bibnamefont
  {Fisher}}, \bibinfo {author} {\bibfnamefont {Jason}\ \bibnamefont {Koeller}},
  \bibinfo {author} {\bibfnamefont {Stefan}\ \bibnamefont {Leichenauer}}, \
  and\ \bibinfo {author} {\bibfnamefont {Aron~C.}\ \bibnamefont {Wall}},\
  }\bibfield  {title} {\enquote {\bibinfo {title} {{Proof of the Quantum Null
  Energy Condition}},}\ }\href {\doibase 10.1103/PhysRevD.93.024017} {\bibfield
   {journal} {\bibinfo  {journal} {Phys. Rev. D}\ }\textbf {\bibinfo {volume}
  {93}},\ \bibinfo {pages} {024017} (\bibinfo {year} {2016})},\ \Eprint
  {http://arxiv.org/abs/1509.02542} {arXiv:1509.02542 [hep-th]} \BibitemShut
  {NoStop}%
\bibitem [{\citenamefont {Son}\ and\ \citenamefont
  {Starinets}(2002)}]{Son:2002sd}%
  \BibitemOpen
  \bibfield  {author} {\bibinfo {author} {\bibfnamefont {Dam~T.}\ \bibnamefont
  {Son}}\ and\ \bibinfo {author} {\bibfnamefont {Andrei~O.}\ \bibnamefont
  {Starinets}},\ }\bibfield  {title} {\enquote {\bibinfo {title} {{Minkowski
  space correlators in AdS / CFT correspondence: Recipe and applications}},}\
  }\href {\doibase 10.1088/1126-6708/2002/09/042} {\bibfield  {journal}
  {\bibinfo  {journal} {JHEP}\ }\textbf {\bibinfo {volume} {09}},\ \bibinfo
  {pages} {042} (\bibinfo {year} {2002})},\ \Eprint
  {http://arxiv.org/abs/hep-th/0205051} {arXiv:hep-th/0205051 [hep-th]}
  \BibitemShut {NoStop}%
\bibitem [{\citenamefont {Glorioso}\ \emph {et~al.}(2018)\citenamefont
  {Glorioso}, \citenamefont {Crossley},\ and\ \citenamefont
  {Liu}}]{Glorioso:2018mmw}%
  \BibitemOpen
  \bibfield  {author} {\bibinfo {author} {\bibfnamefont {Paolo}\ \bibnamefont
  {Glorioso}}, \bibinfo {author} {\bibfnamefont {Michael}\ \bibnamefont
  {Crossley}}, \ and\ \bibinfo {author} {\bibfnamefont {Hong}\ \bibnamefont
  {Liu}},\ }\bibfield  {title} {\enquote {\bibinfo {title} {{A prescription for
  holographic Schwinger-Keldysh contour in non-equilibrium systems}},}\
  }\href@noop {} {\  (\bibinfo {year} {2018})},\ \Eprint
  {http://arxiv.org/abs/1812.08785} {arXiv:1812.08785 [hep-th]} \BibitemShut
  {NoStop}%
\bibitem [{\citenamefont {Stanford}\ and\ \citenamefont
  {Susskind}(2014)}]{Stanford:2014jda}%
  \BibitemOpen
  \bibfield  {author} {\bibinfo {author} {\bibfnamefont {Douglas}\ \bibnamefont
  {Stanford}}\ and\ \bibinfo {author} {\bibfnamefont {Leonard}\ \bibnamefont
  {Susskind}},\ }\bibfield  {title} {\enquote {\bibinfo {title} {{Complexity
  and Shock Wave Geometries}},}\ }\href {\doibase 10.1103/PhysRevD.90.126007}
  {\bibfield  {journal} {\bibinfo  {journal} {Phys. Rev. D}\ }\textbf {\bibinfo
  {volume} {90}},\ \bibinfo {pages} {126007} (\bibinfo {year} {2014})},\
  \Eprint {http://arxiv.org/abs/1406.2678} {arXiv:1406.2678 [hep-th]}
  \BibitemShut {NoStop}%
\bibitem [{\citenamefont {Brown}\ \emph {et~al.}(2016)\citenamefont {Brown},
  \citenamefont {Roberts}, \citenamefont {Susskind}, \citenamefont {Swingle},\
  and\ \citenamefont {Zhao}}]{Brown:2015bva}%
  \BibitemOpen
  \bibfield  {author} {\bibinfo {author} {\bibfnamefont {Adam~R.}\ \bibnamefont
  {Brown}}, \bibinfo {author} {\bibfnamefont {Daniel~A.}\ \bibnamefont
  {Roberts}}, \bibinfo {author} {\bibfnamefont {Leonard}\ \bibnamefont
  {Susskind}}, \bibinfo {author} {\bibfnamefont {Brian}\ \bibnamefont
  {Swingle}}, \ and\ \bibinfo {author} {\bibfnamefont {Ying}\ \bibnamefont
  {Zhao}},\ }\bibfield  {title} {\enquote {\bibinfo {title} {{Holographic
  Complexity Equals Bulk Action?}}}\ }\href {\doibase
  10.1103/PhysRevLett.116.191301} {\bibfield  {journal} {\bibinfo  {journal}
  {Phys. Rev. Lett.}\ }\textbf {\bibinfo {volume} {116}},\ \bibinfo {pages}
  {191301} (\bibinfo {year} {2016})},\ \Eprint
  {http://arxiv.org/abs/1509.07876} {arXiv:1509.07876 [hep-th]} \BibitemShut
  {NoStop}%
\bibitem [{\citenamefont {Caputa}\ \emph {et~al.}(2017)\citenamefont {Caputa},
  \citenamefont {Kundu}, \citenamefont {Miyaji}, \citenamefont {Takayanagi},\
  and\ \citenamefont {Watanabe}}]{Caputa:2017urj}%
  \BibitemOpen
  \bibfield  {author} {\bibinfo {author} {\bibfnamefont {Pawel}\ \bibnamefont
  {Caputa}}, \bibinfo {author} {\bibfnamefont {Nilay}\ \bibnamefont {Kundu}},
  \bibinfo {author} {\bibfnamefont {Masamichi}\ \bibnamefont {Miyaji}},
  \bibinfo {author} {\bibfnamefont {Tadashi}\ \bibnamefont {Takayanagi}}, \
  and\ \bibinfo {author} {\bibfnamefont {Kento}\ \bibnamefont {Watanabe}},\
  }\bibfield  {title} {\enquote {\bibinfo {title} {{Anti-de Sitter Space from
  Optimization of Path Integrals in Conformal Field Theories}},}\ }\href
  {\doibase 10.1103/PhysRevLett.119.071602} {\bibfield  {journal} {\bibinfo
  {journal} {Phys. Rev. Lett.}\ }\textbf {\bibinfo {volume} {119}},\ \bibinfo
  {pages} {071602} (\bibinfo {year} {2017})},\ \Eprint
  {http://arxiv.org/abs/1703.00456} {arXiv:1703.00456 [hep-th]} \BibitemShut
  {NoStop}%
\bibitem [{\citenamefont {Almheiri}\ \emph {et~al.}(2021)\citenamefont
  {Almheiri}, \citenamefont {Hartman}, \citenamefont {Maldacena}, \citenamefont
  {Shaghoulian},\ and\ \citenamefont {Tajdini}}]{Almheiri:2020cfm}%
  \BibitemOpen
  \bibfield  {author} {\bibinfo {author} {\bibfnamefont {Ahmed}\ \bibnamefont
  {Almheiri}}, \bibinfo {author} {\bibfnamefont {Thomas}\ \bibnamefont
  {Hartman}}, \bibinfo {author} {\bibfnamefont {Juan}\ \bibnamefont
  {Maldacena}}, \bibinfo {author} {\bibfnamefont {Edgar}\ \bibnamefont
  {Shaghoulian}}, \ and\ \bibinfo {author} {\bibfnamefont {Amirhossein}\
  \bibnamefont {Tajdini}},\ }\bibfield  {title} {\enquote {\bibinfo {title}
  {{The entropy of Hawking radiation}},}\ }\href {\doibase
  10.1103/RevModPhys.93.035002} {\bibfield  {journal} {\bibinfo  {journal}
  {Rev. Mod. Phys.}\ }\textbf {\bibinfo {volume} {93}},\ \bibinfo {pages}
  {035002} (\bibinfo {year} {2021})},\ \Eprint
  {http://arxiv.org/abs/2006.06872} {arXiv:2006.06872 [hep-th]} \BibitemShut
  {NoStop}%
\bibitem [{\citenamefont {Berglund}\ \emph {et~al.}(2022)\citenamefont
  {Berglund}, \citenamefont {Freidel}, \citenamefont {Hubsch}, \citenamefont
  {Kowalski-Glikman}, \citenamefont {Leigh}, \citenamefont {Mattingly},\ and\
  \citenamefont {Minic}}]{Berglund:2022qcc}%
  \BibitemOpen
  \bibfield  {author} {\bibinfo {author} {\bibfnamefont {Per}\ \bibnamefont
  {Berglund}}, \bibinfo {author} {\bibfnamefont {Laurent}\ \bibnamefont
  {Freidel}}, \bibinfo {author} {\bibfnamefont {Tristan}\ \bibnamefont
  {Hubsch}}, \bibinfo {author} {\bibfnamefont {Jerzy}\ \bibnamefont
  {Kowalski-Glikman}}, \bibinfo {author} {\bibfnamefont {Robert~G.}\
  \bibnamefont {Leigh}}, \bibinfo {author} {\bibfnamefont {David}\ \bibnamefont
  {Mattingly}}, \ and\ \bibinfo {author} {\bibfnamefont {Djordje}\ \bibnamefont
  {Minic}},\ }\bibfield  {title} {\enquote {\bibinfo {title} {{Infrared
  Properties of Quantum Gravity: UV/IR Mixing, Gravitizing the Quantum --
  Theory and Observation}},}\ }in\ \href@noop {} {\emph {\bibinfo {booktitle}
  {{2022 Snowmass Summer Study}}}}\ (\bibinfo {year} {2022})\ \Eprint
  {http://arxiv.org/abs/2202.06890} {arXiv:2202.06890 [hep-th]} \BibitemShut
  {NoStop}%
\bibitem [{\citenamefont {Bena}\ \emph {et~al.}(2022)\citenamefont {Bena},
  \citenamefont {Martinec}, \citenamefont {Mathur},\ and\ \citenamefont
  {Warner}}]{Bena:2022ldq}%
  \BibitemOpen
  \bibfield  {author} {\bibinfo {author} {\bibfnamefont {Iosif}\ \bibnamefont
  {Bena}}, \bibinfo {author} {\bibfnamefont {Emil~J.}\ \bibnamefont
  {Martinec}}, \bibinfo {author} {\bibfnamefont {Samir~D.}\ \bibnamefont
  {Mathur}}, \ and\ \bibinfo {author} {\bibfnamefont {Nicholas~P.}\
  \bibnamefont {Warner}},\ }\bibfield  {title} {\enquote {\bibinfo {title}
  {{Snowmass White Paper: Micro- and Macro-Structure of Black Holes}},}\
  }\href@noop {} {\  (\bibinfo {year} {2022})},\ \Eprint
  {http://arxiv.org/abs/2203.04981} {arXiv:2203.04981 [hep-th]} \BibitemShut
  {NoStop}%
\bibitem [{\citenamefont {Dedushenko}(2022)}]{Dedushenko:2022zwd}%
  \BibitemOpen
  \bibfield  {author} {\bibinfo {author} {\bibfnamefont {Mykola}\ \bibnamefont
  {Dedushenko}},\ }\bibfield  {title} {\enquote {\bibinfo {title} {{Snowmass
  White Paper: The Quest to Define QFT}},}\ }\href@noop {} {\  (\bibinfo {year}
  {2022})},\ \Eprint {http://arxiv.org/abs/2203.08053} {arXiv:2203.08053
  [hep-th]} \BibitemShut {NoStop}%
\bibitem [{\citenamefont {Bourjaily}\ \emph {et~al.}(2022)\citenamefont
  {Bourjaily} \emph {et~al.}}]{Bourjaily:2022bwx}%
  \BibitemOpen
  \bibfield  {author} {\bibinfo {author} {\bibfnamefont {Jacob~L.}\
  \bibnamefont {Bourjaily}} \emph {et~al.},\ }\bibfield  {title} {\enquote
  {\bibinfo {title} {{Functions Beyond Multiple Polylogarithms for Precision
  Collider Physics}},}\ }in\ \href@noop {} {\emph {\bibinfo {booktitle} {{2022
  Snowmass Summer Study}}}}\ (\bibinfo {year} {2022})\ \Eprint
  {http://arxiv.org/abs/2203.07088} {arXiv:2203.07088 [hep-ph]} \BibitemShut
  {NoStop}%
\bibitem [{\citenamefont {Poland}\ and\ \citenamefont
  {Simmons-Duffin}(2022)}]{Poland:2022qrs}%
  \BibitemOpen
  \bibfield  {author} {\bibinfo {author} {\bibfnamefont {David}\ \bibnamefont
  {Poland}}\ and\ \bibinfo {author} {\bibfnamefont {David}\ \bibnamefont
  {Simmons-Duffin}},\ }\bibfield  {title} {\enquote {\bibinfo {title}
  {{Snowmass White Paper: The Numerical Conformal Bootstrap}},}\ }in\
  \href@noop {} {\emph {\bibinfo {booktitle} {{2022 Snowmass Summer Study}}}}\
  (\bibinfo {year} {2022})\ \Eprint {http://arxiv.org/abs/2203.08117}
  {arXiv:2203.08117 [hep-th]} \BibitemShut {NoStop}%
\bibitem [{\citenamefont {Hartman}\ \emph {et~al.}(2022)\citenamefont
  {Hartman}, \citenamefont {Mazac}, \citenamefont {Simmons-Duffin},\ and\
  \citenamefont {Zhiboedov}}]{Hartman:2022zik}%
  \BibitemOpen
  \bibfield  {author} {\bibinfo {author} {\bibfnamefont {Thomas}\ \bibnamefont
  {Hartman}}, \bibinfo {author} {\bibfnamefont {Dalimil}\ \bibnamefont
  {Mazac}}, \bibinfo {author} {\bibfnamefont {David}\ \bibnamefont
  {Simmons-Duffin}}, \ and\ \bibinfo {author} {\bibfnamefont {Alexander}\
  \bibnamefont {Zhiboedov}},\ }\bibfield  {title} {\enquote {\bibinfo {title}
  {{Snowmass White Paper: The Analytic Conformal Bootstrap}},}\ }in\ \href@noop
  {} {\emph {\bibinfo {booktitle} {{2022 Snowmass Summer Study}}}}\ (\bibinfo
  {year} {2022})\ \Eprint {http://arxiv.org/abs/2202.11012} {arXiv:2202.11012
  [hep-th]} \BibitemShut {NoStop}%
\bibitem [{\citenamefont {Bah}\ \emph {et~al.}(2022)\citenamefont {Bah},
  \citenamefont {Freed}, \citenamefont {Moore}, \citenamefont {Nekrasov},
  \citenamefont {Razamat},\ and\ \citenamefont
  {Sch\"afer-Nameki}}]{Bah:2022xfv}%
  \BibitemOpen
  \bibfield  {author} {\bibinfo {author} {\bibfnamefont {Ibrahima}\
  \bibnamefont {Bah}}, \bibinfo {author} {\bibfnamefont {Daniel}\ \bibnamefont
  {Freed}}, \bibinfo {author} {\bibfnamefont {Gregory~W.}\ \bibnamefont
  {Moore}}, \bibinfo {author} {\bibfnamefont {Nikita}\ \bibnamefont
  {Nekrasov}}, \bibinfo {author} {\bibfnamefont {Shlomo~S.}\ \bibnamefont
  {Razamat}}, \ and\ \bibinfo {author} {\bibfnamefont {Sakura}\ \bibnamefont
  {Sch\"afer-Nameki}},\ }\bibfield  {title} {\enquote {\bibinfo {title}
  {{Snowmass Whitepaper: Physical Mathematics 2021}},}\ }\href@noop {} {\
  (\bibinfo {year} {2022})},\ \Eprint {http://arxiv.org/abs/2203.05078}
  {arXiv:2203.05078 [hep-th]} \BibitemShut {NoStop}%
\bibitem [{\citenamefont {Witten}(2021)}]{Witten:2021unn}%
  \BibitemOpen
  \bibfield  {author} {\bibinfo {author} {\bibfnamefont {Edward}\ \bibnamefont
  {Witten}},\ }\bibfield  {title} {\enquote {\bibinfo {title} {{Gravity and the
  Crossed Product}},}\ }\href@noop {} {\  (\bibinfo {year} {2021})},\ \Eprint
  {http://arxiv.org/abs/2112.12828} {arXiv:2112.12828 [hep-th]} \BibitemShut
  {NoStop}%
\bibitem [{\citenamefont {Hughes}\ \emph {et~al.}(2021)\citenamefont {Hughes},
  \citenamefont {Finke}, \citenamefont {German}, \citenamefont {Merzbacher},
  \citenamefont {Vora},\ and\ \citenamefont {Lewandowski}}]{Hughes:2021lyh}%
  \BibitemOpen
  \bibfield  {author} {\bibinfo {author} {\bibfnamefont {Ciaran}\ \bibnamefont
  {Hughes}}, \bibinfo {author} {\bibfnamefont {Doug}\ \bibnamefont {Finke}},
  \bibinfo {author} {\bibfnamefont {Dan-Adrian}\ \bibnamefont {German}},
  \bibinfo {author} {\bibfnamefont {Celia}\ \bibnamefont {Merzbacher}},
  \bibinfo {author} {\bibfnamefont {Patrick~M.}\ \bibnamefont {Vora}}, \ and\
  \bibinfo {author} {\bibfnamefont {H.~J.}\ \bibnamefont {Lewandowski}},\
  }\bibfield  {title} {\enquote {\bibinfo {title} {{Assessing the Needs of the
  Quantum Industry}},}\ }\href@noop {} {\  (\bibinfo {year} {2021})},\ \Eprint
  {http://arxiv.org/abs/2109.03601} {arXiv:2109.03601 [physics.ed-ph]}
  \BibitemShut {NoStop}%
\bibitem [{\citenamefont {Perry}\ \emph {et~al.}(2019)\citenamefont {Perry},
  \citenamefont {Sun}, \citenamefont {Hughes}, \citenamefont {Isaacson},\ and\
  \citenamefont {Turner}}]{Perry:2019bqg}%
  \BibitemOpen
  \bibfield  {author} {\bibinfo {author} {\bibfnamefont {Anastasia}\
  \bibnamefont {Perry}}, \bibinfo {author} {\bibfnamefont {Ranbel}\
  \bibnamefont {Sun}}, \bibinfo {author} {\bibfnamefont {Ciaran}\ \bibnamefont
  {Hughes}}, \bibinfo {author} {\bibfnamefont {Joshua}\ \bibnamefont
  {Isaacson}}, \ and\ \bibinfo {author} {\bibfnamefont {Jessica}\ \bibnamefont
  {Turner}},\ }\bibfield  {title} {\enquote {\bibinfo {title} {{Quantum
  Computing as a High School Module}},}\ }\href@noop {} {\  (\bibinfo {year}
  {2019})},\ \Eprint {http://arxiv.org/abs/1905.00282} {arXiv:1905.00282
  [physics.ed-ph]} \BibitemShut {NoStop}%
\bibitem [{\citenamefont {Hughes}\ \emph {et~al.}(2020)\citenamefont {Hughes},
  \citenamefont {Isaacson}, \citenamefont {Perry}, \citenamefont {Sun},\ and\
  \citenamefont {Turner}}]{Hughes:2020ngh}%
  \BibitemOpen
  \bibfield  {author} {\bibinfo {author} {\bibfnamefont {Ciaran}\ \bibnamefont
  {Hughes}}, \bibinfo {author} {\bibfnamefont {Joshua}\ \bibnamefont
  {Isaacson}}, \bibinfo {author} {\bibfnamefont {Anastasia}\ \bibnamefont
  {Perry}}, \bibinfo {author} {\bibfnamefont {Ranbel}\ \bibnamefont {Sun}}, \
  and\ \bibinfo {author} {\bibfnamefont {Jessica}\ \bibnamefont {Turner}},\
  }\bibfield  {title} {\enquote {\bibinfo {title} {{Teaching Quantum Computing
  to High School Students}},}\ }\href@noop {} {\  (\bibinfo {year} {2020})},\
  \Eprint {http://arxiv.org/abs/2004.07206} {arXiv:2004.07206 [physics.ed-ph]}
  \BibitemShut {NoStop}%
\end{thebibliography}%

\end{document}